\documentclass[twocolumn, amsmath,amssymb,
aps,]{revtex4}
\usepackage{dcolumn}
\usepackage{bm}
\usepackage{graphicx}
\usepackage{float}
\usepackage{url}
\usepackage{hyperref}
\usepackage{xstring}

\def\SdS{Schwarzschild--de~Sitter}
\def\dS{de~Sitter}

\def\KdS{Kerr--de~Sitter}
\def\KNdS{Kerr--Newman--de~Sitter}
\def\be{\begin{equation}} \def\ee{\end{equation}}
\def\bea{\begin{eqnarray}} \def\eea{\end{eqnarray}}
\def\din{\,\mathrm{d}} \def\dbe{\mathrm{d}}
\def\bet{\begin{tabular}} \def\ent{\end{tabular}}
 
\def\sign{\,\mathrm{sign\,}}

\begin{document}

\title{Photon motion in \KdS\, spacetimes}

\author{Daniel Charbul\'{a}k}
\email	{daniel.charbulak@fpf.slu.cz}
\author{Zden\v{e}k Stuchl\'{i}k}
\email	{zdenek.stuchlik@fpf.slu.cz}

\address{Institute of Physics and Research Centre of Theoretical Physics and Astrophysics, 
	Faculty of Philosophy and Science, 
	Silesian university in Opava, 
	Bezru\v{c}ovo n\'{a}m. 13, CZ-746 01 Opava, Czech Republic}

\date{\today}

\begin{abstract}
We study general motion of photons in the Kerr-de Sitter black hole and naked singularity spacetimes. The motion is governed by the impact parameters $X$, related to axial symmetry of the spacetime, and $q$, related to its hidden symmetry. Appropriate 'effective potentials' governing the latitudinal and radial motion are introduced and their behaviour is examined by 'Chinese boxes' technique giving regions allowed for the motion in terms of the impact parameters. Restrictions on the impact parameters $X$ and $q$ are established in dependence on the spacetime parameters $M, \Lambda, a$. The motion can be of orbital type (crossing the equatorial plane, $q>0$) and vortical type (tied above or bellow the equatorial plane, $q<0$). It is shown that for negative values of $q$, the reality conditions imposed on the latitudinal motion yield stronger constraints on the parameter $X$ than that following from the reality condition of the radial motion, excluding existence of vortical motion of constant radius. Properties of the spherical photon orbits of the orbital type are determined and used along with properties of the effective potentials as criteria of classification of the KdS spacetimes according to the properties of the photon motion.    
\end{abstract}

\maketitle

\section{Introduction}\label{intro}

In the framework of inflationary paradigm~\cite{Lin:1990:InfCos:}, recent cosmological observations indicate that a very small relict vacuum energy (equivalently, repulsive cosmological constant $\Lambda > 0$), or, generally, a dark energy demonstrating repulsive gravitational effect, has to be introduced to explain dynamics of the recent Universe~\cite{ArP-Muk-Ste:2000:PHYRL:,Bah-etal:1999:SCIEN:,Cal-Dav-Ste:1998:PHYRL:,Kra:1998:ASTRJ2:,Kra-Tur:1995:GENRG2:,Ost-Ste:1995:NATURE:,Wan-etal:2000:ASTRJ2:}. These conclusions are supported strongly by the observations of distant Ia-type supernova explosions indicating that starting at the cosmological redshift $z \approx 1$ expansion of the Universe is accelerated~\cite{Rie-etal:2004:ASTRJ2:}. The total energy density of the Universe is very close to the critical energy density $\rho_{\mathrm{crit}}$ corresponding to almost flat universe predicted by the inflationary scenario~\cite{Spe-etal:2007:ASTJS:3yrWMAP}, and the dark energy represents about $70\%$ of the energy content of the observable universe \cite{Cal-Kam:2009:NATURE:CosDarkMat, Spe-etal:2007:ApJSuppl:}. These conclusions are confirmed by recent measurements of cosmic microwave background anisotropies by the space satellite observatory PLANCK~\cite{Ada-etal:2013:ASTRA:DifLiYoClG,Ade-etal:2014:ASTRA:}.

The dark energy equation of state is very close to those corresponding to the vacuum energy~\cite{Cal-Kam:2009:NATURE:CosDarkMat}. Therefore, it is relevant to study the astrophysical consequences of the effect of the observed cosmological constant implied by the cosmological tests to be $\Lambda \approx 1.3\times 10^{-56}\,\mathrm{cm^{-2}}$, and the related vacuum energy $\rho_{\mathrm{vac}} \sim 10^{-29}\,\mathrm{g/cm^{3}},$ close to the critical density of the universe. The repulsive cosmological constant changes significantly the asymptotic structure of \mbox{black-hole}, naked singularity, or any compact-body backgrounds as such backgrounds become asymptotically \dS{} spacetimes, and an event horizon (cosmological horizon) always exists, behind which the geometry is dynamic. 

Substantial influence of the repulsive cosmological constant has been demonstrated for astrophysical situations related to active galactic nuclei and their central supermassive black holes~\cite{Stu:2005:MODPLA:}. The black hole spacetimes with the $\Lambda$ term are described in the spherically symmetric case by the vacuum \SdS{}(SdS)
 geometry \cite{Kot:1918:ANNPH2:PhyBasEinsGr,Stu-Hle:1999:PHYSR4:}, while the internal, uniform density SdS spacetimes are given in~\cite{Boh:2004:GENRG2:, Stu:2000:ACTPS2:}. The axially symmetric, rotating black holes are determined by the Kerr-de Sitter (KdS) geometry \cite{Car:1973:BlaHol:,Gib-Haw:1977:PHYSR4:,Lak-Zan:2016:PHYSR4:}. In the spacetimes with the repulsive cosmological term, motion of photons was extensively investigated in many papers~\cite{Kra:2011:CLAQG:,Kra:2014:GRG:,Bak-etal:2007:CEURJP:,Lak:2002:PHYSR4:BendLiCC,Mul:2008:GENRG2:FallSchBH,Sch-Zai:2008:0801.3776:CCTimeDelay,Ser:2008:PHYSR4:CCLens,Stu-Cal:1991:GENRG2:,Stu-Hle:2000:CLAQG:,Vil-etal:2013:ASTSS1:PhMoChgAdS:}. The motion of massive test particles was studied in~\cite{Ali:2007:PHYSR4:EMPropKadS,Che:2008:CHINPB:DkEnGeoMorSchw,Cru-Oli-Vil:2005:CLAQG:GeoSdSBH,Hac-etal:2010:PHYSR4:KerrBHCoStr:,Ior:2009:NEWASTR:CCDGPGrav,Kag-Kun-Lam:2006:PHYLB:SolarSdS,Kra:2004:CLAQG:, Kra:2005:DARK:CCPerPrec,Kra:2007:CLAQG:Periapsis,Oli-etal:2011:MODPLA:ChaParRNadS:,Cha-Har:2012:PHYSR4:BEConGRStar,Zhou-Chen:2011:AstrSpSc:,Stu:1983:BULAI:,Stu:1984:BULAI:,Stu-Hle:1999:PHYSR4:,Stu-Hle:2002:ACTPS2:,Stu-Sla:2004:PHYSR4:}. The KdS geometry can be relevant also for the so called Kerr superspinars representing an alternative to black holes~\cite{Boy-etal:2003:PHYSR4:HoloProtChron,Gim-Hor:2004:hep-th0405019:GodHolo,Gim-Hor:2009:PHYLB:AstVioSignStr, 0264-9381-29-6-065002}, breaking the black hole bound on the dimensionless spin and exhibiting a variety of unusual physical phenomena~\cite{deFel:1974:ASTRA:,deFel:1978:NATURE:InstabNS, Hio-Mae:2009:PHYSR4:KerrSpinMeas,Stu:1980:BULAI:,Stu-Hle-Tru:2011:CLAQG:,Stu-Sch:2010:CLAQG:AppKepDiOrKerrSSp, 0264-9381-29-6-065002, Stu-Sch:2013:CLAQG:UHEKerrGeo}. It is worth to note that the SdS and KdS spacetimes are equivalent to some solutions of the f(R) gravity representing black holes and naked singularities \cite{Per-Rom-PeB:2013:ASTRA:AccDiBHModGra:, Stu-Sla-Kov:2009:CLAQG:PseNewSdS}. 

The role of the cosmological constant can be significant for both the geometrically thin Keplerian accretion discs \cite{Mul-Asch:2007:CLAQG:NonMonoVel,Sla-Stu:2008:CLAQG:CmtNoMonKadS, Stu:2005:MODPLA:,Stu-Hle:1999:PHYSR4:,Stu-Sla:2004:PHYSR4:} and the toroidal accretion discs~\cite{Asc:2008:CHIAA:MassSpinBHQPO, Kuc-Sla-Stu:2011:JCAP:ToroPerFlRNadS:, Pug-Stu:2015:ApJS:,Rez-Zan-Fon:2003:ASTRA:,Sla-Stu:2005:CLAQG:,Stu-Sla-Hle:2000:ASTRA:,Stu-etal:2005:PHYSR4:AschenUnexpTopo:} orbiting supermassive black holes (Kerr superspinars) in the central parts of giant galaxies. Both high-frequency quasiperiodic oscillations and jets originating at the accretion discs can be reflected by current carrying string loops in SdS and KdS spacetimes~\cite{Gu-Cheng:2007:GENRG2:CircLoopKdS,Kol-Stu:2010:PHYSR4:CurCarStrLoops,Stu-Kol:2012:PHYSR4:AccStringLoops,Stu-Kol:2012:JCAP:StringLoops:, Wan-Che:2012:PHYLB:CirLoopPerTens}. In the spherically symmetric spacetimes, the Keplerian and toroidal disc structures can be precisely described the Pseudo-Newtonian potential of Paczynski type~\cite{Stu-Kov:2008:INTJMD:PsNewtSdS:,Stu-Sla-Kov:2009:CLAQG:PseNewSdS} that appears to be useful also in studies of motion of interacting galaxies~\cite{Sche-Stu-Pet:2013:JCAP:, Stu-Sch:2011:JCAP:CCMagOnCloud:,Stu-Sch:2012:INTJMD:GRvsPsNewtMagClou:} demonstrating relation of the gravitationally bound galactic systems to the so called static radius of the SdS or KdS spacetimes  \cite{Stu:1983:BULAI:,Stu:1984:BULAI:,Arra:2014:PHYSR4:,Arra:2017:Universe:,Far:2016:PDU:,Far-Lap-Pra:2015:JCAP:}. This idea has been confirmed by the recent study of general relativistic static polytropic spheres in spacetimes with the repulsive cosmological constant \cite{Stu-Hle-Nov:2016:PHYSR4:,Stu-etal:2017:JCAP:}. 

The present paper is devoted to detailed study of properties of the photon motion in the KdS black hole and naked singularity spacetimes. We concentrate attention to the behavior of the effective potentials determining the regions allowed for the photon motion. Such a study is necessary for full understanding of the optical phenomena occuring in the black hole or naked singularity spacetimes with the repulsive cosmological constant. We generalize the previous work concentrated on the properties of the photon motion in the equatorial plane \cite{Stu-Hle:2000:CLAQG:}, discussing properties of the effective potential of the latitudinal motion in terms of the motion constant related to the equatorial plane, and then continuing by study of the effective potential of the radial motion. We concentrate our study on the spherical photon orbits representing a natural generalization of the photon circular geodesics that enables a natural classification of the KdS spacetimes according to the properties of the null geodesics representing the photon motion.

\section{\KdS\ spacetime and Carter's equations of geodesic motion}

\subsection{\KdS\ geometry}
The line element describing the \KdS\ geometry is in the standard Boyer-Lindquist coordinates, using geometric system of units ($c = G = 1$), given by
\bea
\dbe s^2 =
  &-&\frac{\Delta_r}{I^2 \rho^2}\left(\dbe t - a \sin^2\theta \din \phi\right)^2\nonumber \\
  &+& \frac{\Delta_\theta \sin^2\theta}{I^2\rho^2}\left[a \din t - \left(r^2 +   a^2\right)\din \phi \right]^2 \nonumber \\
&+&\frac{\rho^2}{\Delta_r}\din r^2 + \frac{\rho^2}{\Delta_\theta}\din \theta^2, \label{line}
\eea
where
\bea
\Delta_r&=&\left(1 - \frac{1}{3}\Lambda r^2\right)\left(r^2 + a^2\right) - 2Mr,\\
\Delta_\theta&=&1 + \frac{1}{3}\Lambda a^2 \cos^2\theta,\\
I&=&1 + \frac{1}{3}\Lambda a^2,\\
\rho^2&=&r^2 + a^2\cos^2\theta. \label{rho}
\eea
 Here, as usual, we denoted by $M$ the mass of the central gravitating body, by $a$ its specific angular momentum ($a = J/M$) and by $\Lambda$ the cosmological constant. In order to simplify the discussion of the following equations, it is convenient to introduce a new cosmological parameter $y = \frac{1}{3}\Lambda  M^2,$ and use dimensionless quantities, redefining them such that $s/M \rightarrow s$, $t/M \rightarrow t, r/M \rightarrow r, a/M \rightarrow a$, which is equivalent to putting $M = 1.$ The above expressions then read
\bea
\Delta_r&=&\left(1 - y r^2\right)\left(r^2 + a^2\right) - 2r,\\
\Delta_\theta&=&1 + a^2 y \cos^2\theta,\\
I&=&1 + a^2 y,\eea
with equation (\ref{rho}) being left unchanged.\\
The physical singularity is located, as in the Kerr geometry, at the ring $r=0,$ $\theta=\pi/2.$ \\
The black hole horizons are determined by the condition
\be
\Delta_r=0 \label{event.hor.}
\ee
and their loci can be determined by the relation

\be
y = y_{h}(r;\:a^2)\equiv \frac{r^2-2r+a^2}{r^2(r^2+a^2)}.\label{yh}
\ee
The zeros of  $y_{h}(r;\:a^2),$ determining the loci of black hole horizons in pure Kerr spacetimes, are given by the relation
\be
a^2=a^2_{z(h)}(r)\equiv 2r-r^2, \label{az(h)}
\ee
the loci of its extrema are given by the functions
\be
a^2=a^2_{ex(h)\pm}(r)\equiv \frac{r(1-2r\pm \sqrt{1+8r})}{2},\label{aex(h)}
\ee
where the function $a^2_{ex(h)-}(r)<0$ in its whole definition range, hence is irrelevant. The functions $y_{h}(r;\:a^2),$ $a^2_{z(h)}(r)$ and $a^2_{ex(h)\pm}(r)$ will be needed in the section devoted to the discussion of the radial motion.\\
Three event horizons, two black hole $r_-,$ $r_+,$ and the cosmological horizon $r_c,$ ($r_- < r_+ < r_c $) exist for $y_{min(h)}(a^2) < y < y_{max(h)}(a^2),$ where the limits $y_{min/max(h)}(a^2)$ correspond to local minimum or local maximum of the function $y_{h}(r;\:a^2),$ respectively, for given rotational parameter $a.$ For $0<y<y_{min(h)}(a^2)$ or $y>y_{max(h)}(a^2)$ \KdS\ naked singularity spacetimes exist. The limit case $y=y_{min(h)}(a^2)$ corresponds to an extreme black hole spacetime, when the two black hole horizons coalesce. If $y=y_{max(h)}(a^2),$ the outer black hole and cosmological horizon merge. There exists critical value of the rotational parameter $a^2_{crit}=1.212 02,$ for which the two local extrema of the function $y_{h}(r;\:a^2)$ coalesce in an inflection point at $r_{crit}=1.616 03$ with the critical value $y_{crit}=0.0592.$ Thus, for $a^2>a^2_{crit}$ only \KdS\ naked singularity can exist for any $y>0.$

Properties of the event horizons for the more general case of the \KNdS{} spacetimes can be found in \cite{Stu-Hle:2000:CLAQG:}.

\subsection{Carter's equations of geodesic motion}       
The motion of test particles and photons following its geodesics in the \KdS\ spacetime is described by the well known Carter equations \cite{Car:1973:BlaHol:}
\bea
\rho^2 \frac{\din \theta}{\dbe \lambda} &=& \pm \sqrt{W(\theta;\:E,\:\Phi,\:\mathcal{K},\:y,\:a)}, \label{CarterL} \\
\rho^2 \frac{\din r}{\dbe \lambda} &=& \pm \sqrt{R(r;\:E,\:\Phi,\:\mathcal{K},\:y,\:a)},\label{CarterR} \\
\rho^2 \frac{\din \varphi}{\dbe \lambda} &=& \frac{aI^2[E(r^2+a^2)-a\Phi]}{\Delta_r}\\ \nonumber
&-&\frac{I^2[aE\sin^2\theta-\Phi]}{\Delta_\theta \sin^2\theta},\label{CarterPhi}\\
\rho^2 \frac{\din t}{\dbe \lambda} &=&\frac{I^2(r^2+a^2)[E(r^2+a^2)-a\Phi]}{\Delta_r}\\ \nonumber
&-&\frac{aI^2[aE\sin^2\theta-\Phi]}{\Delta_\theta },\label{CarterT}
\eea
where
\be
W(\theta;\:E,\:\Phi,\:\mathcal{K},\:y,\:a) = \mathcal{K}\Delta_\theta-\frac{I^2(aE\sin^2\theta-\Phi)^2}{\sin^2\theta}\label{Theta}\ee
and
\be R(r;\:E,\:\Phi,\:\mathcal{K},\:y,\:a) = \left[IE\left(r^2 + a^2\right) - Ia\Phi\right]^2 - \Delta_r \mathcal{K}.\label{R} \ee
Here $E$ and $\Phi $ are the constants of motion connected respectively with the time and axial symmetry of the \KdS\ geometry, and $\mathcal{K}$ is the fourth Carter constant of motion connected with the hidden symmetry of the \KdS\ geometry. Another constant of motion is the rest mass $m$ (energy) of the test particle; for  photons $m = 0$. Recall that $E$ and $\Phi$ cannot be interpreted as energy and axial component of the angular momentum of the test particle at infinity, because, due to the presence of the cosmological $\Lambda$ term, the geometry is not asymptotically flat, but de~Sitter \cite{Stu-Hle:1999:PHYSR4:}.\par
Detailed discussion of the equatorial motion of photons in the \KNdS\ spacetimes has been published in \cite{Stu-Hle:2000:CLAQG:}. Circular motion of test particles in the \KdS\ spacetimes has been presented in \cite{Stu-Sla:2004:PHYSR4:}. Here we restrict our attention on the general motion of photons in the \KdS\ spacetimes.

 In fact, the motion of photons is independent of the constant of motion $E$ and depends only on the ratio $\Phi/E$ ($E \neq 0$), usually referred to as the impact parameter $\ell$, and on the parameter $\mathcal{K}/E^2.$ For our general discussion it is convenient to use $Q = \mathcal{K} - I^2(\Phi - a E)^2$ that vanishes for the equatorial motion. For our purposes it is, however, following the paper \cite{Stu-Hle:2000:CLAQG:}, convenient to introduce a new constants of motion $X \equiv \ell - a.$ Further the constant $q \equiv Q/I^2E^2$ will be applied. Then the relations (\ref{Theta}) and (\ref{R}) simplify to the form
 \be
 W(\theta;\:X,\:q,\:y,\:a) \equiv I^2E^2[(X^2+q)\Delta_\theta-\frac{(a\cos^2\theta+X)^2}{\sin^2\theta}],\label{Thetatheta}
 \ee
 \be
 R(r;\:X,\:q,\:y,\:a)\equiv I^2E^2\left[\left(r^2 - aX\right)^2 -\Delta_r\left(X^2 + q\right)\right].\label{RXq} 
 \ee
 Following the work \cite{Stu-Hle:2000:CLAQG:} we study the general photon motion in terms of the parameter X. However, since we consider the non-equatorial motion here, it is also necessary to find out the restrictions to be imposed on parameter X that follows from the reality conditions of the latitudinal motion.
 The latitudinal motion in the \KdS\ spacetimes has been investigated yet \cite{Stu:1983:BULAI:}; however, the discussion has been related to the motion constant $\mathcal{K}.$ Here we give the discussion of the effective potential of the latitudinal motion related to the motion constant Q, as it is convenient for the purposes of our study.
 \section{Latitudinal motion}
 Because it is more convenient to work with algebraic functions instead of trigonometric ones, we introduce a new variable 
 $$m = \cos ^2 \theta,$$
 $$\din m=2\mbox{sign}(\theta-\pi/2)\sqrt{m(1-m)}\mathrm{d}\theta.$$
 This implies replacing the equation (\ref{CarterL}) by
\be
\rho^2 \frac{\din m}{\dbe \lambda} = \pm 2 \sqrt{M(m;\:X,\:q,\:y,\:a)},\label{CarterM} \ee
where
\be
M(m;\:X,\:q,\:y,\:a)\equiv I^2E^2m[(1-m)(X^2+q)\Delta_m-(am+X)^2] \label{M}\ee
with notation \be\Delta_m=1+a^2ym.\ee Note that $\din m/\dbe \lambda =0$ does not necessarily imply $\din \theta/\dbe \lambda=0,$ since it can mean just transit through the equatorial plane or polar axis. Therefore, in some cases, in order to avoid any doubts, we rather discuss the behaviour of the function (\ref{Thetatheta}).\\
The reality condition $M(m;\:a,\:y,\:X,\:q)\geq0$ can be expressed by the relations
\be
 X^\theta_{-}(m;\:q,\:y,\:a)\leq X\leq X^\theta_{+}(m;\:q,\:y,\:a) \label{realityM1} 
 \ee
  in regions where $\Delta_m-a^2y > 0,$ i. e., equivalently, $$m > m_d,$$ where 
  \be
  m_d = 1-1/a^2y \label{md}
  \ee
   is the solution of the equation
  \be
  \Delta_m-a^2y = 0,
  \ee
  and by the relations
 \be
  X\leq X^\theta_{+}(m;\:q,\:y,\:a),\quad X^\theta_{-}(m;\:q,\:y,\:a)\leq X, \label{realityM2}
 \ee
 in regions where $\Delta_m-a^2y < 0,$ i. e., $m < m_d,$ which requires $y > 1/a^2$. The functions $X^\theta_{\pm}(m;\:q,\:y,\:a),$ regarded as 'effective potentials' governing the latitudinal motion, are defined by
 \bea 
 & & X^\theta_{\pm}(m;\:q,\:y,\:a) \equiv \\ \nonumber
 & & \frac{-am \pm  \sqrt{m(1-m)\Delta_m [a^2m+q(\Delta_m-a^2y)]}}{m(\Delta_m-a^2y)}.\label{X(m)}
 \eea
 The functions $X^\theta_{\pm}(m;\:q,\:y,\:a)$ thus determine the regions allowed for the latitudinal motion, conditions $X=X^\theta_{\pm}(m;\:q,\:y,\:a)$ give the turning points.
 In order to understand the behaviour of the functions $X^\theta_{\pm}(m;\:q,\:y,\:a),$ it is necessary to find the reality regions, and loci of its local extrema and divergencies. Following \cite{Stu-Hle:2000:CLAQG:}, we shall perform this analysis using the well known procedure called 'Chinese boxes technique'  and adopting labelling of the appropriate characteristic functions in similar way. The parameters are of various significance - $q$ is a constant of motion, whereas $a,y$ govern the geometry. The natural choice is therefore to give the properties of the potentials $X^\theta_\pm(m;\:q,\:y,\:a)$ by family of functions $q(m;\:y,\:a),$ and properties of these functions by another families of functions of variable $m$ with parameters lowered by one, with spacetime parameters excluded at last.\par
 In the following analysis the relevant range of variable $m$ is, of course, $0\leq m\leq1,$ but somewhere, in order to better understand the behaviour of the characteristic functions, we formally permit $m\in R.$\\
 First we shall determine the reality region of $X_\pm(m).$ It is given by
 \bea
 q&\geq& q^\theta_r(m;\:y,\:a^2)\quad\mbox{if}\quad \Delta_m-a^2y > 0,\label{realityX(m)1}\\
 q&\leq& q^\theta_r(m;\:y,\:a^2)\quad\mbox{if}\quad \Delta_m-a^2y < 0, \label{realityX(m)2}
 \eea
 where
 \be
 q^\theta_r(m;\:y,\:a^2)\equiv \frac{a^2m}{a^2y-\Delta_m}.\ee
 Of course, this function also determines the common points of the potentials $X^\theta_-(m;\:q,\:y,\:a)$ and  $X^\theta_+(m;\:q,\:y,\:a),$ which values are then
 \be
 X^\theta_c=X_{(\pm)}(m;\:q=q^\theta_r,\:y,\:a) = X_{(\pm)}(m;\:y,\:a)\equiv\frac{a}{a^2y-\Delta_m}.
 \ee
 Of particular importance, if defined, is the value $X_{(\pm)}(1;\:y,\:a)=-a$ (see bellow).\par
 The divergency points of the functions $q^\theta_r(m;\:y,\:a^2),$ $X_{(\pm)}(m;\:y,\:a)$ and $X^\theta_-(m;\:q,\:y,\:a)$ are determined by
 \be
 y=y^\theta_{d}(m;\:a^2)\equiv \frac{1}{a^2(1-m)}.\label{yd}
 \ee
 Both the functions $X^\theta_{\pm}(m;\:q,\:y,\:a)$ can diverge, if well defined, for $m=0,$ another divergencies are given by the function $y^\theta_{d}(m;\:a^2)$ for the potential $X^\theta_{-}(m;\:q,\:y,\:a),$ but there are no other divergencies for the potential $X^\theta_{+}(m;\:q,\:y,\:a),$ as can be seen, if we rewrite the definition (\ref{X(m)}) in an alternative form
 \bea
 & & X^\theta_{\pm}(m;\:q,\:y,\:a)=\\ \nonumber
 & & \frac{a^2m^2-q(1-m)\Delta_m}{-am\mp\sqrt{m(1-m)\Delta_m[a^2m+q(\Delta_m-a^2y)]}}.
 \eea \par
 The function $y^\theta_{d}(m;\:a^2)\to \infty$ for $m\to 1$ from the left. There are no local extrema of this function and for $0\leq m<1$ it is increasing. For $m=0$ we get $y^\theta_{d}(0;\:a^2)=1/a^2.$ \\
 The point $m_d$ given by the definition (\ref{md}) determines the loci where the functions $q^\theta_r(m;\:y,\:a^2),$ $X_{(\pm)}(m;\:y,\:a)$ and $X^\theta_-(m;\:q,\:y,\:a)$ diverge; it occurs at relevant interval $(0;1)$ for $y>1/a^2$ and $m_d\to 1$ for $a^2y\to \infty.$ In such case, $q^\theta_r(m;\:y,\:a^2)\to +\infty\;(-\infty)$ for $m\to m_d$ from the left (right). \par
 From the equality
 \be
 \partial q^\theta_r/\partial m = \frac{a^2(a^2y-1)}{(\Delta_m-a^2y)^2}
 \ee
 one can see that the function $q^\theta_r(m;\:y,\:a^2)$ has no local extrema and is decreasing for $y<1/a^2,$ or piecewise increasing with the discontinuity point $m_d$ for $y>1/a^2,$ i. e., $q^\theta_r(m;\:y,\:a^2) \to +\infty\;(-\infty)$ for $m \to m_d$ from the left (right). It always holds $q^\theta_r(m=0;\:y,\:a^2)=0$ and $q^\theta_r(m=1;\:y,\:a^2)=-a^2.$\\
 In the special case $y=1/a^2$ we get
 $$q^\theta_r(m;\:y=1/a^2,\:a^2)=const.= -a^2\quad \mbox{for}\quad m\neq 0$$
 with
 $$\lim_{m\to 0} q^\theta_r(m;\:y=1/a^2,\:a^2)= -a^2.$$ \par
 Based on the conditions (\ref{realityX(m)1}), (\ref{realityX(m)2}) and the above characteristic functions, we can complete setting the definition range of the potentials $X^\theta_\pm(m;\:q,\:a,\:y),$ which we leave to the end of this section.\par

 Now we shall determine the loci of local extrema of the effective potentials $X^\theta_{\pm}(m;\:q,\:y,\:a).$ They can be derived from the condition $\partial X^\theta_{\pm}/\partial m =0,$ which implies the equation
 \be
 (a^2m^2+q)(\Delta_m-a^2y)[a^2m^2I^2+q(1-a^2y+2a^2my)^2]=0.
 \label{dXpm}\ee
 It can be verified that the function $X^\theta_+(m;\:q,\:y,\:a)$ has local extrema given by the relation
 \be q=q^\theta_{ex(+)}(m;\:a^2)\equiv -a^2m^2. \label{qexp} \ee
  A discussion of this function is trivial, so we only note that it is independent of the cosmological parameter $y$ and renders the loci of extrema only for $-a^2\leq q \leq 0,$ while, as we shall see below, they can exist even for $q<-a^2.$
 The character of these extrema reveals inserting this expression into the second derivative, which yields
 \be
 \partial^2X^\theta_+/\partial m^2(m;\:q=q_{ex(+)},\:y,\:a)=\frac{-a}{m(1-m)\Delta_m},
 \ee
 clearly they must be maxima. \par
 From the equation (\ref{dXpm}) we find that another extrema of the potentials $X^\theta_{\pm}(m;\:q,\:y,\:a)$  are determined by the condition
 \be q=q^\theta_{ex(\pm)}(m;\:y,\:a^2)\equiv \frac{-a^2m^2I^2}{[\Delta_m-a^2y(1-m)]^2}.\label{qexpm} \ee
 The divergencies of the functions $q^\theta_{ex(\pm)}(m;\:y,\:a^2)$ are determined by the relation
 \be
 y=y^\theta_{d(ex\pm)}(m;\:a)\equiv \frac{1}{a^2(1-2m)}.\ee
 The function $y^\theta_{d(ex\pm)}(m;\:a^2)$ is positively valued at $0\leq m<0.5,$ where $y^\theta_{d(ex\pm)}(m;\:a^2)\to +\infty$ for $m\to 0.5$ from the left. For $m=0,$ there is $y^\theta_{d(ex\pm)}(0;\:a^2)=y^\theta_{d(r)}(0;\:a^2)=1/a^2.$

 From the properties of the function $y^\theta_{d(ex\pm)}(m;\:a^2)$ we deduce that the function $q^\theta_{ex(\pm)}(m;\:y,\:a^2)$ can diverge only if $y>1/a^2,$ at $$m=m_{d(ex)}\equiv 0.5\;(1-1/a^2y)=0.5\;m_d$$
 located such that $0\leq m_{d(ex)} <0.5.$ Obviously
 \be
 q^\theta_{ex(\pm)}(m;\:y,\:a^2) \to -\infty \quad \mbox{for} \quad m\to m_{d(ex)}. \label{lim_qex}
 \ee  \par
 In the following we shall decide about the monotony and  possible existence of local extrema of the function $q_{ex(\pm)}.$ From 
 \be
 \partial q^\theta_{ex(\pm)}/\partial m = \frac{2m(a^2y-1)a^2I^2}{[\Delta_m-a^2y(1-m)]^3}
 \ee
 it is clear that there are no local extrema of $q^\theta_{ex(\pm)}(m;\:y,\:a^2)$ in the interval $m\in (0;1).$ The derivative changes its sign at the divergent point $m_{d(ex)},$ which reflects the behaviour given by (\ref{lim_qex}).\\
 For $y<1/a^2,$ there is
 $$\partial q^\theta_{ex(\pm)}/\partial m < 0\quad\mbox{for}\quad m\in \langle0;1\rangle,$$
 that is, $q^\theta_{ex(\pm)}(m;\:y,\:a^2)$ is decreasing.\\
 In the limit case $y=1/a^2,$ we get
 $$q^\theta_{ex(\pm)}(m;\:y=1/a^2,\:a^2)=-a^2=q^\theta_r(m;\:y=1/a^2,\:a^2).$$
 Comparing both the functions $q^\theta_{ex+}(m;\:a^2)$ and $q^\theta_{ex(\pm)}(m;\:y,\:a^2),$ we find that
 $$q^\theta_{ex(\pm)}(m;\:y,\:a^2)\leq q^\theta_{ex+}(m;\:a^2)\leq 0$$
 and have common points at $m = 0, 1$ with
 $$q^\theta_{ex+}(0;\:a^2)= q^\theta_{ex(\pm)}(0;\:y,\:a^2)=0$$
 and
 $$ q^\theta_{ex+}(1;\:a^2)= q^\theta_{ex(\pm)}(1;\:y,\:a^2)=-a^2.$$ \par
 In the next step we shall characterize the extrema given by $q^\theta_{ex(\pm)}(m;\:y,\:a^2)$.\\
 First we find that 
 \bea
 & & \frac{\partial X^\theta_{\pm}}{\partial m}(m;\:q=q^\theta_{ex(\pm)},\:y,\:a)=\\ \nonumber
 & & \frac{a^3y\{\sign[1-a^2y(1-2m)]\pm \sign(1-a^2y)\}}{\sign{[1-a^2y(m-1)]^3}}. \label{dXpm(qex_pm)}
 \eea
 If we now require
 $$\partial X^\theta_{+}/\partial m(m;\:q=q^\theta_{ex(\pm)},\:y,\:a) =0 $$
 somewhere at $0 < m < 1,$
 we obtain a condition
 $$m>m_{d(ex)}\quad \mbox{for}\quad y>1/a^2,$$
 which ensures
 \bea
 \partial^2X^\theta_+/\partial m^2(m;\:q=q^\theta_{ex(\pm)},\:y,\:a)&=& \\
 \frac{aI^2}{m(1-m)\Delta_m(1-a^2y)[\Delta_m-a^2y(1-m)]}&<&0. \nonumber \label{d2X(m)} \eea
 Considering the previous results, we can conclude that the functions $q^\theta_{ex(\pm)}(m;\:y,\:a^2)$ determine local maxima of the potential $X^\theta_+(m;\:q,\:y,\:a)$ for $q<-a^2$ that occur on this curve in the case $y>1/a^2$ in the interval $m \in (m_{d(ex)};1).$\\
 Proceeding the same way with the function $X^\theta_-(m;\:q,\:y,\:a),$ we first find that the equation
 $$\partial X^\theta_{-}/\partial m(m;\:q=q^\theta_{ex(\pm)},\:y,\:a) =0$$
 has always solution for some $m\in (0; 1)$ in the case $y<1/a^2,$ but for $y\geq 1/a^2$ this solution must fulfil $m<m_{d(ex)}.$\\
 Substituting $q=q^\theta_{ex(\pm)}$ into the second derivative of $X^\theta_-(m;\:q,\:y,\:a)$ yields the same expression as that in (\ref{d2X(m)}), but now with the above conditions we have
 $$ \partial^2X^\theta_-/\partial m^2(m;\:q=q^\theta_{ex(\pm)},\:y,\:a)>0,$$
 indicating local minima. Therefore, the function $q^\theta_{ex(\pm)}$ gives local minima of $X^\theta_-(m;\:q,\:y,\:a)$ for $y<1/a^2$ at the whole interval $m \in (0;1),$ and for $y>1/a^2$ at $m \in (0;m_{d(ex)}).$ \\
 In the special case $y=1/a^2,$ the function $q^\theta_{ex(\pm)}$ reduces to the form
 \be
 q^\theta_{ex(\pm)}(m;\:y=1/a^2,\:a^2)=-a^2;
 \ee
 \par
 we can easily convince ourself that the function $X^\theta_-$ has no local extremum in such case, and the extrema of $X^\theta_+$ are given by the function $q^\theta_{ex+}(m;\:y,\:a).$ \par
 The conditions (\ref{realityM1}), (\ref{realityM2}) ensuring the allowance of the latitudinal motion must be complemented by case when the functions $X^\theta_\pm$ are not defined. Their definition range is given by relations (\ref{realityX(m)1}), (\ref{realityX(m)2}), but it can be shown that the violation of the latter one imply $M(m;\: X,\:q,\:y,\:a)>0.$ In such case, the latitudinal motion is allowed for any impact parameter $X$ (see the details in the discussion bellow). 
  All characteristic functions are depicted in Fig. 1. and the graphs of the potentials in Fig. 2 for selected representative values of parameters.

  \begin{figure}[htbp]
   \centering
   \begin{tabular}{cc}
   	\includegraphics[width=0.23\textwidth]{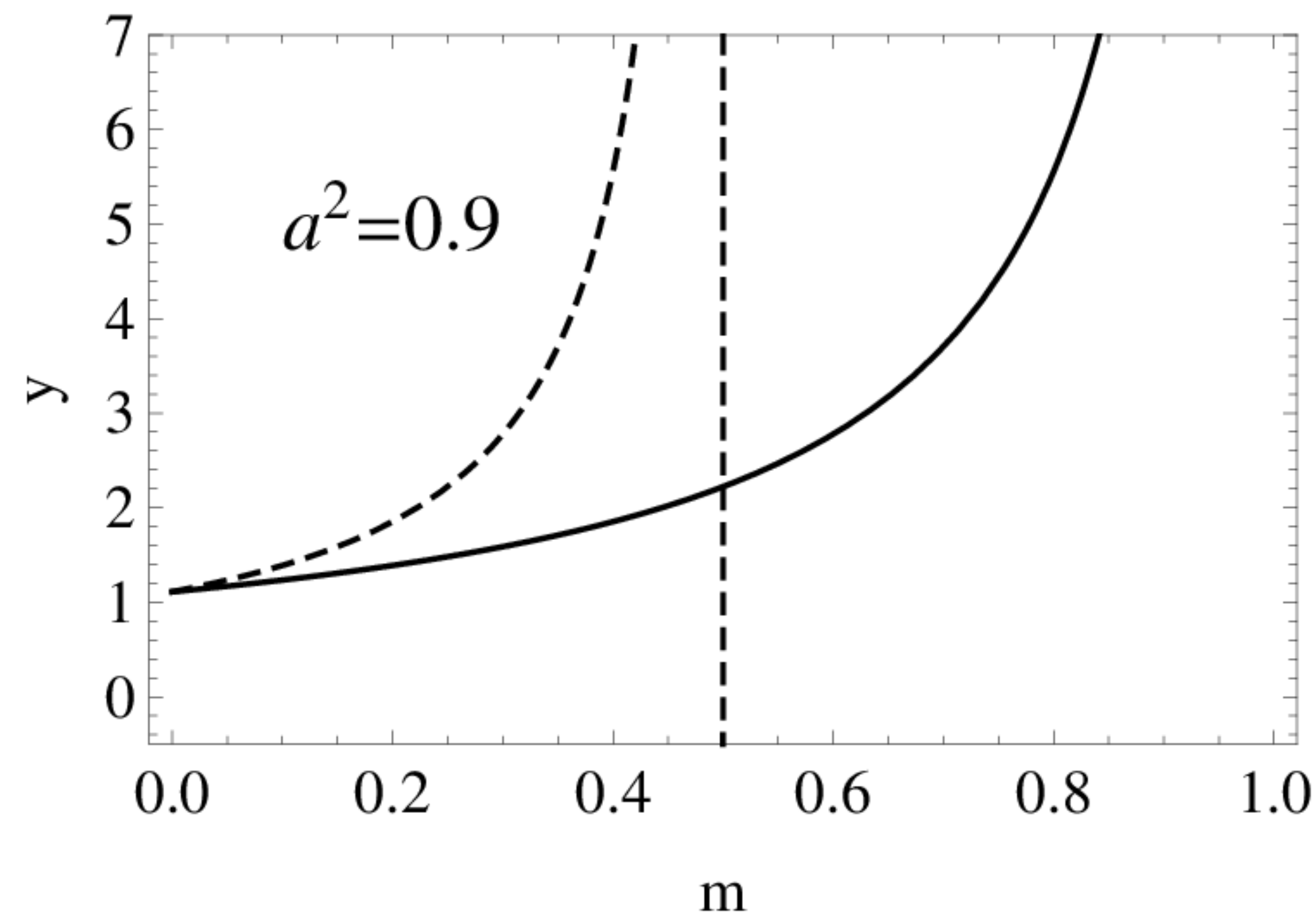}&\includegraphics[width=0.23\textwidth]{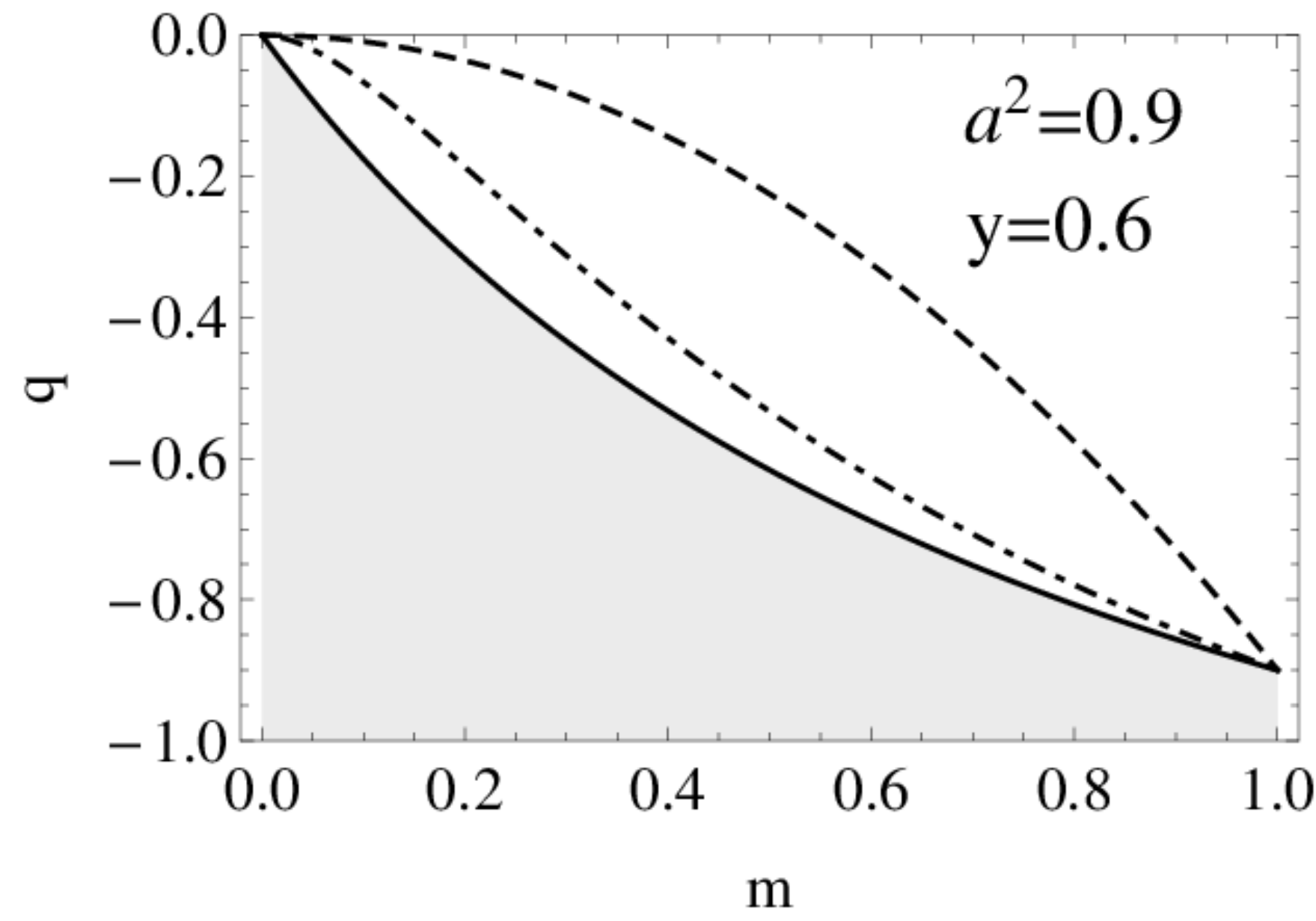}\\
   	(a)&(b)\\
   	\includegraphics[width=0.23\textwidth]{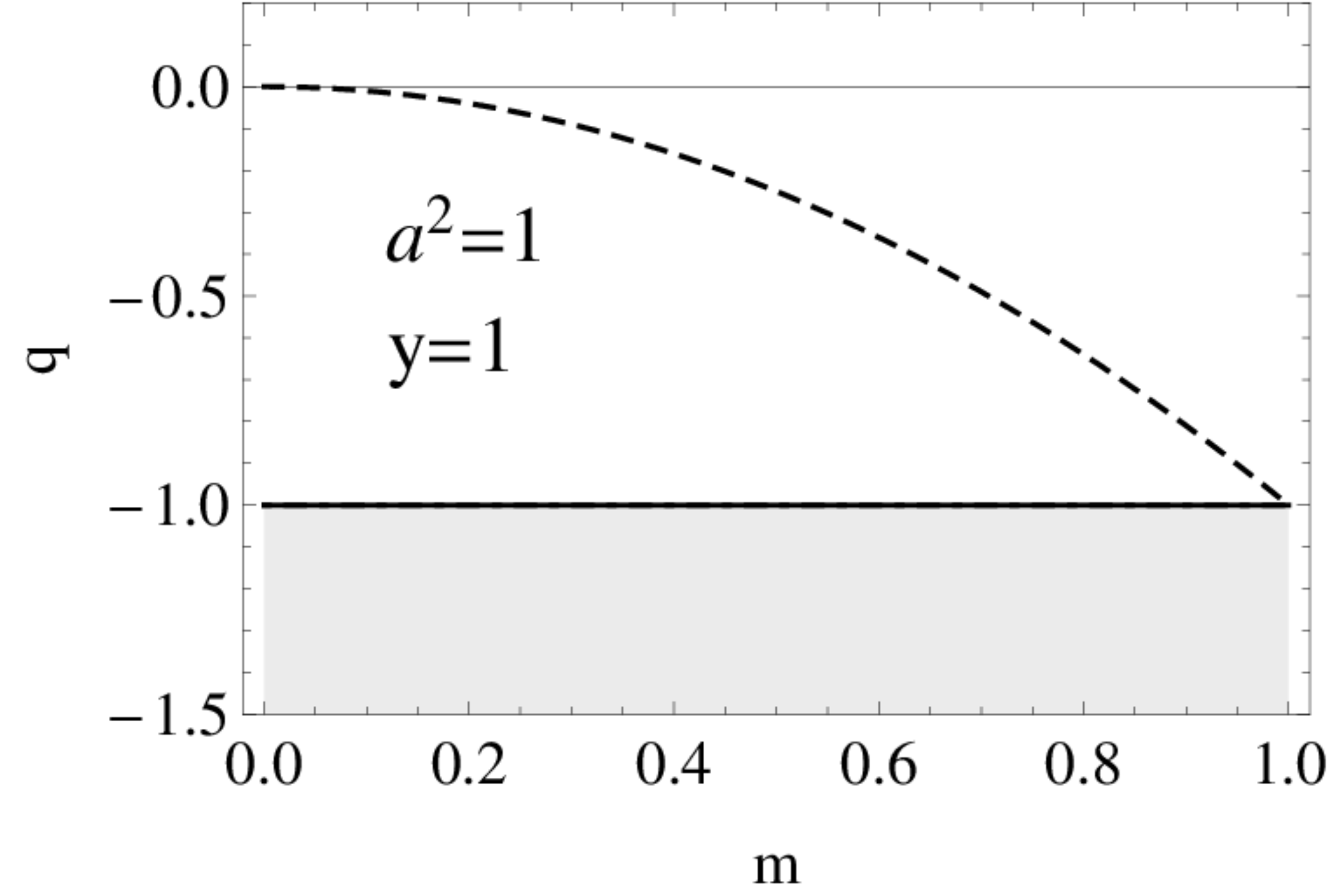}&\includegraphics[width=0.23\textwidth]{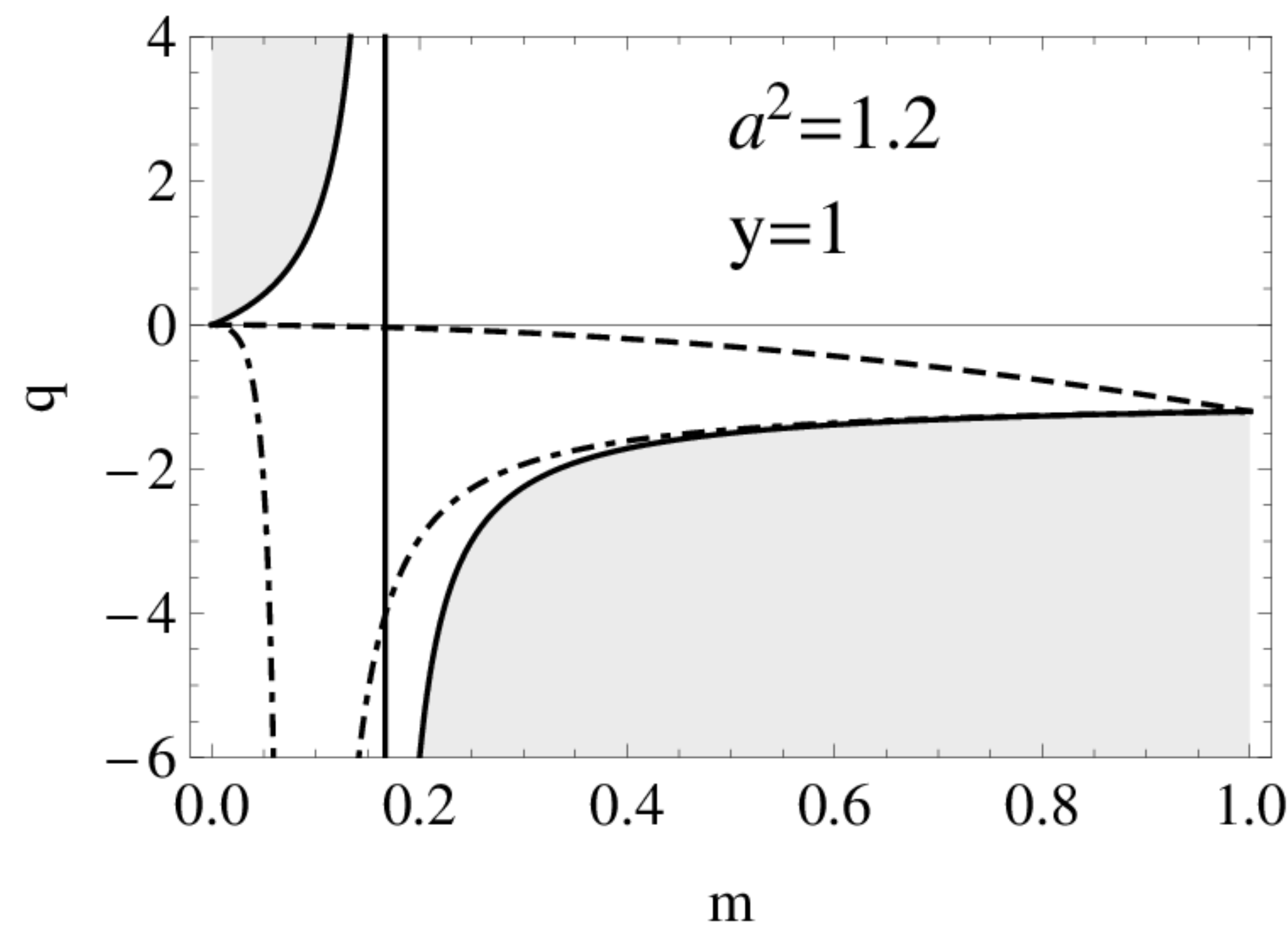}\\
   	(c)&(d)
   \end{tabular}
   \caption{The graphs of characteristic functions depicted for given parameters: \textbf{(a)}$y^\theta_{d(r)}(m;\:a^2)$ (full curve) and $y^\theta_{d(ex\pm)}(m;\:a^2)$ (dashed curve), the vertical line is the asymptote; \textbf{(b-d)}$q^\theta_r(m;\:y,\:a^2)$ (full curve), $q^\theta_{ex(+)}(m;\:a^2)$ (dashed curve), $q^\theta_{ex(\pm)}(m;\:y,\:a^2)$ (dash-dotted curve) successively corresponding to cases $y<1/a^2,\;y=1/a^2,\;y>1/a^2.$ The unshaded region demarcates the definition range of the potentials $X^\theta_{\pm}(m;\:q,\:y,\:a^2)$ given by conditions (\ref{realityX(m)1}), (\ref{realityX(m)2}).}
   \label{Figure 1} \end{figure}

 \begin{figure*}
   \centering
   \begin{tabular}{cccc}
   	\includegraphics[width=0.23\textwidth]{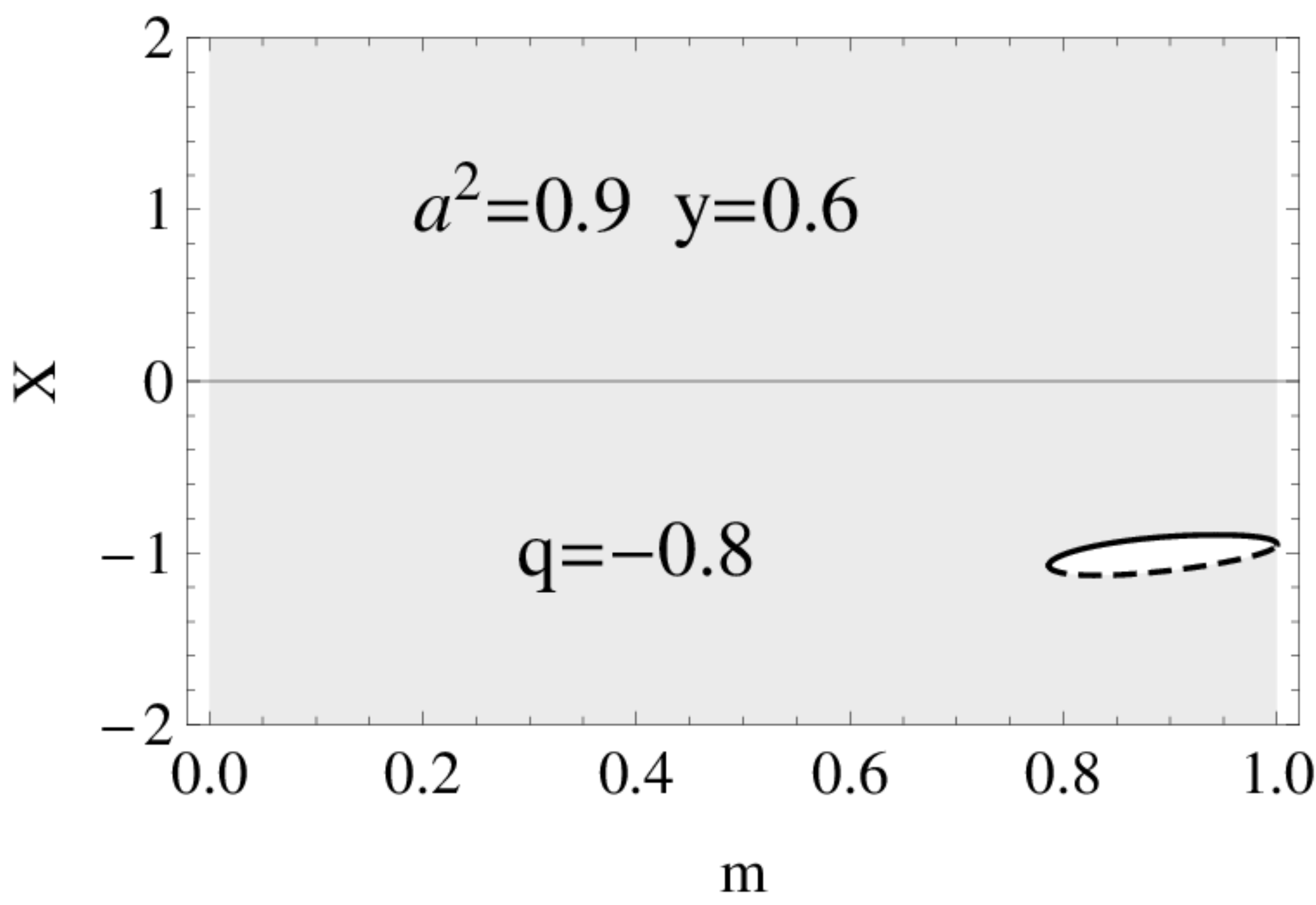}&\includegraphics[width=0.23\textwidth]{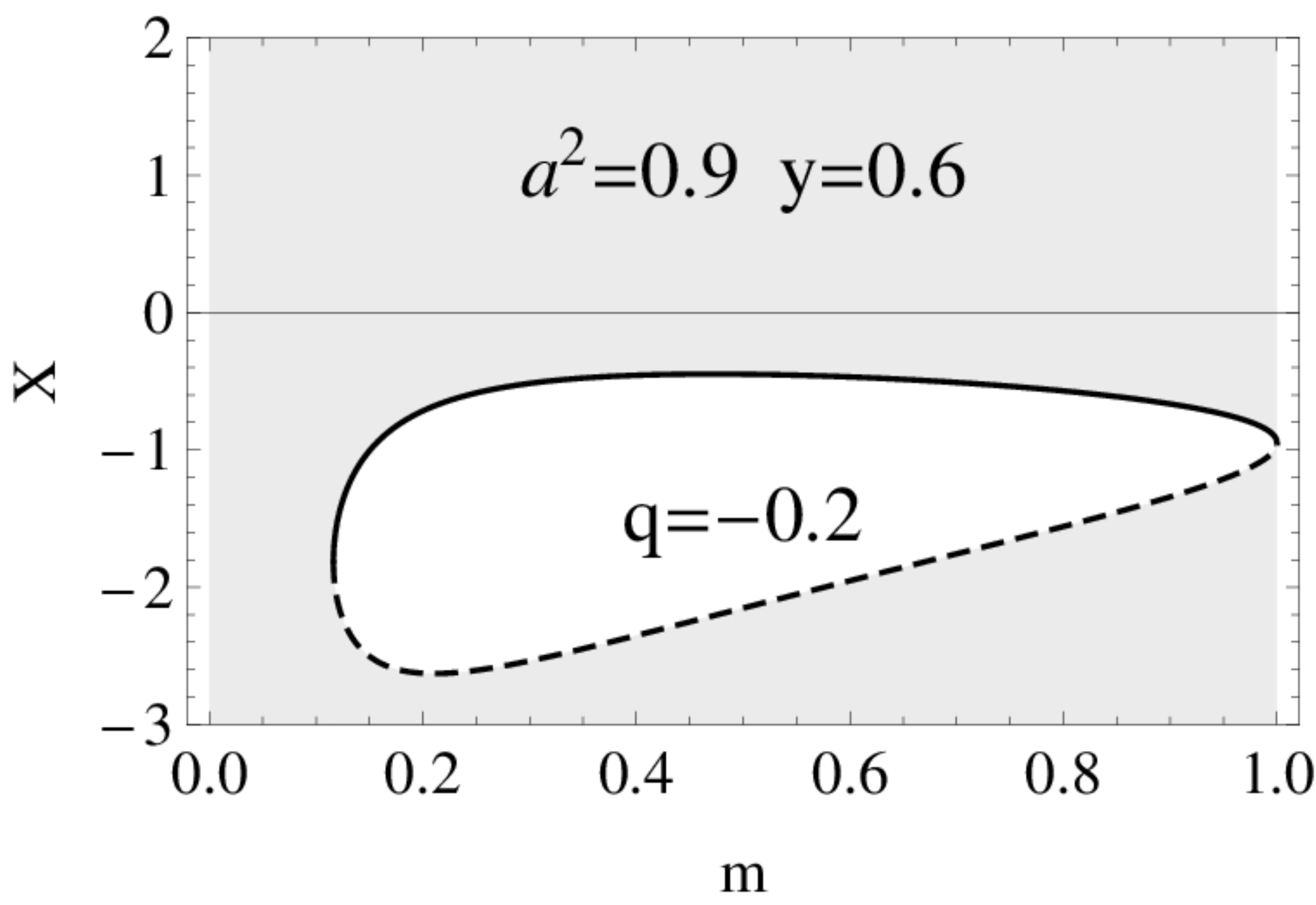}&
   	\includegraphics[width=0.23\textwidth]{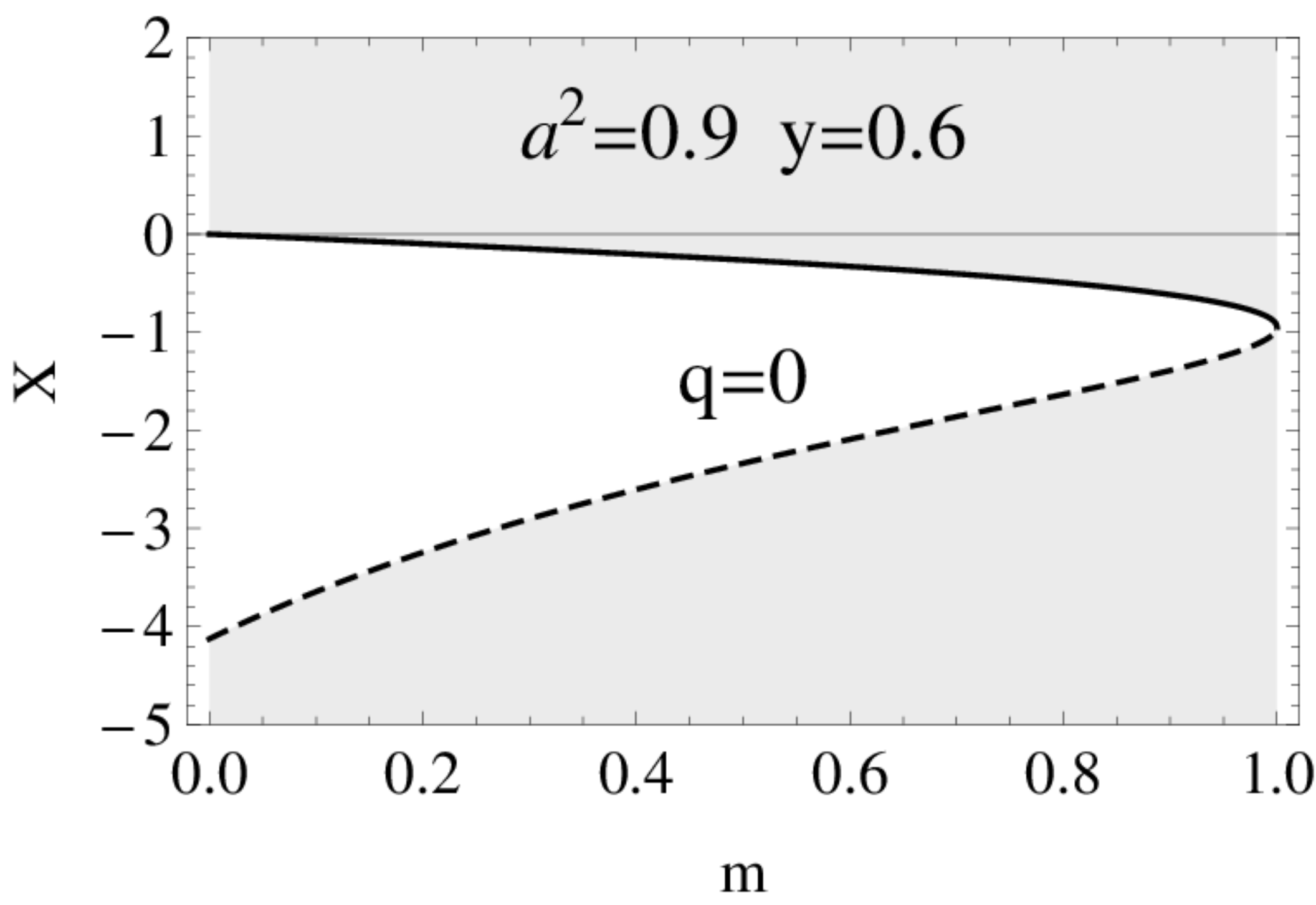}&\includegraphics[width=0.23\textwidth]{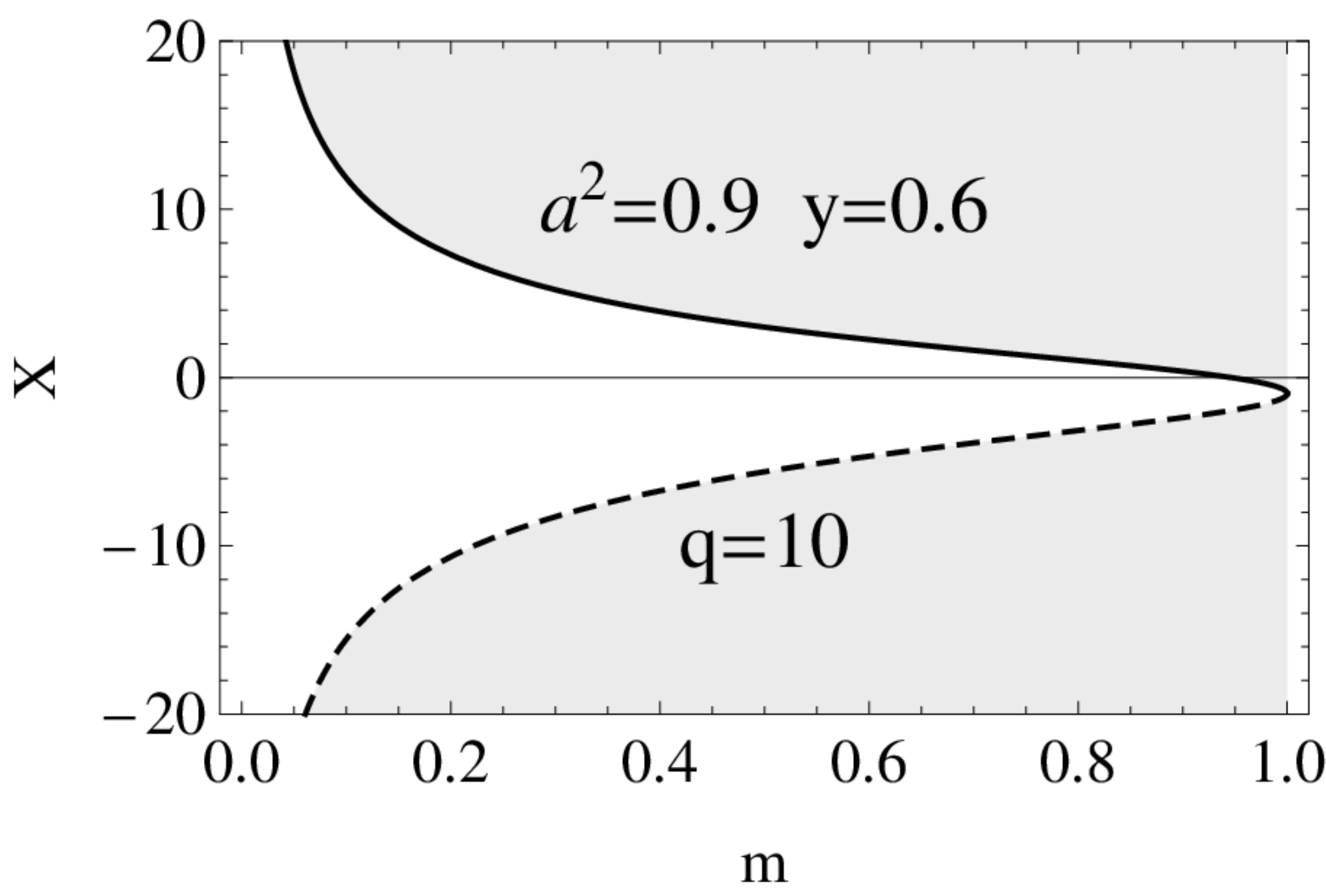}\\
   	(a)&(b)&(c)&(d)\\
   	\includegraphics[width=0.23\textwidth]{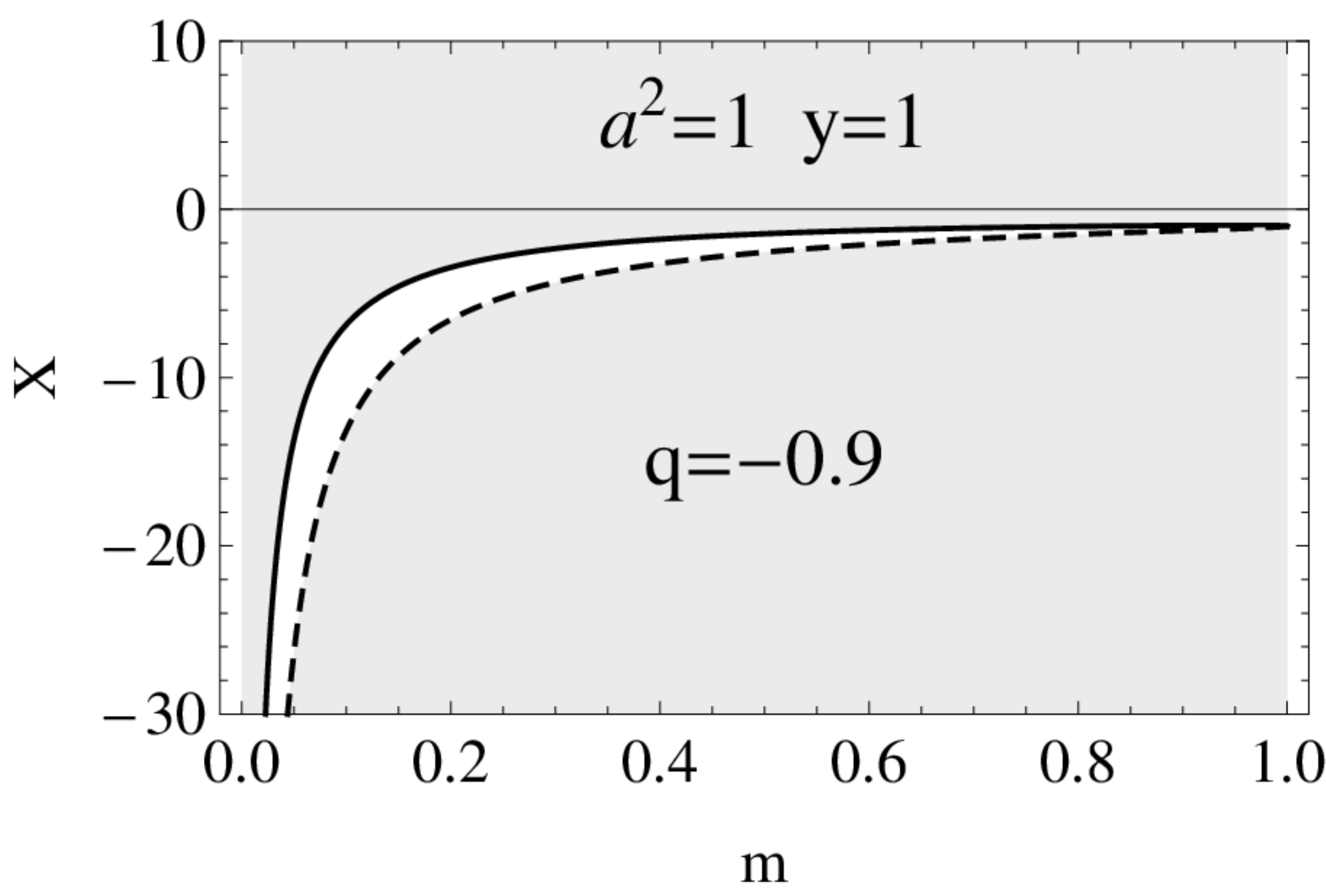}&\includegraphics[width=0.23\textwidth]{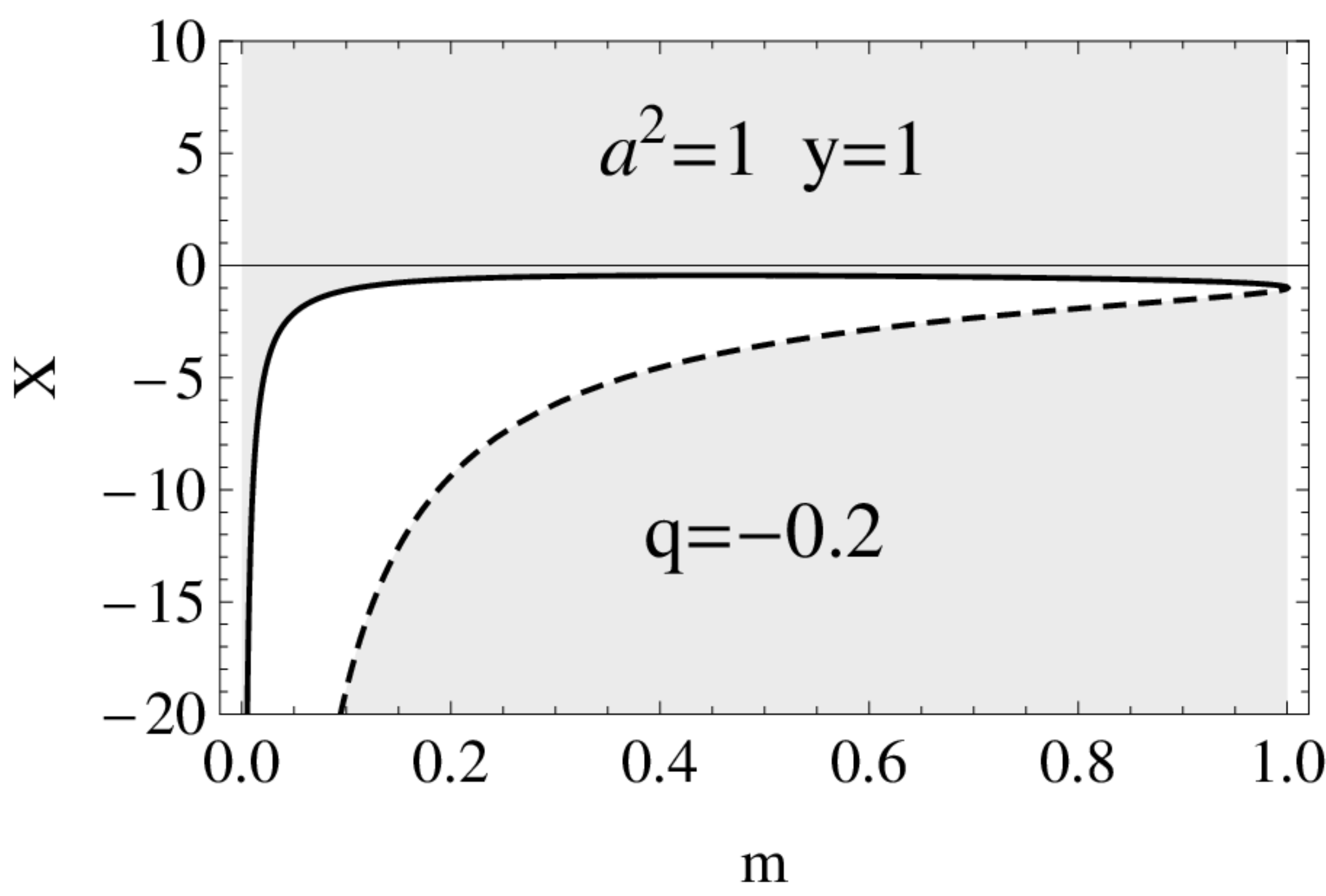}&
   	\includegraphics[width=0.23\textwidth]{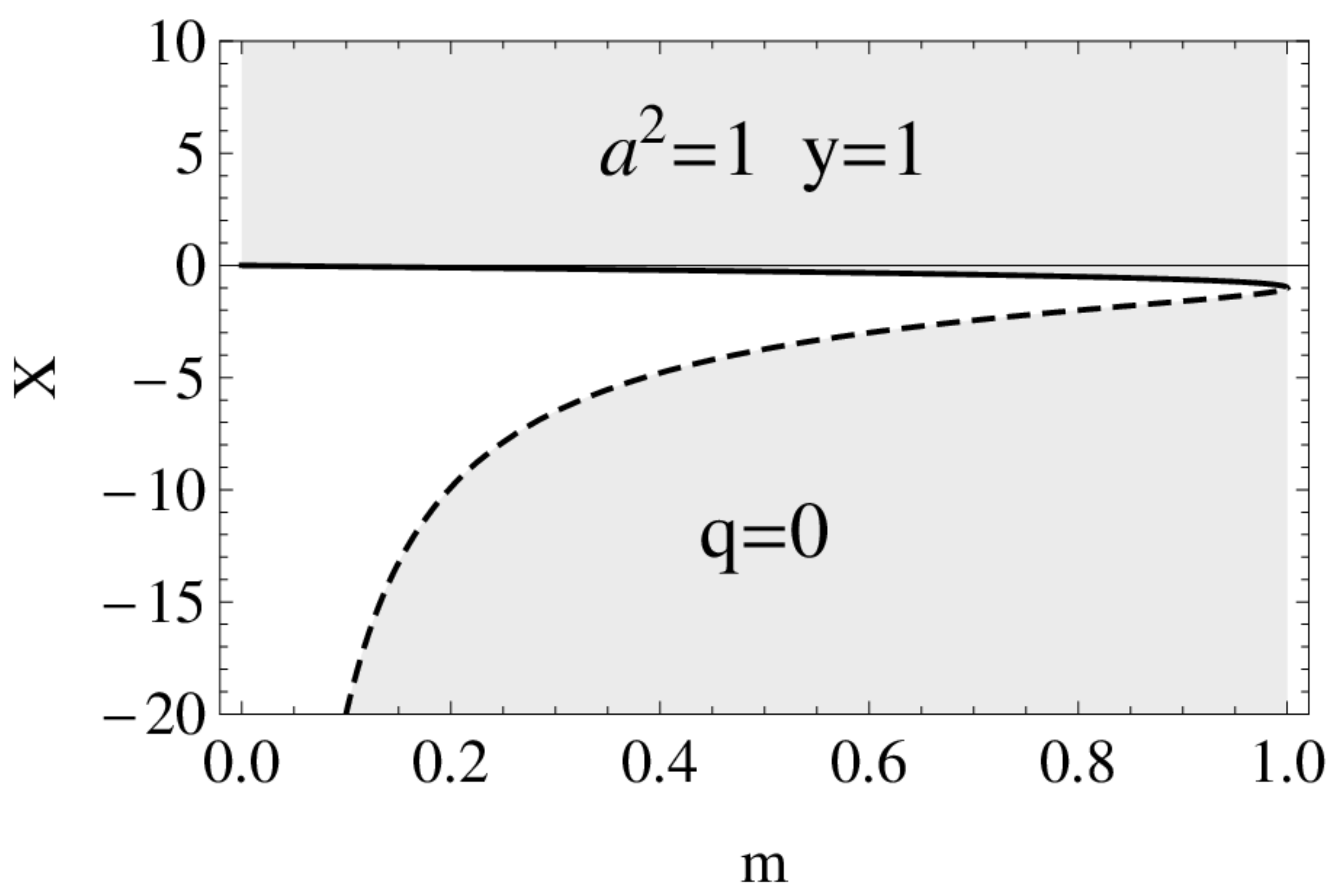}&\includegraphics[width=0.23\textwidth]{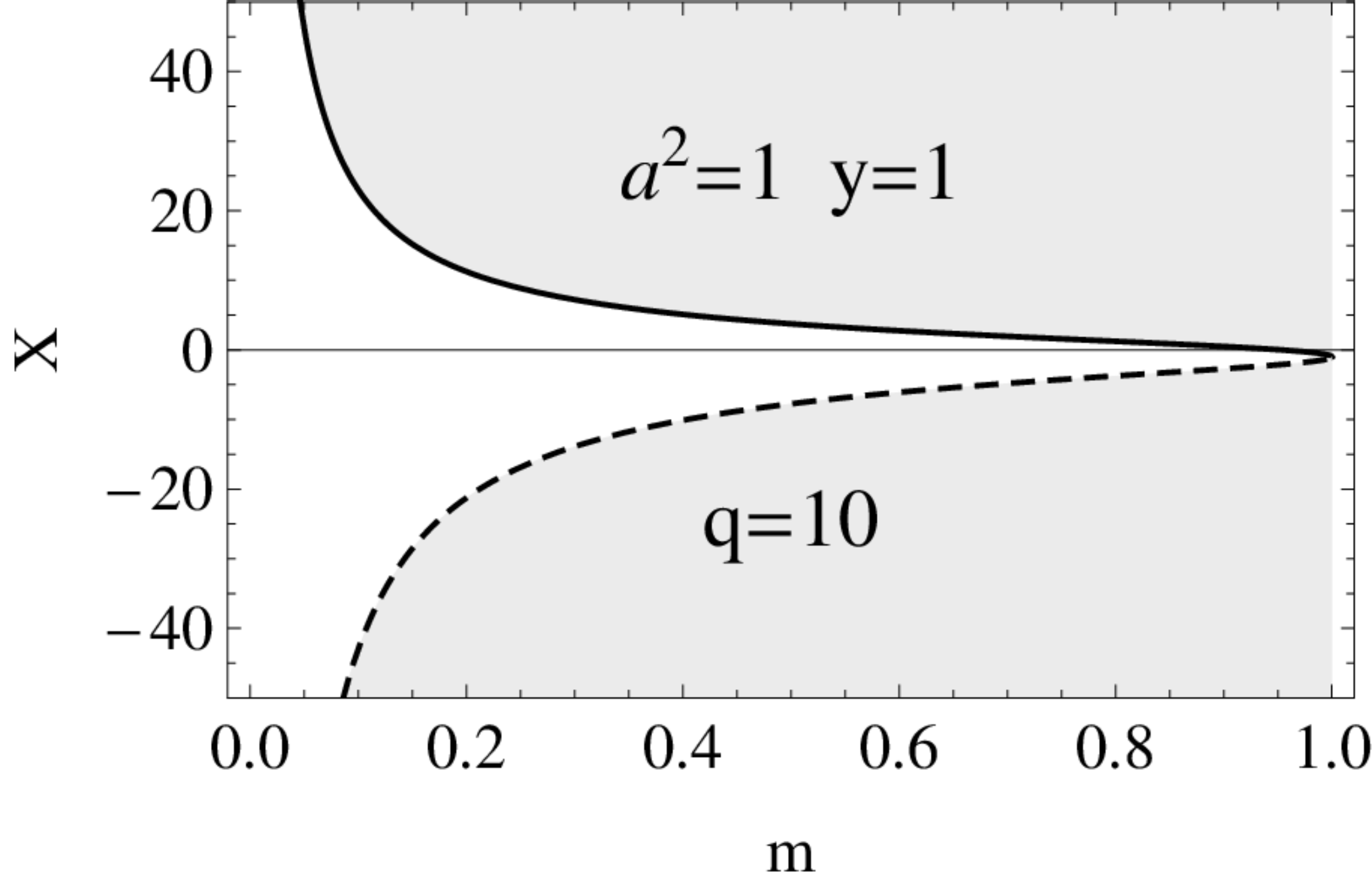}\\
   	(e)&(f)&(g)&(h)\\
   	\includegraphics[width=0.23\textwidth]{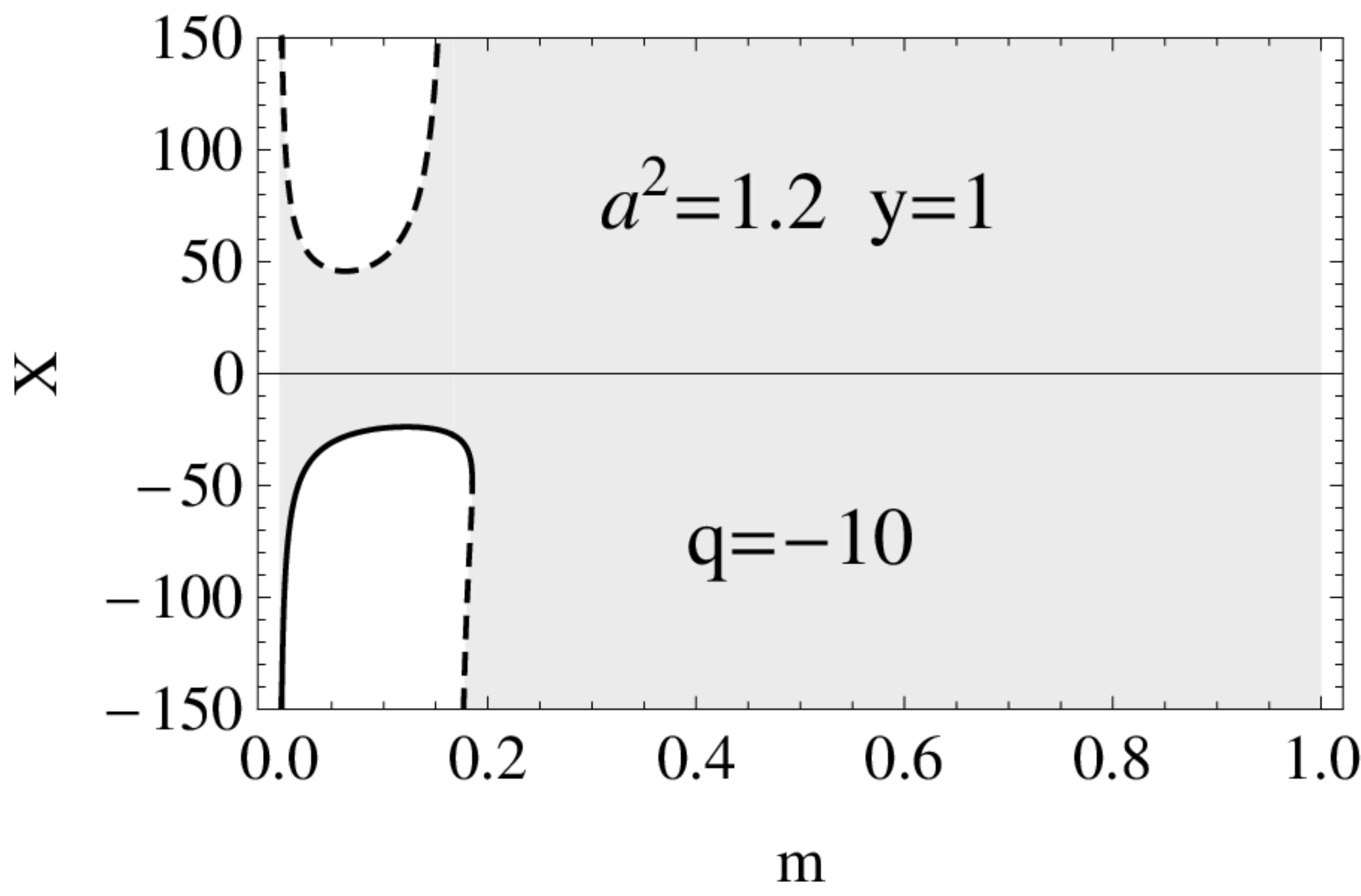}&\includegraphics[width=0.23\textwidth]{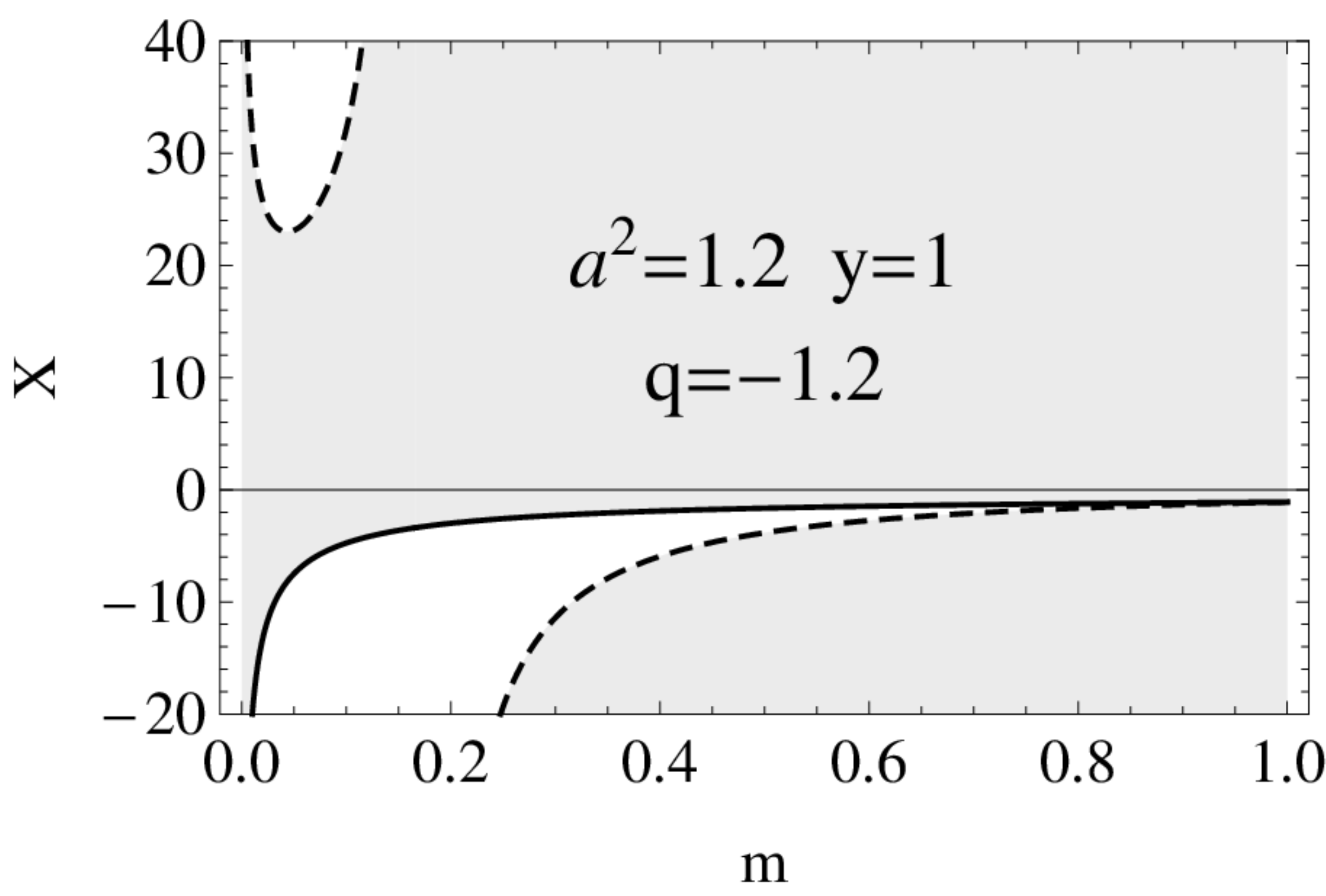}&
   	\includegraphics[width=0.23\textwidth]{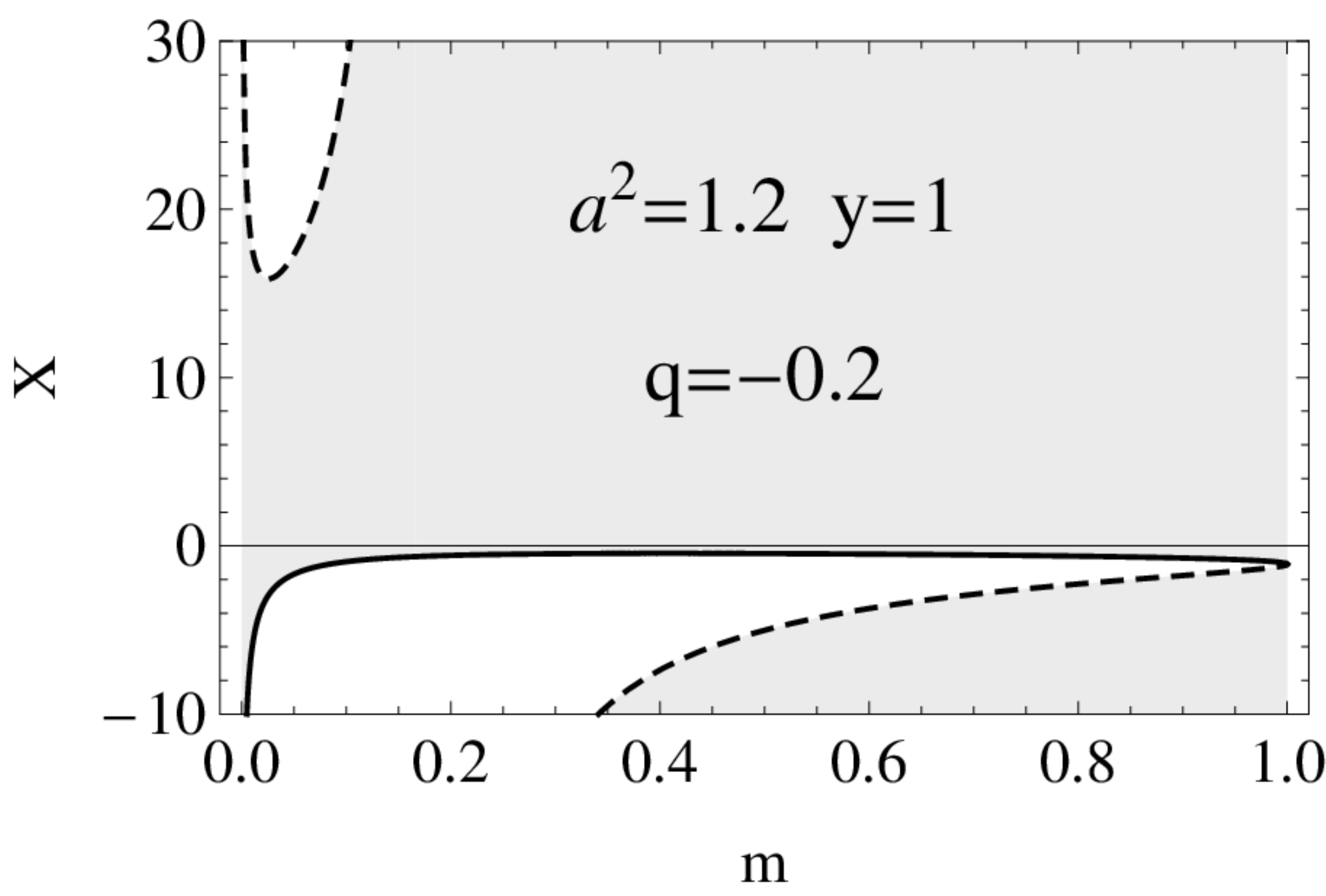}&\includegraphics[width=0.23\textwidth]{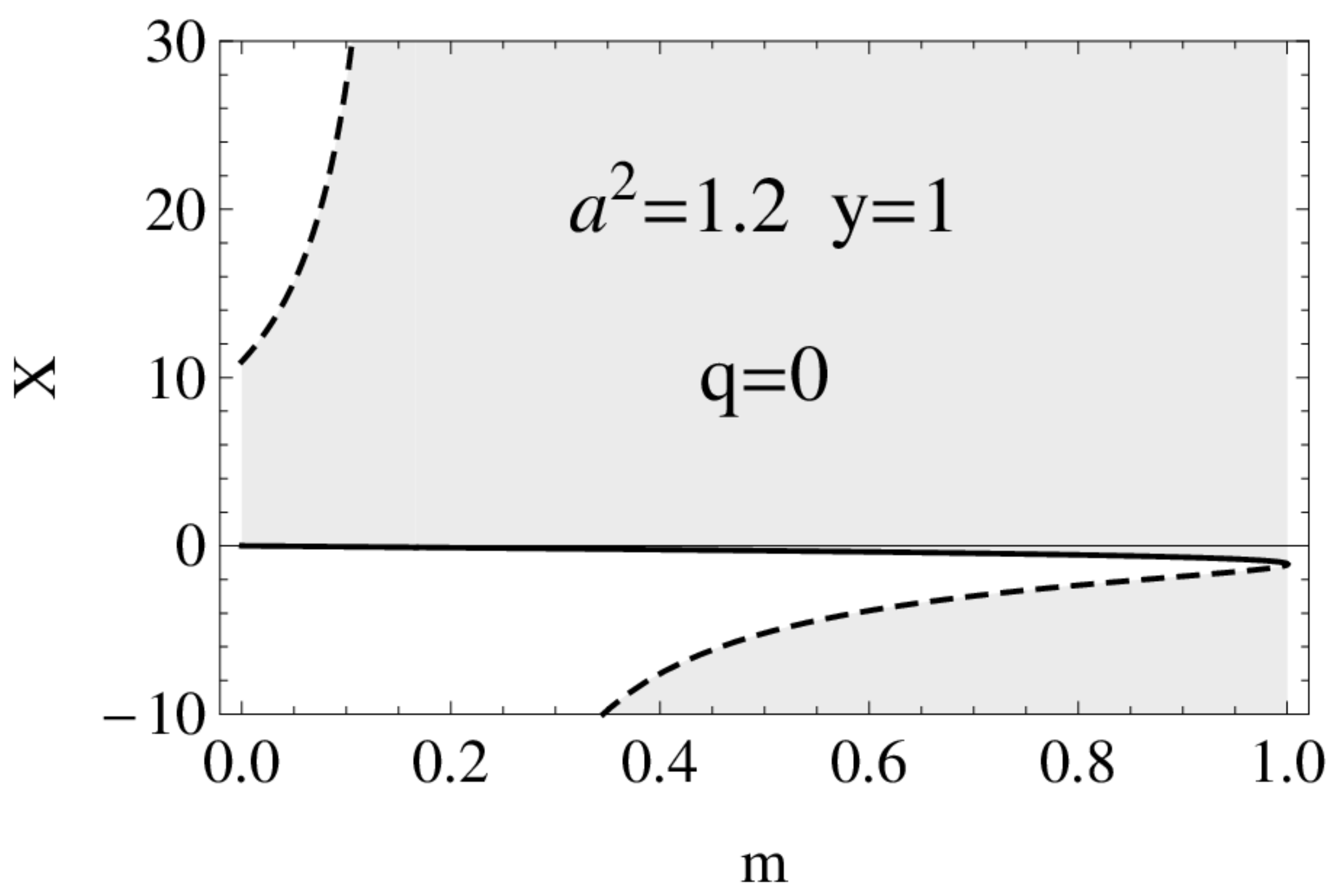}\\
   	(i)&(j)&(k)&(l)\\
   	& \includegraphics[width=0.23\textwidth]{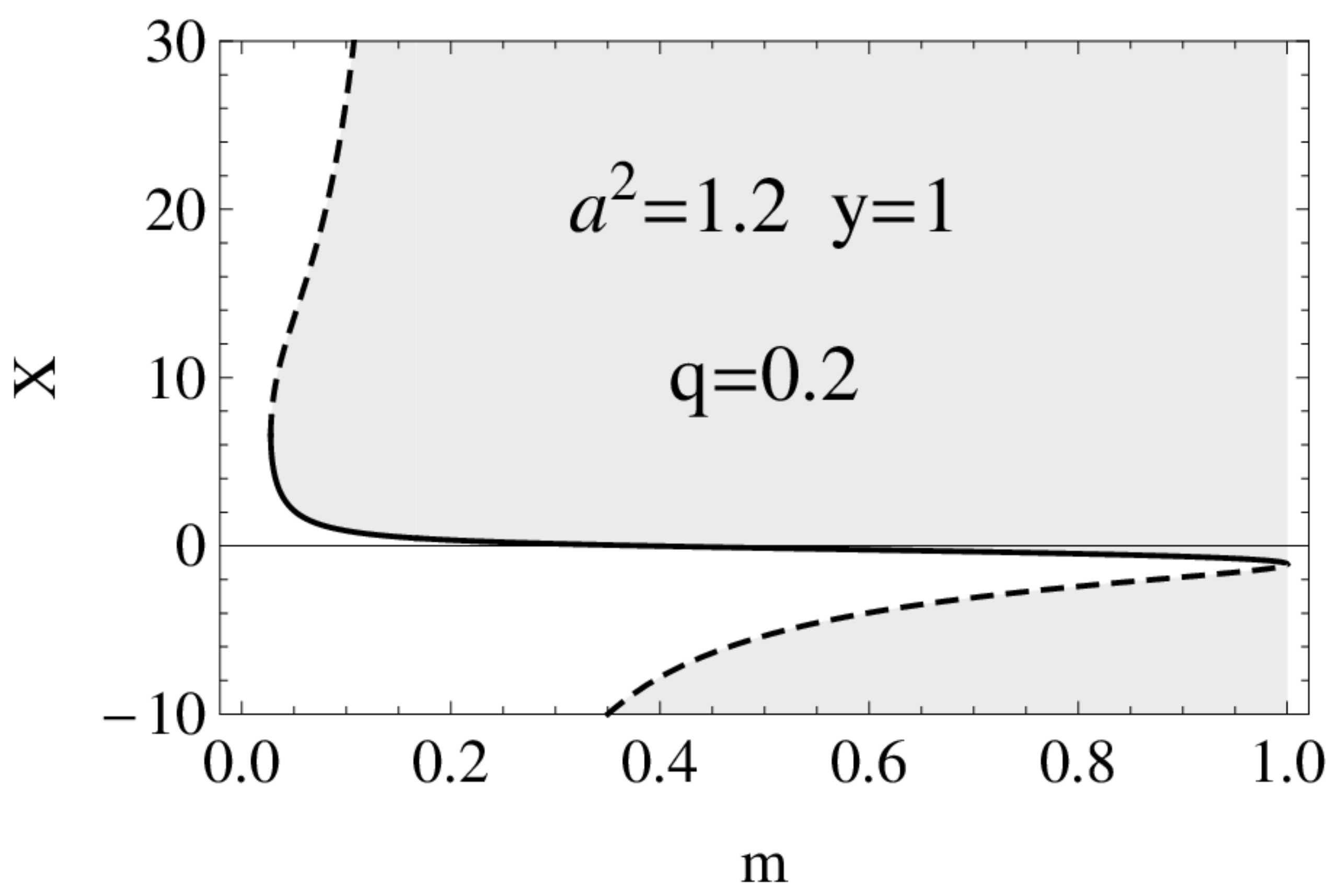}&\includegraphics[width=0.23\textwidth]{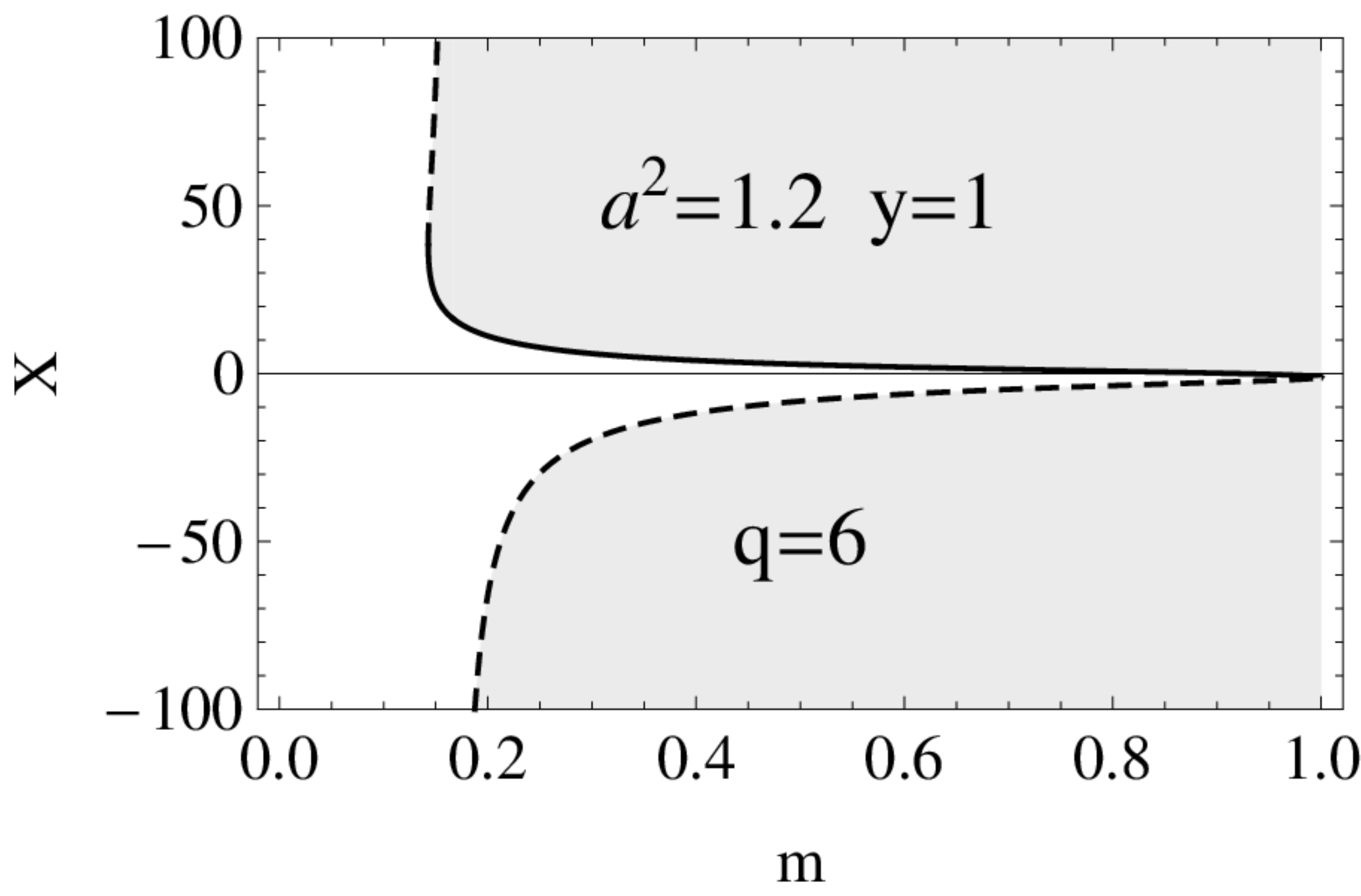} & \\
   	& (m)&(n) & 
   \end{tabular}

   \caption{Graphs of potentials $X^\theta_{+}(m;\:q,\:y,\;a)$ (full curve) and $X^\theta_{-}(m;\:q,\:y,\;a)$ (dashed curve) depicted for given typical values $a^2, y$ corresponding successively to cases $y<1/a^2$ (top row), $y=1/a^2$ (middle row), $y>1/a^2$ (bottom row), and with $q$ representing significant cases $q<0,\;q=0,\;q>0.$   Shading demarcates the region where the latitudinal motion is forbidden.}
   \label{Figure 2} \end{figure*}

    Now we are able to discuss the behaviour of the potentials \(X^\theta_{\pm }(m;\:q,\:y,\:a)\) for various representative values of its parameters.
   The intersections of a line \(X = \mbox{const.}\) with the curves \(X^\theta_{\pm }(m;\:q,\:y,\:a)\) represents the turning points in variable $m.$ From the knowledge of these functions we can thus get qualitative insight into the character of the latitudinal motion. This entitles us to following classification of \KdS\ spacetimes and brief description of the latitudinal motion. The basic division apparently consists of cases $y<1/a^2,\;y=1/a^2,\;y>1/a^2:$
\begin{enumerate}
  \item Case $y<1/a^2$
  \begin{itemize}
    \item $q<-a^2$
    \begin{description}
      \item[--] the definition range of the potentials is an empty set; the latitudinal motion is not possible;
    \end{description}
    \item $q=-a^2$
    \begin{description}
      \item[--] the potentials $X^\theta_\pm(m;\:q,\:y,\:a)$ are defined only for $m=1,$ where
      $$X^\theta_+(1;\:q,\:y,\:a)=X^\theta_-(1;\:q,\:y,\:a)=-a;$$
       photons with such values of parameters are the special
        case of the so called PNC photons 'radially' moving along the spin axis \cite{BS76:};
    \end{description}
    \item $-a^2<q<0$ (Fig. 2a, 2b)
    \begin{description}
    \item[--]both the potentials are defined for $m\in\langle m_l;1\rangle,$ where the lower limit
    \be
    m_l = \frac{q(a^2y-1)}{a^2(qy+1)}>0 \label{ml}
    \ee
    is the solution of the equation
    \[q=q^\theta_r(m;\:y,\:a)\quad \mbox{(see Fig. 1b )};\] 
                
    the limits of the interval are the common points of the potentials, where
    \be
    X^\theta_\pm(m=m_l;\:q,\:y,\:a)=X^\theta_{(\pm)}(m_l)=\frac{a(1+qy)}{a^2y-1}<0; \label{X(ml)}
    \ee
    \item[--]the latitudinal motion is allowed for values of the parameter $X$ between some local minimum $X^\theta_{min(-)}=X^\theta_-(m_{min(-)};\:q,\:y,\:a)$ and maximum $X^\theta_{max(+)}=X^\theta_+(m_{max(+)};\:q,\:y,\:a),$ for which
        $$X^\theta_{min(-)}<-a<X^\theta_{max(+)}<0;$$
        the loci $m_{min(-)}$ of minimum $X^\theta_{min(-)}$ is given by the equation (\ref{qexpm}), the loci $m_{max(+)}$ of maximum $X^\theta_{max(+)}$ is determined by relation (\ref{qexp});
      \item[--] if $X$ takes one of these extremal values, then the trajectory of such photon lies entirely on cones $\theta=\arccos \sqrt{m_{ex}},$ $\theta=\pi -\arccos \sqrt{m_{ex}},$ where $m_{ex}\in \{m_{min(-)},m_{max(+)}\};$ such photons are called PNC photons \cite{BS76:};
      \item[--] for $X^\theta_{min(-)}<X<X^\theta_{max(+)}$ there are two solutions $m_1<m_2$ of each of the two equations $X=X^\theta_\pm(m;\:q,\:y,\:a),$ implying that photon executes so called vortical motion, which is restricted between two pairs of cones, symmetrically placed relative to equatorial plane: \[0 < \arccos \sqrt{ m_2} \leq  \theta  \leq  \arccos \sqrt{m_1} < \frac{\pi }{2}\] and \[ \frac{\pi }{2} < \pi  - \arccos \sqrt{m_1} \leq
\theta  \leq  \pi-\arccos \sqrt{m_2} < \pi; \]
      \item[--] in the special case $X=-a$ one of the turning points is $m_2=1,$ which represents transit through the spin axis; such photon therefore oscillate above one of the poles in cone which is delimited by the angle $\theta=\arccos \sqrt{m_{1}};$
      \item[--] from the preceding discussion it follows that we can expect that the case $X=-a$ represents a change in azimuthal direction with respect to some privileged family of observers;
    \end{description}
    \item $q=0$ (Fig. 2c)
    \begin{description}
    \item[--]the expression in the definition (\ref{X(m)}) can be reduced to
    \be
    X^\theta_\pm(m;\:y,\:a)=\frac{-a(1\mp\sqrt{(1-m)\Delta_m})}{\Delta_m-a^2y}, \label{X(m,q=0)}
    \ee
    which validity can be enlarged, without any repercussion on the correctness of the analysis, even for $m=0;$ the definition range of the potentials is thus $\langle0;1\rangle;$
    \item[--] from the equality $W(\theta=\pi/2;\:X,\:q,\:y,\:a)=q$ it follows that at least in the equatorial plane the (radial) motion always exists for $q=0,$ where it can be both stable or unstable (see bellow); for $q>0$ the equatorial plane is crossed, for $q<0$ it can not be reached;
      \item[--]there are no extrema of the potentials - $X^\theta_+(m;\:q,\:y,\:a)$ is decreasing, $X^\theta_-(m;\:q,\:y,\:a)$ is increasing;  the permissible values of $X$ for which $\din \theta/\dbe \lambda>0$ are still confined to an interval with limits
          \bea
          X^\theta_{min(-)}&=&X^\theta_-(m=0;\:q=0,\:y,\:a) \nonumber \\
          &=&\frac{2a}{(a^2y-1)}\\
          X^\theta_{max(+)}&=&X^\theta_+(m=0;\:q=0,\:y,\:a)=0, \label{X(0)}
          \eea
          where $X^\theta_{min(-)}<X^\theta_{max(+)};$
      \item[--]if $X\leq X^\theta_{min(-)}$ or $X\geq X^\theta_{max(+)}$ then the requirement $W(\theta)\geq0$ is fulfilled only if $\theta=\pi/2,$ and in such case $\din \theta/\dbe \lambda=0,$ thus the motion is stably confined to the equatorial plane;
      \item[--]for $X^\theta_{min(-)}<X<X^\theta_{max(+)}$ photon initially released in the direction off the equatorial plane is once reflected at $\theta=\arccos \sqrt{m_{0}}$ or $\theta=\pi-\arccos \sqrt{m_{0}}$ respectively, where $m_0$ denotes the only solution of $X=X^\theta_\pm(m;\:q,\:y,\:a);$ another point where $\din \theta/\dbe \lambda=0$ is now in the equatorial plane, however the equality $\din ^2\theta/\dbe \lambda ^2=0$ implies halting in the latitudinal direction; the function $W(\theta)$ has at $\theta=\pi/2$ local minimum, which indicates, as follows from perturbation analysis, instability in the equatorial plane;
      \item[--]if specially $X=-a$ then $m_0=1,$ thus photon initially directed off the equatorial plane crosses the spin axis and finally is captured in the equatorial plane;
    \end{description}
    \item $q>0$ (Fig. 2d)
    \begin{description}
    \item[--]the potentials are defined for $m\in (0,1\rangle;$ they are monotonous in the same manner as in the case $q=0,$ but $X^\theta_+(m;\:q,\:a,\:y)\to +\infty$ and $X^\theta_-(m;\:q,\:y,\:a)\to -\infty$ as $m\to 0;$
      \item[--]from the behaviour of the potentials it follows that for $X\neq-a$ photon is forced to oscillate in $\theta$-direction through the equatorial plane between two cones governed by $\arccos \sqrt{m_{0}}\leq \theta \leq \pi-\arccos \sqrt{m_{0}},$ with $m_0$ of the same meaning as above;
      \item[--]case $X=-a$ represents the motion above both poles;
      \item[--]the foregoing conclusion is a reason to have a suspicion that cases $X<-a$ and $X>-a$ differ in the azimuthal direction relative to some family of stationary observers, it corresponds to $\ell>0$ and $\ell<0$;
    \end{description}

  \end{itemize}

  \item Case $y=1/a^2$
  \begin{description}
    \item[--]the potentials simplify into the form
    \be
    X^\theta_\pm(m;\;q,\;a)=\frac{-a\pm\sqrt{(1-m^2)(q+a^2)}}{m};\ee
    \begin{itemize}
      \item $q<-a^2$
      \begin{description}
    \item[--]the potentials are not defined, thus the latitudinal motion is not allowed;
    \end{description}
  \item  $q=-a^2$
    \begin{description}
    \item[--]the curves $X=X^\theta_\pm(m;\;q=-a^2,\:y=1/a^2,\:a)$ coalesce, since
    \bea
    X^\theta_+(m;q=-a^2,y=1/a^2,a)&=& \\ \nonumber
    X^\theta_-(m;q=-a^2,y=1/a^2,a)&=& X^\theta_{(\pm)}(m;\;a)\equiv \frac{-a}{m};\eea
    \item[--]for $X\leq-a$ there is one solution of the equation $X=X^\theta_{(\pm)}(m;\;a),$ which gives $m=m_{(\pm)}\equiv -a/X;$ this corresponds to PNC photons moving along cones $\theta=\arccos \sqrt{m_{(\pm)}},$ $\theta=\pi-\arccos \sqrt{m_{(\pm)}};$
    \item[--]for $X\to -\infty$ the cones approach the equatorial plane;
    \item[--]if specially $X=-a$ the cones degenerate to spin axis, therefore, such PNC photons move along the spin axis;
    \item[--]for $X>-a$ there is no motion allowed;
    \end{description}
  \item $-a^2<q<0$ (Fig. 2e, 2f)
  \begin{description}
    \item[--]the potentials are both defined for $m\in (0;1\rangle$; there is one local maximum $X^\theta_{max(+)}$ given by (\ref{qexp}) of the function $X^\theta_+(m;\:q,\:y,\:a)$ and no extremum of $X^\theta_-(m;\:q,\:y,\:a);$ it holds $X^\theta_-(m;\:q,\:y,\:a)<X^\theta_+(m;\:q,\:y,\:a)<0$ and $X^\theta_-(m;\:q,\:y,\:a),X^\theta_+(m;\:q,\:y,\:a)\to -\infty$ as $m \to 0$ from the right;
    \item[--]if $X<-a$ or $-a<X<X^\theta_{max(+)},$ the vortical motion exists;
    \item[--]for $X=-a$ the 'inner' cones coalesce with the spin axis, thus the vortical motion involves crossing the poles;
    \item[--]for $X=X^\theta_{max(+)}$ both the 'inner' and 'outer' cones coalesce, giving thus rise to PNC photons;
    \item[--]if $X>X^\theta_{max(+)},$ no motion is allowed;
  \end{description}
  \item $q=0$ (Fig. 2g)
  \begin{description}
    \item[--]the same discussion holds as in the case $y<1/a^2,$ except that the motion exists for $X$ arbitrarily small;
  \end{description}
  \item $q>0$ (Fig. 2h)
  \begin{description}
    \item[--]the same conclusions holds as in the case $y<1/a^2,;$
    \end{description}
  \end{itemize}
  \end{description}
  \item Case $y>1/a^2$
  \begin{itemize}
    \item $q<-a^2$ (Fig. 2i)
    \begin{description}
                    \item[--]the definition range of both potentials is an interval $(0;m_u\rangle$ (see the purple curve in Fig. 1d), where the upper limit $m_u<1$ is given as $m_l$ in the previous case by (\ref{ml});
                    \item[--]there is $X^\theta_+(m;\:q,\:y,\:a)\to -\infty$ and $X^\theta_-(m;\:q,\:y,\:a)\to +\infty$ as $m\to 0,$ moreover, $X^\theta_-(m;\:q,\:y,\:a)$ now diverge at $m=m_d,$ which is the solution of (\ref{yd}), and $X^\theta_-(m;\:q,\:y,\:a)\to +\infty\;(-\infty)$ as $m\to m_d$ from the left (right);
                    \item[--]there are thus two regions of permissible values $X$ in the $(m,X)$-plane for which the motion can exist; the lower one bounded by the graph of $X^\theta_+$ and the lower branch of $X^\theta_-,$ which at $m=m_u$ join into continuous curve, and the upper region given by the upper branch of $X^\theta_-;$ the motion is therefore allowed for $X\leq X^\theta_{max(+)}<-a$ or $X\geq X^\theta_{min(-)}>0,$ where the loci of local extrema $X^\theta_{max(+)},$ $X^\theta_{min(-)}$ are given by (\ref{qexpm}) (see the blue curve in Fig. 1d);
                    \item[--]if $X<X^\theta_{max(+)}$ or $X>X^\theta_{min(-)}$ photon executes vortical motion, cases $X=X^\theta_{max(+)}$, $X=X^\theta_{min(-)}$ correspond to PNC photons;
                    \item[--]for $X=X^\theta_-(m_u)=X^\theta_+(m_u)=a(1+qy)/(a^2y-1),$ the inner cones delimiting the vortical motion are the narrowest;
                    \item[--]for $X\to -\infty$ or $X\to +\infty$ the outer cones given by angles $$\theta=\arccos \sqrt{m_1},\quad \theta=\pi-\arccos \sqrt{m_1}$$ approach the equatorial plane since $m_1\to 0;$ for the inner cones $$\theta=\arccos \sqrt{m_2},\quad \theta=\pi-\arccos \sqrt{m_2},$$ there is $$m_2\to m_d=1-1/a^2y;$$
   \end{description}
    \item $q=-a^2$ (Fig. 2j)
     \begin{description} 
                     \item[--]there is no local extremum of the function $X^\theta_+(m;\:q,\:y,\:a),$ which is now increasing; it holds $m_u=1,$ $X^\theta_-(m_u)=X^\theta_+(m_u)=X^\theta_{+(max)}=-a,$ hence for $X=-a$ both the inner and outer cones coalesce with the spin axis, which again corresponds to 'axial' PNC photon;
                     \item[--]another PNC photons exist for $X=X^\theta_{min(-)}>0;$
                     \item[--]there are no other qualitative differences from the case $q<-a^2;$
                   \end{description}
    \item $-a^2<q<0$ (Fig. 2k)
     \begin{description} 
                    \item[--]the definition range is an interval $(0;1\rangle$ and the divergencies of the potentials are the same as above;
                      \item[--]the function $X^\theta_+(m;\:q,\:y,\:a)$ has now local maximum $X^\theta_{max(+)},$ $-a<X^\theta_{max(+)}<0,$ $X^\theta_{max(+)}\to 0$ for $q\to 0,$ determined by equation (\ref{qexp});
                      \item[--]case $X=-a$ now corresponds to vortical motion above the poles - the inner cones have coalesced with the spin axis, the outer ones stay open;
                      \item[--]the vortical motion exists as in the previous cases and above that for $-a<X<X^\theta_{max(+)};$
                    \end{description}
    \item $q=0$ (Fig. 2l)
     \begin{description}
                  \item[--]the definition (\ref{X(m,q=0)}) holds, the functions $X^\theta_{+(-)}(m;\:q,\:y,\:a)$ are defined at $\langle0,1\rangle$
                  ($ \langle0,1\rangle\setminus\{m_d\});$ the values for $m=0$ are given by (\ref{X(0)}), but now  $X^\theta_{max(+)}<X^\theta_{min(-)};$
                  \item[--]the potential $X^\theta_+(m;\:y,\:a)$ is decreasing in its whole definition range, $X^\theta_-(m;\:y,\:a)$ is piecewise increasing because of the divergent point $m_d;$
                  \item[--]if $X\leq X^\theta_{max(+)}=0$ or $X\geq X^\theta_{min(-)}$ then the same conclusions can be made as in the case $y<1/a^2$ for $X^\theta_{min(-)}\leq X\leq X^\theta_{max(+)};$
                  \item[--]for $X^\theta_{max(+)}< X< X^\theta_{min(-)}$ it holds $W(\theta=\pi/2;\:X,\:q=0,\:y,\:a) =0$ again, otherwise $W(\theta;\:X,\:q=0,\:y,\:a)<0,$ therefore photons can radially move in the equatorial plane;
                \end{description}
    \item $q>0$ (Fig.2m, 2n)
     \begin{description}
                  \item[--]the function $X^\theta_+(m;\:q,\:y,\:a)$ is defined at $\langle m_l,1\rangle,$ the function $X^\theta_-(m;\:q,\:y,\:a)$ at
                  $\langle m_l,1\rangle\setminus\{m_d\},$ where $m_l$ is given by (\ref{ml}) with the difference that now $X^\theta_{(\pm)}(m_l)>0$; the graphs of both functions now form a single open curve, which intersects a line $X=const.$ at a single point;
                  \item[--]in the interval $m \in \langle 0; m_l \rangle$ the latitudinal motion is allowed for arbitrarily large or small value of the motion constant $X;$
                  \item[--]for arbitrary $X\neq -a$ there exists oscillatory motion through the equatorial plane as described in the case $y<1/a^2, q>0;$
                  \item[--]if $X=X_{(\pm)}(m_l)$ the boundary cones are closest to equatorial plane, they are given by angles
                  $$\theta=\arccos{\sqrt{m_l}},\quad \theta=\pi-\arccos{\sqrt{m_l}};$$
                  \item[--]the case $X=-a$ corresponds to orbits above both poles crossing also the equatorial plane;
                  \item[--]there is no vortical motion or PNC photons;
                \end{description}

  \end{itemize}
\end{enumerate}
\par
We finish this section with setting the allowed region in the $(X,q)$-plane delimiting such combinations of the motion constants, for which the latitudinal motion is possible, in dependence on the spacetime parameters $a,y.$ \\

From the requirement that the function $M(m;\:a,\:y,\:X,\:q)$ defined in the relation (\ref{M}) has to be non-negative somewhere in the interval $m \in \langle0;1\rangle,$ one can derive that the allowed region of the $(X-q)$-plane is determined by the condition
\be 
q\geq q_{min}(X,\:y,\:a),\label{qmin(X,y,a)}
\ee
where $q_{min}(X,\:y,\:a)$ is defined using functions
\be
q_1(X)\equiv -X^2 \label{f_1(X)}
\ee
and 
\be
q_2(X;\:y,\:a)\equiv -I^{-2}[(1-a^2y)X+2a]^{2} \label{f_2(X)}
\ee
as follows (see Fig. 3 and Fig. 4):
\begin{itemize}
  \item Case $y<1/a^2$ (Fig. 3a)
  \begin{equation}
q_{min}(X,\:y,\:a) \equiv \left\{
\begin{array}{l}
0,\quad\mbox{for}\quad X < \frac{2 a}{a^2 y - 1}\quad\mbox{or}\quad X > 0;\\
\\
 q_2(X;\:y,\:a),\quad\mbox{for}\quad\frac{2 a}{a^2y-1} \leq X < -a;\\
\\
 q_1(X),\quad\mbox{for}\quad -a \leq  X \leq  0;\\
  \end{array}\right.
  \label{q_l_plane1}
 \end{equation}
  \item Case $y=1/a^2$ (Fig. 3b)
  \begin{equation}
q_{min}(X,\:y,\:a) \equiv \left\{
\begin{array}{l}
-a^2,\quad\mbox{for}\quad X \leq -a;\\
\\
 q_1(X),\quad\mbox{for}\quad -a \leq X \leq 0;\\
\\
 0,\quad\mbox{for}\quad X\geq 0; \end{array}\right.\label{q_l_plane2}
 \end{equation}
  \item Case $y>1/a^2$ (Fig. 3c)
  \begin{equation}
q_{min}(X,\:y,\:a) \equiv \left\{
\quad \begin{array}{l}
q_2(X;\:y,\:a),\quad\mbox{for}\quad X < -a\\
\\
\qquad \qquad \qquad \mbox{or} \quad \frac{2 a}{a^2 y - 1} < X;\\
\\
 q_1(X),\quad\mbox{for}\quad -a \leq X \leq 0;\\
\\
 0,\quad\mbox{when}\quad 0 < X \leq  \frac{2 a}{a^2 y - 1}; \end{array}\right.\label{q_l_plane3}
 \end{equation}
\end{itemize}
The case $y<1/a^2$ qualitatively corresponds to both black hole and naked singularity spacetimes, the other two cases $y=1/a^2$ and $y>1/a^2$ describe the naked singularity spacetimes (see Fig. \ref{Figure 7} in the next section). The $q=const.$- slices of the function $q_{min}(X,\:y,\:a)$ give for $q<0$ extremal values $X_{min(-)},$ $X_{max(+)}$ of the potentials $X^\theta_{\pm}(m;\:q,\:y,\:a)$ discussed in the text.
\begin{figure}[htbp]
   \centering
   \begin{tabular}{cc}
   	\includegraphics[width=0.23\textwidth]{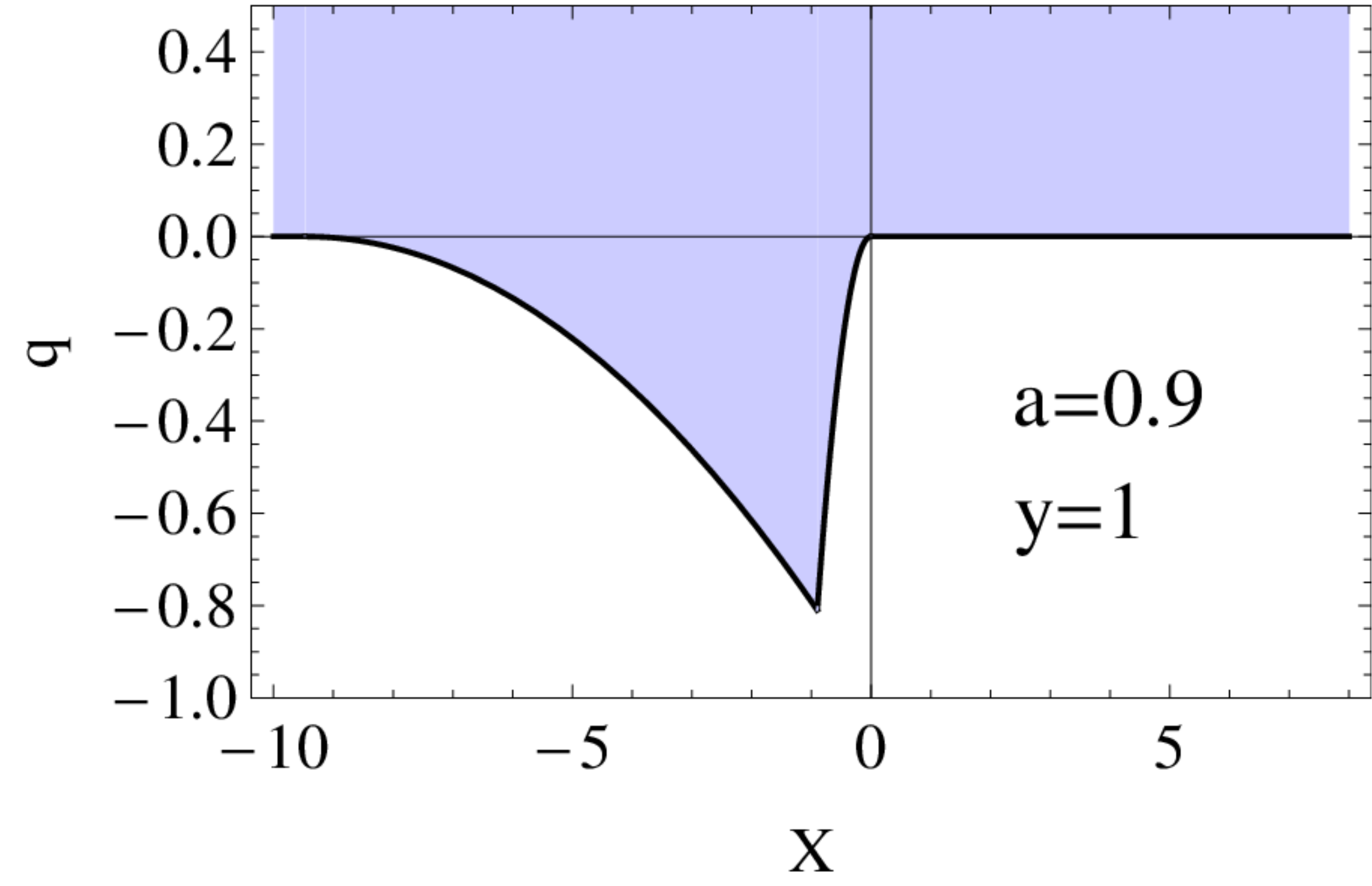}&\includegraphics[width=0.23\textwidth]{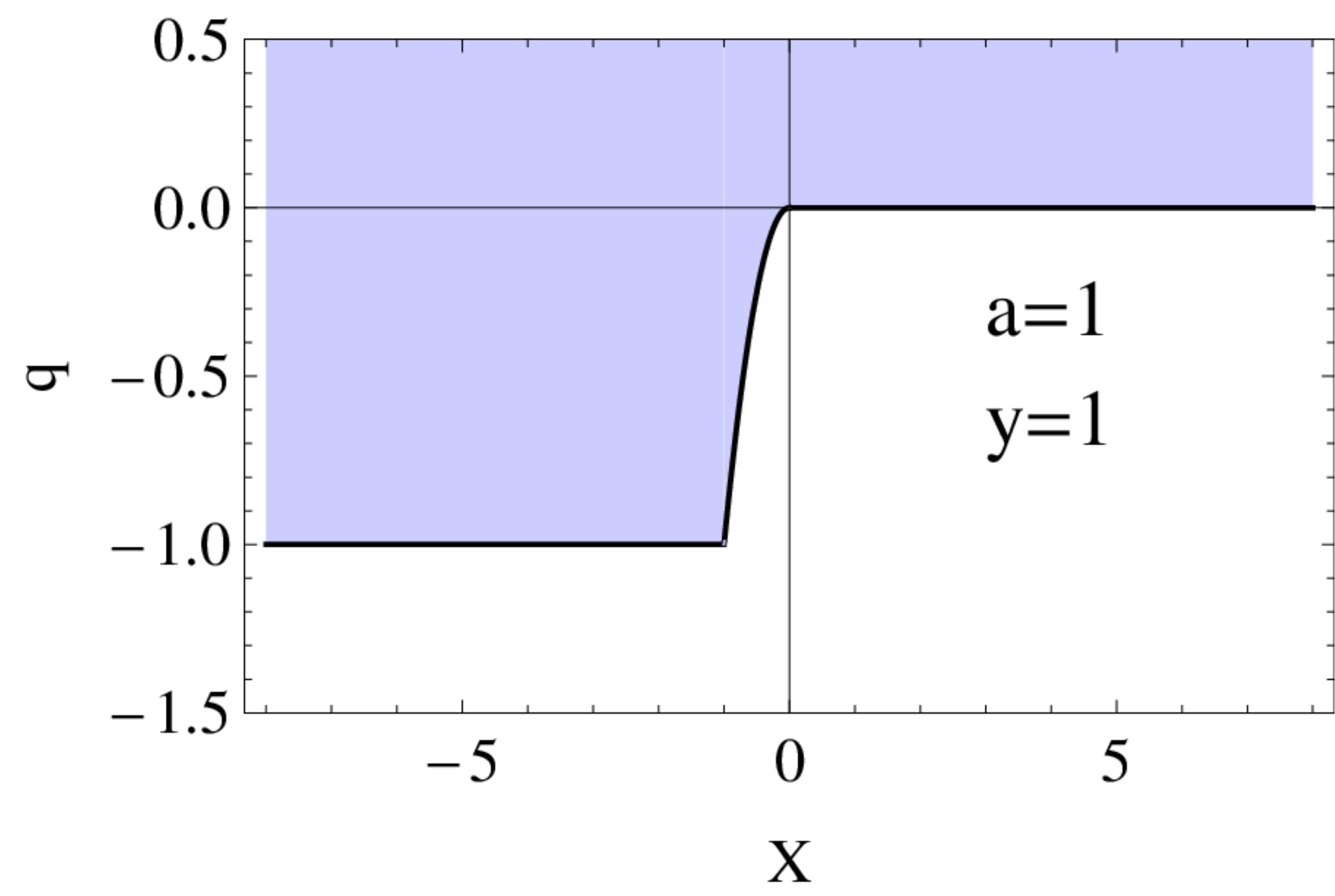}\\
   	(a)&(b)\\
   	\includegraphics[width=0.23\textwidth]{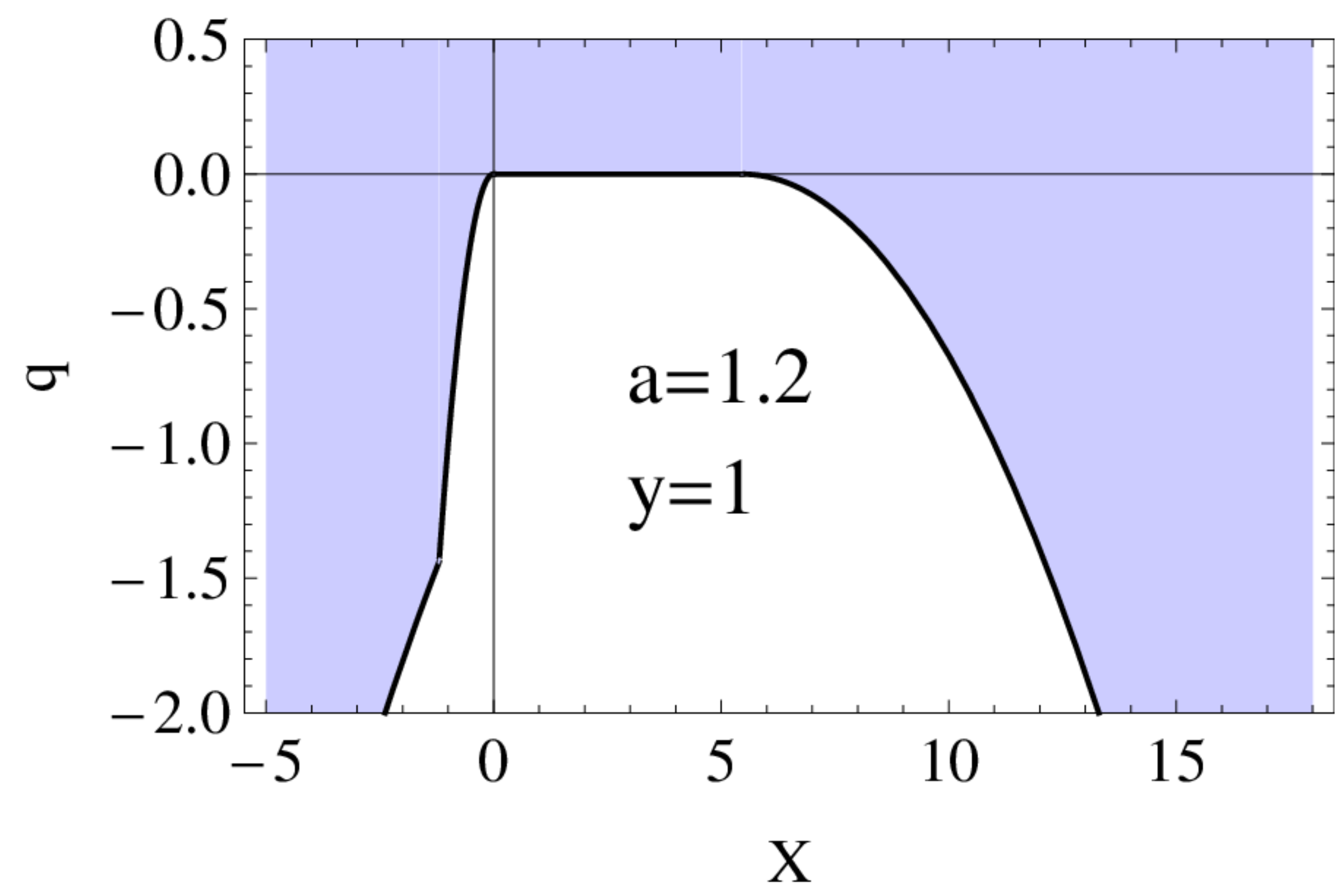}&\\
   	(c)&
   \end{tabular}
   \caption{The allowed region in the motion constant plane $(X-q)$ depicted for some representative values of spacetime parameters $a, y,$ successively corresponding to cases $y<1/a^2,\;y=1/a^2,\;y>1/a^2.$  An intersections of a line $q=const.<0$ with the border curves $q_1(X),$ $q_2(X;\:y,\:a)$ determines the extremal values $X_{min(-)},$ $X_{max(+)}$ of the potentials $X^\theta_{\pm}(m;\:q,\:y,\:a)$ introduced in the discussion above. }
   \label{Figure 3} \end{figure}
   
 \begin{figure}[htbp]
   \centering
   \begin{tabular}{c}
   	\includegraphics[width=0.38\textwidth]{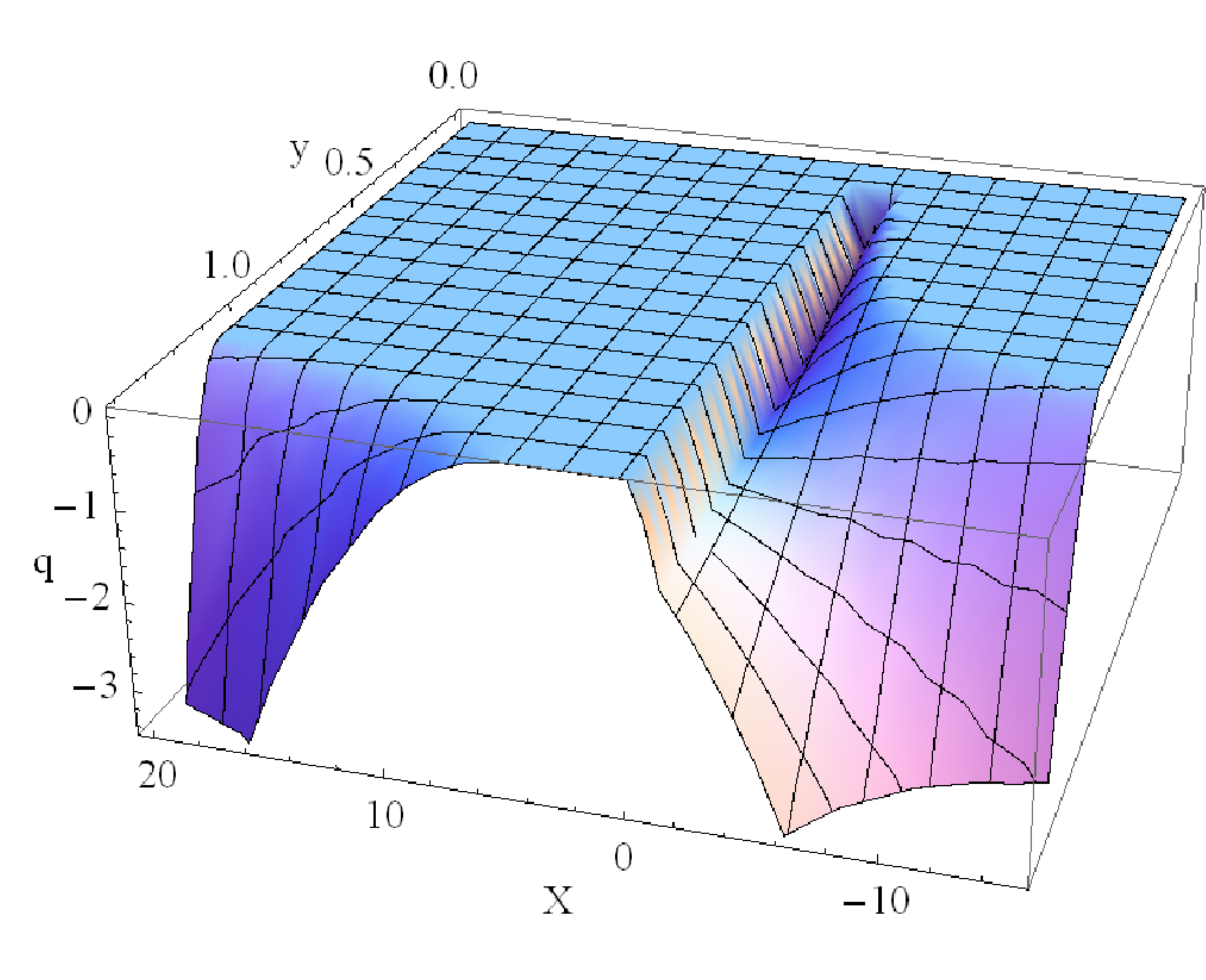}\\
   	(a)\\
   	\includegraphics[width=0.38\textwidth]{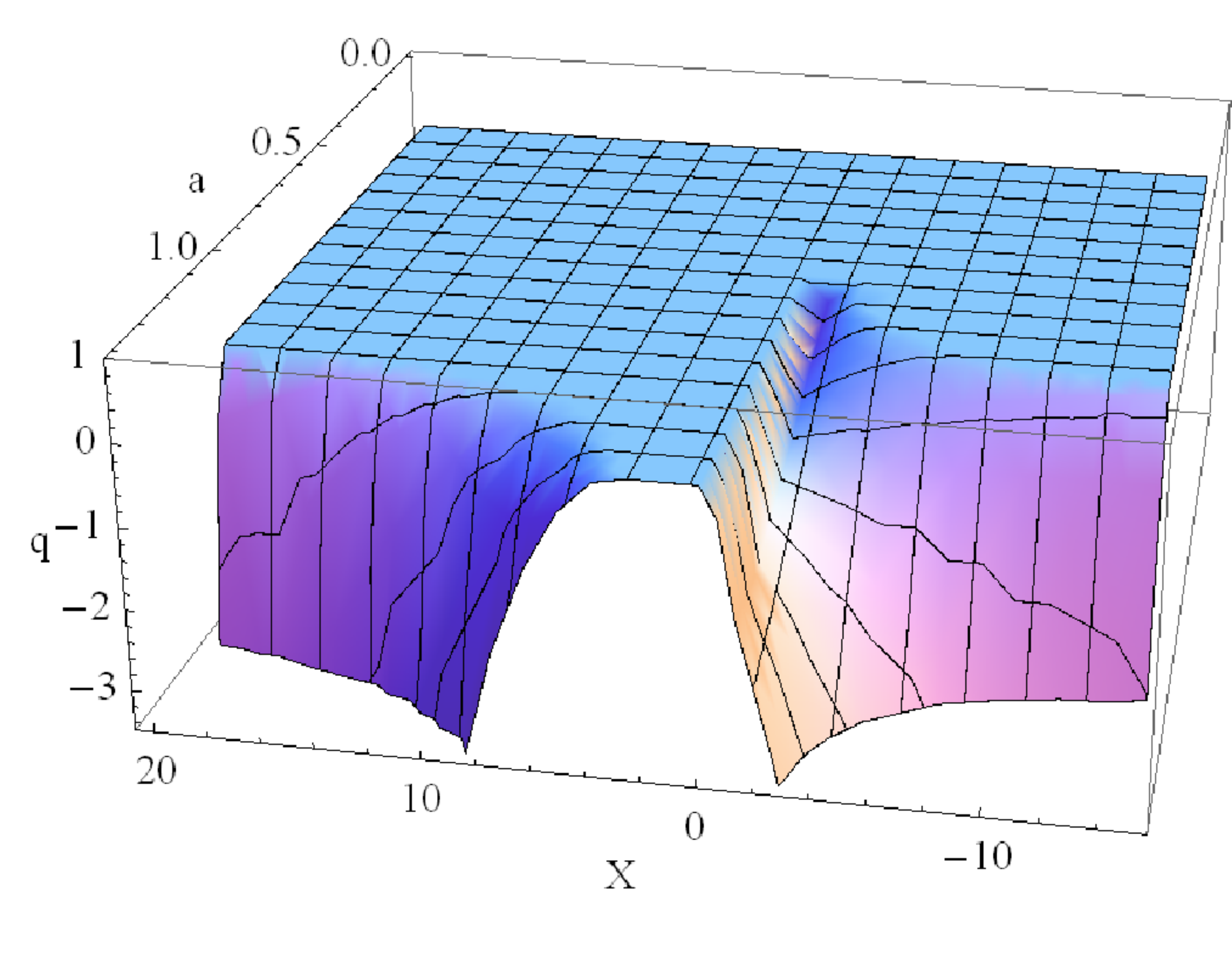}\\
   	(b)
   \end{tabular}
   \caption{Parameter spaces $(q-X-y)$ for $a=1$ \textbf{(a)}, and $(q-X-a)$ for $y=1$ \textbf{(b)} divided by separating surface in two sub-spaces corresponding to allowed (above) and forbidden (beneath) values of kinetic constants $q, X.$ }
   \label{Figure 4} \end{figure}

\section{Radial motion}

 From the equation (\ref{CarterR}) it is clear that the radial motion can exist if $R(r)\geq 0,$ where the equality gives the turning points of the radial motion. This condition can be rewritten in terms of an `effective potential` $X_\pm$ in the form
 \be
 X\leq X_-\quad \mathrm{or}\quad X\geq X_+, \label{realityCR1}
 \ee
 if
 \[
 a^2 - \Delta_r > 0\quad (\mathrm{and}\quad X_- < X_+), 
 \]
 or
 \be
 X_+\leq X \leq X_-\quad \mathrm{if}\quad a^2 - \Delta_r < 0, \label{realityCR2}
 \ee
 where
 \be
 X_\pm(r;q,\:y,\:a)\equiv \frac{ar^2\pm\sqrt{\Delta_r\left[r^4+q(a^2-\Delta_r)\right]}}{a^2-\Delta_r}.\label{Xpm} \ee \par

 We start the analysis by determining the reality region of the effective potential $X_\pm.$ From the expression (\ref{Xpm}) it follows that this function is well defined for
 \be q \left\{
  \begin{array}{l} \leq q_{r}(r;\:y,\:a^2)\quad \mbox{if} \quad a^2 < \Delta_r \quad \mbox{or}\quad \Delta_r \leq 0 \\
  \geq  q_{r}(r;\:y,\:a^2)\quad \mbox{if} \quad 0\leq \Delta_r < a^2 \label{realityX}
  \end{array}\right.,\ee
where we have introduced the reality function
\be
q_{r}(r;\:y,\:a^2)\equiv \frac{r^4}{\Delta_r-a^2}= \frac{r^3}{r-2-yr(r^2+a^2)}.
\ee
There are thus two different types of the boundary points of the definition range of $X_{\pm}(r;\:q,\:a,\:y).$ The points of the first type lie stably for given spacetime parameters on the borders of the static regions determined by the relation \ref{event.hor.},  i. e. at the event horizons $(r=r_h).$ At these horizons, if they exist, for arbitrary parameter $q,$ the functions $X_{\pm}$ have common values \be X_+(r_h) = X_-(r_h) = \frac{r_h^2}{a} \label{Xpm(rh)} \ee (c. f.  \cite{Stu-Hle:2000:CLAQG:}). The points of the second type, which are also common points of $X_{\pm},$ depend on the value of parameter $q$ and are given by the equality $q = q_{r}(r;\:y,\:a^2).$ If we denote them $r = r_q$ then it holds \be X_+(r_q) = X_-(r_q) = -\frac{a q}{r_q^2}.\ee

The divergencies of the function $q_{r}(r;\:y,\:a^2)$, which are incident with divergent points of $X_{+}(r;\:q,\:y,\:a),$ are located at radii where
\be
\Delta_r=a^2, \label{div.q,X}
\ee 
which one can express by the relation
\be
y=y_d(r;\:a^2)\equiv \frac{r-2}{r(r^2+a^2)}. \label{yd(r)}
\ee
The function $X_{-}(r;\:q,\:y,\:a)$ can not diverge at radii $r_d$ given by (\ref{div.q,X}), since using an alternative expression
\be
X_{\pm}(r;\:q,\:y,\:a)\equiv \frac{r^4-q\Delta_{r}}{ar^2\mp \sqrt{\Delta_{r}[r^4+q(a^2-\Delta_{r})]}}
\ee
it can be shown that it has finite value 
\be
X_{-}(r_{d};\:q,\:y,\:a)=\frac{r_{d}^4-qa^2}{2ar_{d}^2}.
\ee
 Another point where the functions $X_{\pm}(r;\:q,\:y,\:a)$ diverge is $r=0,$ with $X_{\pm}(r;\:q,\:y,\:a)\to \pm \infty$ as $r\to 0$ for $q>0,$ but  for $q=0$ it holds $X_{\pm}(r;\:q=0,\:y,\:a)\to 0.$\\
The character of the function $y_d(r;\:a^2)$ has been discussed in \cite{Stu-Ba_Ost:1998:KNdSrest.rep.bar.epm, Stu-Hle:2000:CLAQG:}, therefore we briefly repeat that the only zero of $y_{d}(r;\:a^2)$ is at $r=2,$ the extrema, which for $a^2>0$ must be maxima, yields the relation
\be
a^2=a^2_{max(d)}(r)\equiv r^2(r-3).\label{aex(d)}
\ee.
\\ The only zero of $q_{r}(r;\:y,\:a^2)$ is at $r = 0.$ For $r \to \infty$ it holds $q_{r}(r;\:y,\:a) \to -1/y.$ Its extrema are determined by
\be
y = y_{ex(r)}(r; \:a^2) \equiv \frac{r-3}{a^2r}.
\ee
The divergency of $y_{ex(r)}(r; \:a^2)$ is at $r = 0$ and $y_{ex(r)}(r; \:a^2)\to -\infty$ for $r\to 0.$ For $r \to \infty$ it approaches the line $1/a^2$ from bellow. The zero is at $r = 3$ and its extrema do not exist, the function is purely increasing.\par
Now we shall specify the local extrema of the effective potential, which determine the radii of spherical photon orbits. They are given by the condition $\partial X_{\pm}/\partial r = 0,$ which implies
\be
r^4+a^2q=0,
\ee
or
\bea
&&qa^2[2yr^3+(ya^2-1)r+1]^2+\\ \nonumber
&&r^3[y^2a^4r^3+2ya^2r^2(r+3)+ r(r-3)^2-4a^2]=0.
\eea

This can be rewritten in terms of parameter $q$ as
\be
q=q_{ex1}(r;\:a^2)\equiv -\frac{r^4}{a^2}, \label{qex1}
\ee
and
\bea
q&=&q_{ex}(r;\:y,\:a^2)\label{qex2} \\ \nonumber 
&\equiv& -\frac{r^3}{a^2}\; \frac{y^2a^4r^3+2ya^2r^2(r+3)+r(r-3)^2-4a^2}{[2yr^3+(ya^2-1)r+1]^2}.  
\eea
 Note that the function $q_{ex1}(r;\:a^2)$ is independent of  the cosmological parameter. Both the functions $q_{ex1}(r;\:a^2),$ $q_{ex}(r;\:y,\:a^2)$ have common points determined by
\be
y=\frac{r^2-2r+a^2}{r^2(r^2+a^2)}=y_{h}(r;\:a^2),
\ee
and
\be
y=\frac{1}{r^3},
\ee
i. e., they are located at event horizons and so called static radius $r_{s}=1/\sqrt[3]{y}$, where the gravitational attraction is just compensated by cosmological repulsion \cite{Stu:1983:BULAI:, Stu-Hle:1999:PHYSR4:}. The function $q_{ex1}(r;\:a^2)$ is negative valued and hence, as we shall see bellow, the extrema of the potentials $X_\pm$ determined by this function lie in regions forbidden by conditions for the reality of latitudinal motion.\\
The divergencies of $q_{ex}(r;\:y,\:a^2)$ are determined by the relation
\be
y=y_{d(ex)}(r;\:a^2)\equiv \frac{r-1}{r(2r^2+a^2)},\ee
its asymptotic behaviour is given by $q_{ex}(r;\:y,\:a^2)\to -(I/2ay)^2$ as $r\to \infty.$\\
The function $y_{d(ex)}(r;\:a^2)$ diverges for $r=0$ and $y_{d(ex)}(r;\:a^2)\to -\infty$ as $r\to 0.$ For $r\to \infty$ it holds $y_{d(ex)}(r;\:a^2)\to 0.$ The zero of this function is at $r=1$ and its local extrema are determined by the relation
\be
a^2=a^2_{max(d(ex))}(r)\equiv 2r^2(2r-3),\ee
where the label 'max' indicates that at relevant range $r\geq 3/2,$ these extrema must be maxima.\\
The zero point of the function $q_{ex}(r;\:y,\:a^2)$ is at $r=0,$ another zeros determine the loci of the circular equatorial photon orbits. They are given by the relation
\be
y=y_{z(ex)\pm}(r;\:a^2)\equiv \frac{-r(r+3)\pm 2\sqrt{r(3r^2+a^2)}}{a^2r^2}.\ee
Since the function $y_{z(ex)-}(r;\:a^2) < 0$ for $r>0,$ it is irrelevant in our discussion. The function $y_{z(ex)+}(r;\:a^2)$ is real valued for all $r>0$ and it diverges at $r=0,$ with $y_{z(ex)+}(r;\:a^2)\to \infty$ as $r\to 0.$ For $r\to \infty,$ we find $y_{z(ex)+}(r;\:a^2)\to -1/a^2.$ Its zeros represent the equatorial circular photon orbits in the Kerr spacetimes, being determined by the relation \cite{Stu:1981:BAIC:Rad.mot.ph.Kerr}
\be
a^2=a^2_{z(z(ex)+)}(r)\equiv \frac{r(r-3)^2}{4}.\ee
The extrema of the function $y_{z(ex)+}(r;\:a^2)$ are determined by the equation
\be
a^2=a^2_{ex(z(ex)+)\pm}(r)\equiv \frac{r(1-2r\pm\sqrt{1+8r})}{2}=a^2_{ex(h)\pm}(r),\ee
hence the loci of extrema of the functions $y_{z(ex)+}(r;\:a^2)$ and $y_{h}(r;\:a^2)$ coalesce.\\ 
The function $a^2_{ex(z(ex)+)-}(r)$ should be excluded from further analysis since for $r>0$ there is $a^2_{ex(z(ex)+)-}(r)<0.$ \par
It remains to determine loci of the local extrema of the function $q_{ex}(r;\:y,\:a^2).$ Proceeding the usual way we find that their occurrence is governed by the relations
\be
y=y_{ex(ex)}(r;\:a^2)\equiv \frac{r-3}{a^2r}=y_{ex(r)}(r; \:a^2) \label{yex(ex)_yex(r)}
\ee
and
\[y=y_{ex(ex)\pm}(r;\:a^2)\equiv \]
\be
\frac{3r^2\sqrt{r}-a^2\sqrt{r}(3+2r)\pm \sqrt{(4a^2-3r)(a^4+6a^2r^2-3r^4)}}{2a^4\sqrt{r^3}}.
\ee
Using the relation (\ref{yex(ex)_yex(r)}), one can show that the extrema of both functions $q_{ex}(r;\:y,\:a^2),$ $q_{r}(r;\:y,\:a^2)$ coalesce. At this point let us add that another common points of functions  $q_{ex}(r;\:y,\:a^2),$ $q_{r}(r;\:y,\:a^2)$, as well as of the functions $q_{ex1}(r;\:y,\:a^2),$ $q_{r}(r;\:y,\:a^2),$ are also given by
\be
y=y_{h}(r;\:a^2),
\ee
i. e. they are located at the event horizons.\\
The reality conditions of the functions $y_{ex(ex)\pm}(r;\:a^2)$ read
\be
a^2\leq a^2_{r(ex(ex\pm))+}(r)\quad \mbox{or}\quad a^2_{r(ex(ex\pm))}(r)\leq a^2, \label{reality-yexexpm1}
\ee
\[\mbox{if}\quad 0<r\leq \hat{r},\]
and
\be
a^2\leq a^2_{r(ex(ex\pm))}(r)\quad \mbox{or}\quad a^2_{r(ex(ex\pm))+}(r)\leq a^2,
\ee
\[\mbox{if}\quad \hat{r}\leq r,\]
where
\be a^2_{r(ex(ex)\pm)+}(r)\equiv (+2\sqrt{3}-3)r^2\ee
and
\be a^2_{r(ex(ex)\pm)}(r)\equiv \frac{3}{4}r. \ee
The marginal radius $\hat{r}$ has value
\be \hat{r}=\frac{2\sqrt{3}+3}{4}=1.61603=r_{crit}\ee
and it holds
\be a^2_{r(ex(ex)\pm)+}(\hat{r})=a^2_{r(ex(ex)\pm)}(\hat{r})=a^2_{crit}=1.21202, \label{reality-yexexpm2} \ee
where $a^2_{crit}$ corresponds to local maximum of the function $a^2_{ex(h)+}(r)$ (see e. g. \cite{Stu-Hle:2000:CLAQG:} for details).\\
The functions $y_{ex(ex)\pm}(r;\:a^2)$ have the divergency point at $r=0$ and $y_{ex(ex)\pm}(r;\:a^2)\to \pm\infty$ for $r\to 0.$ For $r\to \infty$ we find that $y_{ex(ex)+}(r;\:a^2)\to \infty$ and $y_{ex(ex)-}(r;\:a^2)\to 0$ from above.\\
The zero point of $y_{ex(ex)}(r;\:a^2)$ is at $r=3$ and the function is increasing for all $r>0.$\\
Zeros of the functions $y_{ex(ex)\pm}(r;\:a^2)$ are given by
\be
a^2=a^2_{z(ex(ex)\pm)}(r)\equiv r(r^2-3r+3).\ee
The condition for stationary points $$\partial y_{ex(ex)\pm}(r;\:a^2)/\partial r=0$$ leads to
$$ a^4+a^2r(2r-1)+r^3(r-3)=0,$$
which can be solved with respect to $a^2$ with the same result as given by (\ref{aex(h)}). However, substitution into the second derivative concurrently with the requirement $y_{ex(ex)\pm}(r;\:a^2)>0$ implies
\be
\partial^2 y_{ex(ex)\pm}(r;\:a^2=a^2_{ex(h)+}(r))=0,
\ee
therefore the function
\be
a^2_{inf(ex(ex)\pm)+}(r)\equiv a^2_{ex(h)+}(r)
\ee
determines the loci of the inflex points of the functions $y_{ex(ex)\pm}(r;\:a^2).$\\
If we compare the asymptotic behaviour of all characteristic functions $y(r;\:a^2),$ we find that following inequality is satisfied:\\
$1/a^2>y_{ex(ex)}(r;\:a^2)>y_{ex(ex)-}(r;\:a^2)>y_h(r;\:a^2)>y_d(r;\:a^2)>y_{d(ex)}(r;\:a^2)$
$>0>y_{z(ex)}(r;\:a^2)>-1/a^2$ as $r\to \infty.$ \\
In  Fig.5 we present all the characteristic functions related to spin parameter governing the effective potential on the lowest level:
\begin{itemize}
  \item $a^2_{z(h)}(r)$
  \item $a^2_{ex(h)+}(r)$=$a^2_{ex(z(ex))+}(r)$=$a^2_{inf(ex(ex)\pm)+}(r)$
  \item $a^2_{max(d)}(r)$
  \item $a^2_{max(d(ex))}(r)$
  \item $a^2_{z(z(ex)+)}(r)$
  \item $a^2_{r(ex(ex)\pm)+}(r)$
  \item $a^2_{r(ex(ex)\pm)}(r)$
  \item $a^2_{z(ex(ex)\pm)}(r).$
\end{itemize}
These functions determine the behaviour of the characteristic functions related to the cosmological parameter, characterizing the functions $q(r;y,a^2)$ and then effective potentials on the higher level:
\begin{itemize}
  \item $y_{h}(r;\:a^2)$
  \item $y_{d}(r;\:a^2)$
  \item $y_{ex(r)}(r;\:a^2)=y_{ex(ex)}(r;\:a^2)$
  \item $y_{d(ex)}(r;\:a^2)$
  \item $y_{z(ex)+}(r;\:a^2)$
  \item $y_{ex(ex)\pm}(r;\:a^2).$
\end{itemize}

\begin{figure}[H]
	\centering
	\includegraphics[width=0.45\textwidth]{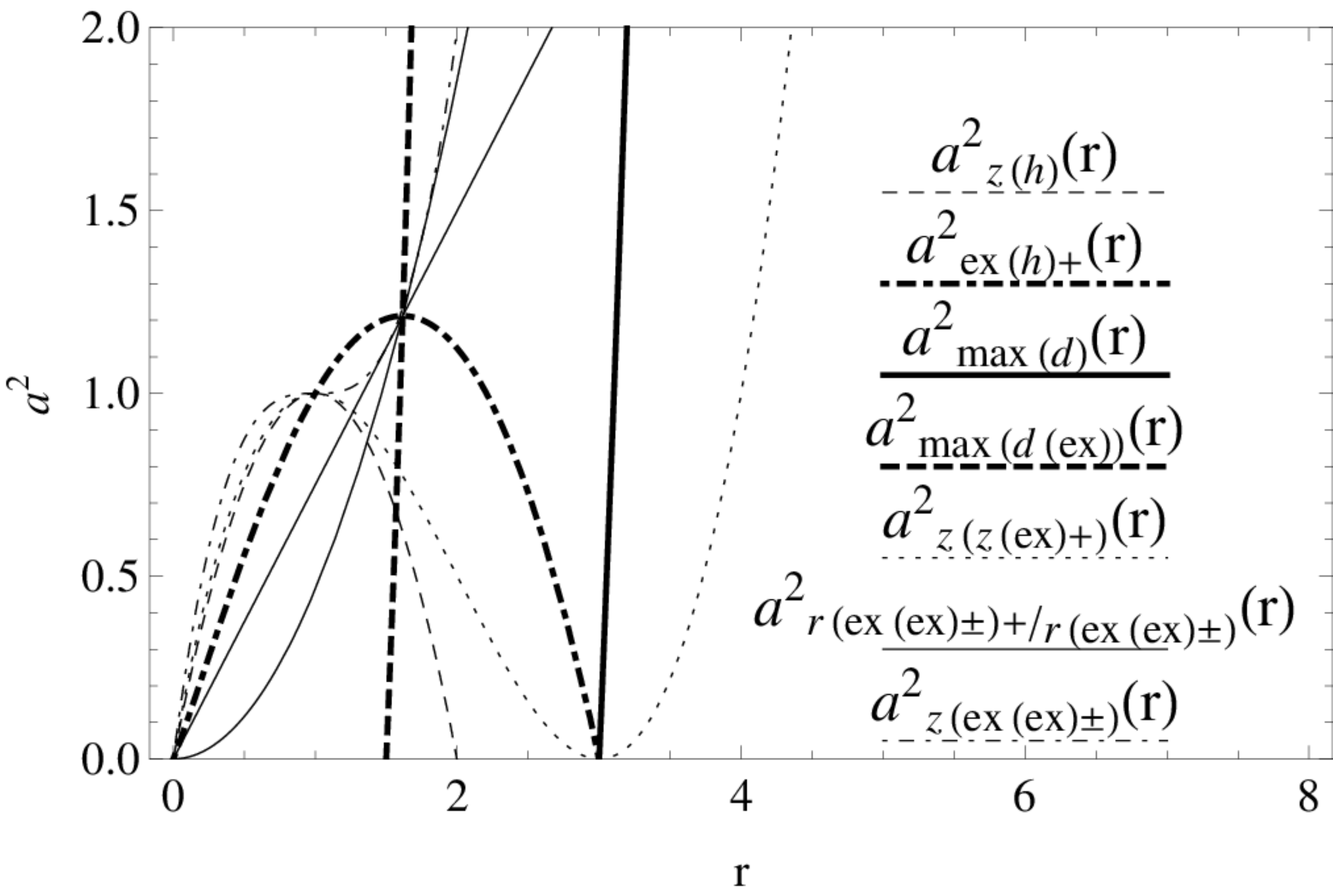}\\
	\includegraphics[width=0.45\textwidth]{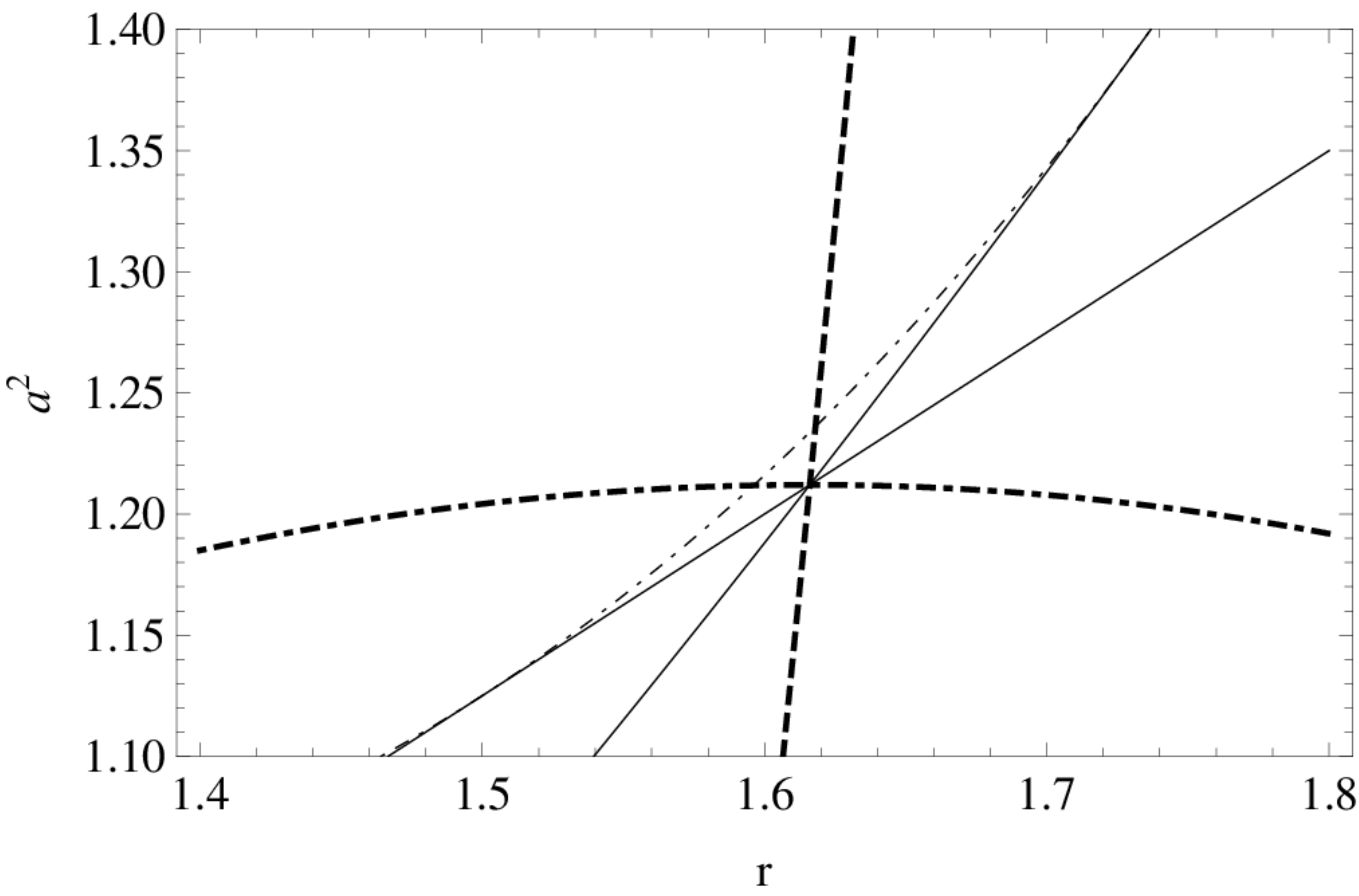}
	\caption{Characteristic functions $a^2(r)$ determining the behaviour of functions the $y(r;\:a^2).$ }\label{Figure 5}
\end{figure}
From the significance of the individual characteristic functions $a^2(r)$ depicted in Fig. 5, one can infer that there are just two values of $a^2$ being of particular importance and leading to qualitatively different behaviour of the functions $y(r;\:a^2):$
\begin{description}
	\item[$a^2=1$]-- the common local maximum of the functions $a^2_{z(h)}(r)$ and $a^2_{z(z(ex))}(r)$ at $r=1,$ which coincides with the inflection point of the function $a^2_{z(ex(ex)\pm)}(r;\:a^2)$ and with the intersection with the curve $a^2_{ex(h)+}(r)$
	\item[$a^2=a^2_{crit}=1.21202$]-- the local maximum $a^2_{ex(h)+}(r)$ which is the intersection of the curves $a^2_{r(ex(ex)\pm)+}(r)$, $a^2_{r(ex(ex)\pm)}(r)$ and $a^2_{max(d(ex))}(r).$
\end{description}

The graphs of characteristic functions $y(r;\:a^2)$ depicted for some values of spin parameter $a$ representing cases $0<a^2<1,$ $1<a^2<a^2_{crit}$ and $a^2_{crit}<a^2$ are presented in Fig. 6.
\begin{figure*}[htbp]
	\centering
	\begin{tabular}{cc}
		\includegraphics[width=0.45\textwidth]{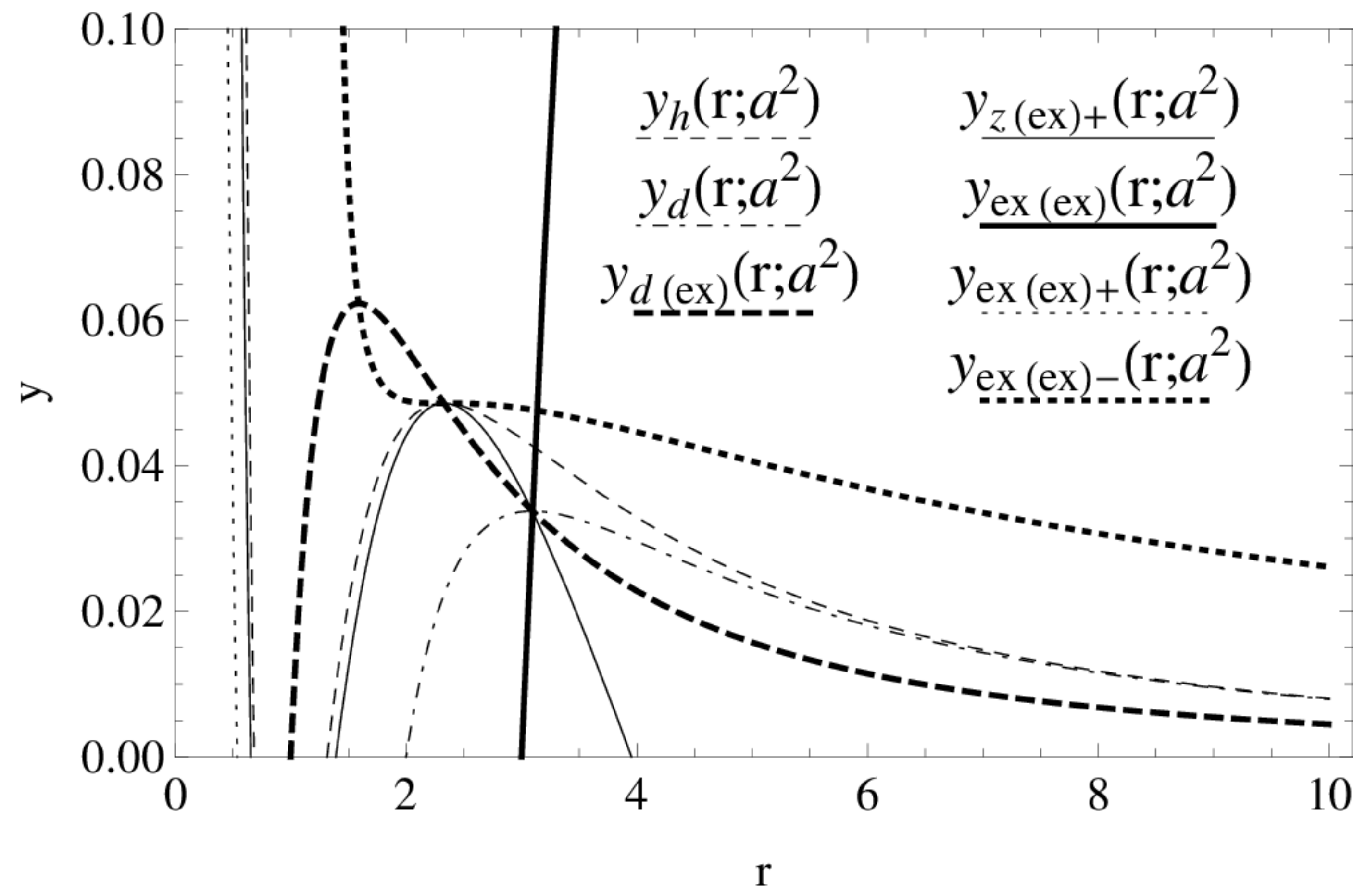} & \includegraphics[width=0.45\textwidth]{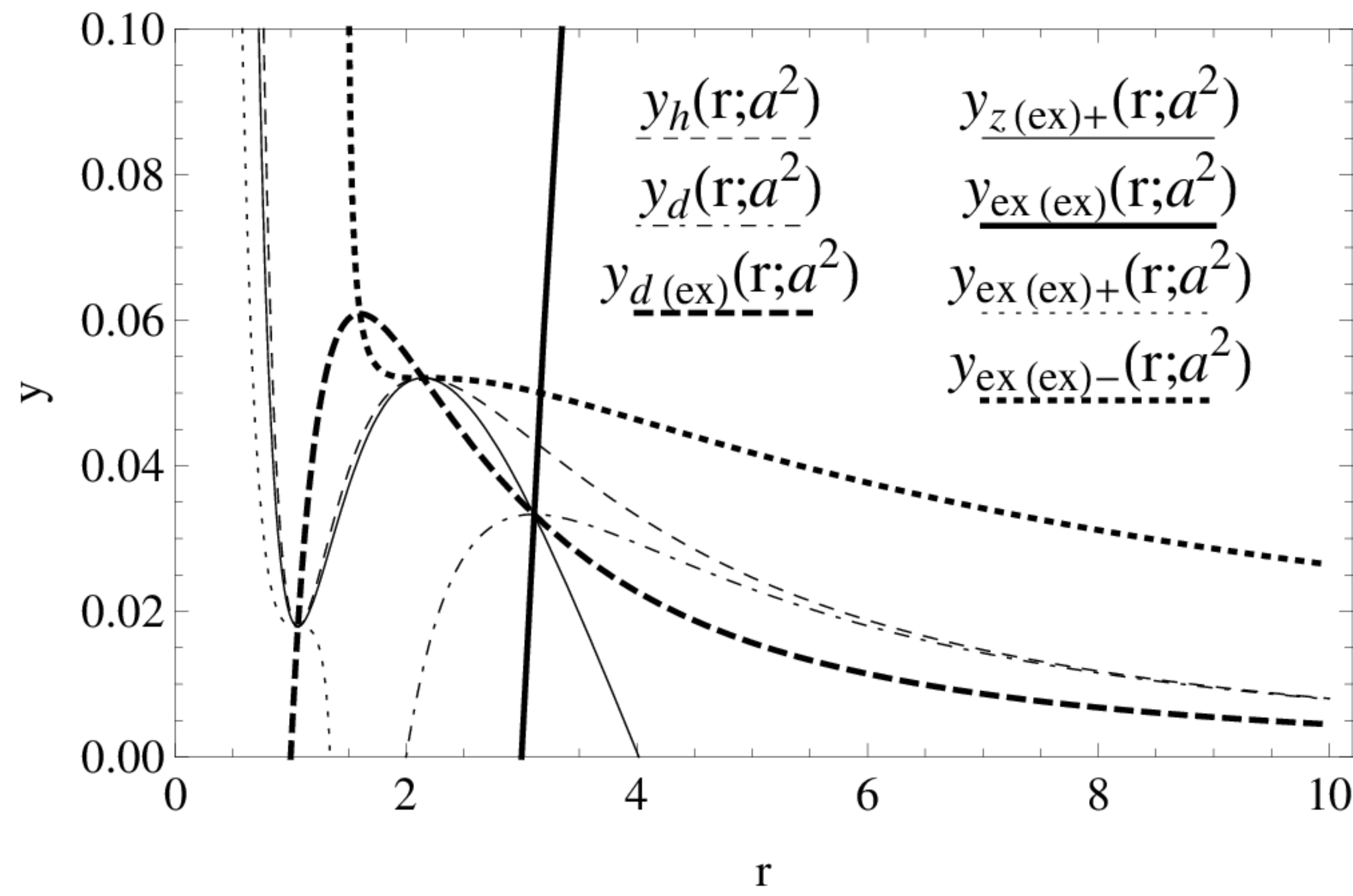}\\
		(a) & (b)\\
		\includegraphics[width=0.45\textwidth]{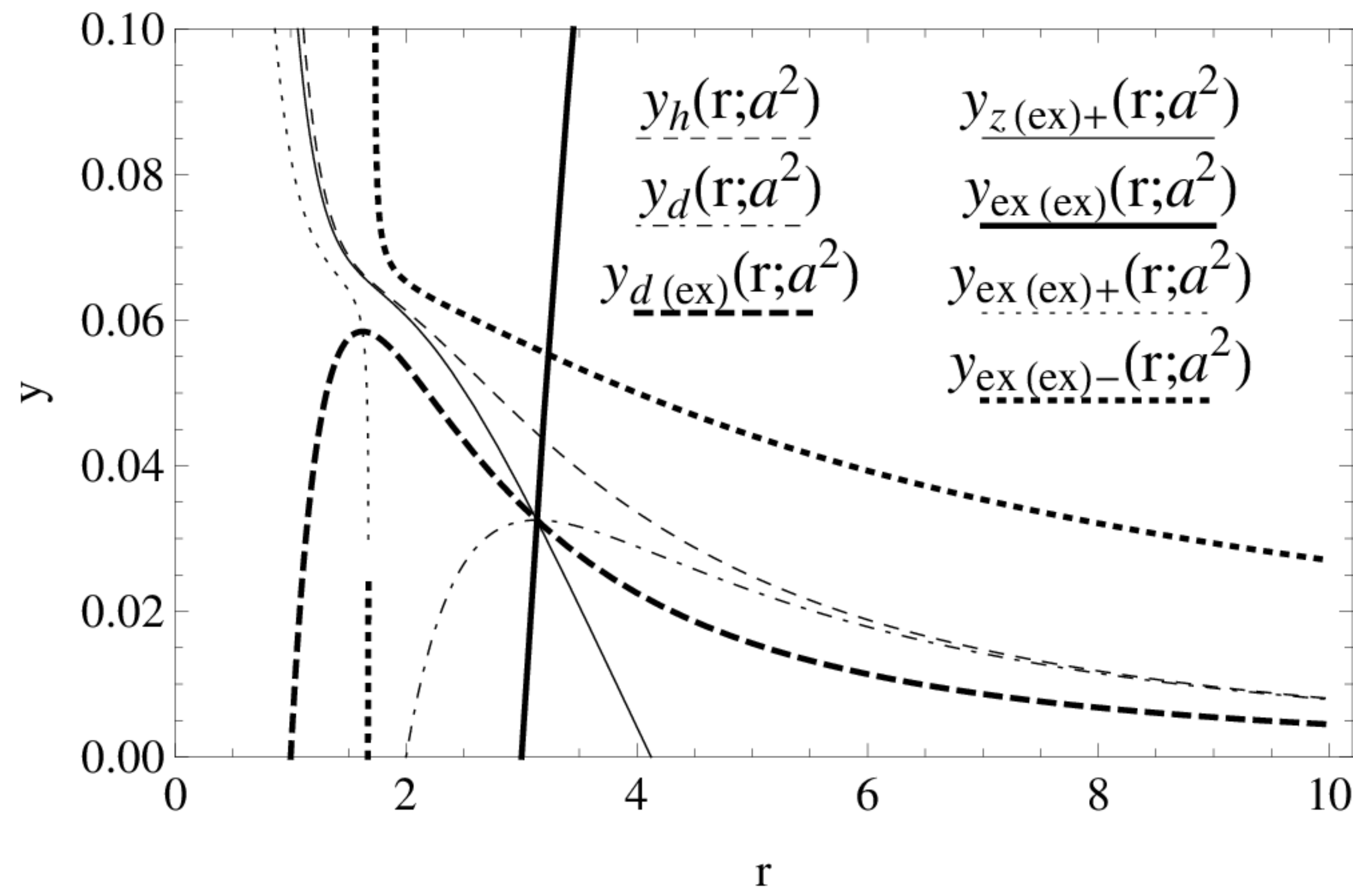} & \includegraphics[width=0.45\textwidth]{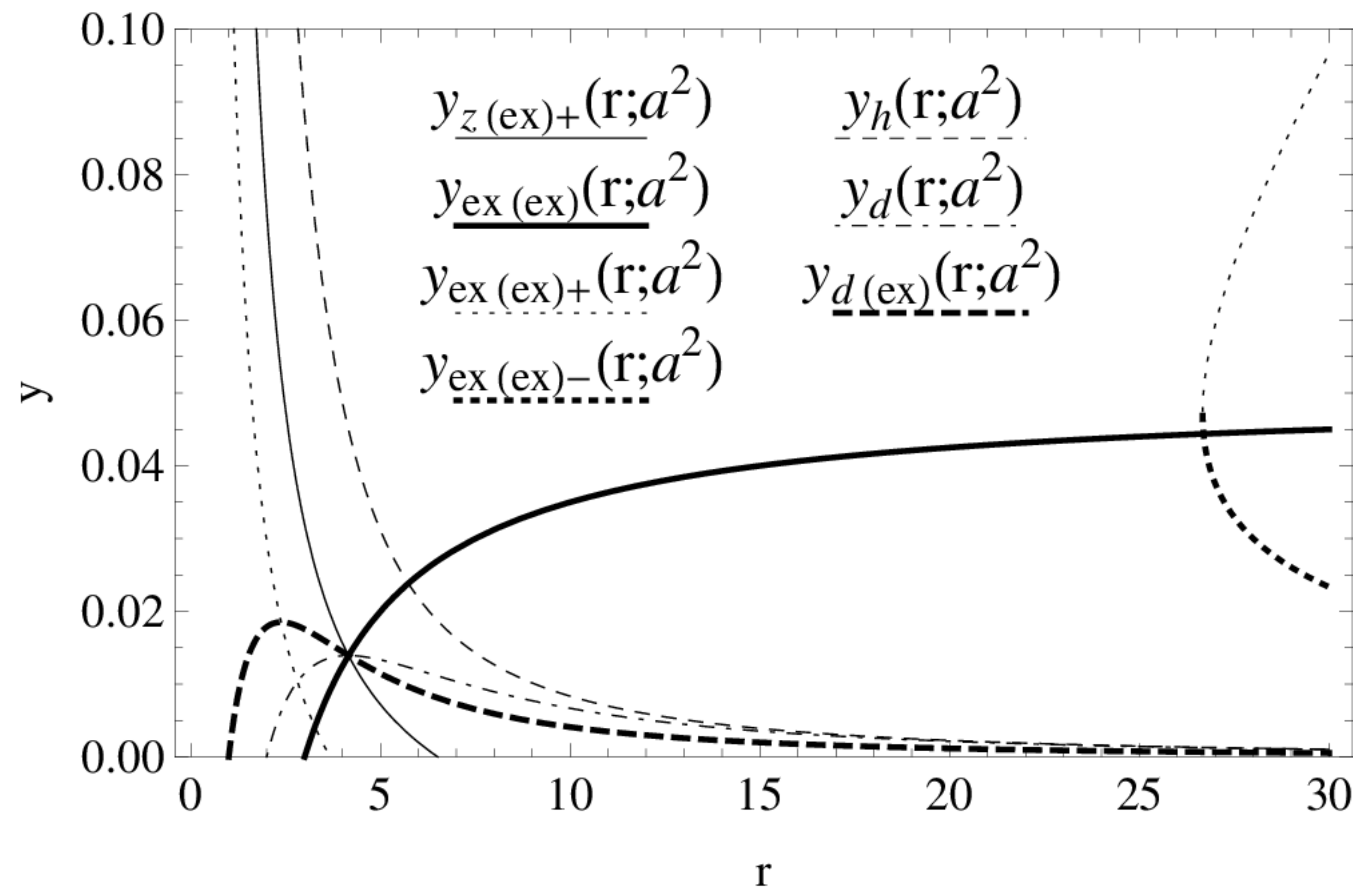}\\
		(c) & (d)
		
	\end{tabular}
	\caption{Characteristic functions $y(r;\:a^2)$ given for $a^2=0.9$ \textbf{(a)}, $a^2=1.04$ \textbf{(b)}, $a^2=1.3$ \textbf{(c)} and $a^2=20$ \textbf{(d)}. }\label{Figure 6}
\end{figure*} 

In general, behaviour of the characteristic functions $q_{r}(r;\:y,\:a^2)$ and $q_{ex}(r;\:y,\:a^2)$ will be qualitatively different, if for the parameter $a$ being fixed we take the $y$- values from different intervals, which are limited by intersections and/or extrema of the characteristic functions $y(r;\:a^2)$ that are demonstrated in Fig. 6. We therefore need to determine curves $y(a^2)$ that separate the $(a^2\mbox{-}y)$-plane into regions that correspond to that different behaviour of the characteristic functions $q_{r}(r;\:y,\:a^2)$ and $q_{ex}(r;\:y,\:a^2)$. The number of these functions is substantially lowered by the fact that all the local extrema are multiple intersections with other curves and coincide with other extrema. Moreover, as explained bellow, the behaviour of the characteristic functions $q_{r}(r;\:y,\:a^2)$ and $q_{ex}(r;\:y,\:a^2)$ in their negative values we can omit as irrelevant for the character of the photon motion. The functions we need are the following:
\begin{itemize}
	\item $y_{max(h)}(a^2)=y_{max(z(ex)+)}(a^2)=y_{inf(ex(ex)-)}(a^2)=
	y_{d(ex)\mbox{-}h\mbox{-}(z(ex)+)\mbox{-}ex(ex)-}(a^2)$
	\item $y_{min(h)}(a^2)=y_{min(z(ex)+)}(a^2)=y_{inf(ex(ex)+)}(a^2)=
	y_{d(ex)\mbox{-}h\mbox{-}(z(ex)+)\mbox{-}ex(ex)+}(a^2)$
	\item $y_{max(d)}(a^2)=y_{d\mbox{-}d(ex)\mbox{-}(z(ex)+)\mbox{-}ex(ex)}(a^2)$
	\item $y_{d\mbox{-}(z(ex)+)}(a^2)$
	\item $y_{max(d(ex))}(a^2)=y_{d(ex)\mbox{-}ex(ex)-}(a^2)$
	\item $y_{ex(ex)\mbox{-}(ex(ex)+)}(a^2)$
\end{itemize}

Here the dashes between two labels denote affiliation to intersection of appropriate functions (it can be proved that there are no other intersections of these functions than that shown in Fig.6). These functions are projections of extremal values or intersections of characteristic functions $y(r;\:a^2)$ into $(a^2\mbox{-}y)$-plane and they are demonstrated in Fig. 7.
\begin{figure*}[htbp]
	\centering
	\begin{tabular}{c}
		\includegraphics[width=0.7\textwidth]{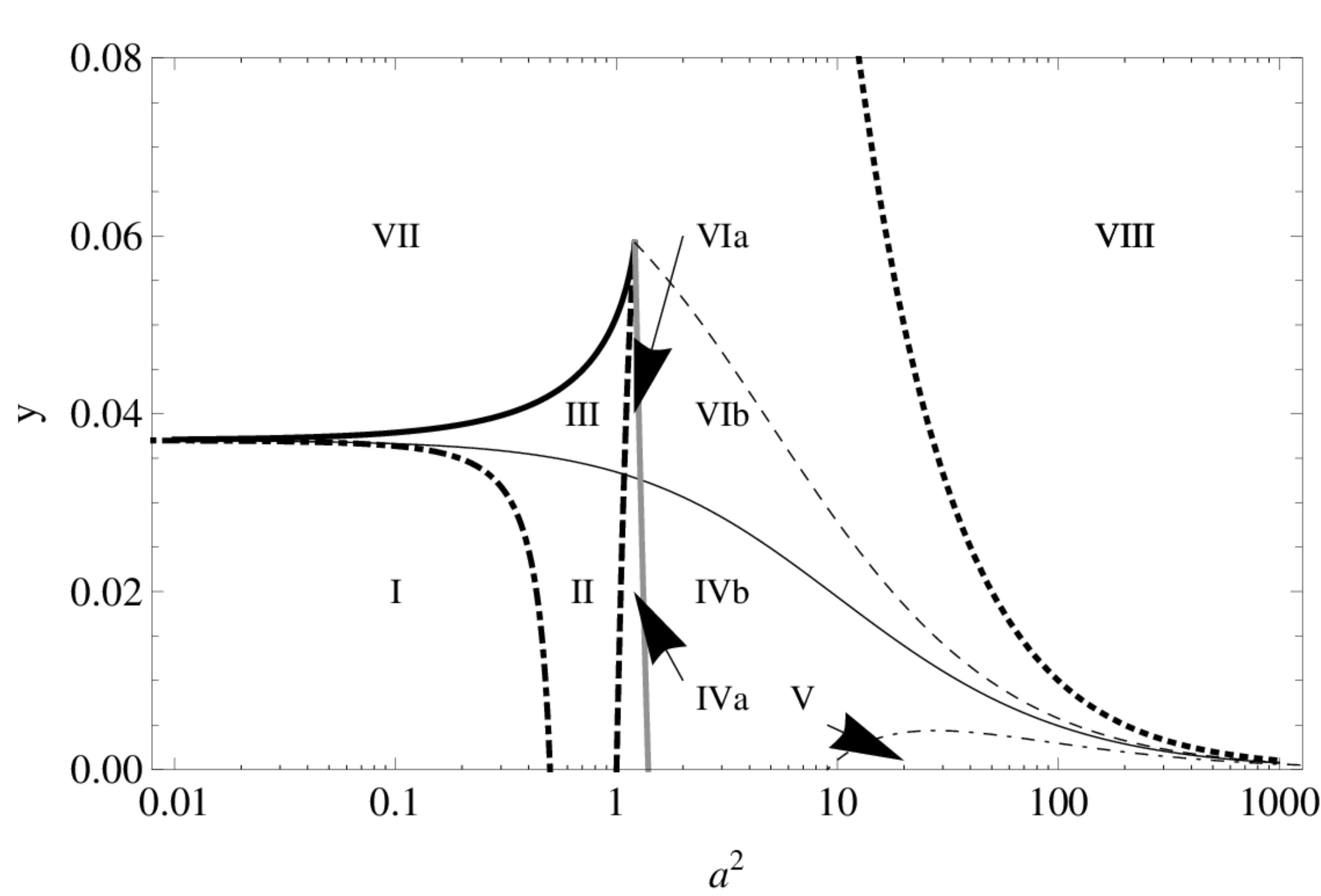}
	\end{tabular}
	\caption{The functions $y_{max(h)}(a^2)$ (bold full curve), $y_{min(h)}(a^2)$ (bold dashed curve), $y_{d\mbox{-}z(ex)+}(a^2)$ (bold dash-dotted curve), $y_{max(d)}(a^2)$ (full curve), $y_{max(d(ex))}(a^2)$ (dashed curve),  $y_{ex(ex)\mbox{-}ex(ex)+}(a^2)$ (dash-dotted) and $y=1/a^2$ (bold dotted curve). In order to clearly display the asymptotic behaviour of these functions we give their plots with the $a^2$-axis in logarithmic scale. These curves represent such qualitative changes in behaviour of the characteristic functions $q_{r}(r;\:y,\:a^2),$ $q_{ex}(r;\:y,\:a^2),$ which are resulting in different character of the photon motion, and divide the $(a^2-y)$-plane into regions distinguished by different Roman numerals. The bold grey line is function  $y_{(ex(ex)+)\mbox{-}(ex(ex)-)2}(a^2)$ that separates spacetimes I-III, IVa, VIa endowed with both prograde and retrograde spherical photon orbits, as seen by family of the locally non-rotating observers (see bellow), that are separated by 'polar' spherical photon orbit, from the spacetimes IVb, V, VIb,VII, VIII possessing just retrograde spherical orbits. The function $y=1/a^2$ represents additional division of parameter plane reflecting different character of the latitudinal motion as discussed in previous section.}\label{Figure 7}
\end{figure*}

The functions $y_{ex(h)}(a^2)$ divide the parameter plane $(a^2\mbox{-}y)$ into regions describing \KdS\ black hole and naked singularity spacetimes, the curve $y_{max(d)}(a^2)$ divides spacetimes with so called divergent and restricted repulsive barrier of photon motion. A detailed discussion of these functions have been performed e. g. in \cite{Stu-Hle:2000:CLAQG:,Stu-Sla:2004:PHYSR4:} and will not be repeated here. The significance of the remaining functions can be understood from the depiction of the characteristic functions $q_{r}(r;\:y,\:a^2),$ $q_{ex}(r;\:y,\:a^2)$ in Fig 8. They are given parametrically by appropriate functions $a^2(r),$ $y(r;\:a^2(r))$ with $r$ being the parameter:
\begin{itemize}
	\item the functions $y_{max(h)}(a^2)$ and  $y_{min(h)}(a^2)$ are both determined by $a^2_{ex(h)+}(r)$ and $y_h(r;\:a^2=a^2_{ex(h)+}(r));$ 
	\item $y_{max(d)}(a^2)$ we obtain from $a^2_{max(d)}(r)$ with $y_d(r;\:a^2=a^2_{max(d)}(r));$ 
	\item the curve $y_{max(d(ex))}(a^2)$ is given by functions $a^2_{max(d(ex))}(r)$ and $y_{d(ex)}(r;\:a^2=a^2_{max(d(ex))}(r));$
	\item  $y_{d\mbox{-}z(ex)+}(a^2)$ is determined by 
	\[a^2_{d\mbox{-}z(ex)+}(r)\equiv \frac{r}{8}(1-4r+\sqrt{40r+1})\]
	and $y_d(r;\:a^2=a^2_{d\mbox{-}z(ex)+}(r)),$ where the function $a^2_{d\mbox{-}z(ex)+}(r)$ is a solution of $ y_{d}(r;\:a^2)=y_{z(ex)+}(r;\:a^2) $ with respect to parameter $a^2;$ all such functions are obtained by analogous manner;	
	\item $y_{ex(ex)\mbox{-}ex(ex)+}(a^2)$ are constructed from 
	\[a^2_{ex(ex)\mbox{-} ex(ex)\pm}(r) \equiv \]
	\[\frac{r}{2}(4r^2-12r+3\pm \sqrt{16r^4-96r^3+156r^2-36r+9})\]
	and $ y_{ex(ex)}(r;\:a^2=a^2_{ex(ex)\mbox{-} ex(ex)+}(r));$ 
\end{itemize}

There exist another functions $y(a^2)$, corresponding to intersections of the characteristic functions $y(r;\:a^2),$ which are not displayed in Fig.7. The reason is that all the functions $y(a^2)$ lie under the curve $y=1/a^2,$ and thus we have to take into account the restriction $q\geq-a^2$ (see Section 1). Therefore, the changes of the characteristic functions $q_{r}(r;\:y,\:a^2),$ $q_{ex}(r;\:y,\:a^2)$  in values under this limit can be omitted as irrelevant. Moreover, we can easily show that in the case $q<0,$ the restrictions (\ref{qmin(X,y,a)}) imposed on the latitudinal motion yields stronger constraints on the value $X$ than that given by the relations (\ref{realityCR1}), (\ref{realityCR2}), (\ref{Xpm}) conditioning the reality of the radial motion. Indeed, for any triad $(q<0,y,a^2)$ there is no intersection of the curves $X=X_\pm(r;\:q,\:y,\:a)$  with the lines $X^\theta_{min(-)},$ $X^\theta_{max(+)},$ where $X^\theta_{min(-)},$ $X^\theta_{max(+)}$ are extrema of the functions $X^\theta(m;\:q,\:y,\:a)$ introduced in Sec.1 (see Fig. 9 e-g,$\omega$). To verify this, it is convenient to regard the curves $X=X_\pm(r;\:q,\:y,\:a)$ as $q=const$-slices of the surface $q=q_{max}(r;\:X,\:a,\:y),$ where
\be
q \leq q_{max}(r;\:X,\:a,\:y)\equiv \frac{(r^2-a X)^2}{\Delta_r}-X^2 \label{qr}
\ee
is an alternative expression of the reality condition $R(r;\:X,\:q,\:y,\:a)\geq0,$,  and search instead for intersections of surfaces $q=q_{max}(r;\:X,\:y,\:a)$ and $q=q_{min}(X,\:y,\:a),$ defined by relations (\ref{f_1(X)}) - (\ref{q_l_plane3}). \\
We therefore solve two equations - $q_{max}(r;\:X,\:y,\:a^2)=q_1(X),$ with the result
\be
X=\frac{r^2}{a}, \label{qr_q1}
\ee
and $q_{max}(r;X,\:a,\:y)=q_2(X;\:a,\:y),$ which gives
\be
X_{1,2}=\frac{I^2r^2+2(a^2y-1)\Delta_r\mp 2I\sqrt{-2r\Delta_r}}{a(I^2-4y\Delta_{r})}. \label{qr_q2}
\ee
The solution (\ref{qr_q1})  yields $X>0$, which, however, does not apply to the case $q<0$ for $y<1/a^2.$  Moreover, this solution represents touching points of the surface $q_r(r;\:X,\:y,\:a^2)$ with the parabolic surface $q_1(X)$ at $X=+\sqrt{-q},$ and hence can be omitted even in the case $y\geq 1/a^2,$ since theses values lie in the region forbidden by the relations (\ref{q_l_plane2})-(\ref{q_l_plane3}). The solutions (\ref{qr_q2}) are evidently irrelevant, since in stationary regions $\Delta_r>0$ they are imaginary.\\
The above analysis shows that in the case $q<0$ the 'potentials' $X_\pm(r;\:q,\:y,\:a)$ have values in regions forbidden by reality conditions of the latitudinal motion and hence play no role at all. The limits for impact parameter $X$ of photons with $q<0$ are thus given by relation (\ref{qmin(X,y,a)}); photons satisfying the relation (\ref{qmin(X,y,a)}) have thus no turning points of the radial motion. In the rest of this treatise we can thus focus on the behaviour of the characteristic functions for $q\geq 0.$ 
In Fig. 8 we present all possible variants of behaviour of the characteristic functions $q(r;\:y,\:a^2).$ These variants involve cases
\begin{description}
	\item[I:] $y \le y_{d\mbox{-}z(ex)+}(a^2)$ for $a^2\le 0.5$
	\item[II:] $y_{d\mbox{-}z(ex)+}(a^2)\le y\le y_{max(d)}(a^2)$ for $a^2\le 0.5,$\\
	or\\
	$y\le y_{max(d)}(a^2)$ for $0.5 \le a^2\le 1,$\\
	or\\
	$y_{min(h)}(a^2)\le y\le y_{max(d)}(a^2)$ for $1\le a^2\le 1.08316;$
	\item[III:] $y_{max(d)}(a^2)\le y \le y_{max(h)}(a^2)$ for $a^2\le 1.08316,$\\
	or\\
	$y_{min(h)}(a^2)\le y \le y_{max(h)}(a^2)$ for $1.08316\le a^2\le 1.21202=a^2_{crit};$ 
	
	\item[IVa:] $y\le y_{min(h)}(a^2)$ for $1\le a^2\le 1.08316,$\\
	or\\
	$y\le y_{max(d)}(a^2)$ for $1.08316\le a^2\le 1.28282,$\\
	or\\
	$y\le y_{(ex(ex)+)\mbox{-}(ex(ex)-)2}(a^2)$ for $1.28282\le a^2\le 6\sqrt{3}-9=1.3923;$
	
	\item[IVb:] $y_{(ex(ex)+)\mbox{-}(ex(ex)-)2}(a^2)\le y\le y_{max(d)}(a^2)$ for $1.28282\le a^2\le 1.3923,$\\
	or\\
	$y\le y_{max(d)}(a^2)$ for $1.3923\le a^2\le 9,$\\
	or\\
	$y_{ex(ex)\mbox{-}ex(ex)+}(a^2)\le y\le y_{max(d)}(a^2)$ for $a^2\ge 9;$
	\item[V:] $y\le y_{ex(ex)\mbox{-}ex(ex)+}(a^2)$ for $a^2\ge 9;$
	\item[VIa:] $y_{max(d)}(a^2)\le y\le y_{min(h)}(a^2)$ for $1.08316\le a^2\le 1.21202,$\\
	or\\
	$y_{max(d)}(a^2)\le y\le y_{(ex(ex)+)\mbox{-}(ex(ex)-)2}(a^2)$ for $1.21202\le a^2\le 1.28282;$
	
	\item[VIb:] $y_{(ex(ex)+)\mbox{-}(ex(ex)-)2}(a^2)\le y\le y_{max(d(ex))}(a^2)$ for $1.21202\le a^2\le 1.28282,$\\
	or\\
	$y_{max(d)}(a^2)\le y\le y_{max(d(ex))}(a^2)$ for $a^2\ge 1.28282;$
	
	\item[VII:] $y_{max(h)}(a^2)\le y\le 1/a^2$ for $a^2\le 1.21202,$
	or\\
	$y_{max(d(ex))}(a^2)\le y\le 1/a^2$ for $a^2\ge 1.21202;$
	\item[VIII:] $y\ge 1/a^2;$	     
\end{description}

Now it remains to assign to each region of the $(a^2\mbox{-}y)$-plane functions $q(y,\:a^2),$ which by themselves represent marginal values of the parameter $q$ corresponding to some qualitative shift in the behaviour of the potentials $X_\pm(r;\:q,\:y,\:a).$\par
In the regions I, II, which describe black hole spacetimes with divergent repulsive barrier, we have to compare the two local maxima $q_{max(ex+)}(y,\:a^2)$ located under the inner horizon and $q_{max(ex)}(y,\:a^2)=q_{min(r)}(y,\:a^2)$ between the outer and cosmological horizons respectively (see Figs. 8a,b). The function $q_{max(ex+)}(y,\:a^2)$ is given parametrically by functions $y_{ex(ex)+}(r;\:y,\:a^2)$ and $q_{ex}(r;\:y=y_{ex(ex)+}(r;\:y,\:a^2),\:a^2)$ with $r$ being the parameter, similarly $q_{max(ex)}(y,\:a^2)$ is given by $y_{ex(ex)}(r;\:y,\:a^2)$ and $q_{ex}(r;\:y=y_{ex(ex)}(r;\:y,\:a^2),\:a^2).$ \\
The extrema function $q_{max(ex)}(y,\:a^2)$ diverges at the curve $y_{max(d)}(a^2)$, which forms boundary between regions II-III and IV--VI, i. e. $q_{max(ex)}(y=y_{max(d)}(a^2),\:a^2) \to +\infty$ (c. f. Figs. 8b, 8c and 8d, 8f). In the region III, corresponding to black hole spacetimes with the restricted repulsive barrier, only local maximum $q_{max(ex+)}(y,\:a^2)$ located under the inner horizon remains.\\
As can be seen from the behaviour of the characteristic functions in Fig. 6b, for $y \to y_{min(h)}(a^2)$ from above, the 'inner' local maximum of $q_{(ex)}(r;\:y,\:a^2),$ determined by  $y_{ex(ex)+}(r;\:y,\:a^2),$ approaches from the left its divergency point given by $y=y_{d(ex)}(r;\:a^2),$ where $q_{ex}\to -\infty$ (Fig. 8b), so that $q_{ex}(r;\: y=y_{min(h)}(a^2),\:a^2)$ becomes continuous.  For $y \le y_{min(h)}(a^2),$ the divergency of the function $q_{ex}(r;\:y,\:a^2)$ appears again with $q_{ex}\to +\infty$ and a local minimum has formed on the right (cf. Figs. 8b, 8d). Hence the curve $y_{min(h)}(a^2)$ forms a boundary on which the local maxima $q_{max(ex+)}(y,\:a^2)$ convert into local minima. We denote them $q_{min(ex\pm)}(y,\:a^2),$ since, as follows from relations (\ref{reality-yexexpm1})-(\ref{reality-yexexpm2}), for $1.125\le a^2 \le a^2_{crit}$ and 
\be y\le y_{(ex(ex)+)\mbox{-}(ex(ex)-)1}(a^2)\equiv \frac{8a^2-9}{8a^4} , \label{yexexpexexm1}
\ee
or $a^2_{crit} \le a^2 \le 1.3923$ and
\be y\le y_{(ex(ex)+)\mbox{-}(ex(ex)-)2}(a^2)\equiv \sqrt{\frac{3(2\sqrt{3}-3)}{a^6}}-\frac{1}{a^2} , \label{yexexpexexm2} 
\ee
they are given by $y_{ex(ex)-}(r;\:y,\:a^2).$ The function $y_{(ex(ex)+)\mbox{-}(ex(ex)-)1}(a^2)$ is given parametrically by $a^2_{r(ex(ex)\pm)}(r)$ and, e.g., by $y_{ex(ex)-}(r;\:y,\:a^2=a^2_{r(ex(ex)\pm)}(r)),$ and the function $y_{(ex(ex)+)\mbox{-}(ex(ex)-)2}(a^2)$ by $a^2_{r(ex(ex)\pm)+}(r)$ and $y_{ex(ex)-}(r;\:y,\:a^2=a^2_{r(ex(ex)\pm)+}(r)).$ The analytical expressions in (\ref{yexexpexexm1}), (\ref{yexexpexexm2}) can be then derived by eliminating the radius $r.$ Both these functions have their relevant parts entirely in regions IV,VI.\\

The function $y_{(ex(ex)+)\mbox{-}(ex(ex)-)2}(a^2)$ corresponds to the local minima of the potential $X_{-}(r;\:q,\:y,\:a)$ reaching the value $X=-a,$ i. e., $\ell =0.$ The photons corresponding to these minima, and having appropriate motion constant $q$, persist on 'spherical' orbits with $r=const$, which are crossing the spacetime rotation axis alternately above both poles. In the next we shall call them 'polar' spherical orbits -- in the following section we shall see that such polar spherical orbits form a border surface between prograde and retrograde spherical photon orbits, as related to the locally non-rotating observers.
The function $y_{(ex(ex)+)\mbox{-}(ex(ex)-)2}(a^2)$ has therefore an important meaning, since it represents a boundary between regions of qualitatively different KdS spacetimes in the $(a^2-y)$-plane. From this point of view, the function $y_{(ex(ex)+)\mbox{-}(ex(ex)-)2}(a^2)$ creates another qualitative shift in the parameter plane $(a^2-y)$ with regard to character of the photon motion, however, no qualitative shift in mathematical properties of characteristic functions  $q_{r}(r;\:y,\:a^2)$ and $q_{ex}(r;\:y,\:a^2)$ in their relevant values $q\geq 0.$ The parts of the $(a^2-y)$-plane corresponding to different behaviour of the characteristic functions are in Fig. 7 distinguished by Roman numerals, the curve $y_{(ex(ex)+)\mbox{-}(ex(ex)-)2}(a^2)$ then induces an additional division a/b.\\
Further we have to relate the minima $q_{min(ex\pm)}(y,\:a^2)$ with the maxima $q_{max(ex)}(y,\:a^2)=q_{min(r)}(y,\:a^2).$ In the region V, the minima of $q_{ex}(r;\:y,\:a^2)$ coalesce with the minima of $q_{r}(r;\:y,\:a^2)$ (Fig. 8e). We therefore have to compare the minima function $q_{min(ex)}(y,\:a^2)$=$q_{min(r)}(y,\:a^2)$ determined by $y_{ex(ex)}(r;\:y,\:a^2)$ and, e.g., $q_{ex}(r;\:y=y_{ex(ex)}(r;\:y,\:a^2),\:a^2),$ with the maxima function $q_{max(ex+)}(y,\:a^2)$ parametrized by $y_{ex(ex)+}(r;\:y,\:a^2)$ and $q_{ex}(r;\:y=y_{ex(ex)+}(r;\:y,\:a^2),\:a^2).$ The boundary curve $y_{ex(ex)\mbox{-}ex(ex)+}(a^2)$ then represents such combinations of parameters $a^2,y$ for which the local extrema of $q_{ex}(r;\:y,\:a^2)$ have coalesced into an inflection point. For parameters from the region VI, corresponding to naked singularity spacetimes with restricted repulsive barrier (as well as from the remaining regions VII, VIII), the function $q_{ex}(r;\:y,\:a^2)$ has one local minimum (Fig. 8f), and we therefore construct a function $q_{min(ex\pm)}(y,\:a^2)$ determined by functions $y_{ex(ex)\pm}(r;\:a^2)$ and $q_{ex}(r;\:y=y_{ex(ex)\pm}(r;\:a^2),\:a^2),$ where the minus sign has to be chosen for $1.17007 \le a^2 \le a^2_{crit}$ and $y_{max(d)}(a^2)\le y \le y_{(ex(ex)+)\mbox{-}(ex(ex)-)1}(a^2),$ or $a^2_{crit} \le a^2 \le 1.2828$ and $y_{max(d)}(a^2) \le y \le y_{(ex(ex)+)\mbox{-}(ex(ex)-)2}(a^2)$ (see Fig.7). For $y\to y_{max(d(ex))}(a^2)$ and $a^2\ge a^2_{crit}$ there is $q_{min(ex+)}(y,\:a^2)\to +\infty,$, and for $y> y_{max(d(ex))}(a^2),$ i.e. in the region VII, it converts into the local maximum (c.f. Figs 8f, 8g). The transition into the region VII from the region III can be inferred from comparison of Fig. 8c with Fig. 8g. Therefore, in the region VII we have to follow up the values of the function $q_{max(ex+)}(y,\:a^2)$ determined by the functions $y_{ex(ex)+}(r;\:a^2)$ and $q_{ex}(r;\:y=y_{ex(ex)+}(r;\:a^2),\:a^2).$ The functions  $q(y,\:a^2)$ are demonstrated in Fig. 9.\par
With the knowledge of the behaviour of the extremal values $q_{min/max(ex)}(y,\:a^2)$ at each region of the $(a^2\mbox{-}y)$-plane, we can finally construct all qualitatively different types of the behavior of the effective potentials $X_{\pm}(r;\:q,\:y,\:a).$ They are presented in Figure 10 for appropriately chosen representative combinations $(q,y,a^2).$
   \begin{figure*}[htbp]
   	\centering
   	
   	\begin{tabular}{cc}
   		\includegraphics[width=0.45\textwidth]{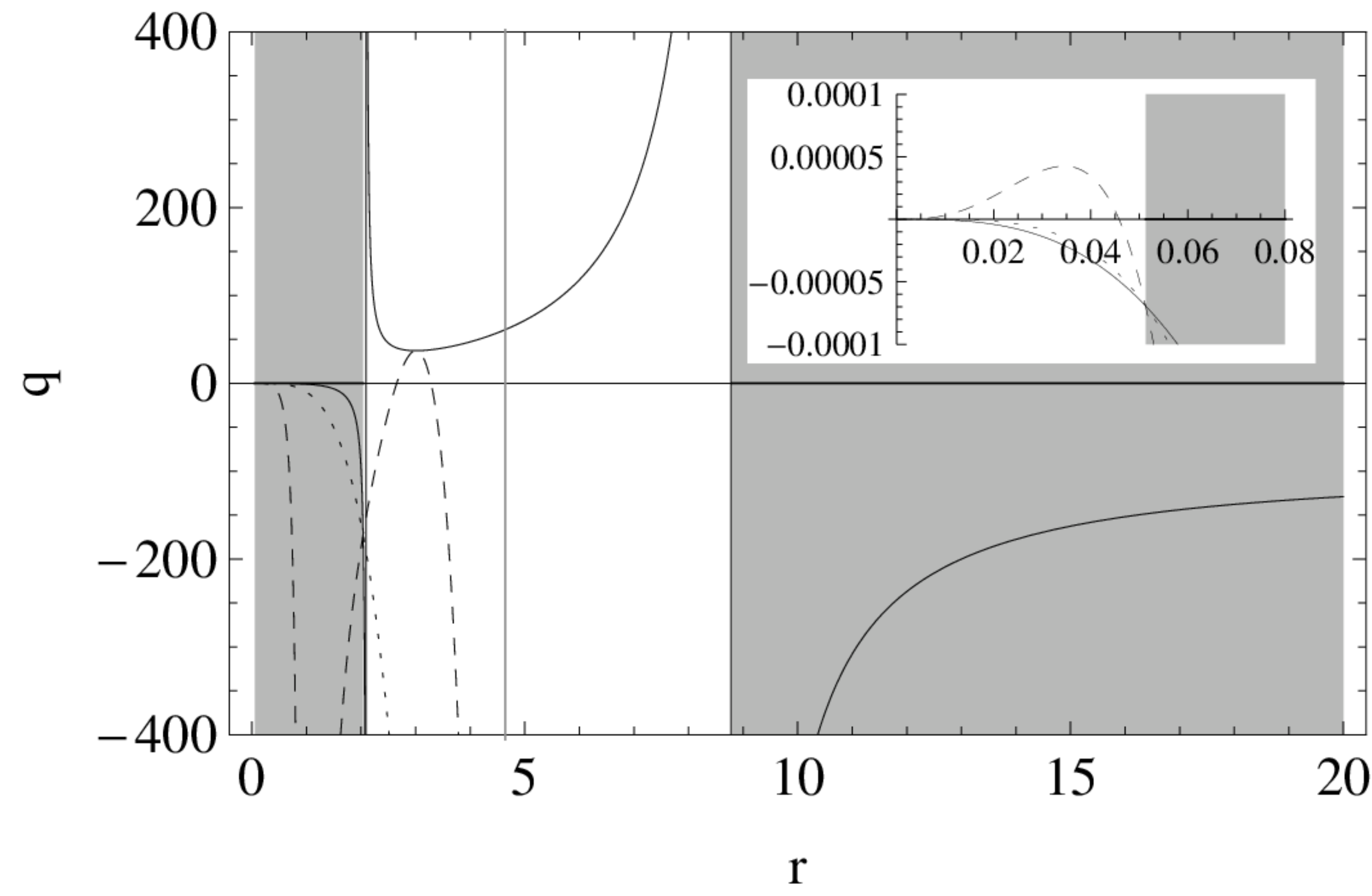} & \includegraphics[width=0.45\textwidth]{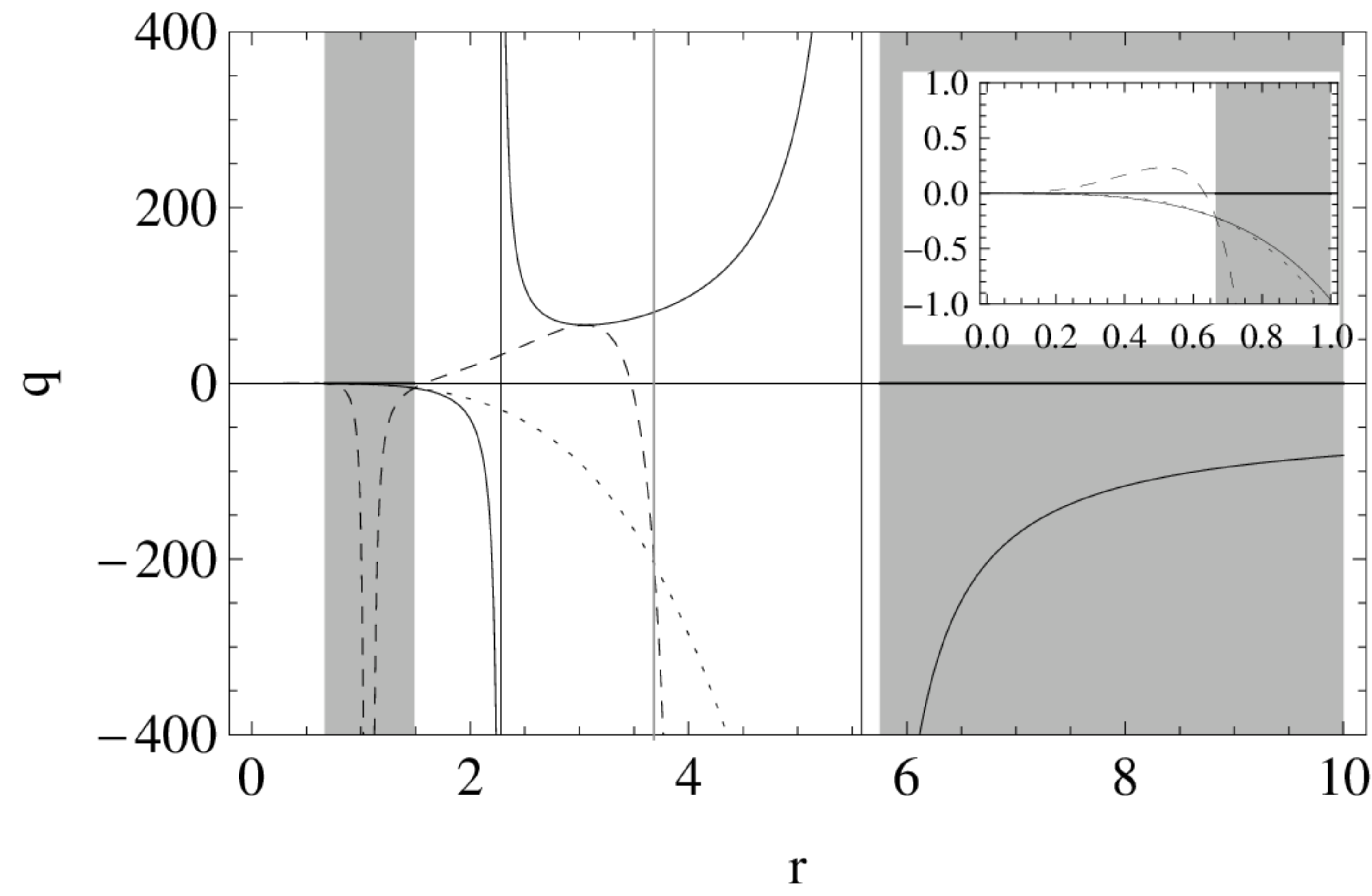} \\
   		(a) I: $y=0.01$, $a^2=0.1$ & (b) II: $y=0.02$, $a^2=0.9$ \\
   		\includegraphics[width=0.45\textwidth]{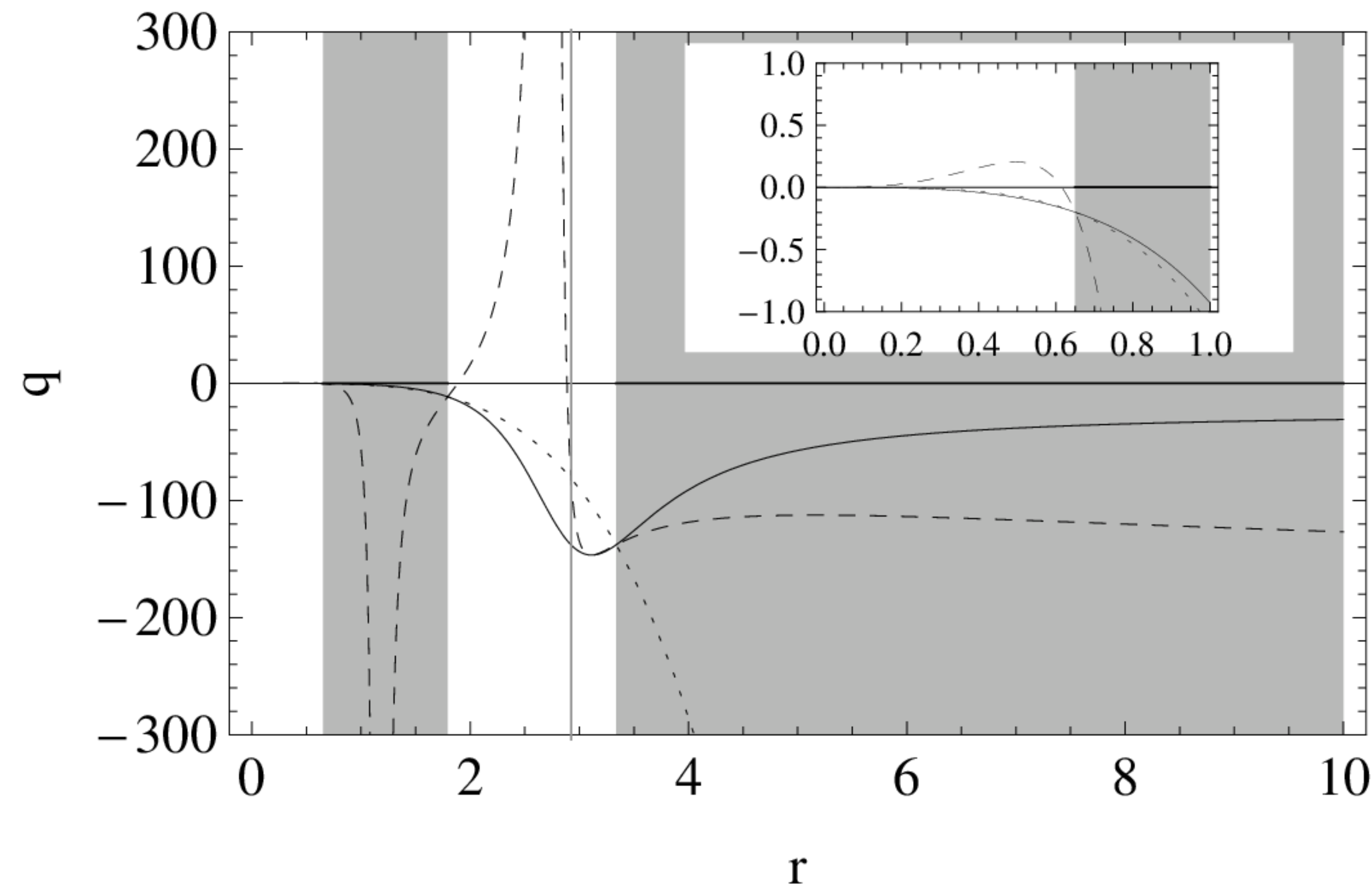} & \includegraphics[width=0.45\textwidth]{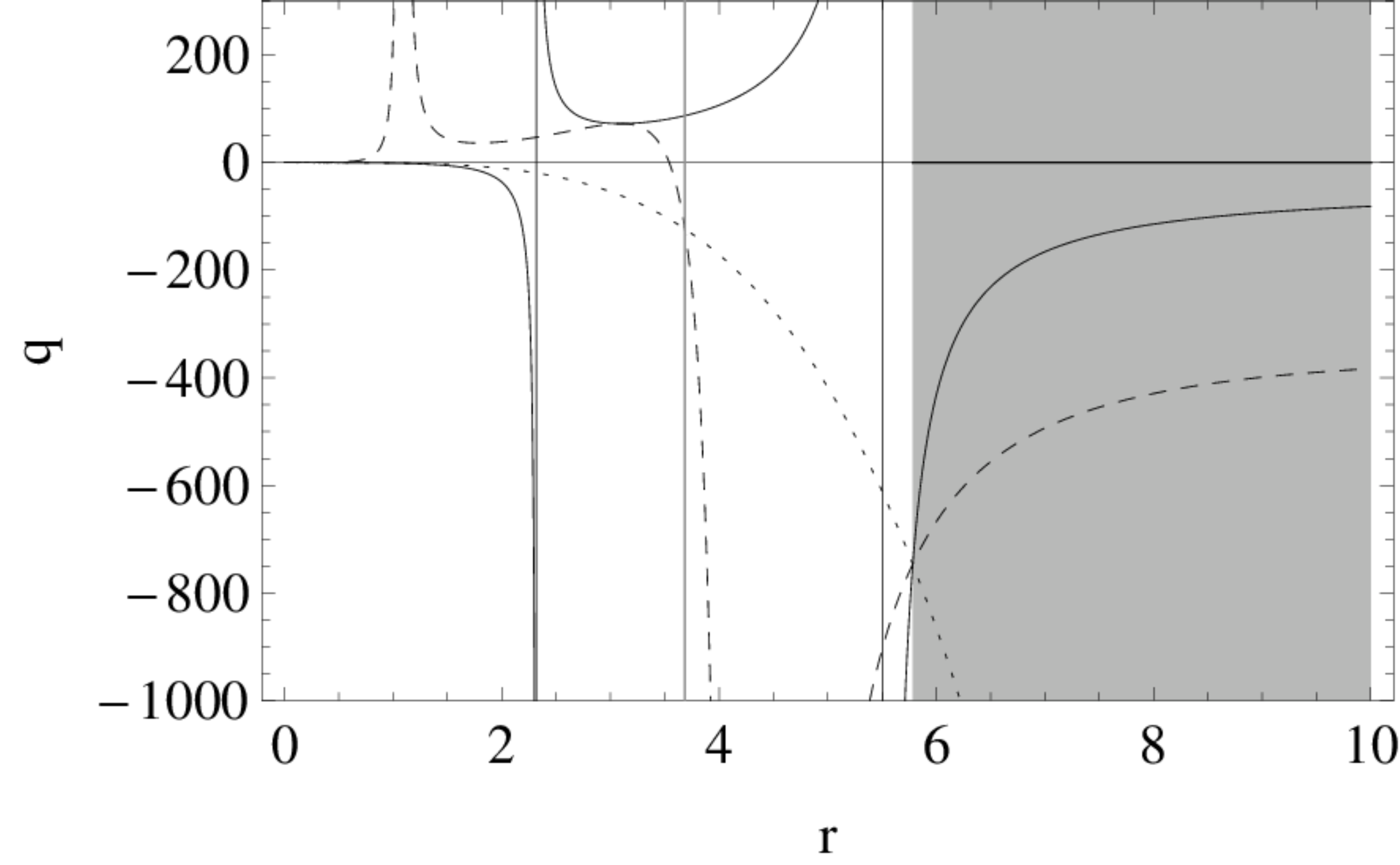} \\
   		(c) III: $y=0.04$, $a^2=0.9$ & (d) IV:  $y=0.02$, $a^2=1.5$ \\
   		\includegraphics[width=0.45\textwidth]{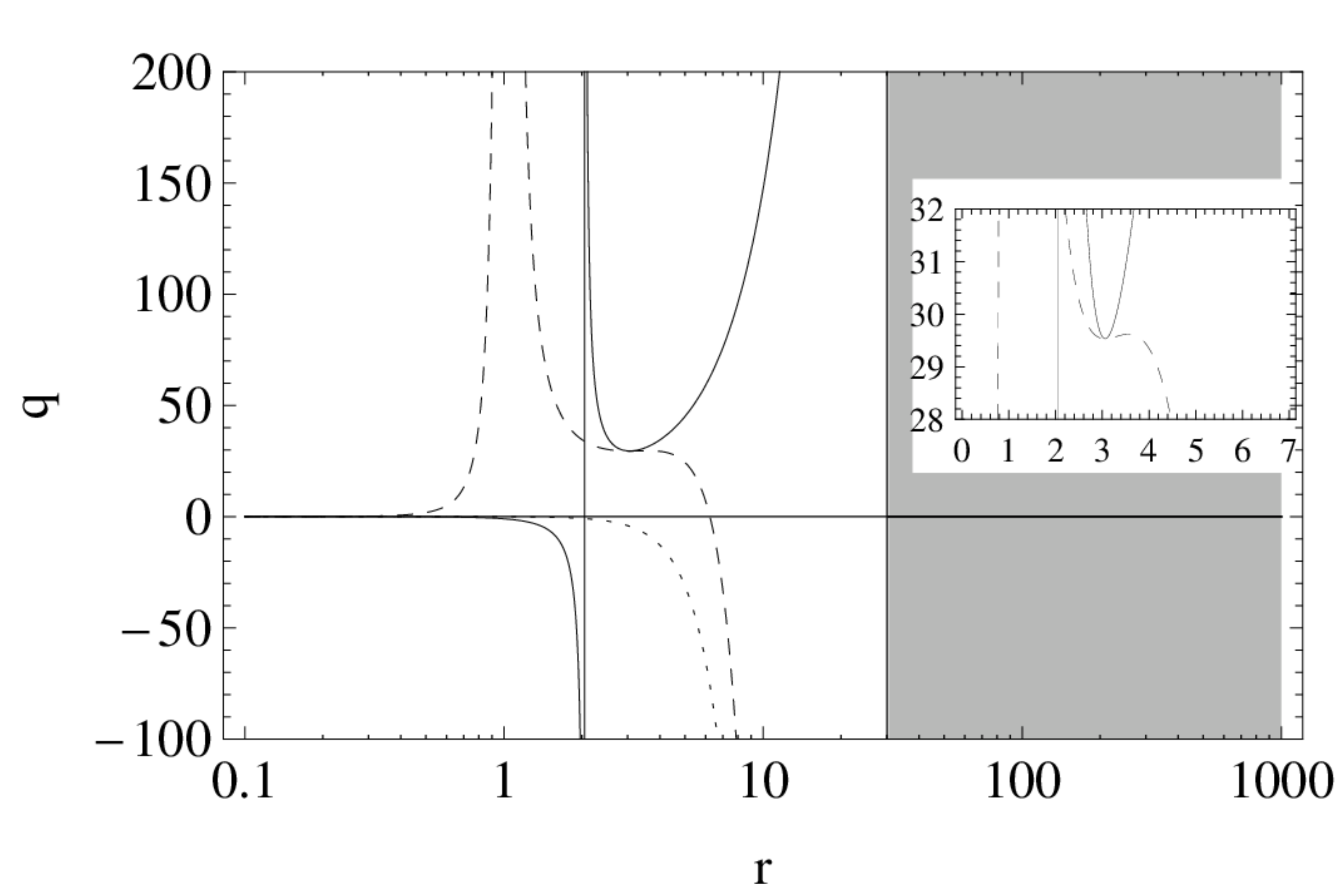} & \includegraphics[width=0.45\textwidth]{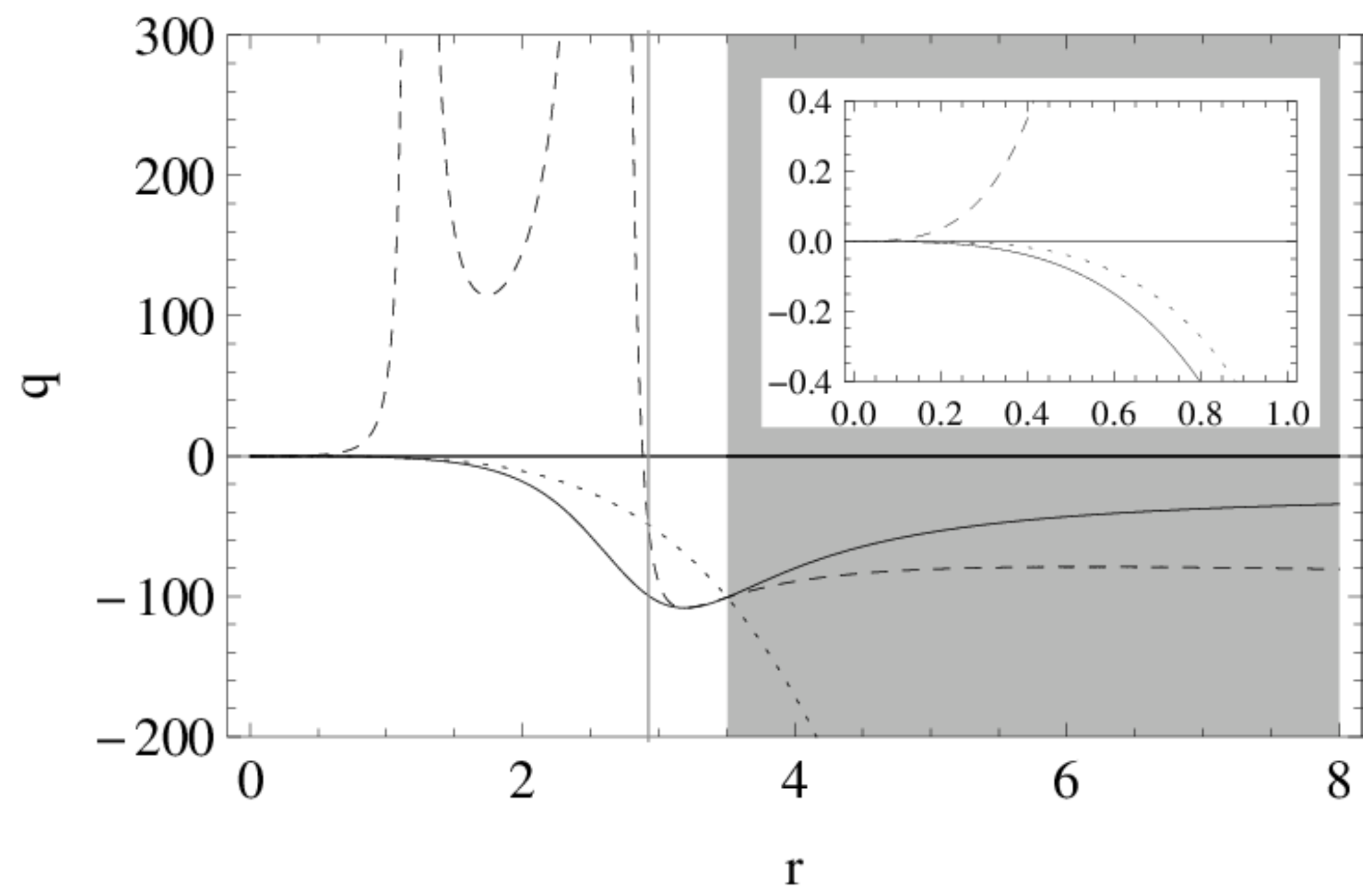} \\
   		(e) V: $y=0.001$, $a^2=20$ & (f) VI: $y=0.04$, $a^2=1.5$ \\
   		\includegraphics[width=0.45\textwidth]{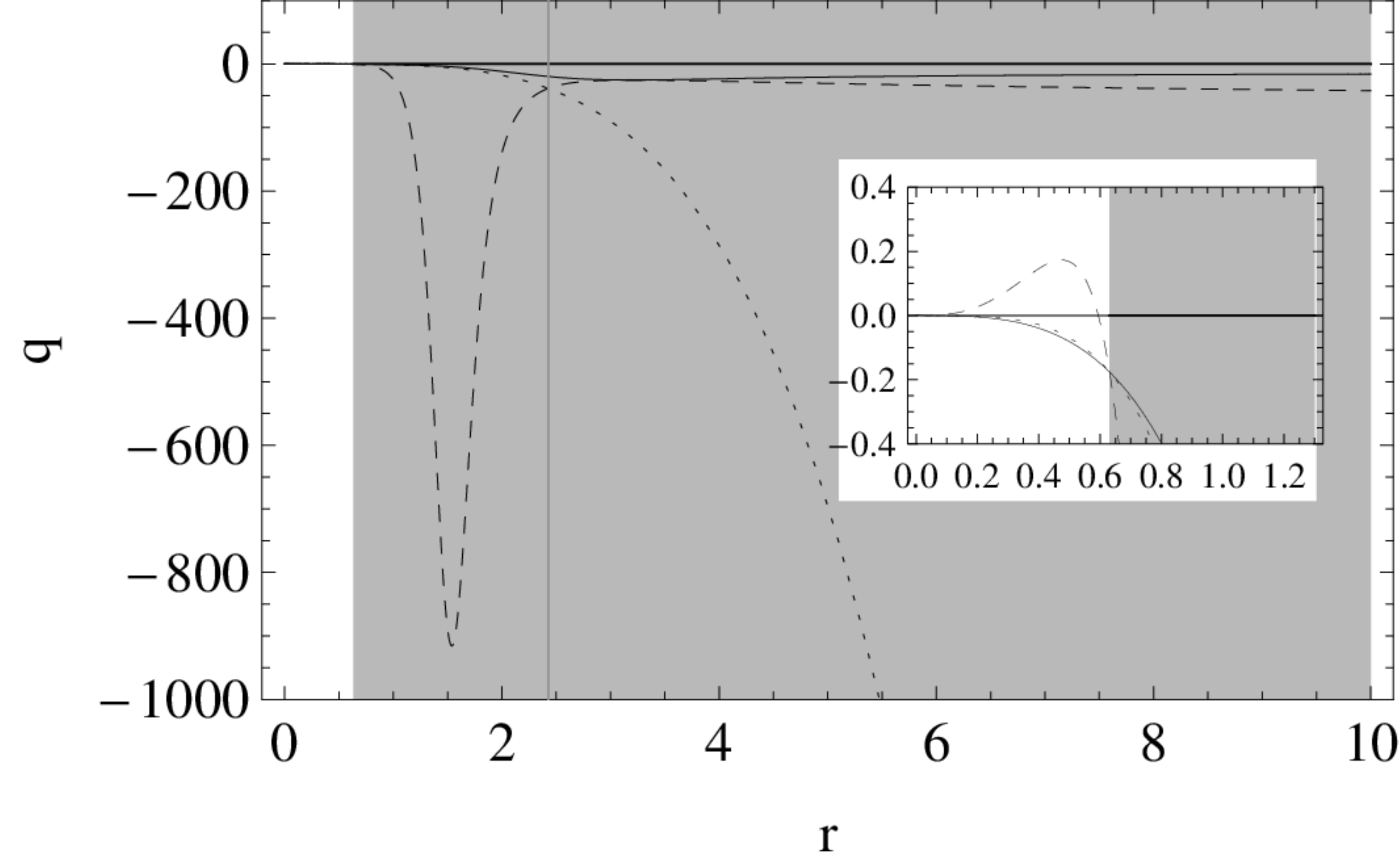} & \includegraphics[width=0.45\textwidth]{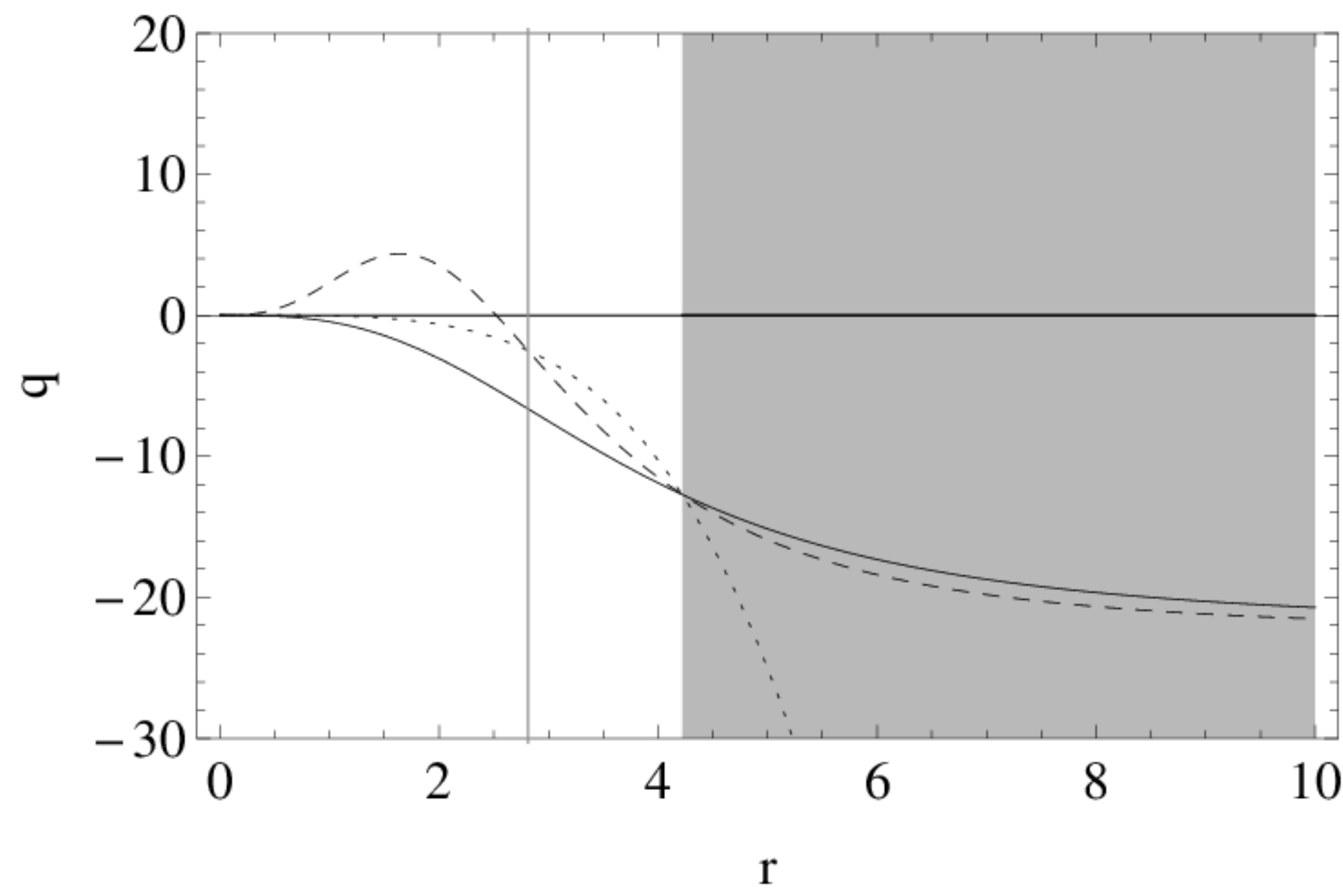} \\
   		(g) VII: $y=0.07$, $a^2=0.9$ & (h) VIII: $y=0.045$ $a^2=25$\\   				
   	\end{tabular}

	 \caption{Characteristic functions $q_{r}(r;\:y,\:a^2)$ (full curve), $q_{ex}(r;\:y,\:a^2)$ (dashed curve) and $q_{ex1}(r;\:y,\:a^2)$ (dotted curve) governing the behaviour of effective potentials $X_{\pm}(r;\:q,\:y,\:a).$ The dynamic region is highlighted by shading, the grey vertical bar demarcates loci of the static radius. Note that the last scheme corresponding to case $y>1/a^2,$ which is introduced due to changes in the latitudinal motion, brings no new features in character of the radial motion, since it qualitatively coincide with case VII and is shown for completeness. }\label{Figure 8}
  \end{figure*}

 \begin{figure*}[htbp]
	\centering
	\begin{tabular}{cc}
		\includegraphics[width=0.48\textwidth]{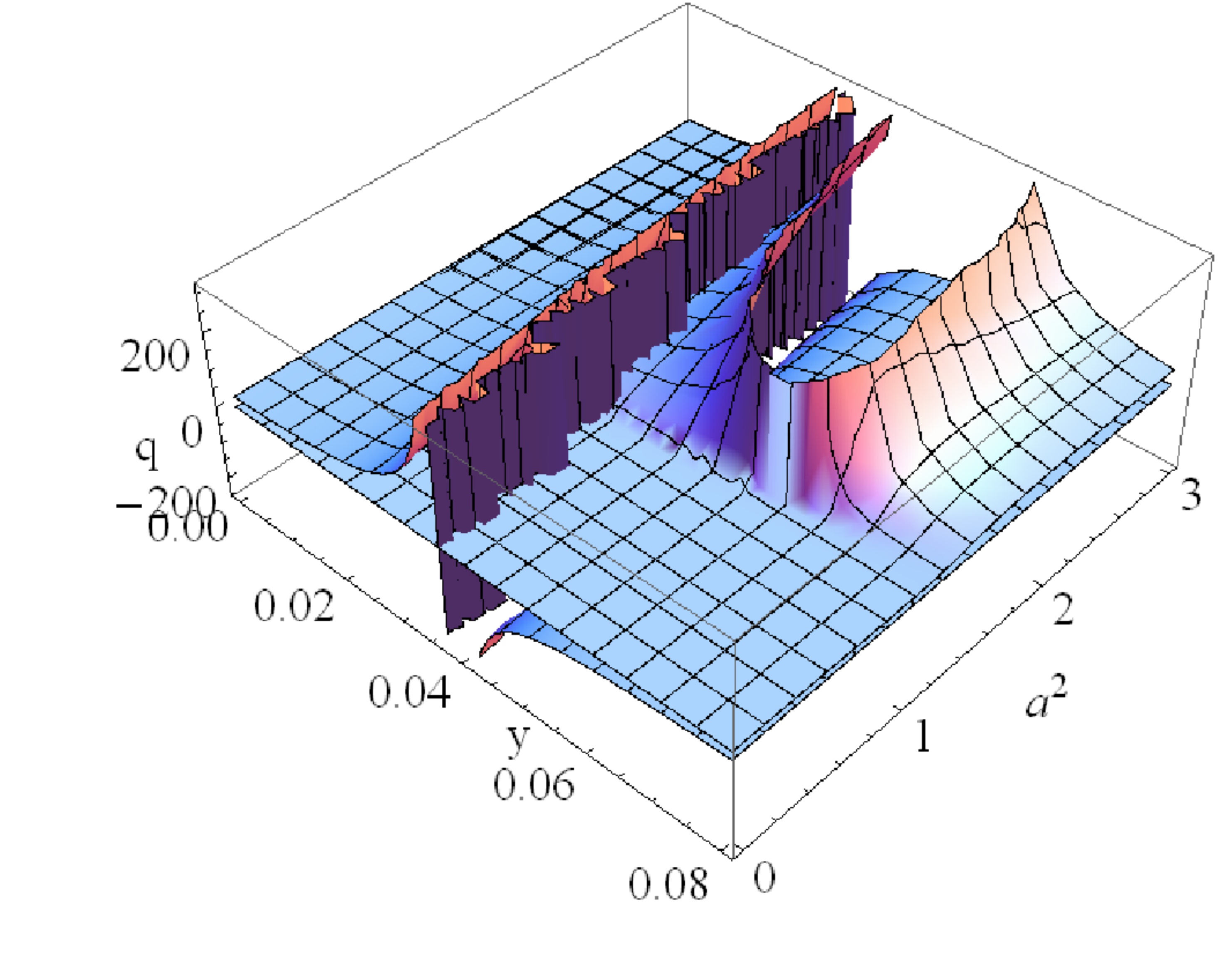}&\includegraphics[width=0.48\textwidth]{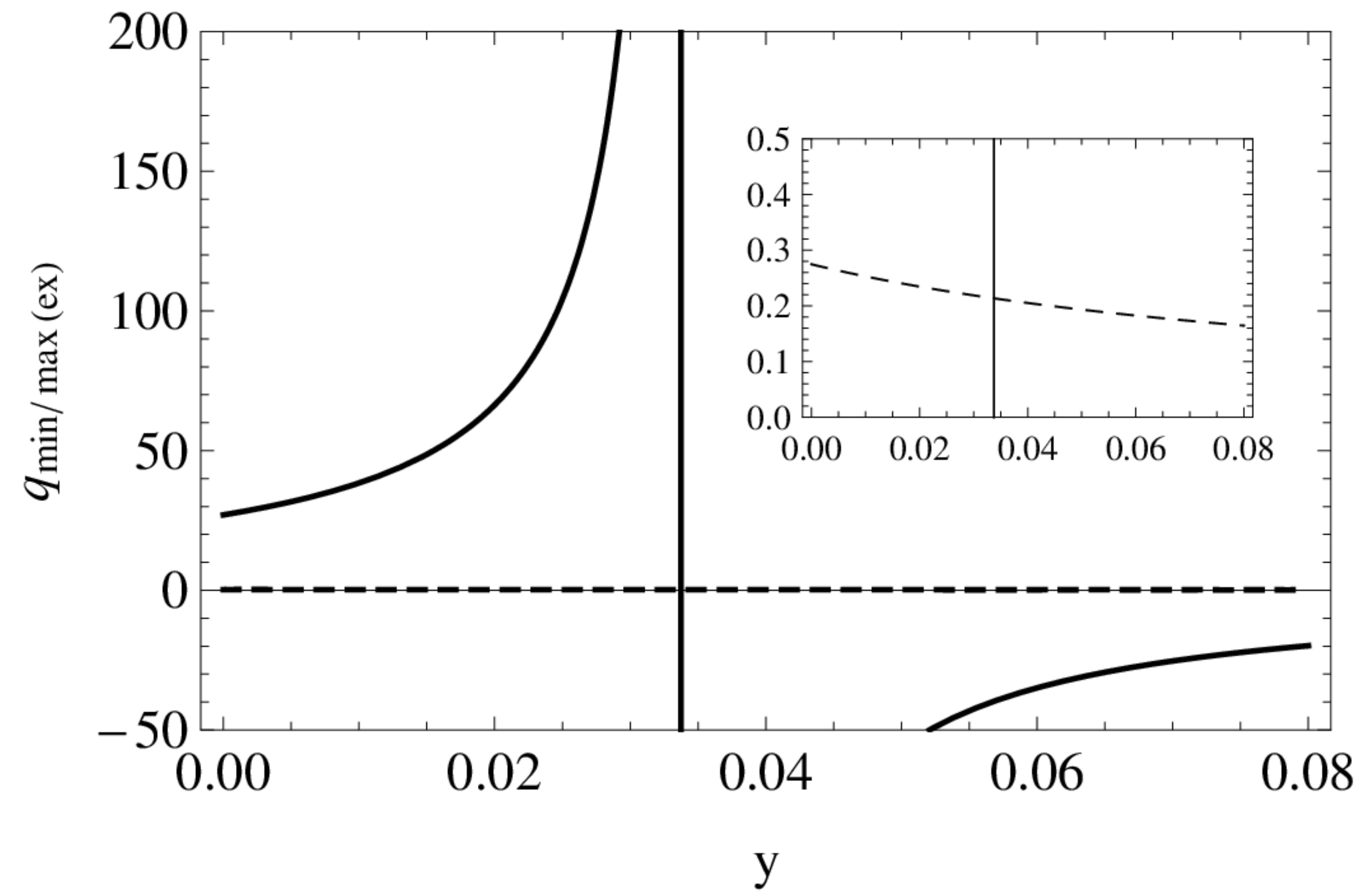}\\
		(a) & (b) $a^2=0.9$\\
		\includegraphics[width=0.48\textwidth]{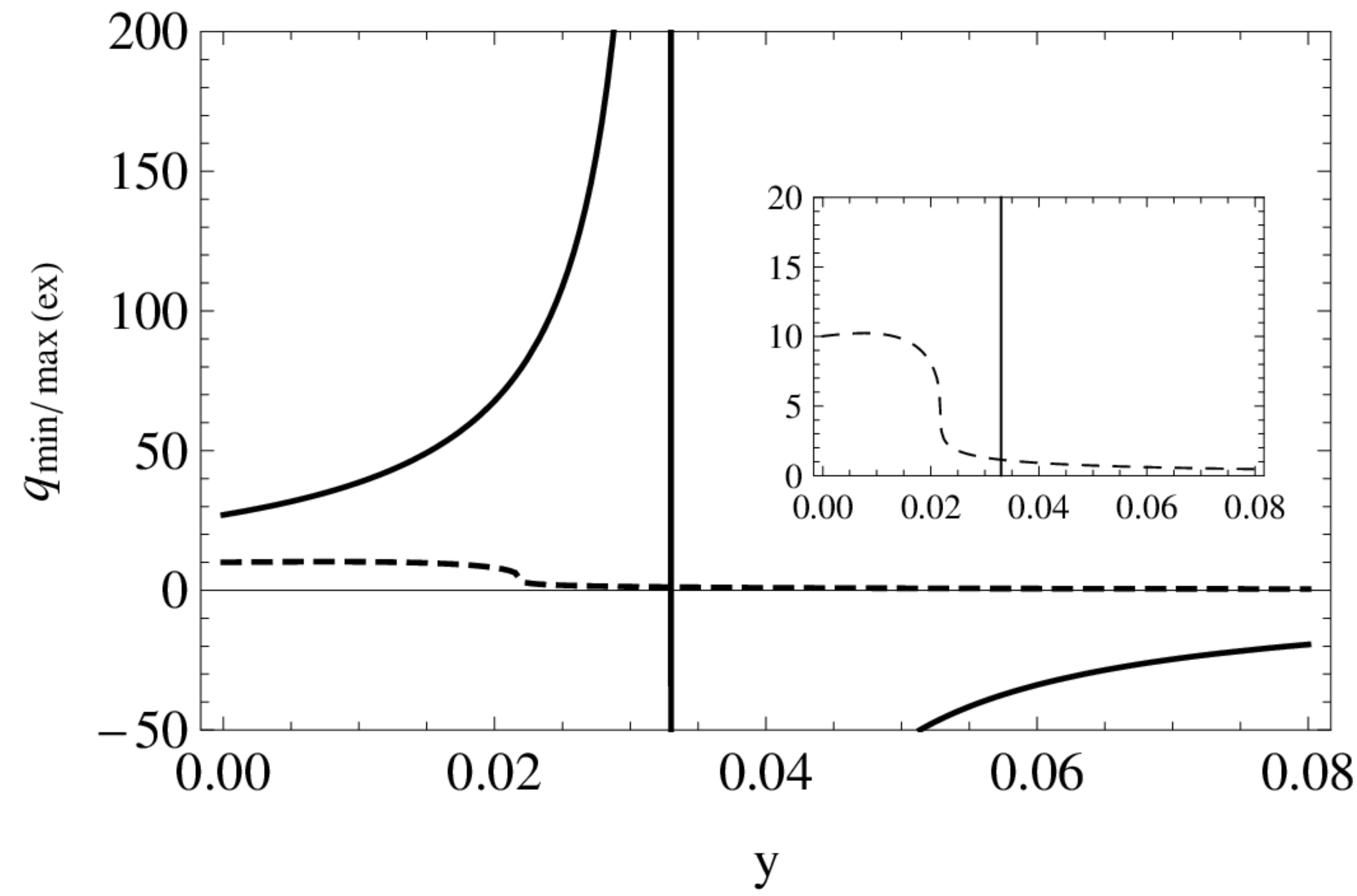}&\includegraphics[width=0.48\textwidth]{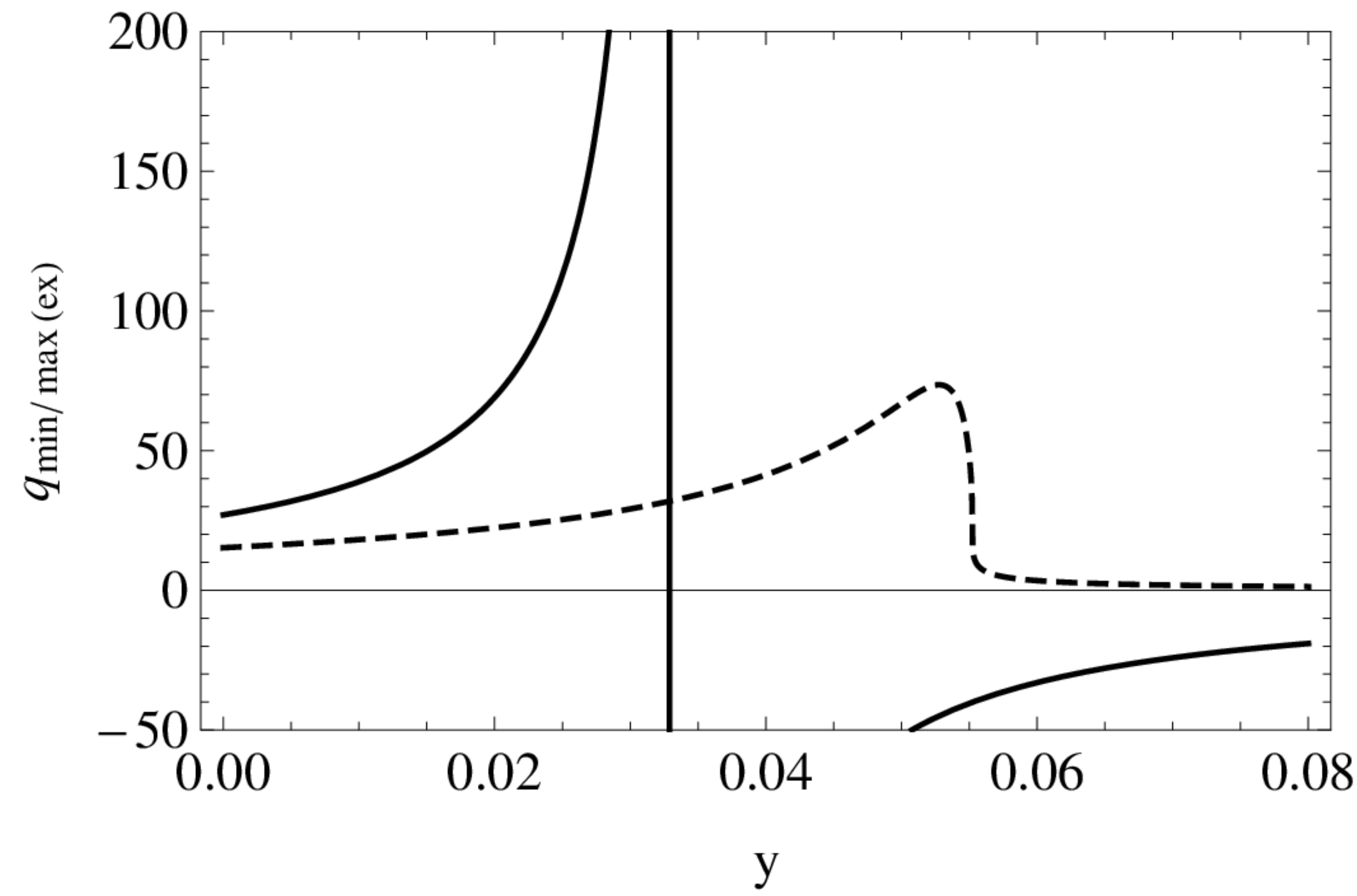}\\
		(c) $a^2=1.05$ & (d) $a^2=1.18$\\
		\includegraphics[width=0.48\textwidth]{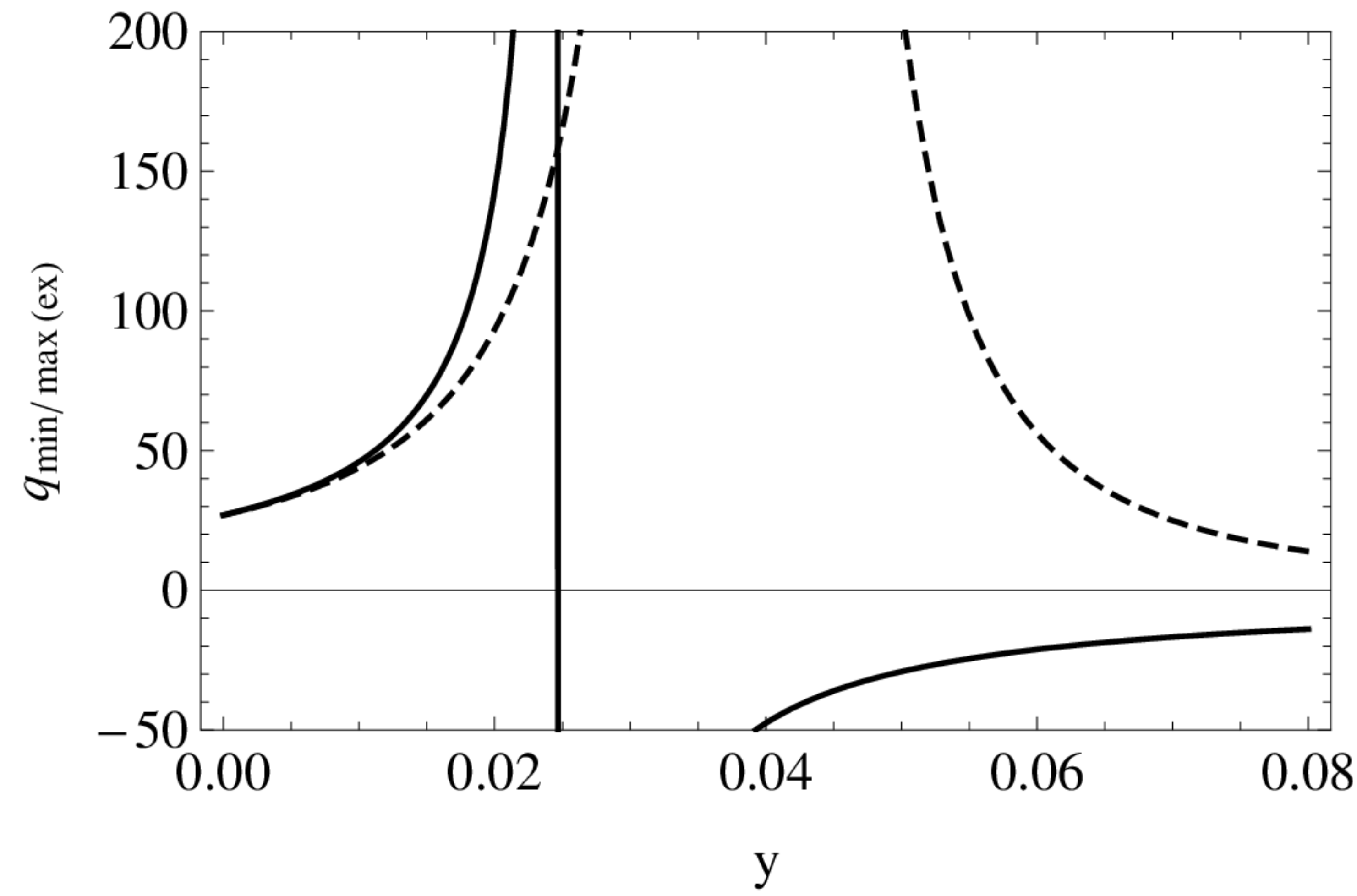}&\includegraphics[width=0.48\textwidth]{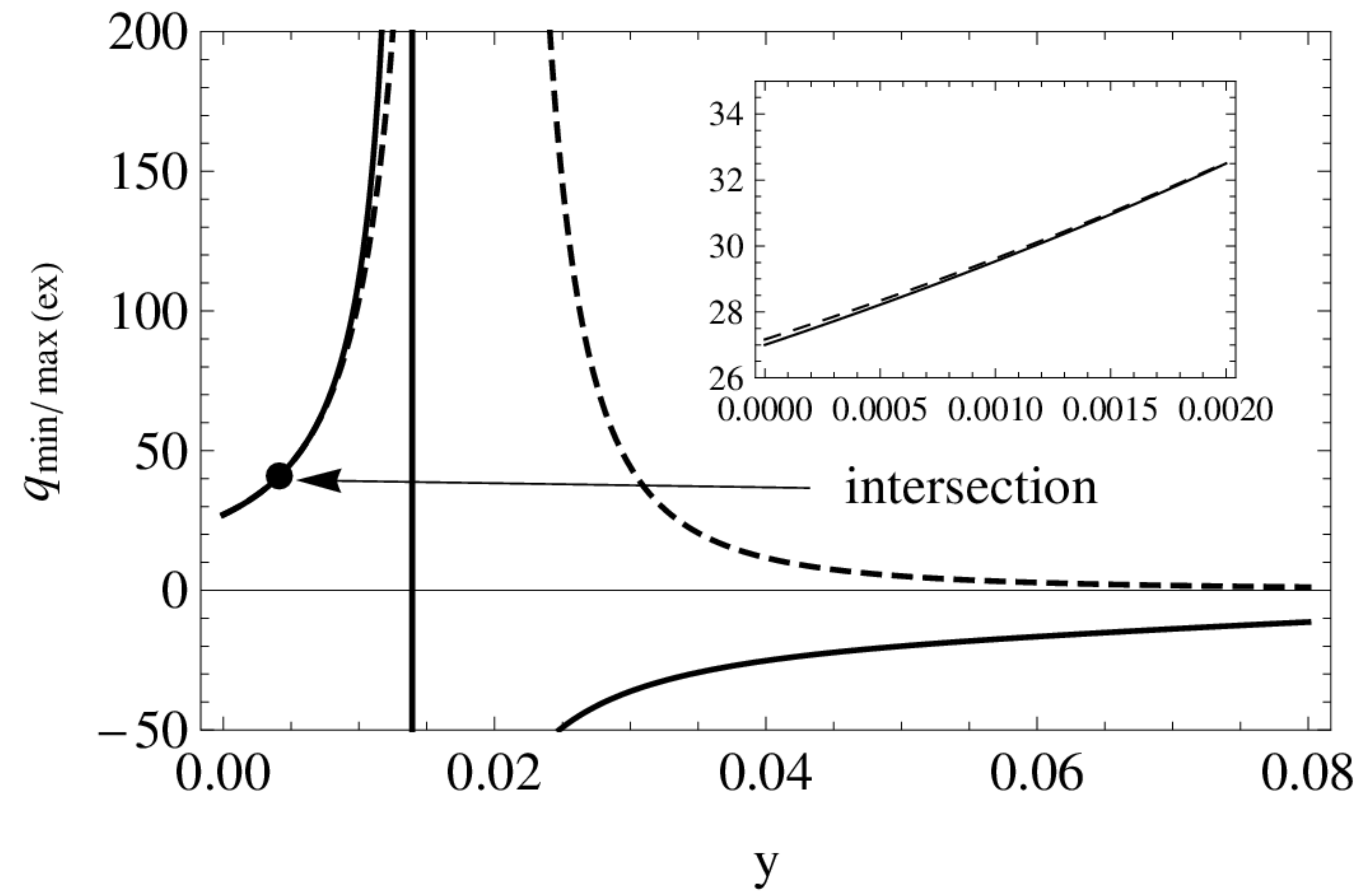}\\
		(e) $a^2=5$ & (f) $a^2=20$	 		 		
	\end{tabular}	 	
	\caption{3D plot of functions $q(y,\:a^2)$ (a) and its $a^2=const.$-slices corresponding to values between outstanding limits $a^2=1, a^2_{rrb(max)}=1.08316, a^2_{crit}=1.21202$ and $a^2_{(ex(ex)\mbox{-}ex(ex)+)min}=9.$ The positive branch of the full curve is the graph of the function $q_{max(ex)}(y,\:a^2),$ its asymptote intersects the $(a^2\mbox{-}y)$-plane in the curve $y_{max(d)}(a^2).$ The dashed curve in Fig. (b) describes the local maxima $q_{max(ex+)}(y,\:a^2)$ under the inner horizon. In Figs. (c), (d) it represents local minima $q_{min(ex+)}(y,\:a^2)$ ($q_{min(ex-)}(y,\:a^2)$ eventually) for $y<y_{min(h)}(a^2),$ and maxima $q_{max(ex+)}(y,\:a^2)$ for $y>y_{min(h)}(a^2),$ where the critical value $q_{max(ex+)}(y=y_{min(h)}(a^2),\:a^2)$ can be identified as the inflection point of this curve. The minima diverge as $y\to y_{max(d(ex))}(a^2)$ (Fig. e), the descending part then describes the local maxima $q_{max(ex+)}(y,\:a^2)$ in region VII. Fig. (f) demonstrates intersection of both curves at $y=y_{ex(ex)\mathrm{-}ex(ex+)}(a^2).$ The full curve for $y<y_{ex(ex)\mathrm{-}ex(ex+)}(a^2)$ then corresponds to the local minima $q_{min(ex)}(y,\:a^2)=q_{min(r)}(y,\:a^2),$ while the dashed one  now matches the maxima $q_{max(ex+)}(y,\:a^2).$  } \label{Figure 9}
\end{figure*} 

\begin{figure*}[htbp]
	\centering
	
	\begin{tabular}{cc}
		\includegraphics[width=0.48\textwidth]{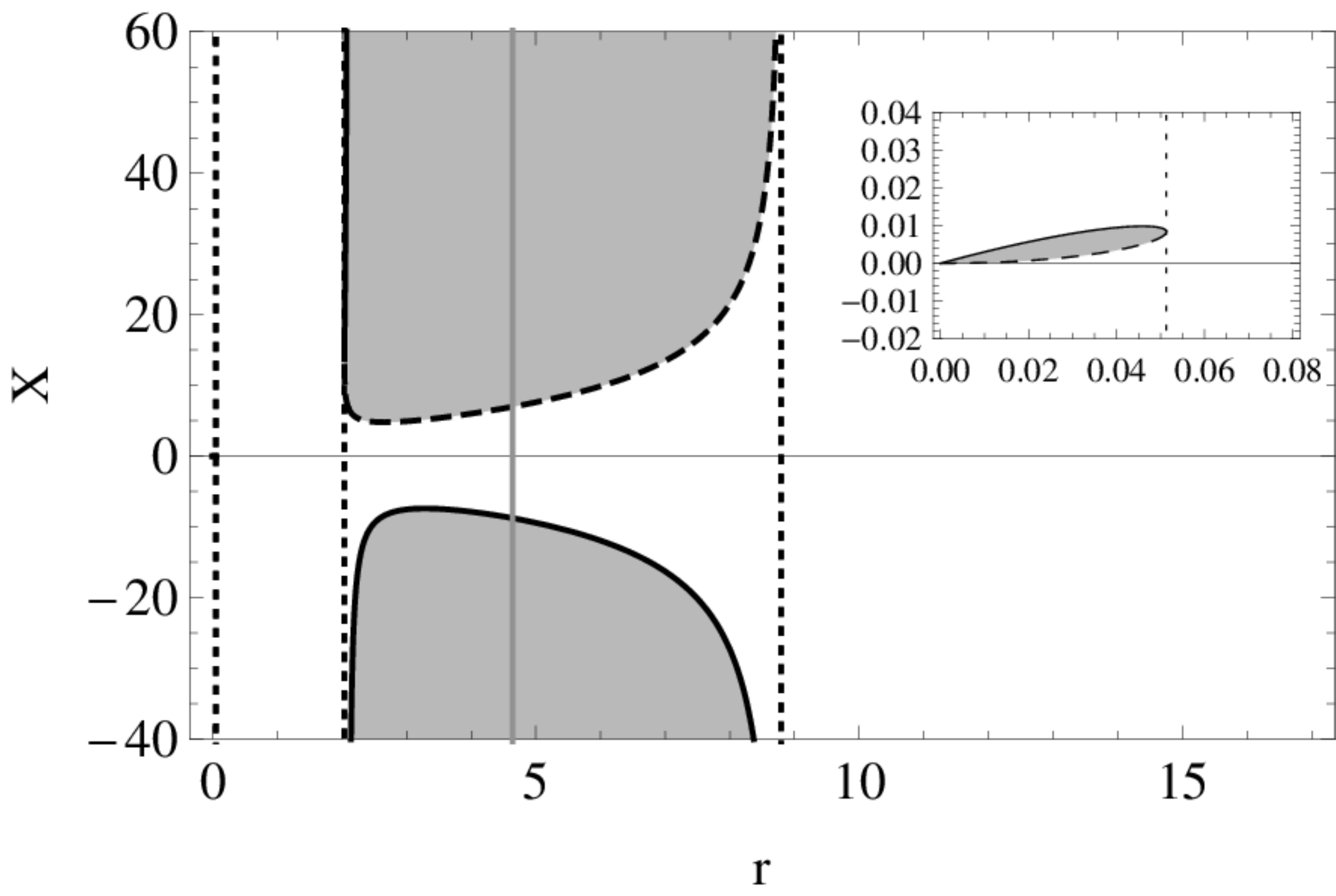} & \includegraphics[width=0.48\textwidth]{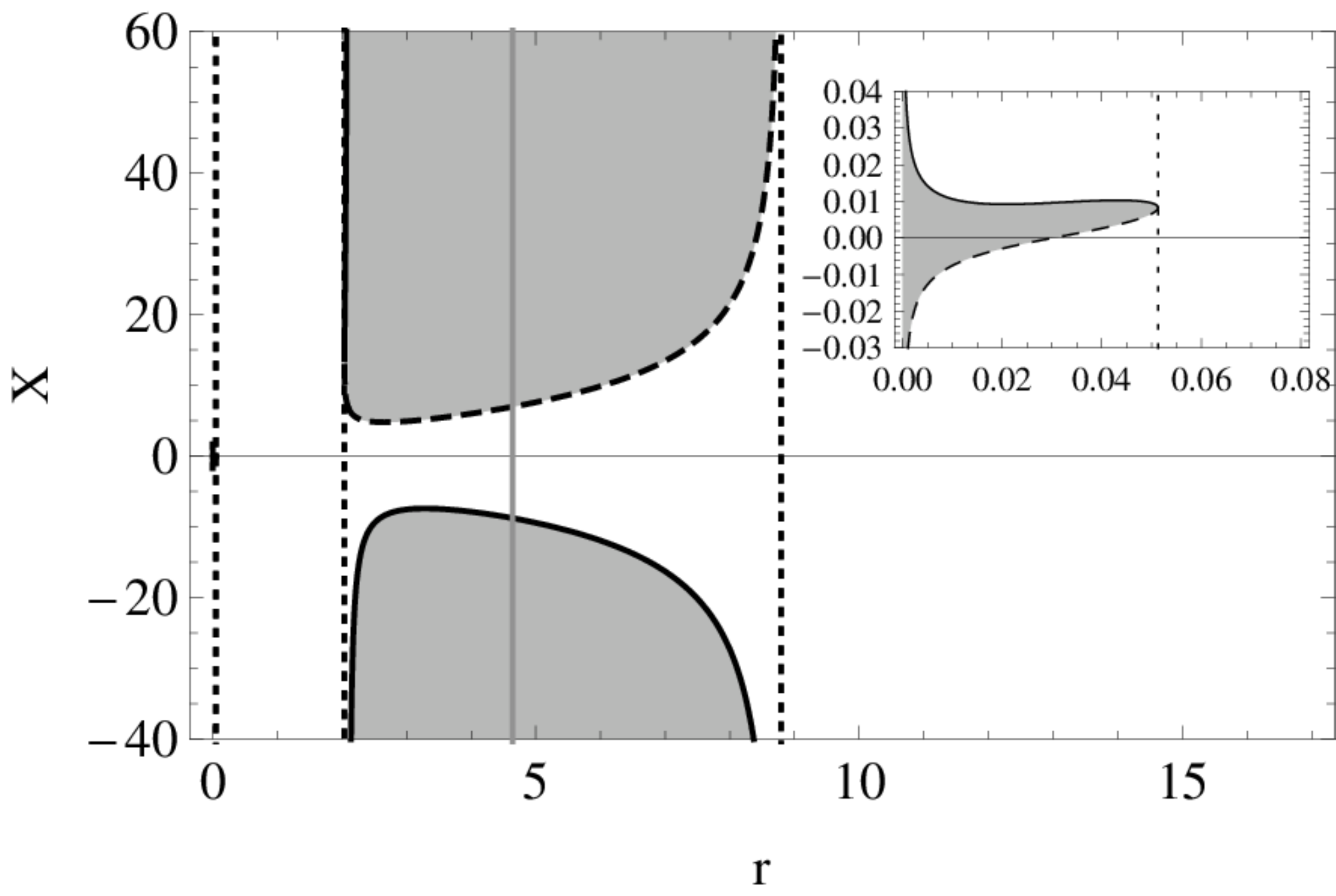} \\
		(a) $q=0,$ I: $y=0.01,$ $a^2=0.1$ & (b) $q=2.10^{-5},$ I: $y=0.01$, $a^2=0.1$ \\
		\includegraphics[width=0.48\textwidth]{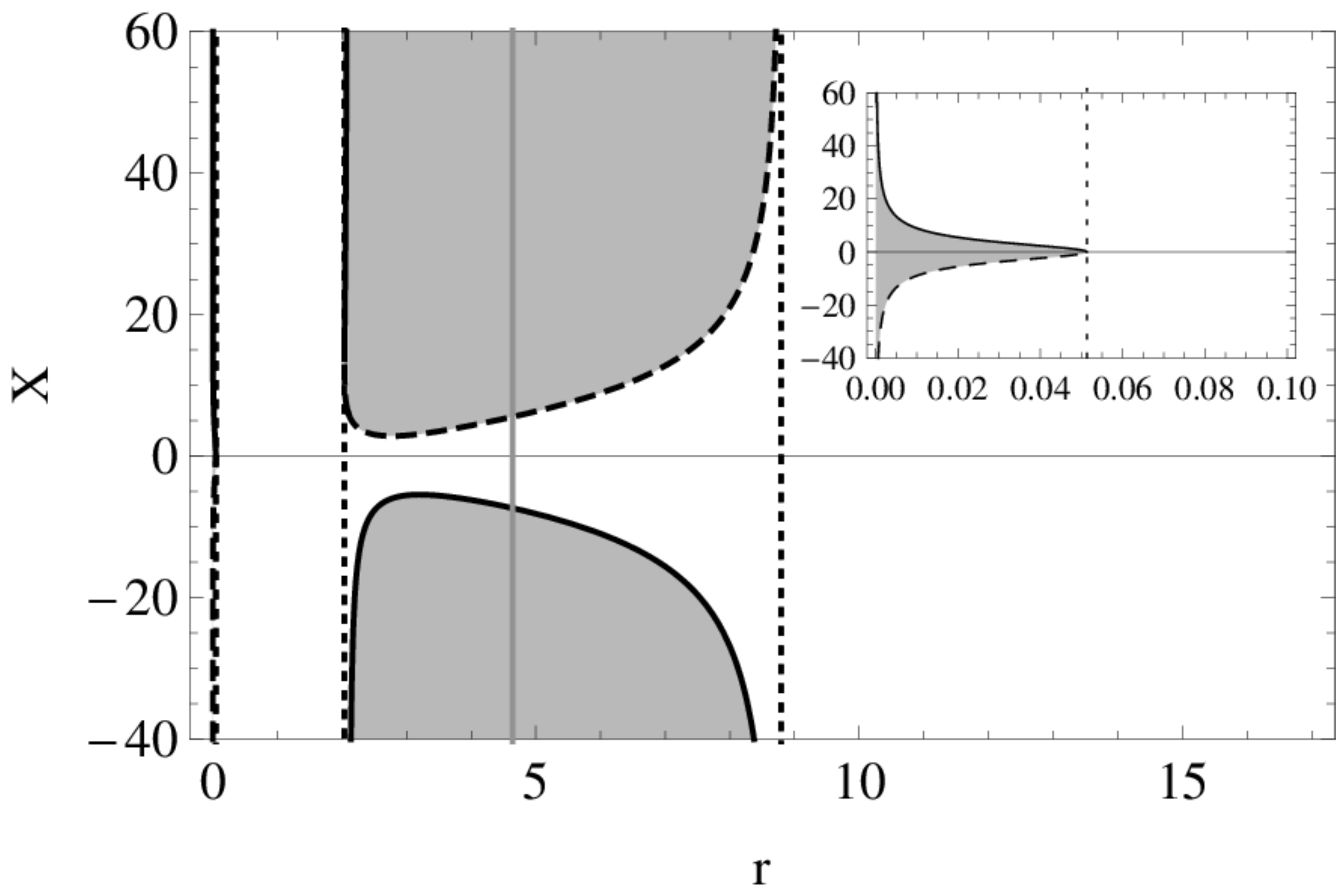} & \includegraphics[width=0.48\textwidth]{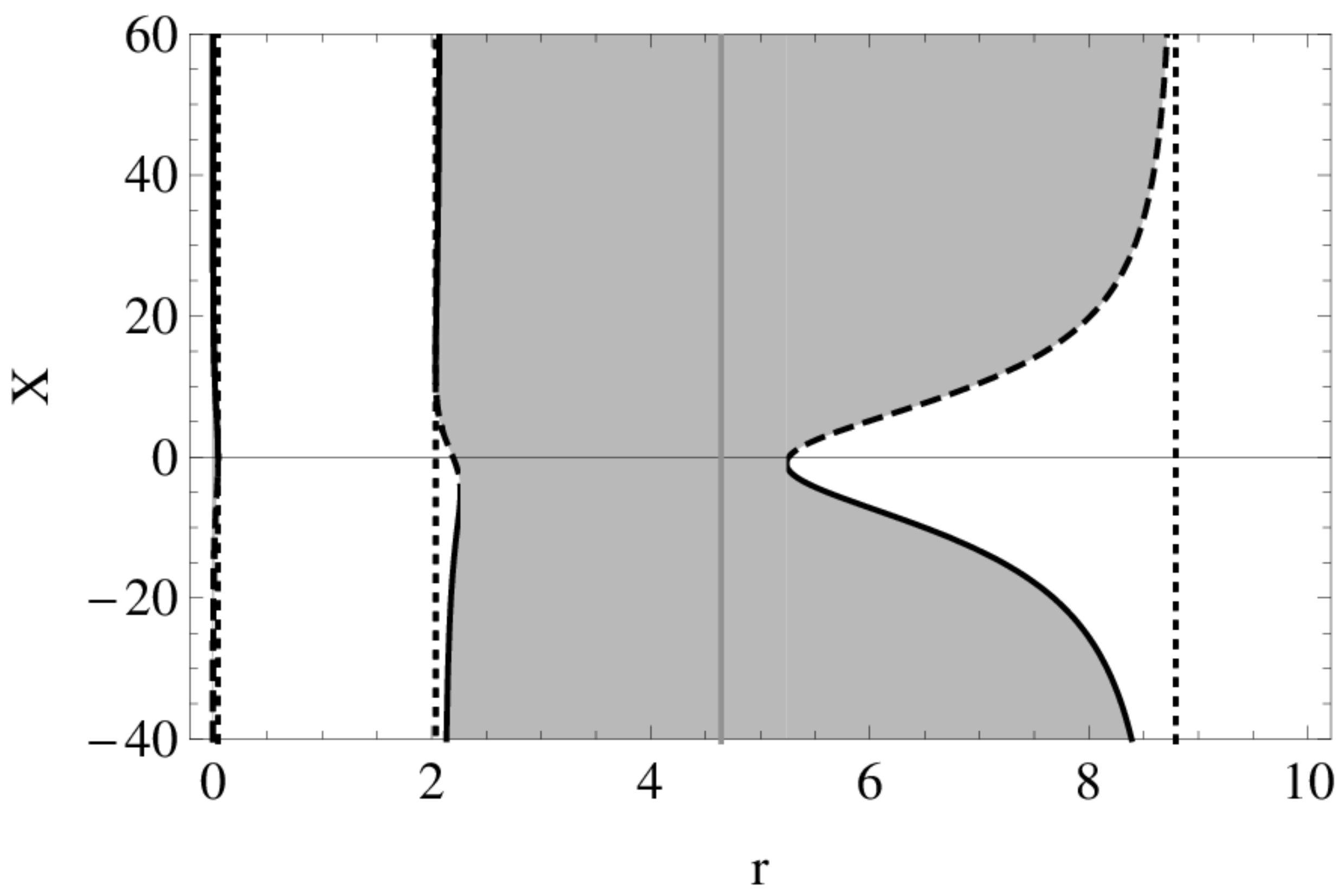} \\
		(c) $q=20,$ I: $y=0.01,$ $a^2=0.1$ & (d) $q=80,$ I: $y=0.01,$ $a^2=0.1$ \\
		\includegraphics[width=0.48\textwidth]{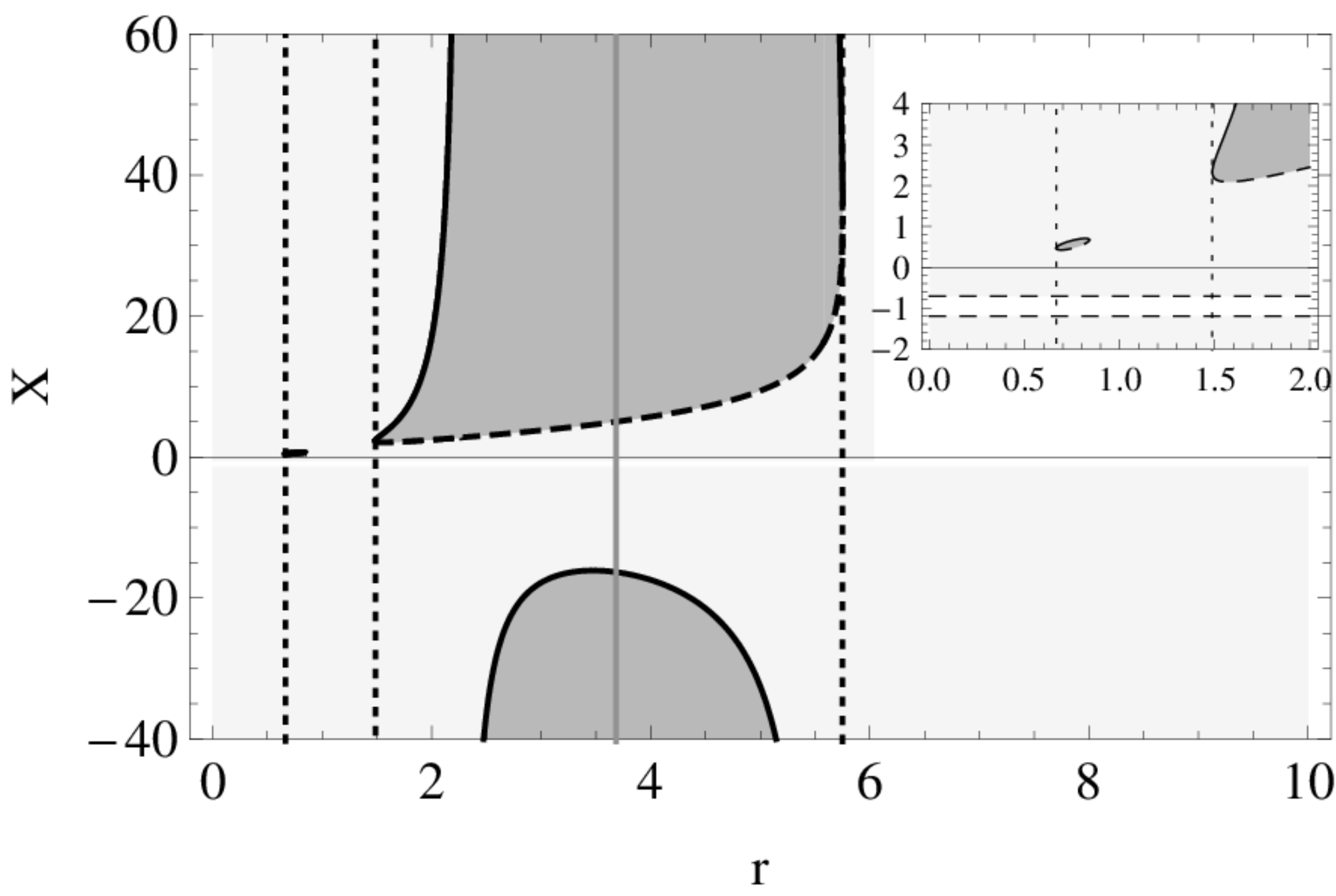} & \includegraphics[width=0.48\textwidth]{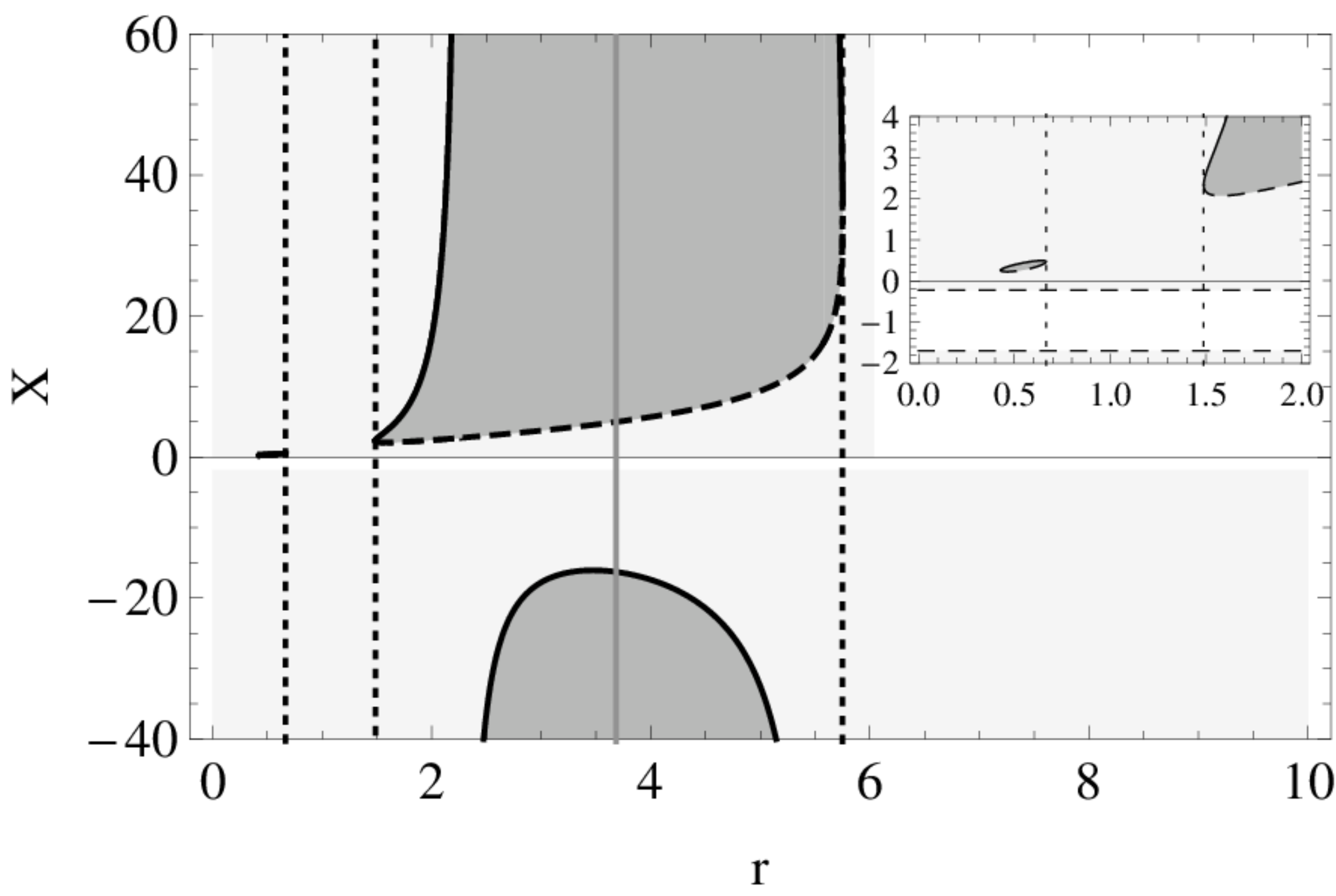} \\
		(e) $q=-0.5,$ II: $y=0.02,$ $a^2=0.9$ & (f) $q=-0.05,$ II: $y=0.02,$ $a^2=0.9$ \\
		\includegraphics[width=0.48\textwidth]{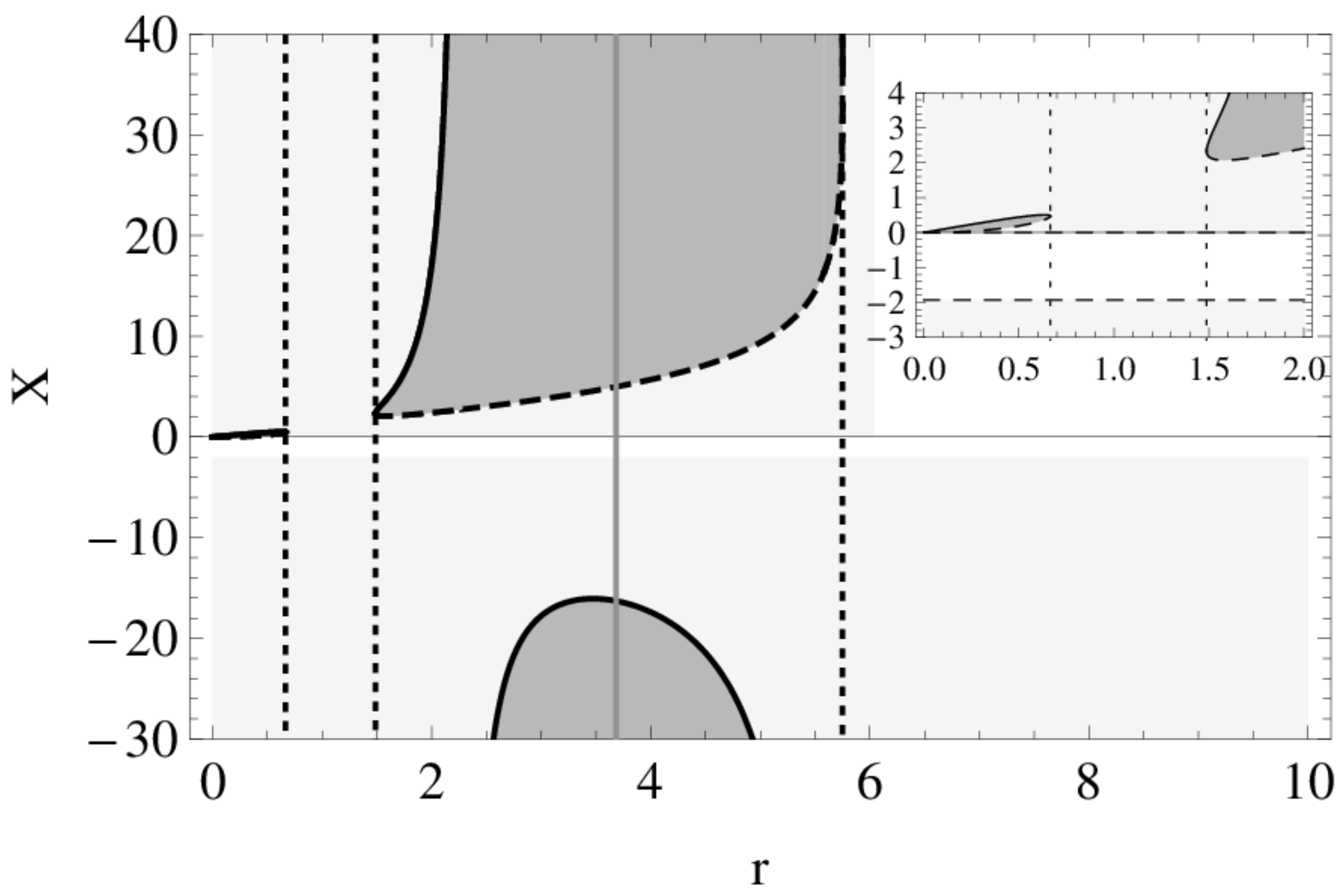} & \includegraphics[width=0.48\textwidth]{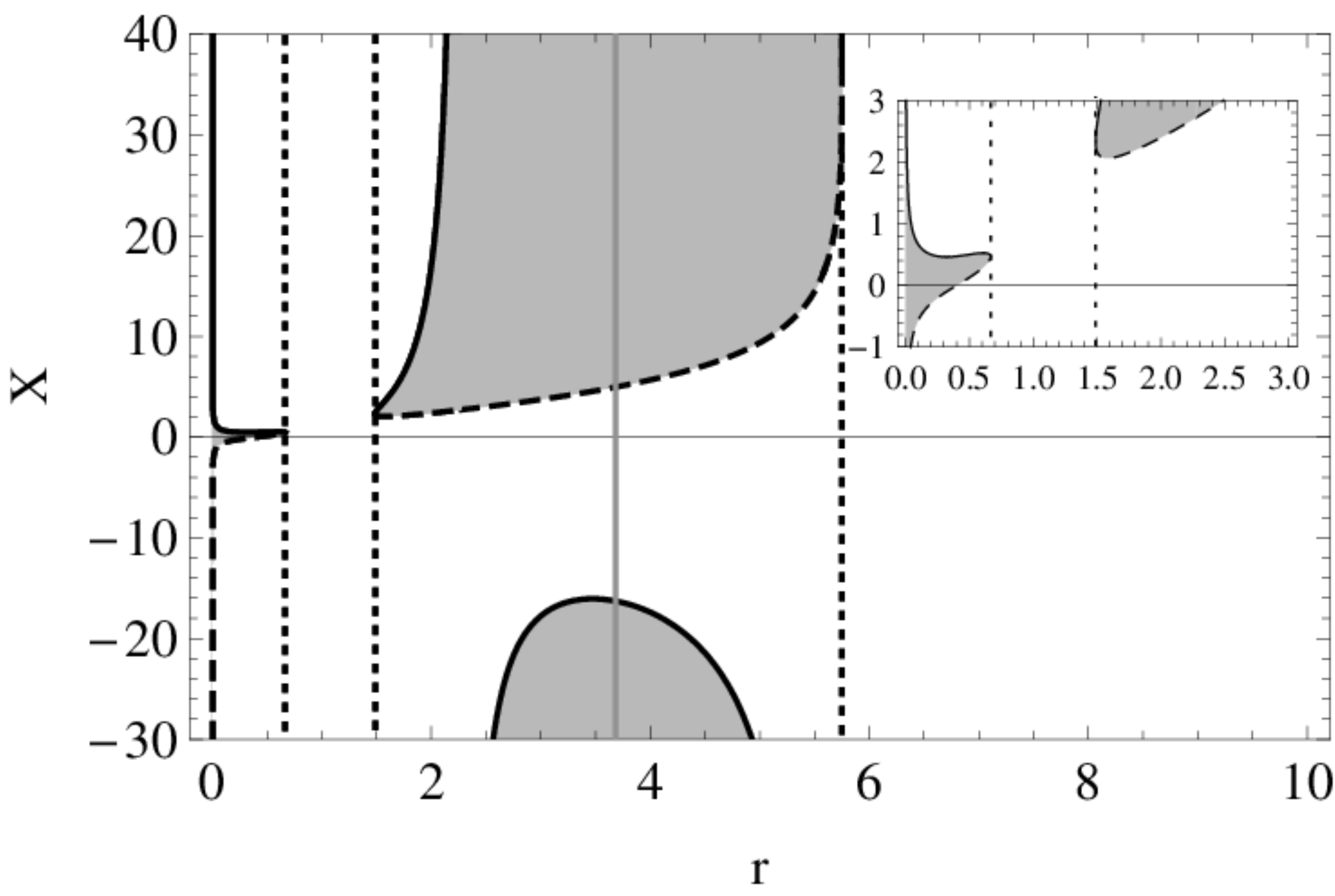} \\
		(g) $q\to 0^{-}$, II: $y=0.02,$ $a^2=0.9$ & (h) $q=0.1,$ II: $y=0.02,$ $a^2=0.9$		
	\end{tabular}
	\center (\textit{Figure continued})
	
\end{figure*}

\begin{figure*}[htbp]
	\centering
	\begin{tabular}{cc}
		\includegraphics[width=0.48\textwidth]{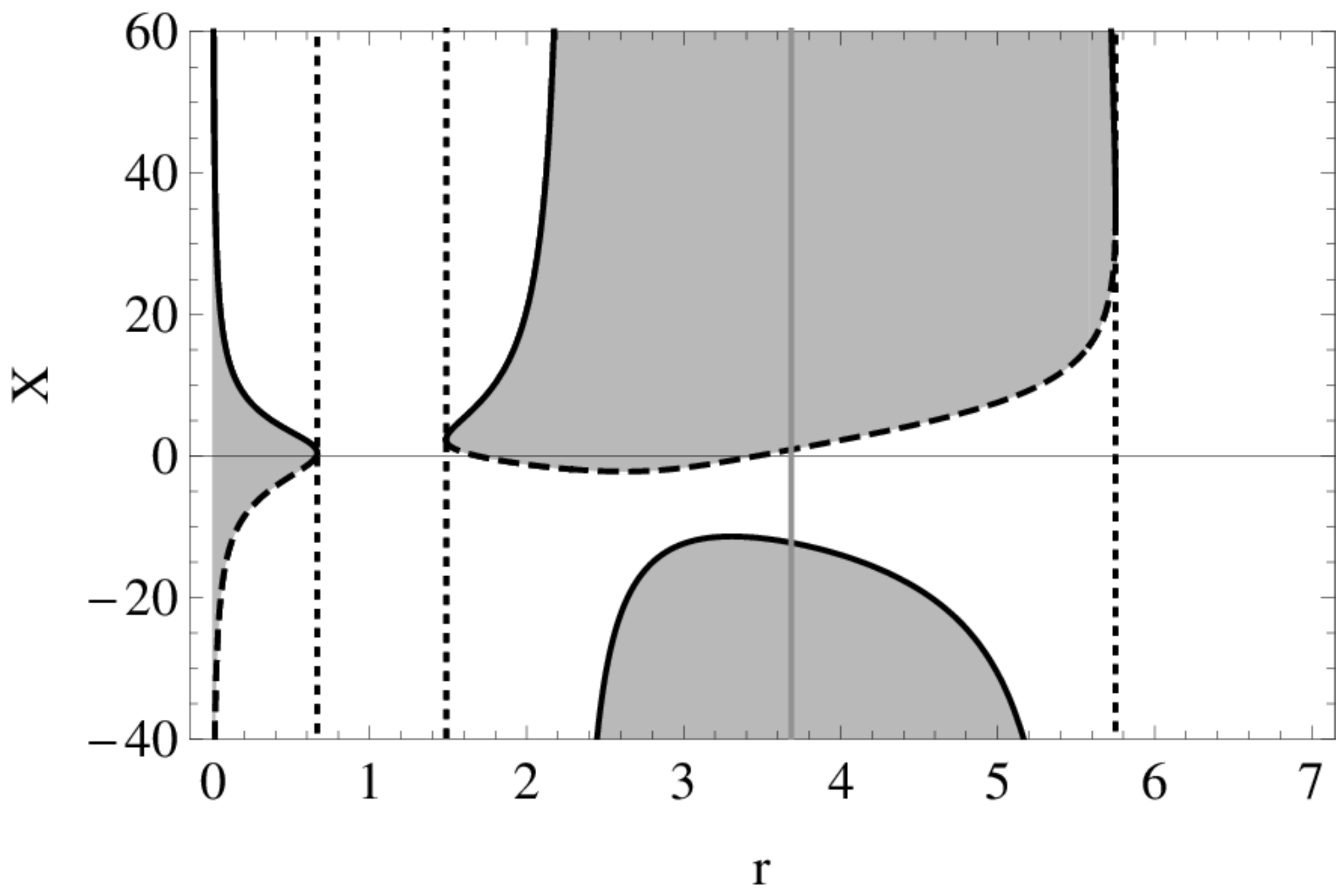} & \includegraphics[width=0.48\textwidth]{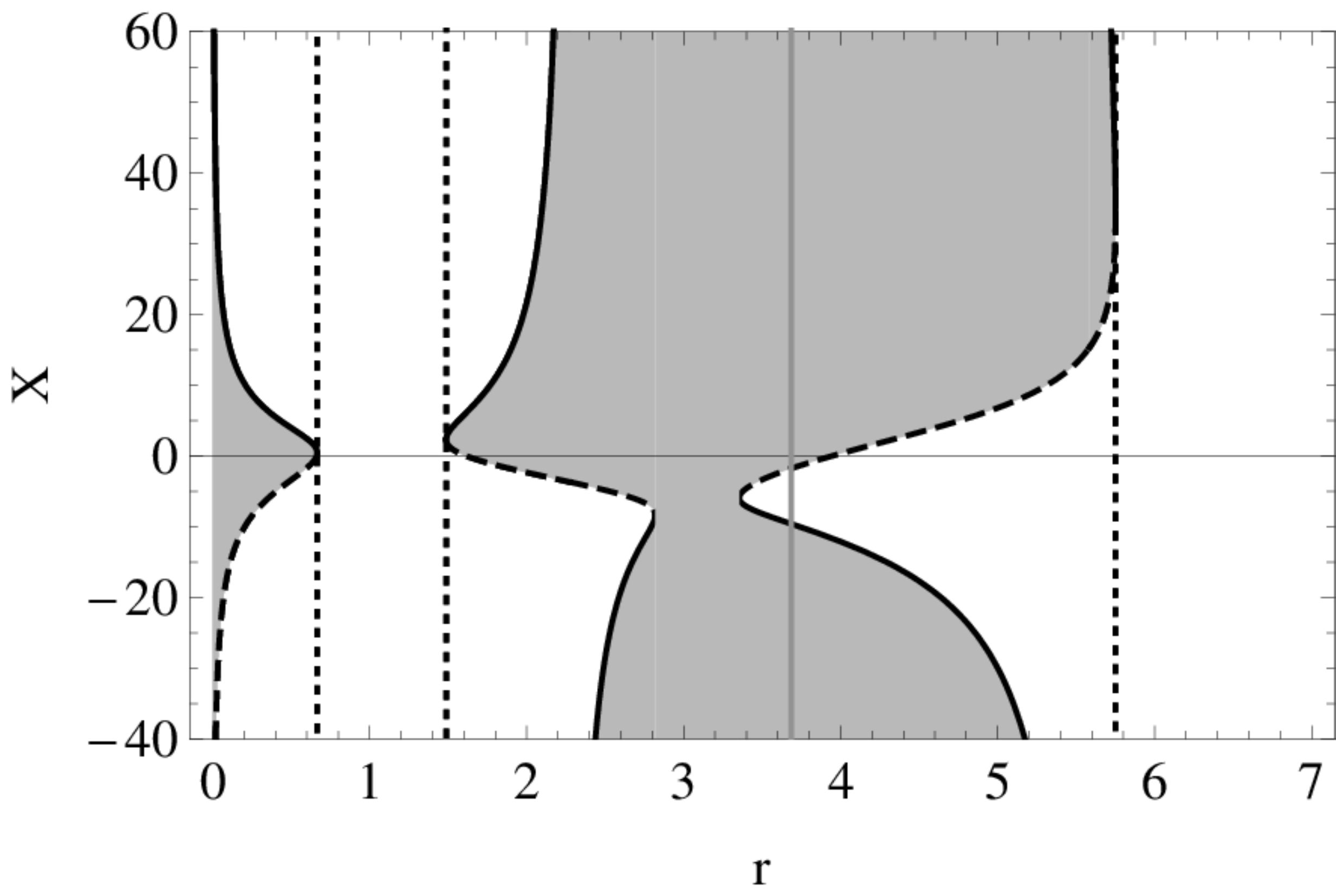} \\
		(i)  $q=50,$ II: $y=0.02,$ $a^2=0.9$ & (j) $q=70,$ II: $y=0.02,$ $a^2=0.9$ \\
		\includegraphics[width=0.48\textwidth]{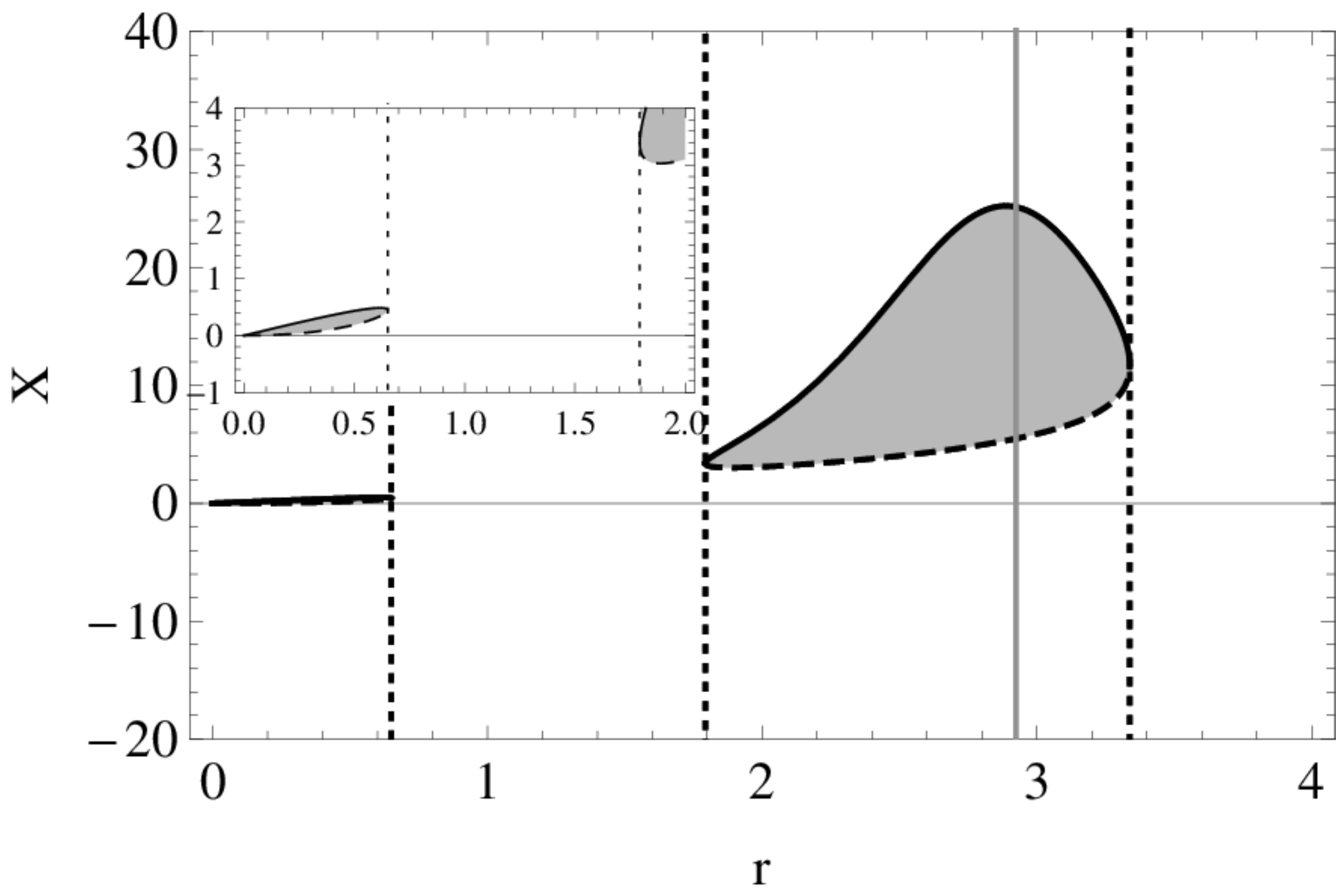}&
		\includegraphics[width=0.48\textwidth]{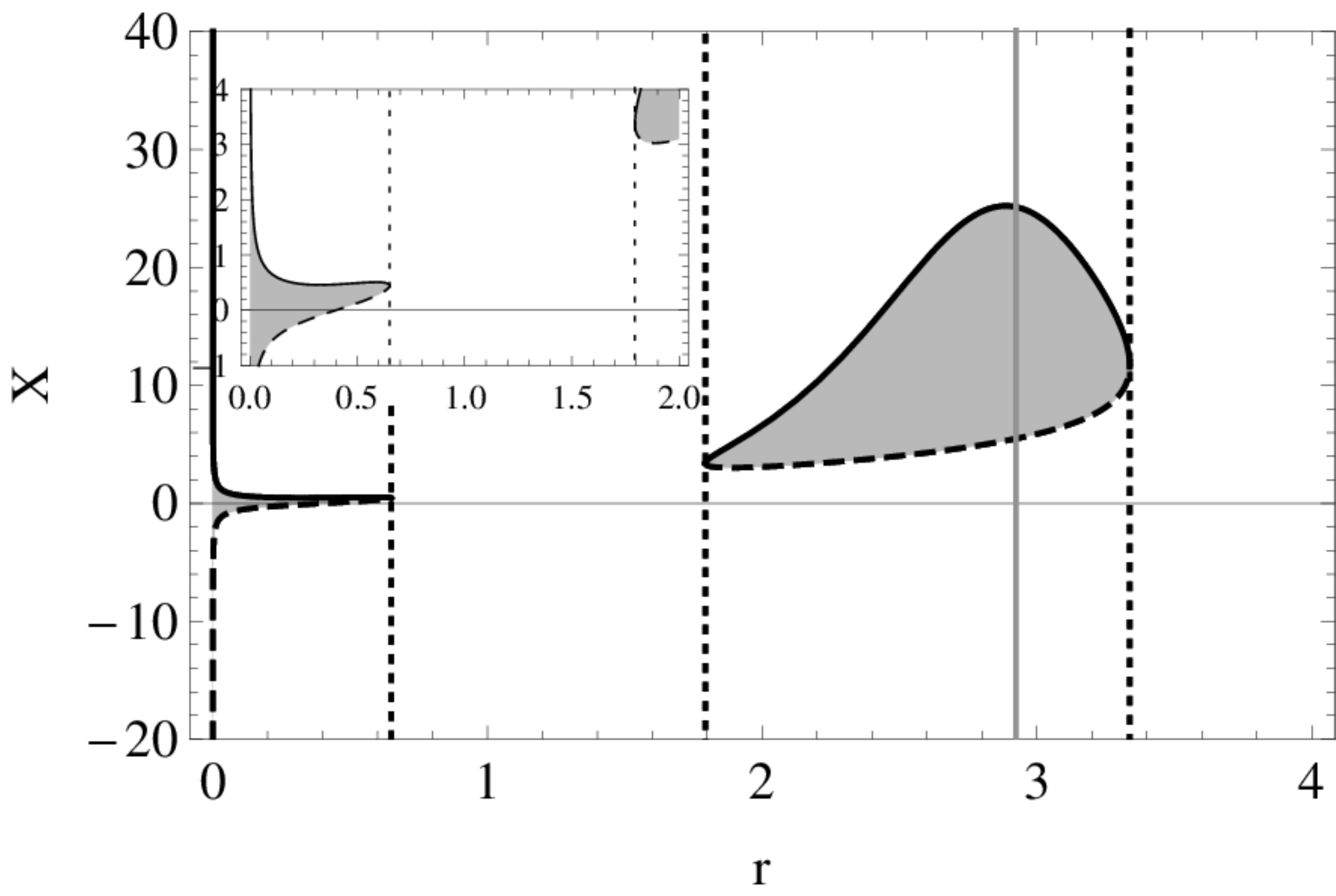} \\
		(k) $q=0,$ III: $y=0.04,$ $a^2=0.9$ & (l)  $q=0.1,$ III: $y=0.04,$ $a^2=0.9$ \\
		\includegraphics[width=0.48\textwidth]{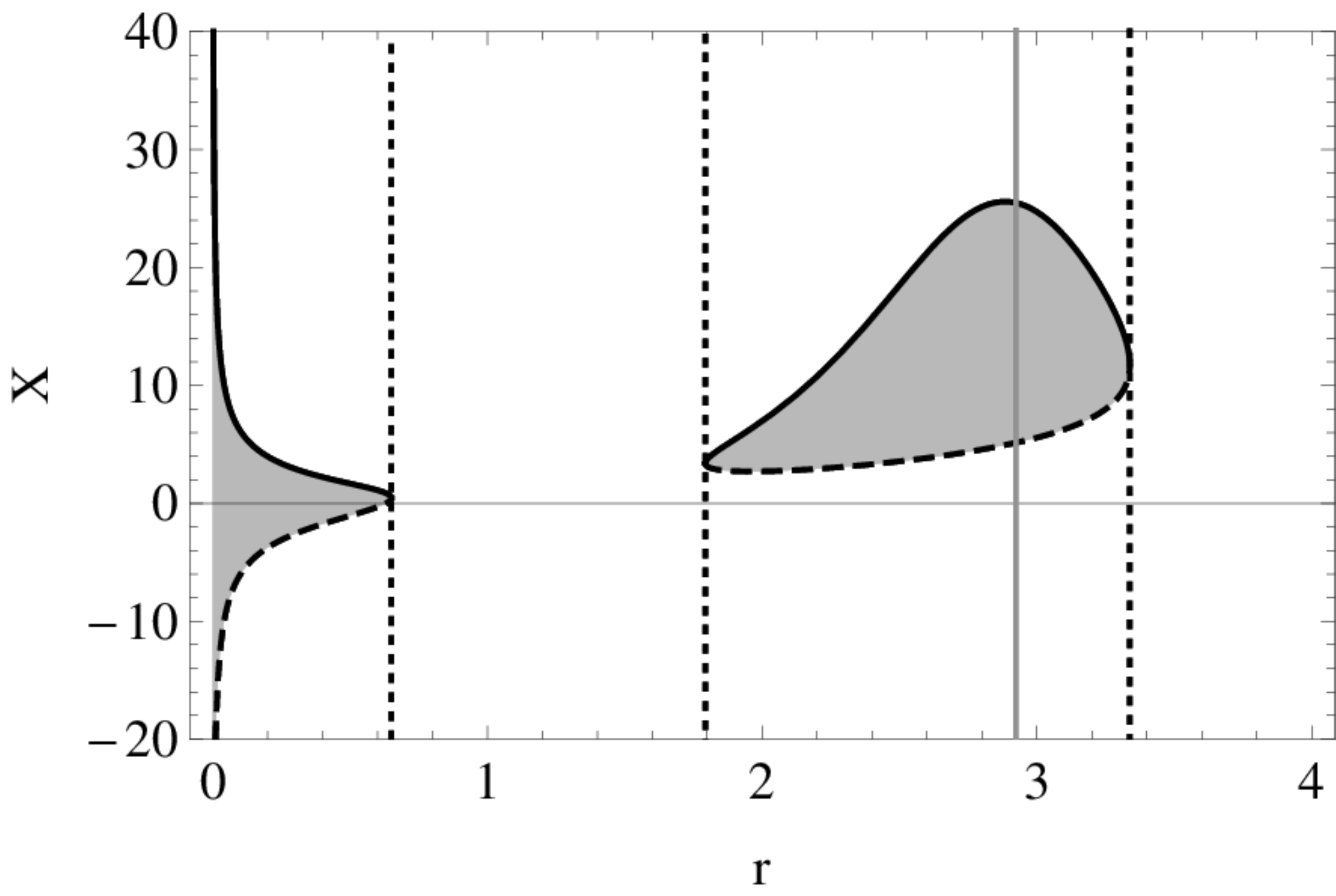} & \includegraphics[width=0.48\textwidth]{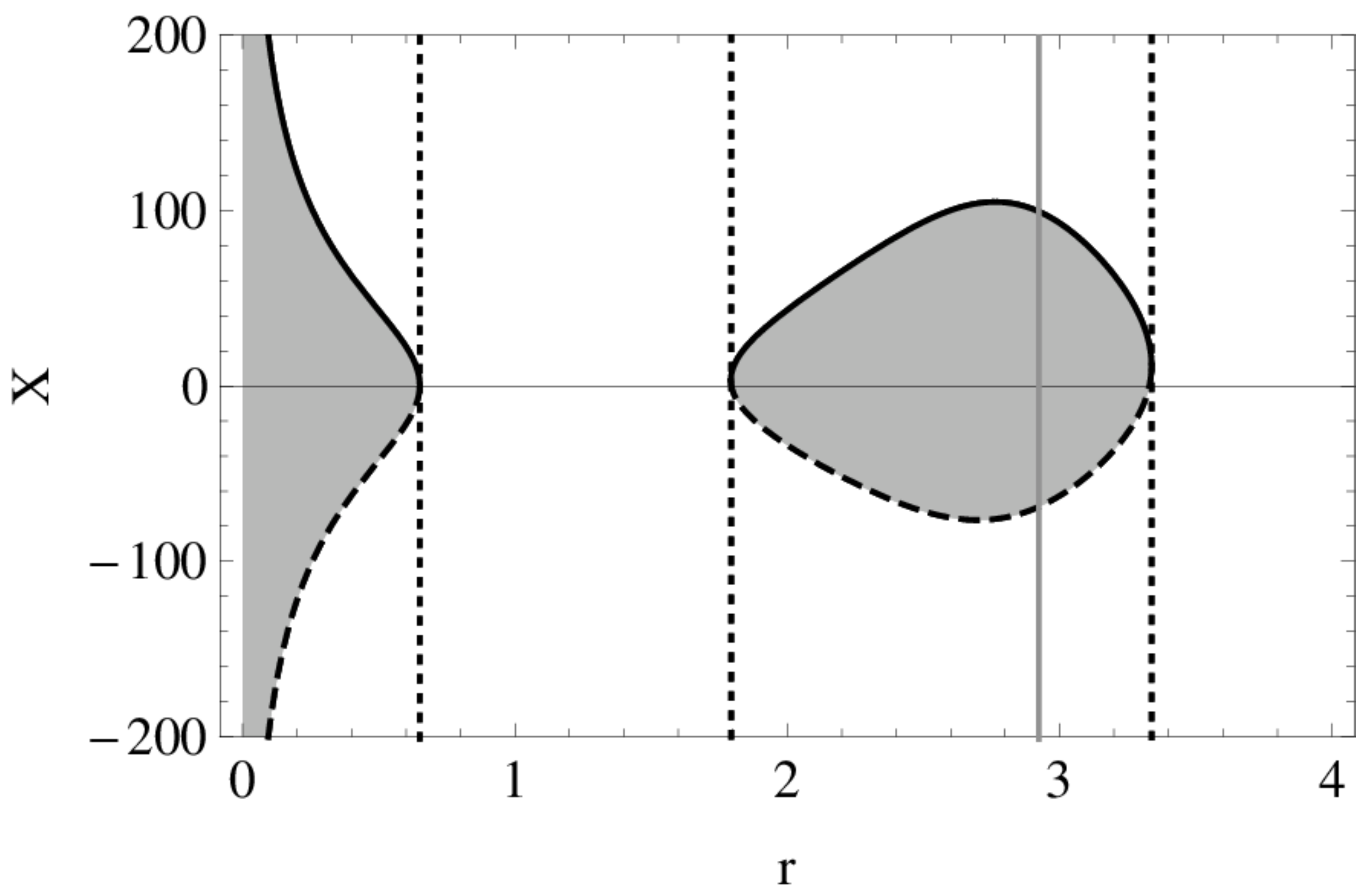} \\
		(m) $q=10,$ III: $y=0.04,$ $a^2=0.9$ & (n) $q=10000,$ III: $y=0.04,$ $a^2=0.9$\\
		\includegraphics[width=0.48\textwidth]{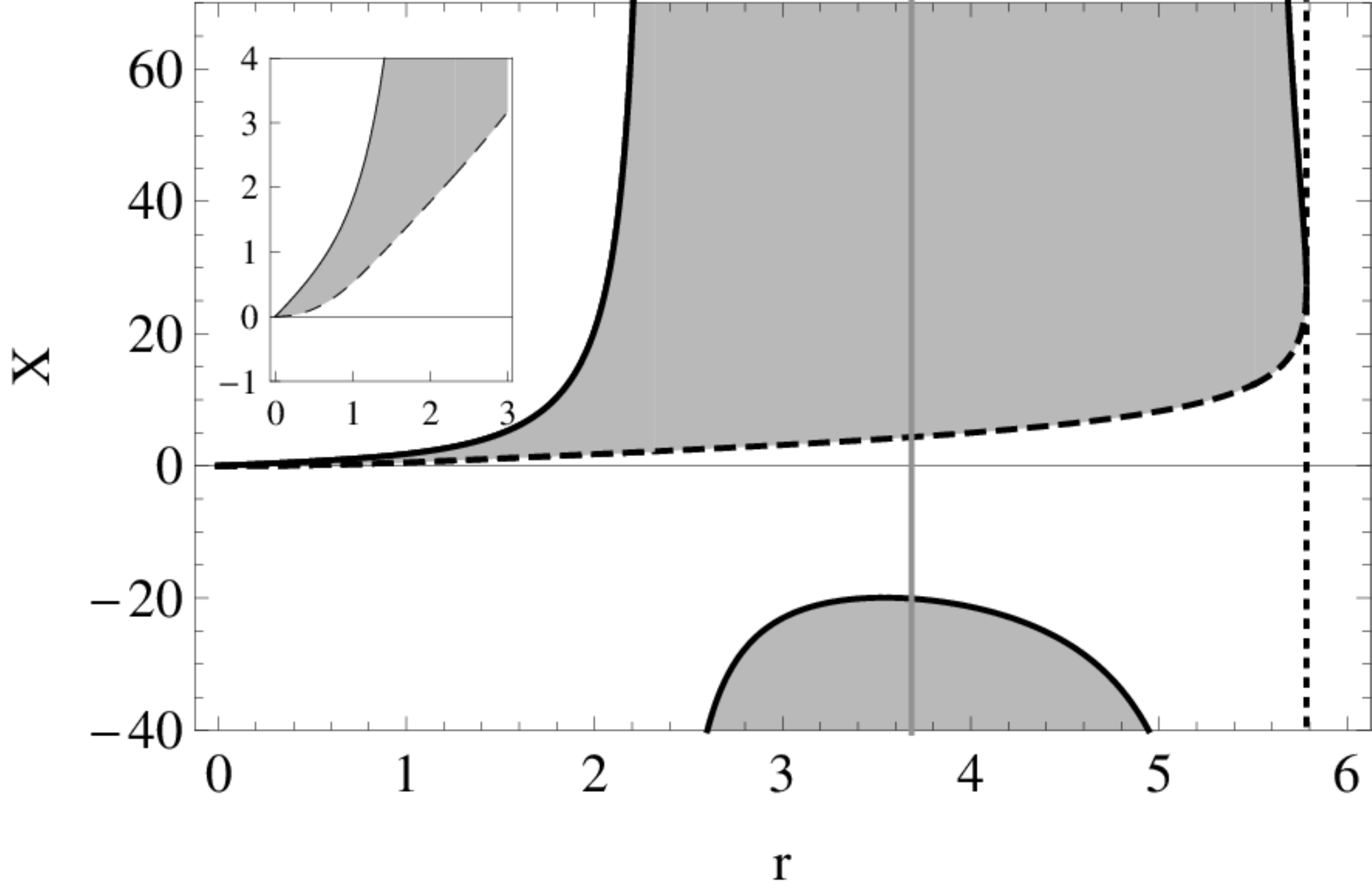} & \includegraphics[width=0.48\textwidth]{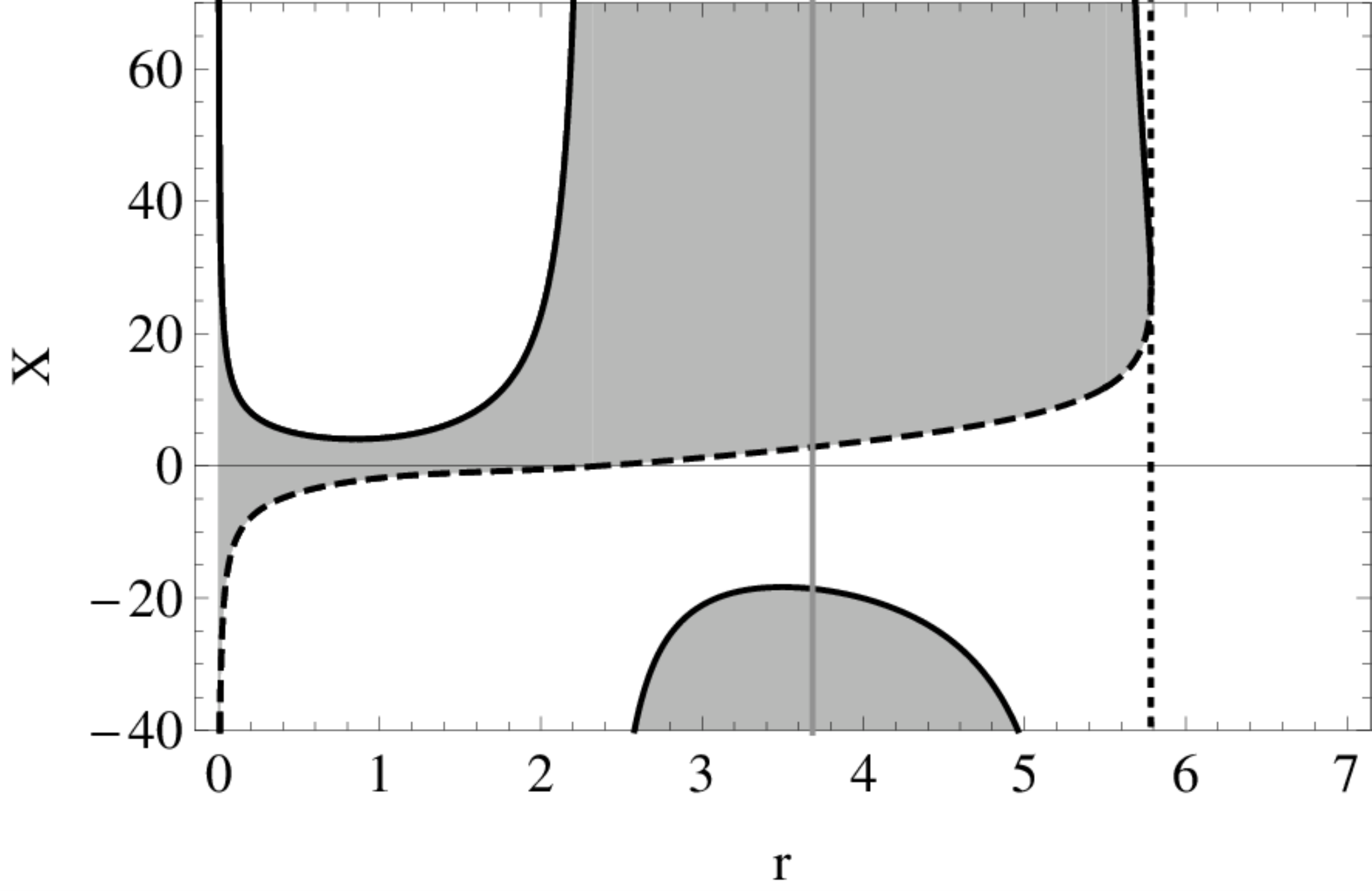}\\
		(o) $q=0,$ IVb: $y=0.02,$ $a^2=1.5$ & (p) $q=20,$ IVb: $y=0.02,$ $a^2=1.5$		
	\end{tabular}
	
\end{figure*}
\begin{figure*}[htbp]
	\centering
	\begin{tabular}{cc}
		\includegraphics[width=0.48\textwidth]{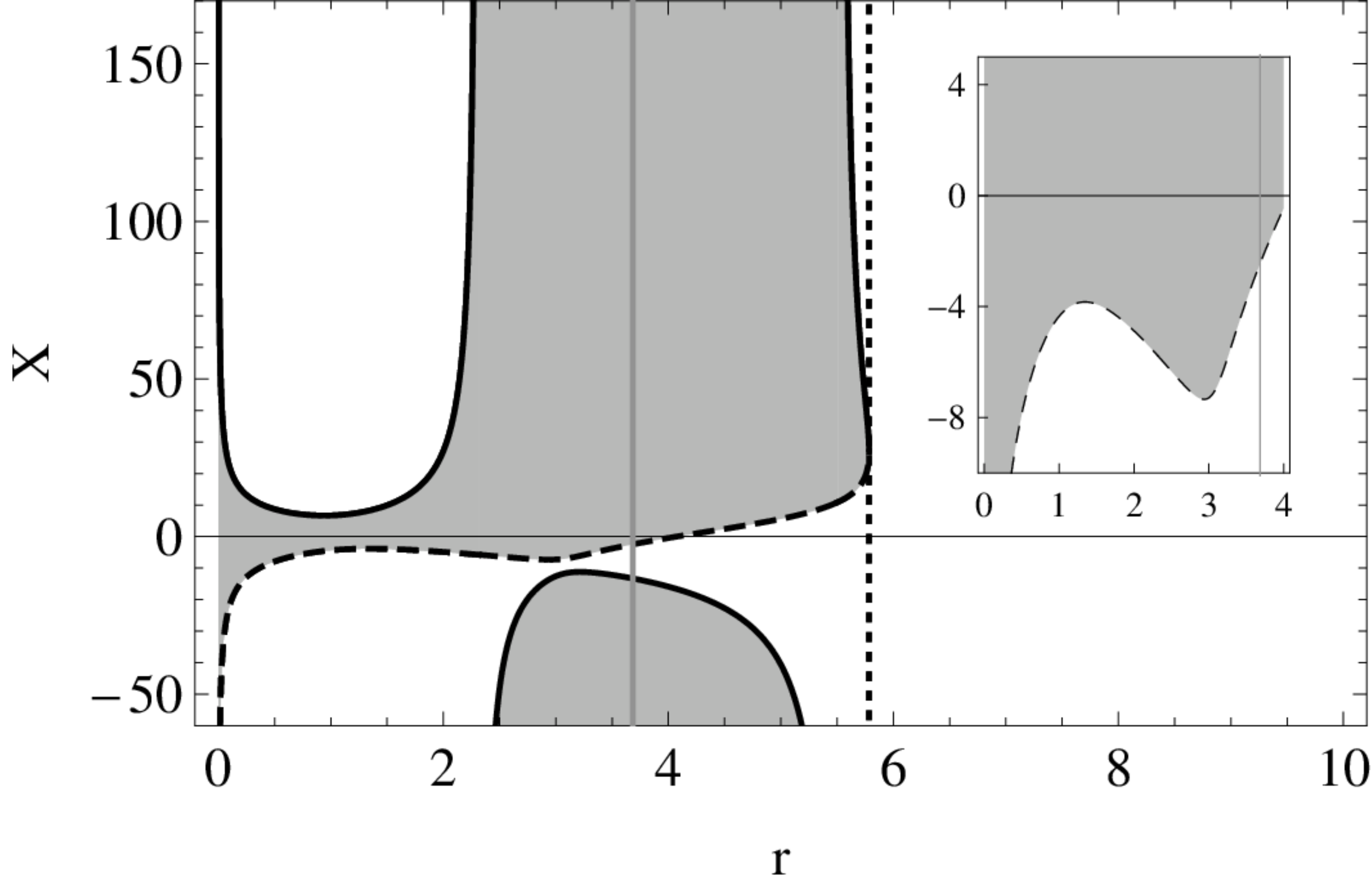} & \includegraphics[width=0.48\textwidth]{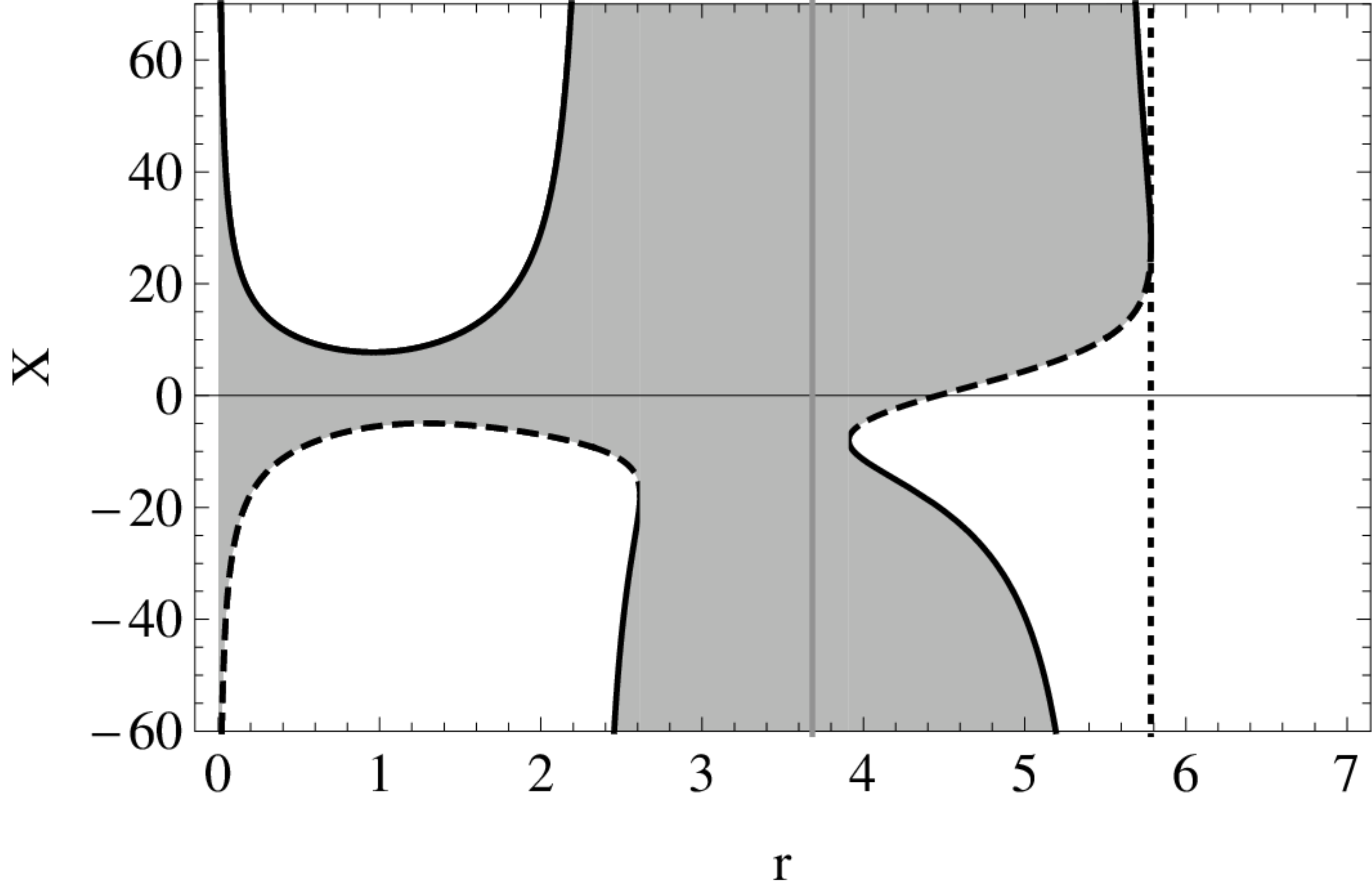}\\
		(q) $q=70,$ IVb: $y=0.02,$ $a^2=1.5$ & (r) $q=100,$ IVb: $y=0.02,$ $a^2=1.5$\\
		\includegraphics[width=0.48\textwidth]{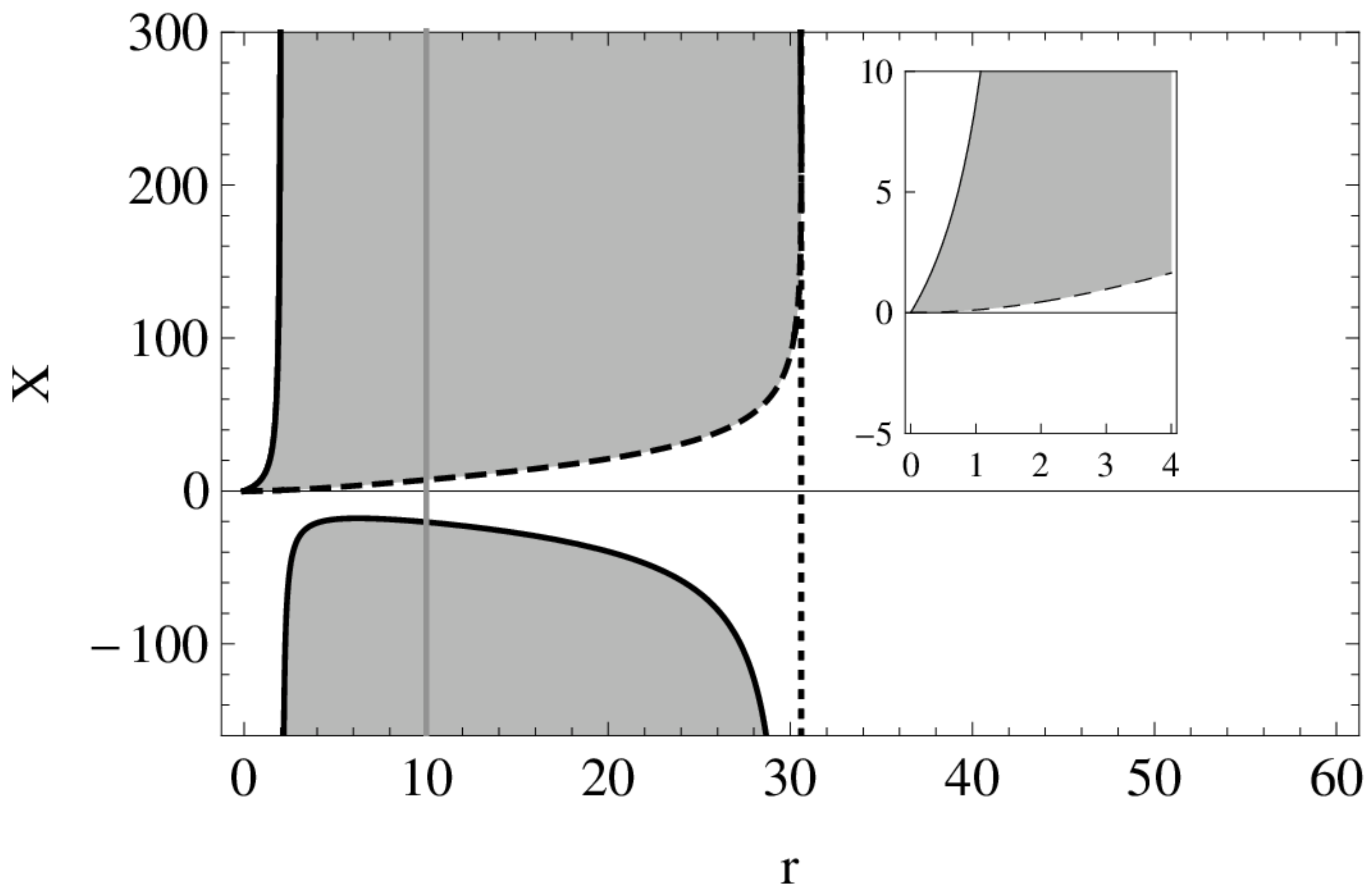} & \includegraphics[width=0.48\textwidth]{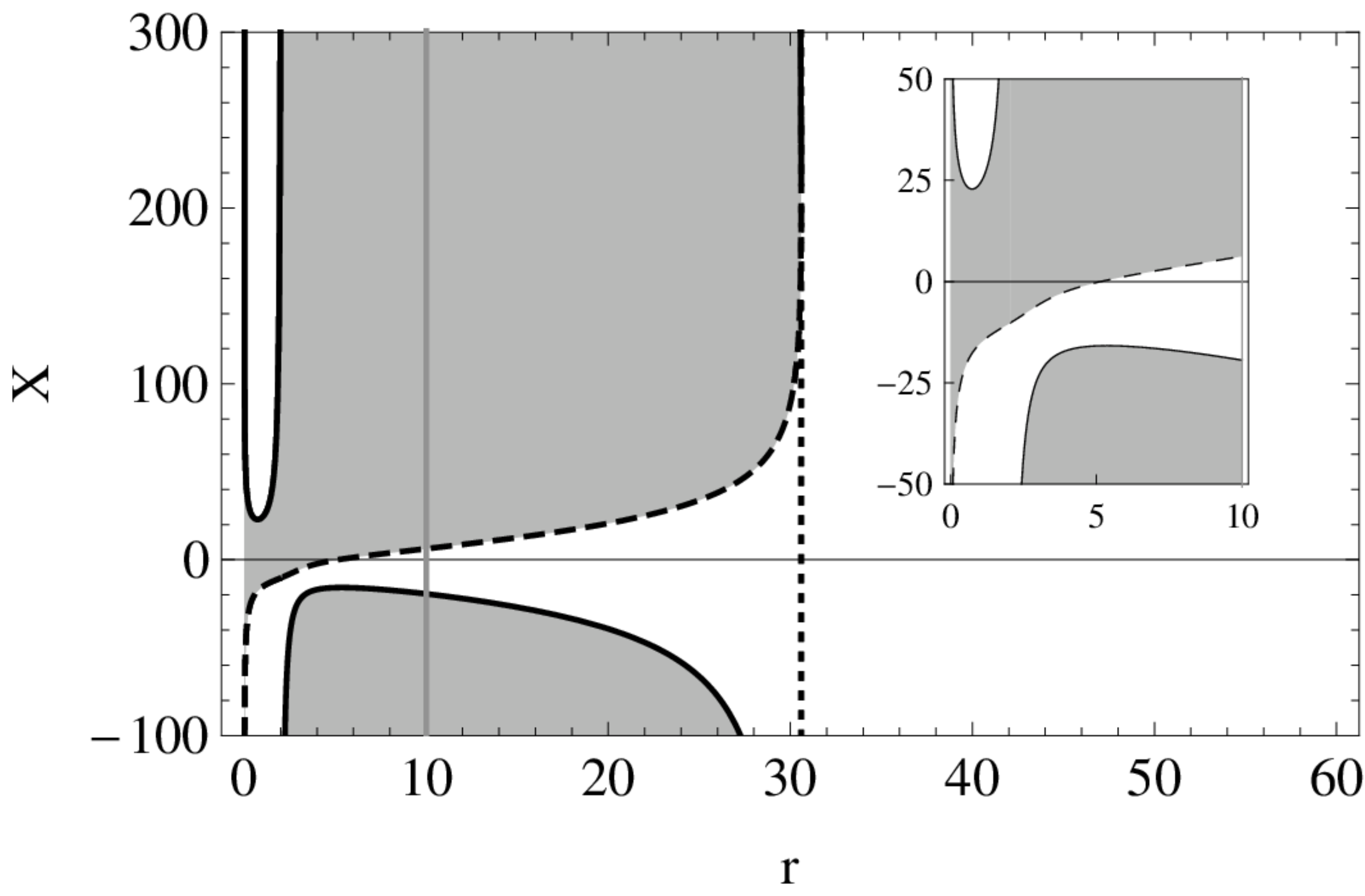}\\
		(s) $q=0,$ V: $y=0.001,$ $a^2=20$ & (t) $q=20,$ V: $y=0.001,$ $a^2=20$\\
		\includegraphics[width=0.48\textwidth]{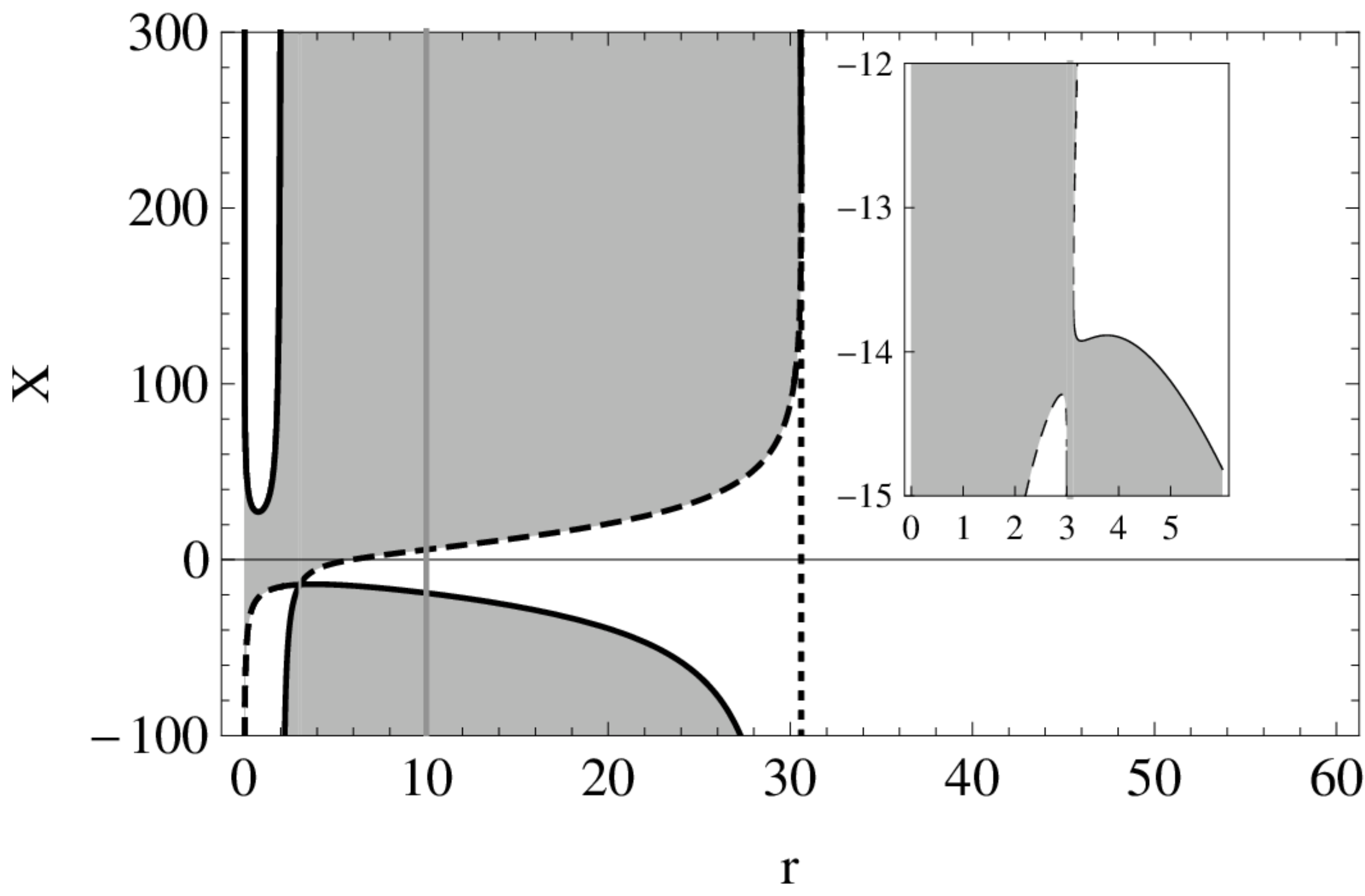} & \includegraphics[width=0.48\textwidth]{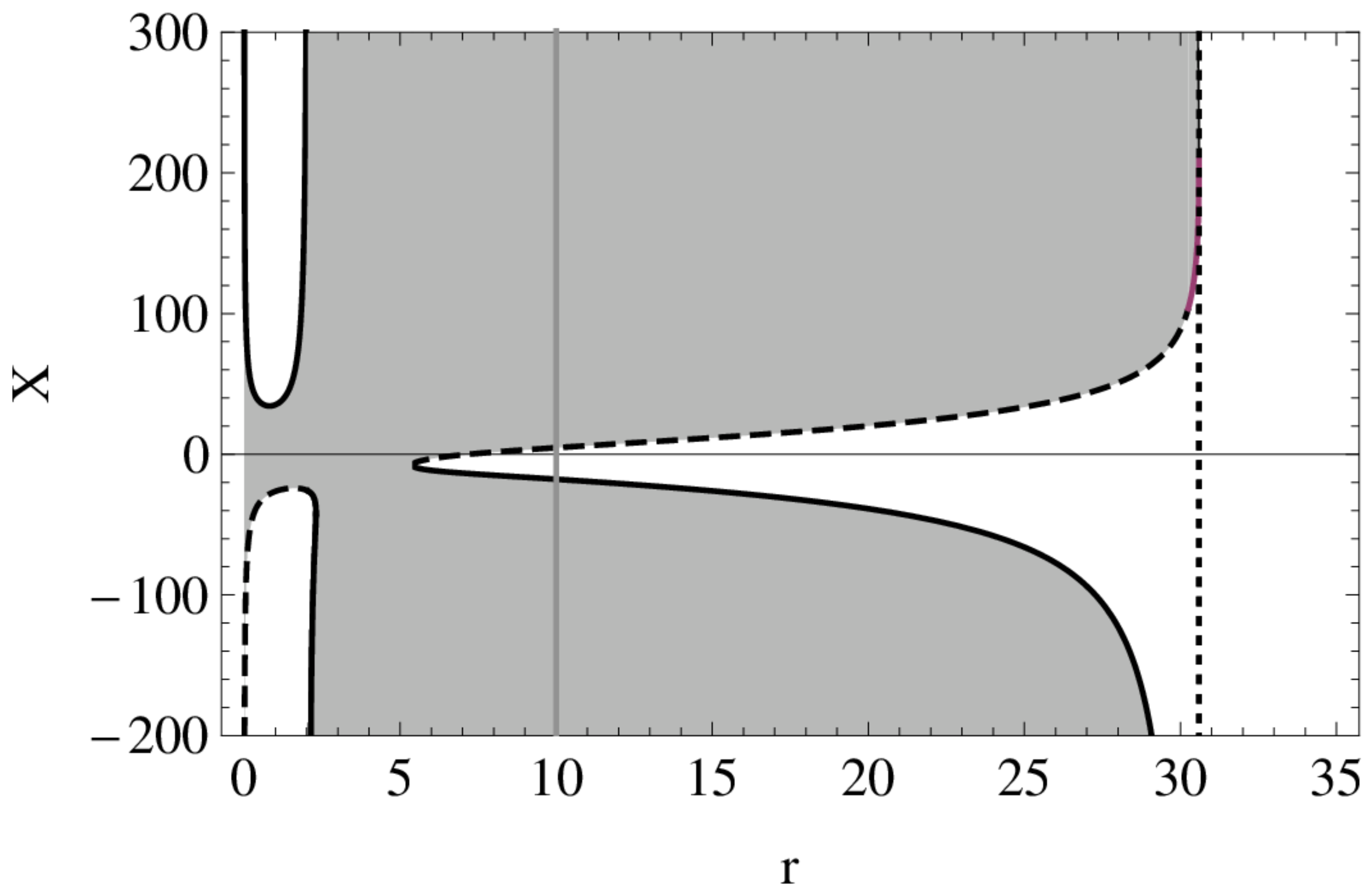}\\
		(u) $q=29.6,$ V: $y=0.001,$ $a^2=20$ & (v) $q=50,$ V: $y=0.001,$ $a^2=20$\\
		\includegraphics[width=0.48\textwidth]{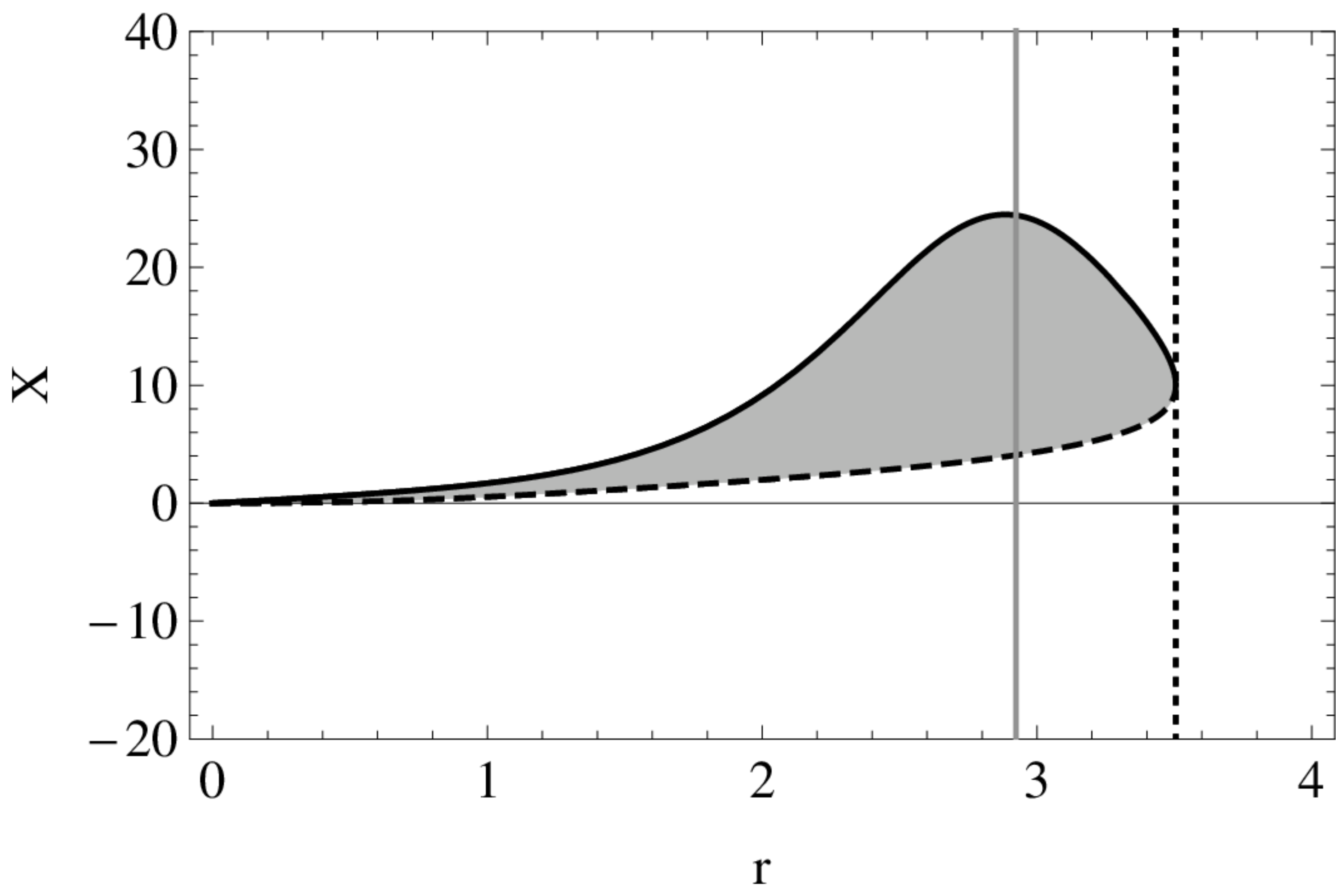} & \includegraphics[width=0.48\textwidth]{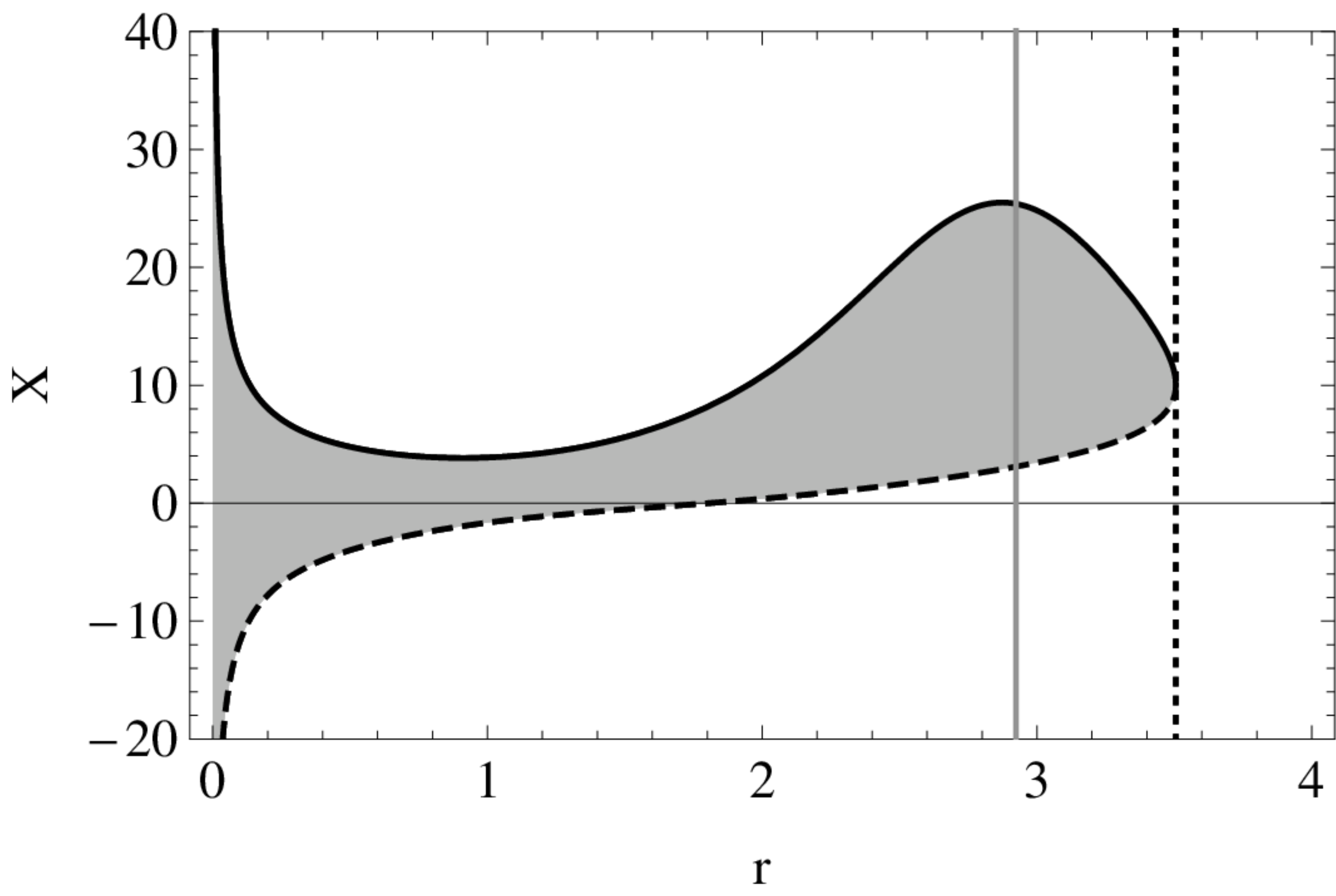}\\
		(w) $q=0,$ VIb: $y=0.04,$ $a^2=1.5$ & (x) $q=20,$ VIb: $y=0.04,$ $a^2=1.5$\\				
	\end{tabular}
	\center (\textit{Figure continued})
\end{figure*}
\begin{figure*}[htbp]
	\centering
	\begin{tabular}{cc}
		
		\includegraphics[width=0.48\textwidth]{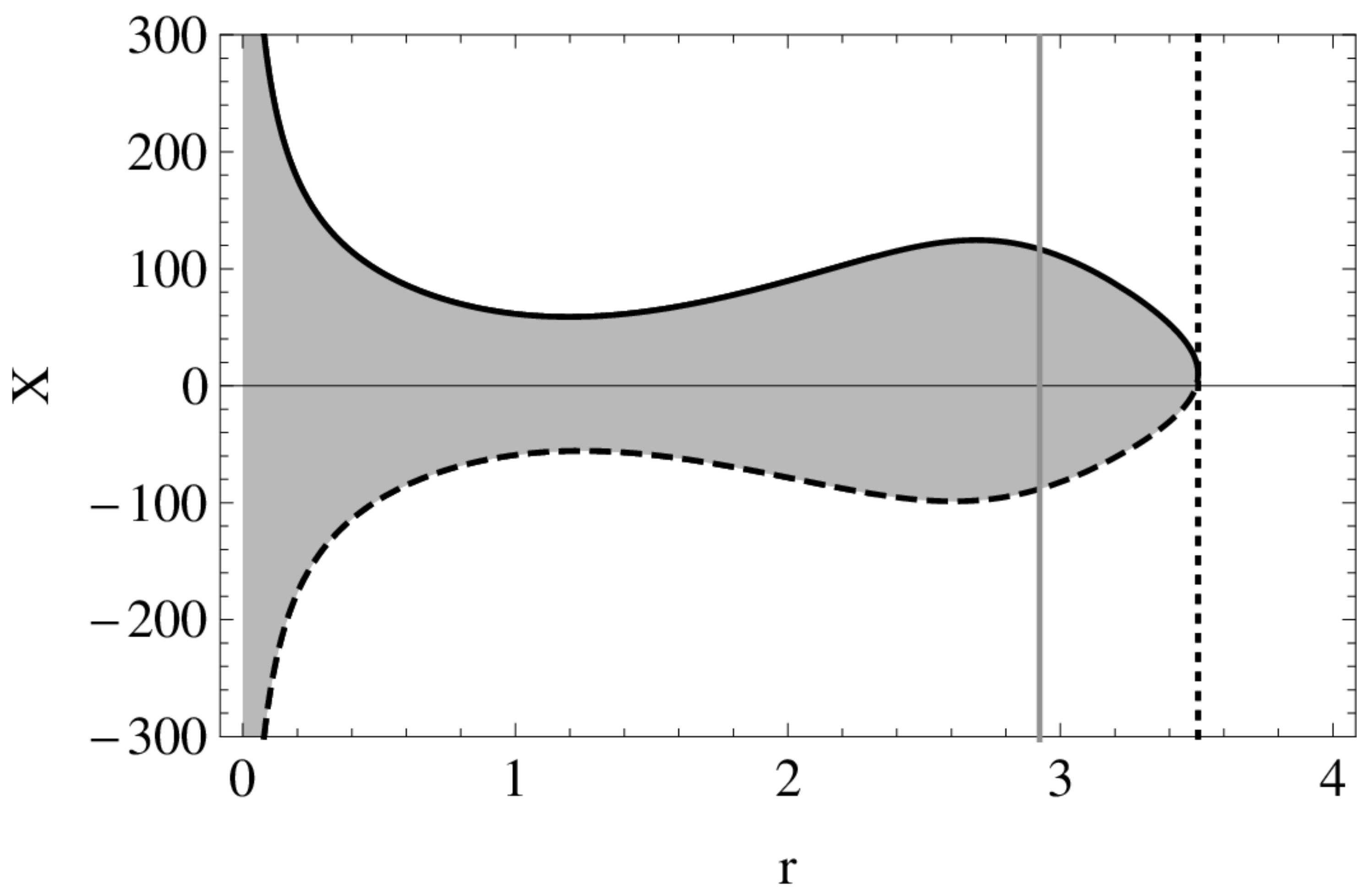} & \includegraphics[width=0.48\textwidth]{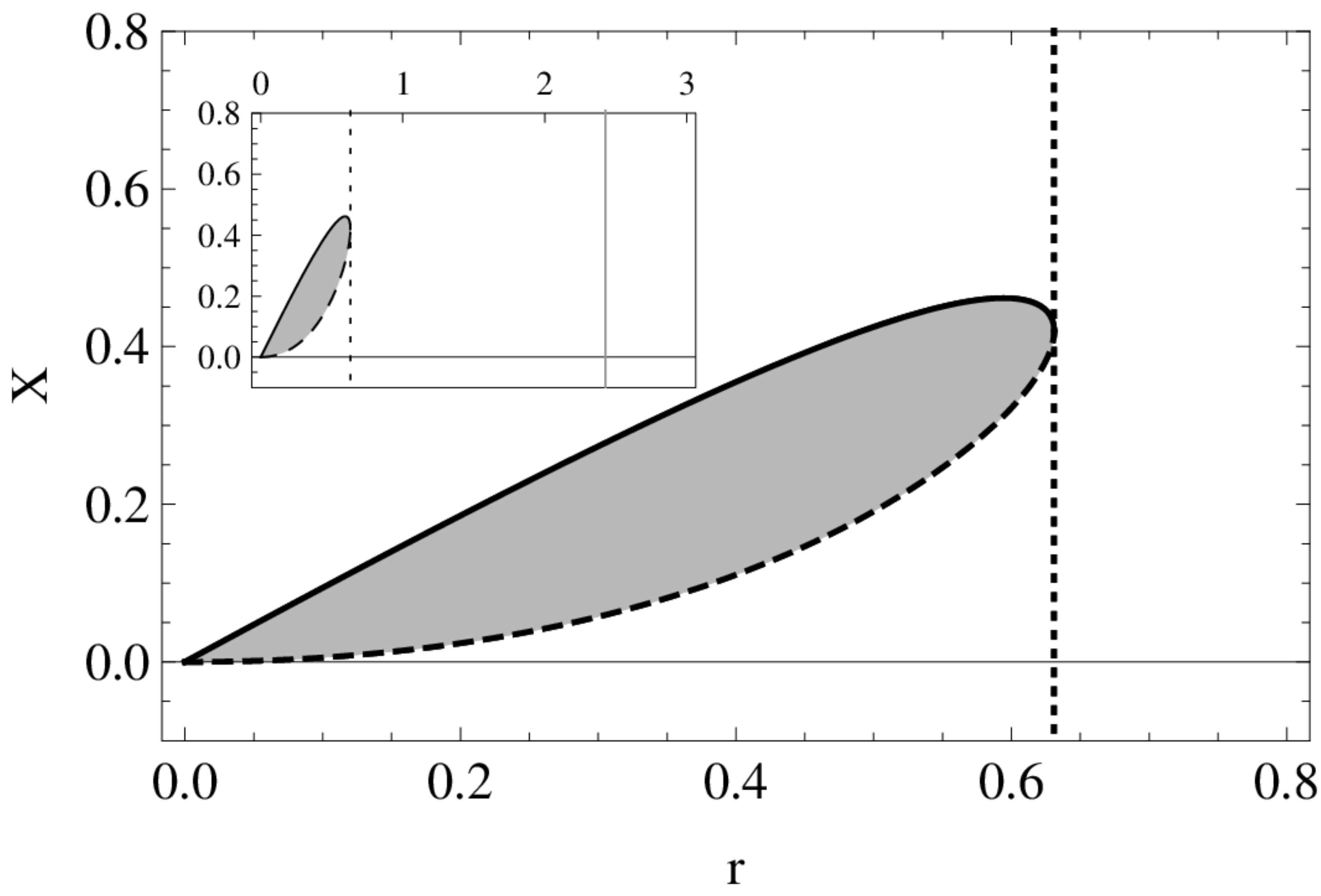}\\
		(y) $q=10000,$ VIb: $y=0.04,$ $a^2=1.5$ & (z)  $q=0,$ VII: $y=0.07,$ $a^2=0.9$\\
		\includegraphics[width=0.48\textwidth]{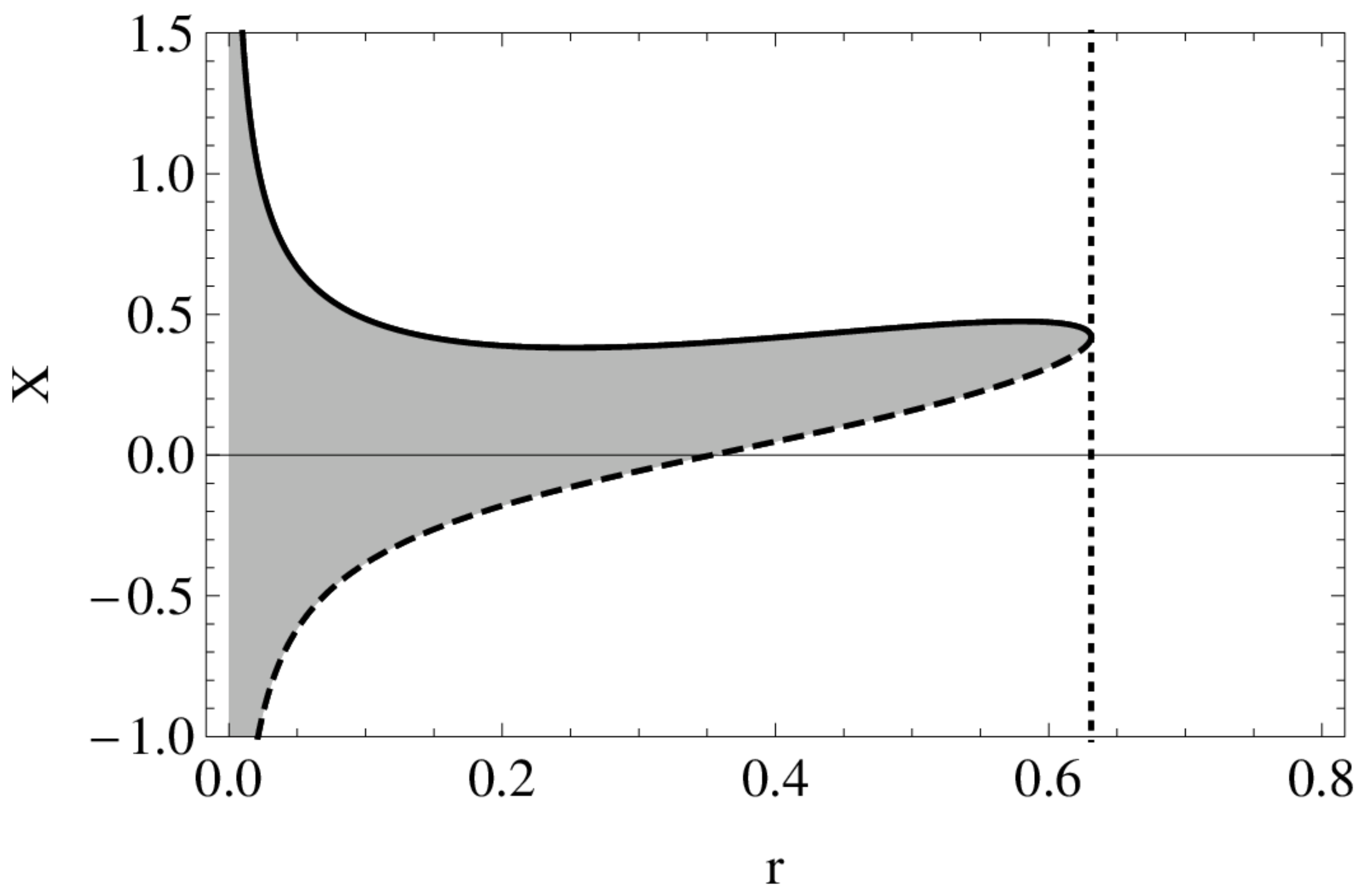} & \includegraphics[width=0.48\textwidth]{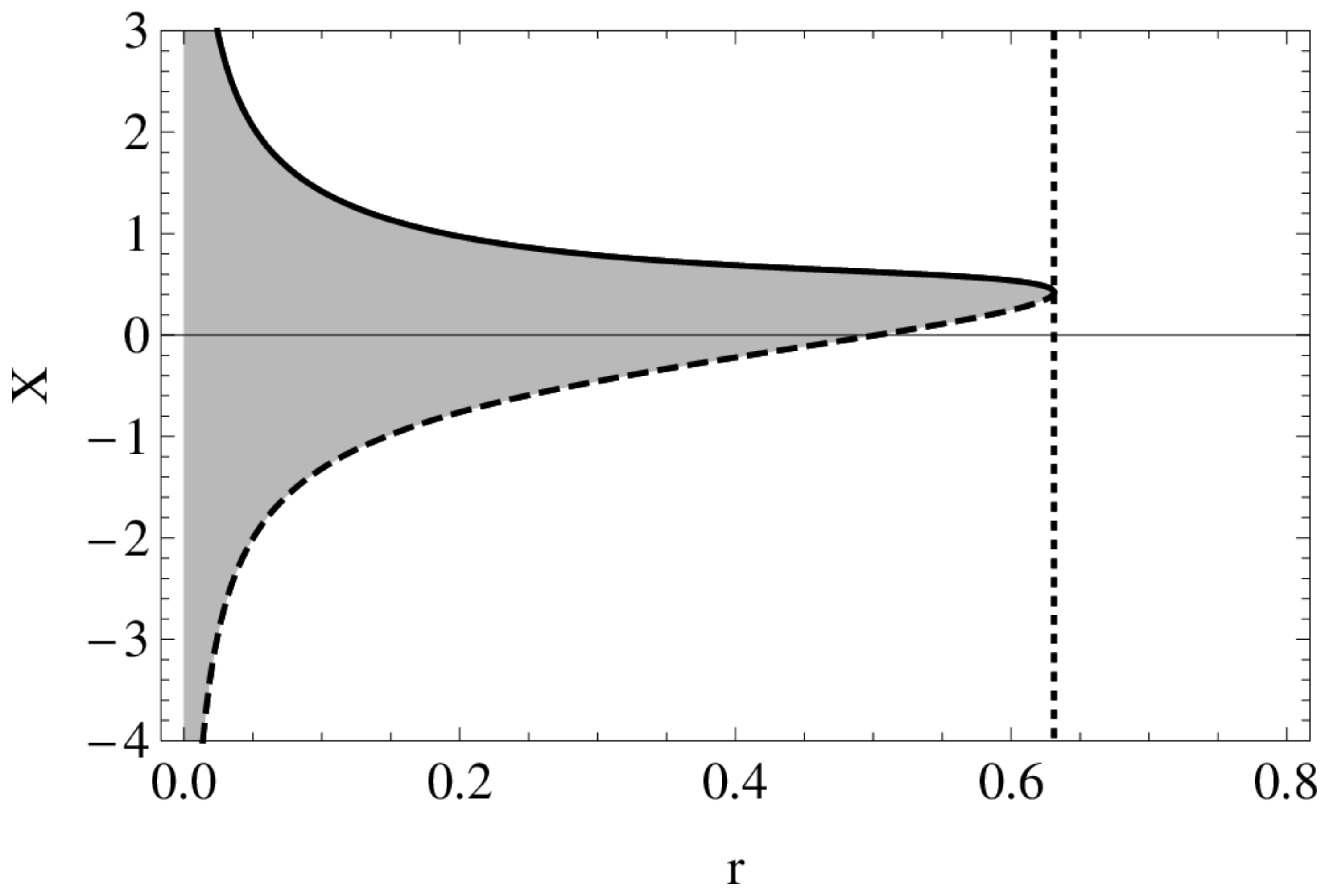}\\
		($\alpha$) $q=0.05,$ VII: $y=0.07,$ $a^2=0.9$ & ($\beta$)  $q=0.5,$ VII: $y=0.07,$ $a^2=0.9$\\
		\includegraphics[width=0.48\textwidth]{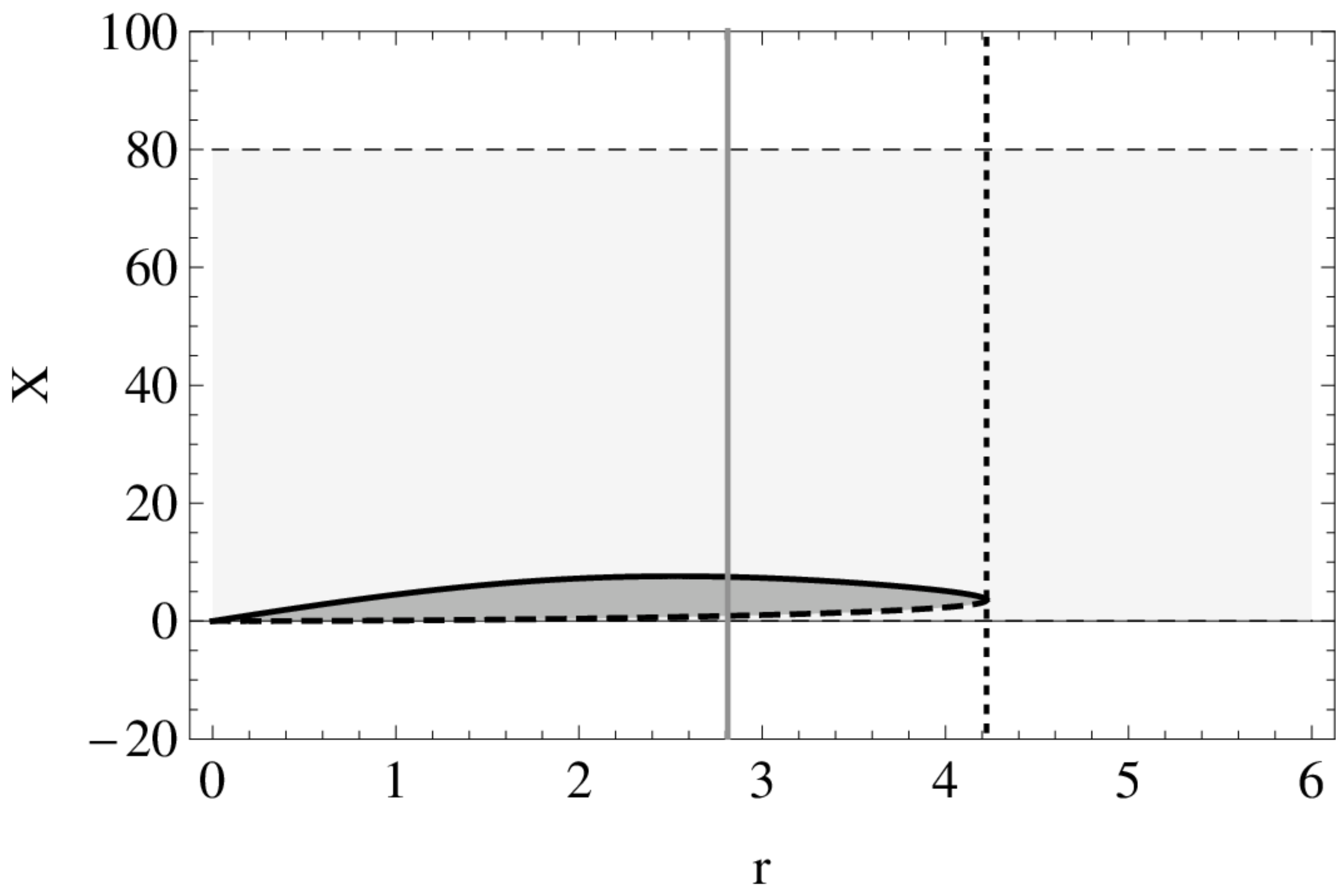} & \includegraphics[width=0.48\textwidth]{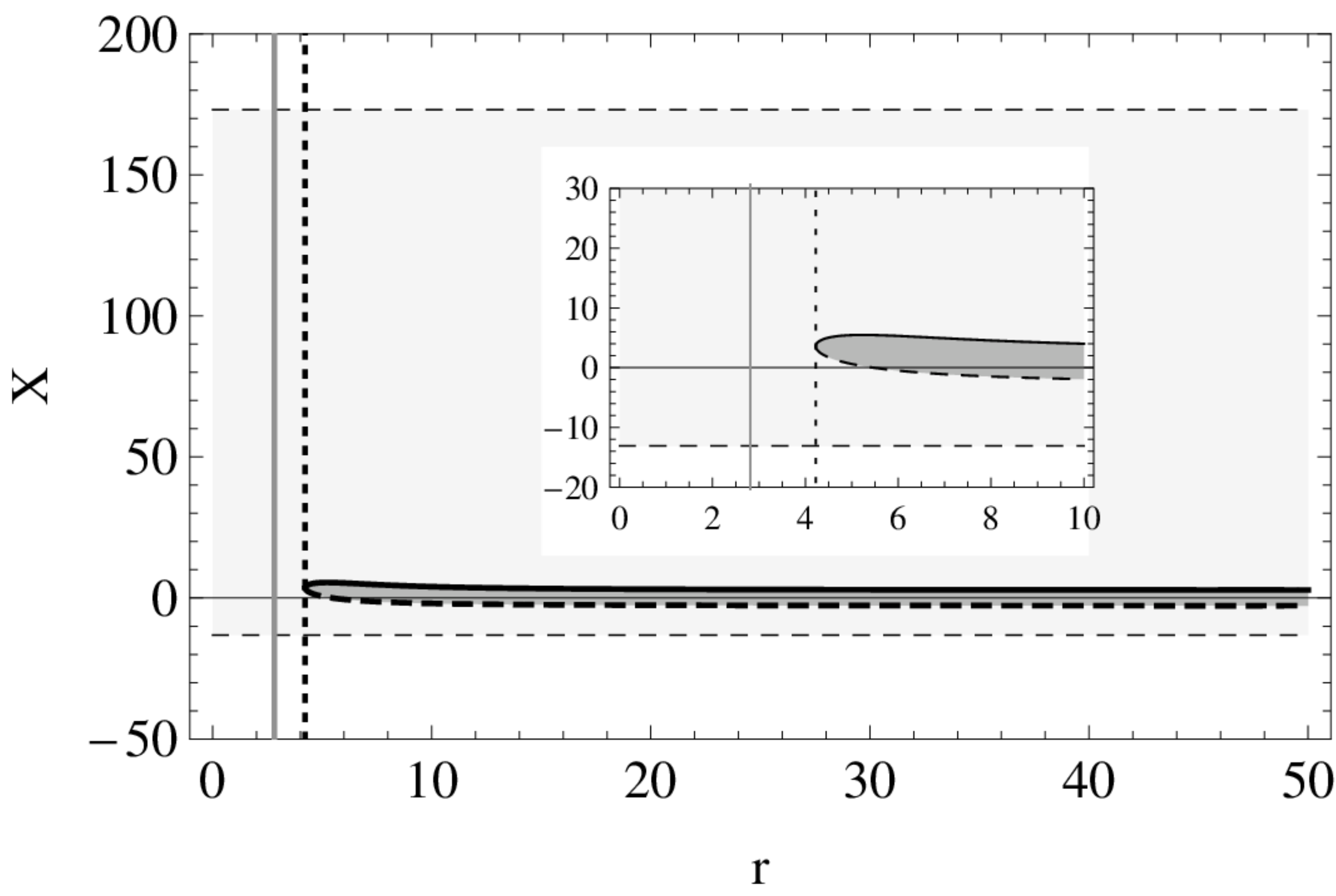}\\
		($\gamma $) $q=-0.04,$ VIII: $y=0.045,$ $a^2=25$ & ($\delta $) $q=-30,$ VIII: $y=0.045,$ $a^2=25$
	\end{tabular}
	
	\caption{Different types of behaviour of the potentials $X_{+}(r;\:q,\:a,\:y)$ (full curve), and $X_{-}(r;\:q,\:a,\:y)$ (dashed curve), with spacetime parameters chosen successively from regions corresponding to the classes I-VIII in the $(a^2-y)$-plane, and some representative values of parameter $q$ divided by the extrema $q(y,\:a^2).$ Vertical dashed lines demarcate loci of the event horizons, the grey line demarcate the static radius, shading highlights the forbidden region. Figs. \textbf{(e)}-\textbf{(g)} depict the case $q<0$ with the limiting values $X^\theta_{min(-)},$ $X^\theta_{max(+)}$ as horizontal dashed lines. The graphs $X_{\pm}(r;\:q,\:a,\:y)$ lie entirely in forbidden region. The behaviour of these functions in other cases differing by values $a,$ $y$ with $q<0$ is qualitatively the same. Figs. \textbf{($\gamma$)}-\textbf{($\delta$)} describes the case $y>1/a^2,$ $q<0.$ The potentials have values in forbidden region again; for $q>0$ the behaviour of the potentials qualitatively corresponds to class VII. In the classes IVb, VIb, the local minima of the potentials $X_{-}(r;\:q,\:a,\:y)$ (see Figs. \textbf{(q)},\textbf{(y)}) have values $X_{min(-)}<-a,$ contrary to the case IVa,VIa, which are not displayed, since there are no other qualitative differences.   }\label{Figure 10}
\end{figure*}
\section{Spherical photon orbits and classification of the \KdS\ spacetimes due to properties of the photon motion}

In the following section we demonstrate by using the behaviour of the effective potentials $X_{\pm}(r;\:q,\:a,\:y)$ that the null geodesics create qualitatively different structures in the various cases of \KdS\ spacetimes with the spacetime parameters chosen from different parts of the $(a^2\mbox{-}y)$-plane labelled by numerals I-VIII. Hence the regions of the spacetime parameter space of these labels can be considered as representatives of the classification of the \KdS\ spacetimes due to the photon motion (null geodesics). Similarly as in \cite{Stu-Hle:2000:CLAQG:}, there are three (four) criteria used -- the main criterion for the classification is the existence (number) of the event horizons. The other differentiating factors follow from the nature of photon motion. First, there is some kind of repulsive barrier defending a light to reach the ring singularity, which is always created in its vicinity for photons with $q>0$. However, a similar barrier can emerge between the outer black hole horizon and the cosmological horizon in black hole and naked singularity spacetimes, repelling photons towards one of these horizons. In the naked singularity spacetimes, occurrence of an additional barrier, which reflects photons towards the ring singularity, leads to occurence of the phenomenon of bound photon orbits. Such bound photon orbits are not present in the case of the black hole spacetimes. The presence and character of this barrier we take as another criterion in the following classification. The other aspect that authorizes us to make such distinction between the KdS spacetimes will be the existence and character of the spherical photon orbits. In the KdS naked singularity spacetimes the bound orbits are concentrated around the stable spherical photon orbits. \\
\newpage

\subsection{Spherical photon orbits}
The spherical photon orbits are determined by the conditions $R(r)=0$ and $dR/dr=0$ that have to be solved simultaneously. The physically acceptable solution is governed by the relations for the photon motion constants $X$ and $q$ that are expressed as functions of the radius $r$ and the spacetime parameters $a,y$, and take the form 
 \be
   X =X_{sph}(r)\equiv \frac{r[(1-a^2y)r^2-3r+2a^2]}{a[yr(2r^2+a^2)-r+1]}, 
\ee   
\be
   q = q_{sph}(r)\equiv q_{ex}(r;\:y,\:a^2),
\ee
where the function $q_{ex}(r;\:y,\:a^2)$ is defined by the relation (\ref{qex2}).
These solutions governing the spherical photon orbits are allowed in the interval of radii limited by the equatorial photon circular orbits. Stability of the spherical photon orbits relative to radial perturbations is determined by the sign of the expression 
\be
      \frac{d^{2}R}{d^{2}r} = 12r^2[1+y(X^2+q)]+2(X^2+q)(a^2y-1)-4aX, \label{ddR} 
\ee
evaluated at appropriate radii. It can be shown that local maxima of the potential $X_{+}$ and local minima of $X_{-}$ correspond to unstable orbits in the black hole spacetimes. However, local minima of $X_{+}$ and local maxima of $X_{-}$ represent stable orbits, which occur in the naked singularity spacetimes.

It is useful to relate the parameters (wave vector components) of photons orbiting along the spherical null geodesics to the locally non-rotating frames (LNRFs) that are the most convenient frames for description of physical processes in the Kerr(dS) spacetimes \cite{Bardeen:1973}.
The LNRF tetrad of differential one-forms is given by the relations

\bea
\omega^{(t)}&=&\sqrt{\frac{\Delta_{r} \Delta_{\theta} \rho^2}{I^2 A}} \din t, \\
\omega^{(r)}&=&\sqrt{\frac{\rho^2}{\Delta_{r}}} \din r,\\
\omega^{(\theta)}&=&\sqrt{\frac{\rho^2}{\Delta_{\theta}}} \din \theta,\\
\omega^{(\phi)}&=&\sqrt{\frac{A \sin ^2 \theta}{I^2 \rho^2}}(\din \phi-\Omega_{LNRF} \din t)\label{o_i_LNRF}, 
\eea
the corresponding tetrad of dual vectors reads 
\bea
     e_{(t)}&=&\sqrt{\frac{I^2A}{\Delta_r \Delta_{\theta} \rho^2}}\left( \frac{\partial}{\partial t}+\Omega_{LNRF} \frac{\partial}{\partial \phi}\right) ,\\
     e_{(r)}&=&\sqrt{\frac{\Delta_{r}}{\rho^2}}\frac{\partial}{\partial r},\\
     e_{(\theta)}&=&\sqrt{\frac{\Delta_{\theta}}{\rho^2}}\frac{\partial}{\partial \theta},\\
     e_{(\phi)}&=&\sqrt{\frac{I^2 \rho^2}{A \sin^2\theta}}\frac{\partial}{\partial \phi}. \label{e_i_LNRF}
\eea
The wave-vector components related to the LNRFs are then determined by the relations 
\be k^{(a)}= \omega^{(a)}_{\mu}k^{\mu}, \quad k_{(b)}=k_{\nu} e_{(b)}^{\nu}, \ee
hence
\bea
k^{(t)}&=&I E \sqrt{\frac{A}{\Delta_r \Delta_{\theta} \rho^2}} (1-\Omega_{LNRF}(X+a)),\label{k(t)} \\ 
k^{(r)}&=&\pm \frac{I E}{\sqrt{\Delta_{r}\rho^2}}\left[(r^2-aX)^2-\Delta_{r}(X^2+q)\right],\label{k(r)} \\ 
k^{(\phi)}&=&I E \sqrt{\frac{\rho^2}{A \sin^2\theta}}(X+a), \label{k(phi)} \\
k^{(\theta)}&=&\pm \frac{I E}{\sqrt{\Delta_{\theta}\rho^2}}\times \\ \nonumber
&&\sqrt{(X^2+q)\Delta_\theta-\frac{(a\cos^2\theta+X)^2}{\sin^2\theta}},
\eea
where
\be
A = (r^2+a^2)^2-a^2\Delta_{r}\sin^2\theta,
\ee
and
\be
\Omega_{LNRF} = \frac{a[(r^2+a^2)\Delta_{\theta}-\Delta_{r}]}{A}
\ee
is the angular velocity of the LNRFs related to distant static observers.

In order to determine the orientation of the spherical orbits, we have chosen as the azimuthal direction indicator the sign of the ratio $k^{(\phi)}/k^{(t)}.$ If we define the directional angle $\Psi$ in such a way that $\Psi=0$ for motion in the direction of the latitudinal tetrad vector $e_{(\theta)},$ while $\Psi=\pi/2$ for motion in the direction of the azimuthal tetrad vector $e_{(\phi)},$ then $k^{(\phi)}/k^{(t)}=\sin \Psi$ and we find the relation 
\be
sin\Psi = \frac{\rho^2\sqrt{\Delta_{r}\Delta_{\theta}}}{A \sin\theta}\frac{X+a}{1-\Omega_{LNRF}(X+a)} \label{sinpsi(X)}.
\ee 
If the sign of $sin\Psi$ is positive, we call the spherical orbit prograde, if it is negative, we call the spherical orbit retrograde. Special case of limiting spherical orbits corresponds to the equatorial circular orbits that are again co-rotating (prograde), respectively counter-rotating (retrograde). It can be shown that the sign of the directional angle remains fixed at any latitude of any particular spherical orbit, i.e., the locally non-rotating observers see the photon motion in fixed azimuthal direction. \footnote{However, we have to note that, similarly to the case of Kerr black holes \cite{Teo:2003:GenRelGrav:}, the sign of the variation of the azimuthal coordinate can be changed at some latitude, if related to distant observers.} Since the functions $A, \Omega_{LNRF}$ are positive \cite{Stu-Hle:2000:CLAQG:}, it is clear that all photons with $X<-a$ ($\ell<0$) are retrograde. However, photons with 
\be
X>\frac{1}{\Omega_{LNRF}}-a \quad (\ell>1/\Omega_{LNRF}>0) \label{positel}
\ee
can be retrograde as well. Considering in such a case the relation for the tetrad LNRF component (\ref{k(t)}), we can see that in order to keep for $k^{(t)}$ the standard physical meaning, i.e., $k^{(t)}>0,$ we have to put $E<0.$ \footnote{Keeping $E>0$ means $k^{(t)}<0,$ i.e., a photon in negative-root state with time evolution directed to past -- for details see \cite{Bic-Stu-Bal:1989:BAC:}.} In order to find conditions under which such a situation occurs, it is convenient to express from the alternate relation  
\be
\sin\Psi = \frac{\rho^2\sqrt{\Delta_{r}\Delta_{\theta}}}{A \sin\theta}\frac{\ell}{1-\Omega_{LNRF}\ell} \label{sinpsi(el)}
\ee 
the impact parameter $\ell$ in the form
\be
\ell=\frac{A\sin\psi \sin\theta}{\sqrt{\Delta_{r}\Delta_{\theta}}\rho^2+A\Omega_{LNRF} \sin\psi \sin\theta} \label{ell(psi)}
\ee 
and reverse the problem by searching for conditions, under which there is $\ell>0.$ Such a relation is evidently fulfilled if $\sin \psi >0,$ i.e., the positive impact parameters pertain to prograde photons. However, there is another possibility, to have $\sin \psi <0$ together with 
\be
-1\le \sin \psi < -\frac{\rho^2 \sqrt{\Delta_{r}\Delta_{\theta}}}{A \Omega \sin \theta},
\ee
from which it follows
\be
A\Omega \sin \theta - \rho^2 \sqrt{\Delta_{r}\Delta_{\theta}}\ge0.
\ee	
However, the last inequality can be written in the form
\be
 A I^2\rho^2 g_{tt}\ge0,
\ee
which implies 
\be
g_{tt}\ge0.
\ee
Hence, such a situation can occur only in the ergosphere. Of course, the impact parameter of such photons must fulfil the condition (\ref{positel}). The function
\be
\frac{1}{\Omega}-a=\frac{r^2(r^2+a^2)\Delta_{\theta}+a^2\Delta_{r}\cos\theta^2}{a[(r^2+a^2)\Delta_{\theta}-\Delta_{r}]}
\ee
has for $\Delta_{r}=0$ common points with the potentials $X_{\pm}$ given by the relation (\ref{Xpm(rh)}). There are no other intersections with the potentials, hence, the reality condition of the radial motion together with (\ref{positel}) imply $X>X_{+}>0.$ Therefore, the motion of photons with negative energy $E,$ which appear to be retrograde in the LNRFs, is governed by effective potentials $X_{+}(r;\:q,\:a,\:y)$ with positive values.\\
	
Using the properties of the effective potentials $X_{\pm}(r;\:q,\:a,\:y)$, we can identify the radii $r=r_{0}$ of the spherical photon orbits as loci of the local extrema of the effective potentials and determine their stability and orientation as described above. At each allowed radius $r_{0}$, located between the radii of the equatorial photon circular orbits, we can assign corresponding limits $\theta_{min},$ $\theta_{max}$ on the latitudinal motion by solving the equation
\be
M(m;\:X_{sph}(r_{0}),\:q_{sph}(r_{0}),\:y,\:a)=0, \label{marg.m}
\ee
which due to he results of Section 3 has one real positive root $m_{0},$ since $q_{sph}(r_{0})\geq 0$ ($q_{sph}(r_{0})=0$ for $r_{0}=r_{ph\pm},$ i.e., equatorial circular co-rotating or counter-rotating photon orbit). The marginal latitudes (turning points of the latitudinal motion) then read
\be
\theta_{min}=\arccos \sqrt{m_0}, \quad \theta_{max}=\pi-\arccos \sqrt{m_0} ,
\ee
for details see the discussion of the latitudinal motion in Section 3. We can thus easily determine for a spherical orbit at an allowed radius $r_0$ the impact parameters of the orbit, the extension of the latitudinal motion, and the orientation of the azimuthal motion. 

\subsection{Classification}
\begin{figure*}[htbp]
	\centering
	
	\begin{tabular}{c}
		\includegraphics[width=0.67\textwidth]{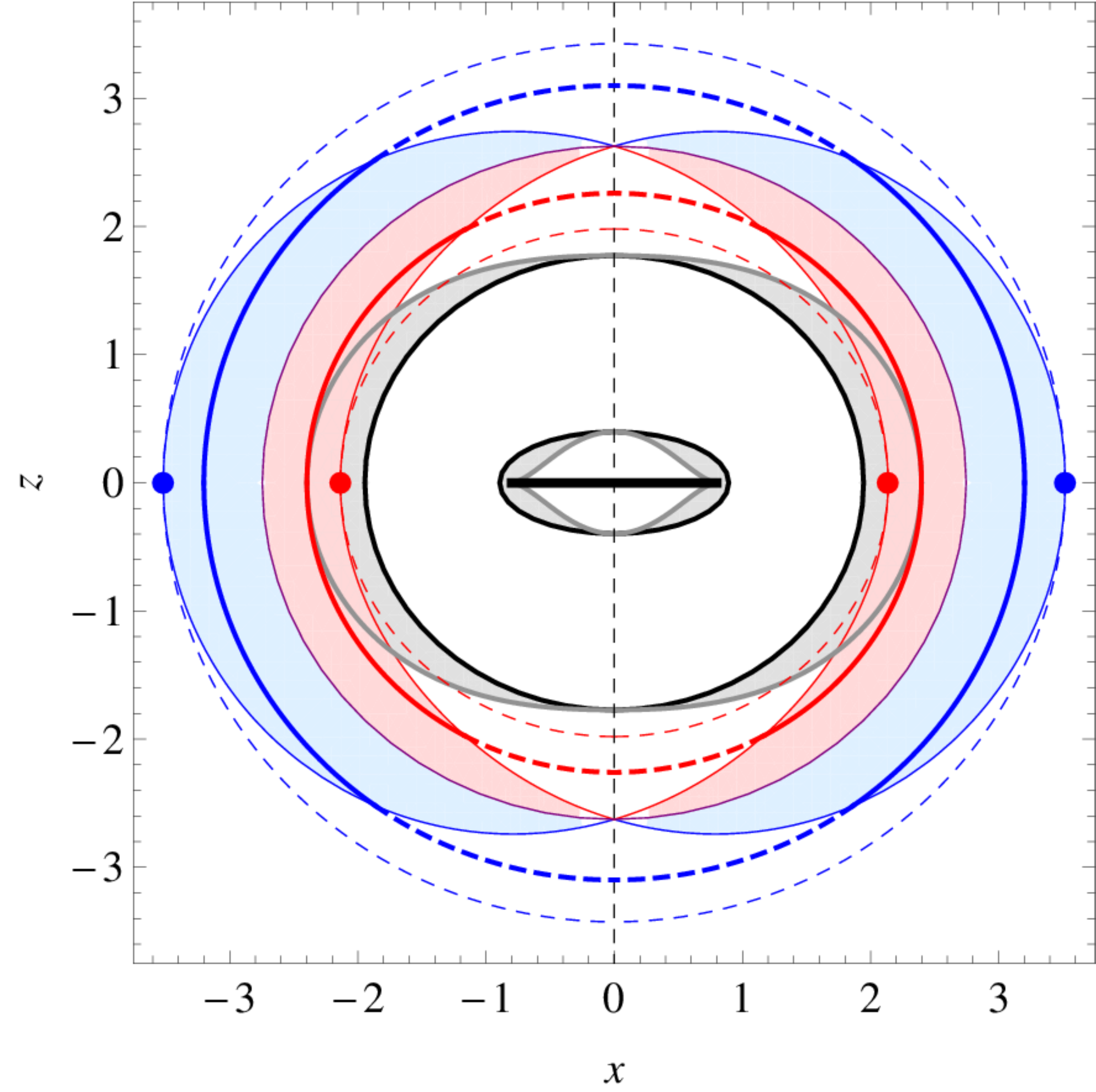}
		
	\end{tabular}
	\caption{Illustration of some characteristics of the KdS spacetime with parameters $a^2=0.64,$ $y=0.02.$ The labels $x,z$ are the Kerr-Schild coordinates, $y$-coordinate is suppressed. Horizontal abscissa is the ring singularity, black ellipses mark the black hole event horizons, grey curves represent the static limit surface (ergosurface), black dashed vertical line is the rotary axis. The ergosphere is denoted by shading; note however that the shading is canceled by colours corresponding to various types of the spherical photon orbits located at the ergosphere (here and in the following figures representing the classification). The outer ergosphere and the cosmological horizon is outside of the drawing in this figure. Thin dashed red ellipse denotes the sphere with radius $r=r_{ph+}=1.98,$ which intersects the equatorial plane in the co-rotating photon circular orbit, displayed as red bold points. Similarly, thin dashed blue ellipse represents the $r=r_{ph2}=3.43$-sphere, and the bold blue points mark the equatorial counter-rotating circular orbit. Bold red arcs represent a prograde spherical photon orbit at $r=2.26,$ selected just to touch the ergosurface in the equatorial plane. Thick red dashed arcs denote rest of the selected sphere. Accordingly, full parts of the bold blue ellipse mark some spherical retrograde photon orbit, selected at $r=3.1.$ The endpoints of the bold arcs then correspond to marginal latitudes, among which the photons oscillate. All such endpoints thus create a surface, here depicted by full thin red or blue curves, limiting regions filled by the prograde or retrograde, respectively, spherical photon orbits, coloured in light red or blue, respectively. In this case, the regions of opposite orientations are separated by purple ellipse representing the polar spherical orbit at which a photon with $X=-a$ ($\ell=0$) crosses the spaxcetime rotation axis alternately above both poles. Light/full hues we reserved to regions of unstable/stable orbits, the latter ones not present in this case. In some KdS spacetimes, there exist retrograde spherical orbits of photons with $X>-a.$ They are confined exclusively inside the ergosphere and occupied by photons with negative energy $E.$ These regions will be depicted in green.       }\label{Sphorbs}
\end{figure*}

In the following classification we introduce ten classes of the \KdS\ spacetimes and demonstrate properties of the photon motion using the spherical photon orbits that serve as crucial characteristic for the classification. We give the loci of the spherical photon orbits and their extension in latitude, stability against radial perturbations, and orientation of their azimuthal motion. The classification is represented by family of characteristic figures corresponding to the separated classes of the KdS spacetimes. For easy interpretation of the family of the figures representing the classification, we introduce an auxiliary Fig.\ref{Sphorbs} commented with detailed explanatory notes. In order to fully and clearly characterize the KdS spacetimes and their horizon and ergosphere structure, and to demonstrate the spheroidal character of the applied coordinate system, we use now the so called Kerr-Schild coordinates $x,y,z$ that are connected to the Boyer-Lindquist coordinates $r,\theta$ by the relations 
\be
     x^2 + y^2 = (r^2 + a^2)\sin^2\theta ,\quad z^2 = r^2\cos^2\theta.
\ee 
In the figures we, of course, use the meridional sections of $y=0$. The characteristics of the classes of the KdS spacetime according to the photon orbits are presented as follows.  

\begin{description}
	
		\item[Class I]:		
		Black hole spacetimes with the divergent repulsive barrier of the radial photon motion, having one equatorial counter-rotating circular unstable orbit with negative energy located under the inner black hole horizon ($0<r<r_{-}$), which is limiting the range of the spherical photon orbits with negative energy. There exist stable orbits, corresponding to local minima $X_{min(+)}$ of the effective potential $X_{+}$ at $0<r<r_{max(ex)1},$ and unstable orbits, corresponding to local maxima $X_{max(+)}$ of $X_{+}$ at $r_{max(ex)1}<r<r_{z(ex)1}$ for $0<q<q_{max(ex+)}(y,\:a^2)$ (Fig. \ref{Classes}a). We denote as  $r_{min/max(ex)}$ the local extrema, and as $r_{z(ex)}$ the zero point, of the function $q_{ex}(r;\:y,\:a^2)$ hereafter. Such a structure is present under the inner horizon of any KdS black hole spacetime. Outside the ergosphere, one unstable co-rotating equatorial circular orbit, located at $r=r_{ph+}=r_{z(ex)2},$ and polar spherical orbit with $r=r_{pol},$ $r_{ph+}<r_{pol},$ limit the range of unstable prograde spherical orbits given by the local minima $X_{min(-)}$ of the effective potential $X_{-},$ for which $X_{min(-)}>-a$. The radius of the polar spherical orbit is found by solving $X_{-}(r_{pol};q_{ex}(r_{pol}))=-a.$ The counter-rotating equatorial circular orbit at $r=r_{ph-}=r_{z(ex)3}$ gives the limit of region of unstable retrograde spherical orbits, given by the local minima $X_{min(-)}<-a$, and maxima $X_{max(+)}<-a$ for $0<q<q_{max(ex)}(y,\:a^2),$ such that $r_{pol}<r_{ph-}.$	
		
		\item[Class II]:		 
		 Black hole spacetimes with the same features as in the class I, but now the ergosphere enters the region of the spherical photon orbits (Fig. \ref{Classes}b). No spherical orbit is fully immersed in the ergosphere and photons at all the spherical orbits have positive energy. The presence of the ergosphere in region of the spherical photon orbits influences character of the light escape cones \cite{l_e_cones:}.
	 
		\item[Class III]:	  
		Black hole spacetimes with the restricted repulsive barrier of the radial photon motion. The ergosphere spreads over all radii. The prograde spherical orbits are given by the local minima $X_{min(-)}>-a$ at $r_{ph+}<r<r_{pol},$ while the retrograde spherical orbits with $E>0$ are given by the minima $X_{min(-)}<-a$ at $r_{pol}<r<r_{d(ex)}.$ The spherical orbits given by the local maxima $X_{max(+)}$ (see Figs 10k-n) at $r_{d(ex)}<r<r_{ph-},$ where $r_{d(ex)}$ denotes the divergence point of $q_{ex}(r;\:y,\:a^2)$ (see Fig. 8c), are fully immersed in the ergosphere. Such areas are drawn in Figs \ref{Classes} in green and the spheres with $r=r_{d(ex)}$ as full/dashed green ellipses. Photons in such regions have $E<0$.
		
		\item[Class IVa]:
		Naked singularity spacetimes with divergent repulsive barrier of the photon motion. At radii $0<r<r_{d(ex)}$ (Fig.8d can be used for illustration), there are local minima of the potential $X_{+}$ (for illustration use Figs. 10p-r) corresponding to the stable retrograde spherical orbits with negative energy ($E<0$) (Fig. \ref{Classes}d). The stable retrograde orbits with positive energy corresponding to the local maxima of $X_{-}$ are at $r_{d(ex)}<r<r_{pol1}$. These maxima exceed the value $X_{max(-)}=-a$ at $r_{pol1}<r<r_{min(ex)},$ where they yield stable prograde spherical orbits. At radii $r_{min(ex)}<r<r_{pol2}$, there are the local minima of $X_{-}$ with values $X_{min(-)}>-a,$ -- these radii are thus occupied by the unstable prograde orbits. The local minima of $X_{-}$, and the local maxima of $X_{+}$ at $r>r_{pol2}$, correspond to the unstable retrograde orbits. There are thus two polar spherical orbits enclosing region of prograde orbits -- the inner at the radius $r=r_{pol1}$ being stable, the outer at the radius $r=r_{pol2}$ being unstable. 
	
        \item[Class IVb]: 
        Naked singularity spacetimes with the same features as in the class IVa, but the two polar orbits have coalesced, therefore, there are no prograde spherical orbits (Fig. \ref{Classes}e). 
        
        \item[Class V]:
        Naked singularity spacetimes having the structure of the spherical orbits corresponding to the previous case (Fig. \ref{Classes}f), but with is a small region of bound orbits for photons with motion constants $q_{min(ex)}(y,\:a^2)<q<q_{max(ex+)}(y,\:a^2)$ and $X$ between the appropriate local extrema of $X_{+}$ (c. f. Figs. 10q, u), which is not contained in the other cases. 
        
        \item[Class VIa]: Naked singularity spacetimes with the restricted repulsive barrier of the radial photon motion. For $0<r<r_{d(ex)1}$ (the function $q_{ex}(r;\:y,\:a^2)$ has two divergence points $r_{d(ex)1}, r_{d(ex)2},$ -- see Fig. 8f for preview) the minima of $X_{+}$ correspond to stable retrograde orbits with $E<0$; for $r_{d(ex)1}<r<r_{pol1}$, there are the local maxima of $X_{-}$ with values $X_{max(-)}<-a$ giving retrograde orbits with $E>0.$ The local minima of $X_{-}$ at $r_{pol1}<r<r_{min(ex)}$ give the stable prograde orbits. At radius $r=r_{pol1}$ the stable polar orbit is located. For $r_{min(ex)<r<r_{pol2}}$, the function $X_{-}$ has minima with values $X_{min(-)}>-a$ giving the unstable prograde orbits. For $r_{pol2}<r<r_{d(ex)2}$, they correspond to the unstable retrograde orbits. At radius $r=r_{pol2}$, the unstable polar orbit exists. The local maxima of the function $X_{+}$ at $r_{d(ex)2<r<r_{ph-}}$ correspond to the retrograde unstable spherical orbits with $E<0.$
        
        \item[Class VIb]:
        The structure of the spherical orbits corresponds to the class VIa with an exception that the local extrema of the potential $X_{-}$ have values $X<-a,$, implying that there are no polar spherical orbits, neither the prograde spherical orbits (Fig. \ref{Classes}h). 

        \item[Class VII]:
        Naked singularity spacetimes with the restricted repulsive barrier of the radial photon motion having stable retrograde spherical orbits at $0<r_{max(ex)}$ corresponding to local minima of $X_{+}$ (Fig. 10$\alpha$), and unstable retrograde spherical orbits at $r_{max(ex)}<r<r_{ph-}$ corresponding to local maxima of $X_{+}.$ All these spherical orbits, including the counter-rotating equatorial circular orbit at $r=r_{ph-}$, correspond to photons with $E<0.$
        
        \item[Class VIII]: Special class of the naked singularity spacetimes demonstrating the same features of the radial motion of photons with $q\geq 0$ as the class VII, but differing from all previous cases by the existence of null geodesics for arbitrary $q<0.$ The allowed values of the impact parameter $X$ are for $q<0$ confined to the intervals $X<X^\theta_{max(+)}<0$ or $X>X^\theta_{min(-)}>0$ (see Section 3). The potentials governing the radial photon motion are fully immersed in the forbidden region (Figs. 10($\gamma$), ($\delta$)), thus in the radial direction the photons with such parameters move freely in the whole range between the ring singularity and the cosmological horizon. 
\end{description}

\begin{figure*}[htbp]
	\centering
	\begin{tabular}{cc}
		
		\includegraphics[width=0.44\textwidth]{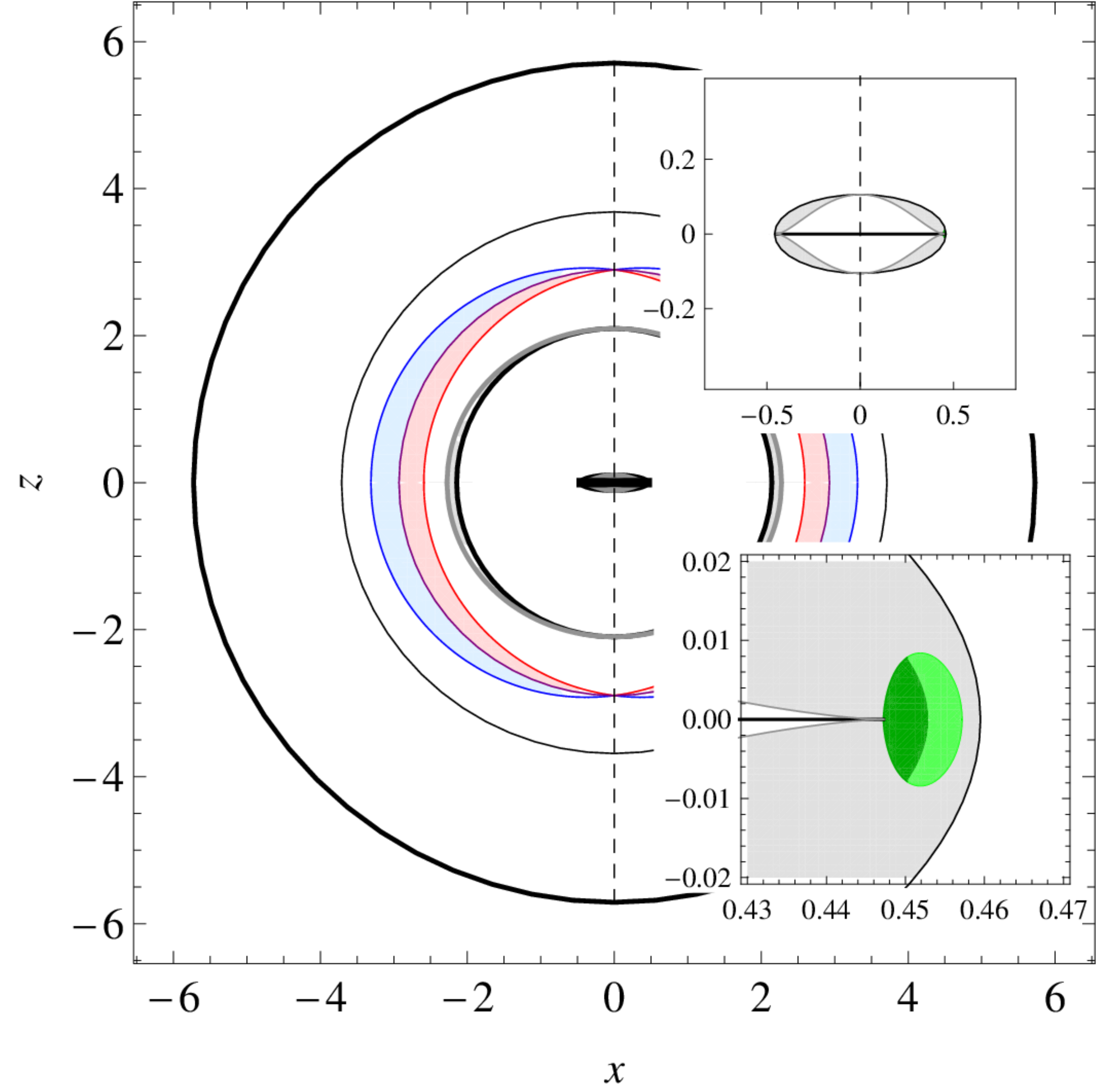} & \includegraphics[width=0.44\textwidth]{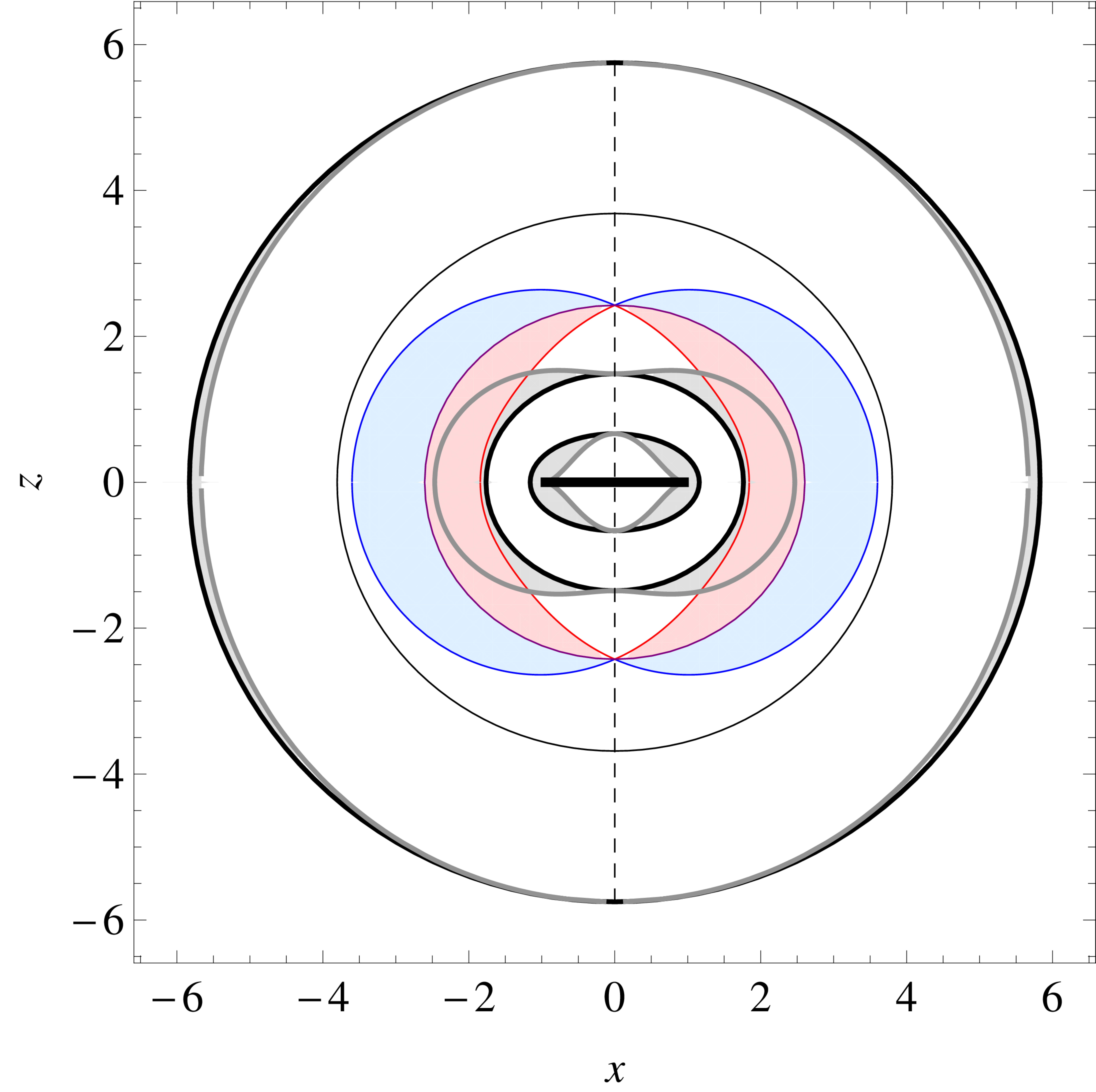}\\ 
		(a) Class I: $y=0.02,$ $a^2=0.2$ & (b)  Class II: $y=0.02,$ $a^2=0.9$\\
		\includegraphics[width=0.44\textwidth]{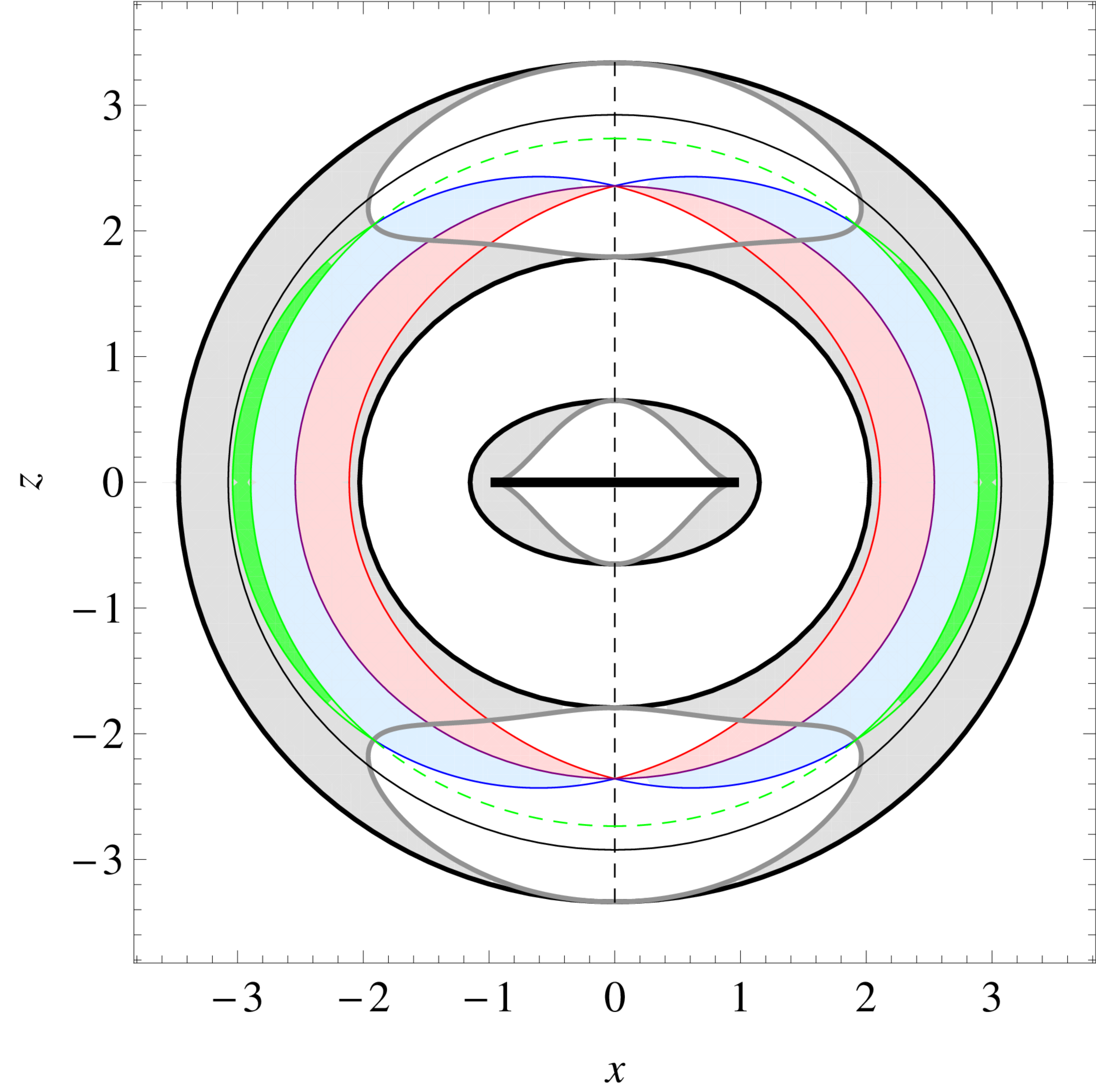} & \includegraphics[width=0.44\textwidth]{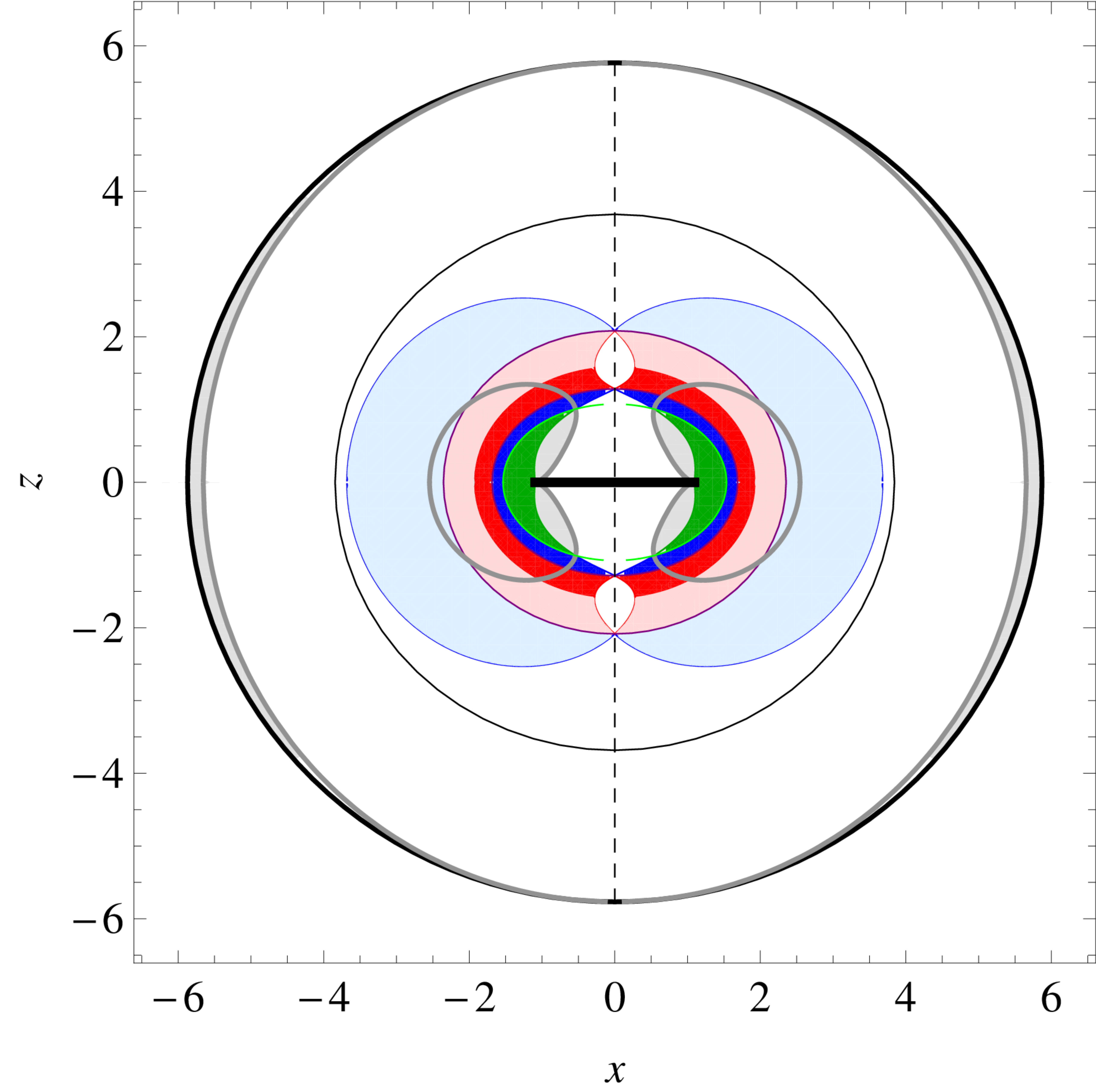}\\
		(c) Class III: $y=0.04,$ $a^2=0.9$ & (d)  Class IVa: $y=0.02,$ $a^2=1.2$\\
		\includegraphics[width=0.44\textwidth]{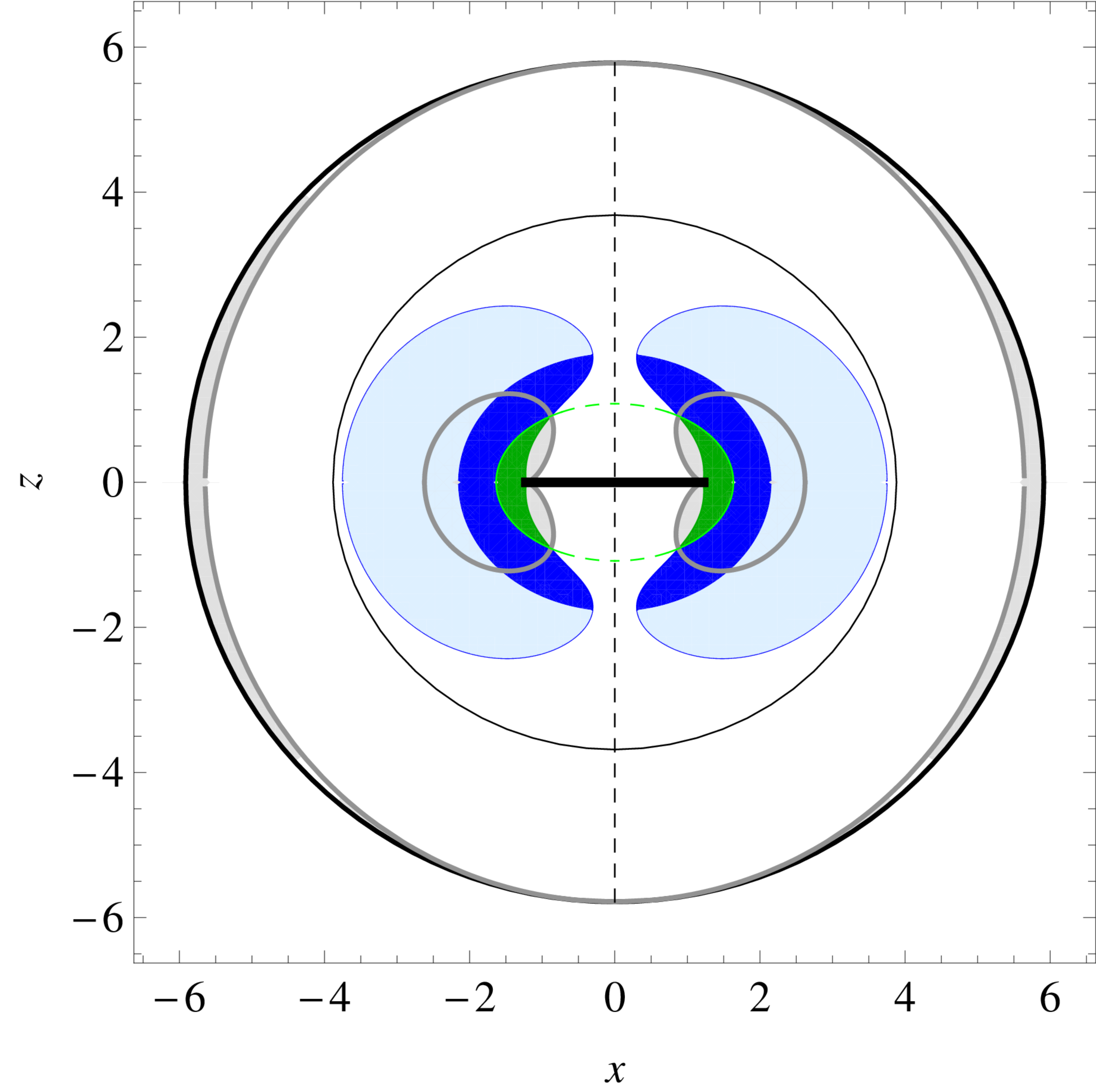} & \includegraphics[width=0.44\textwidth]{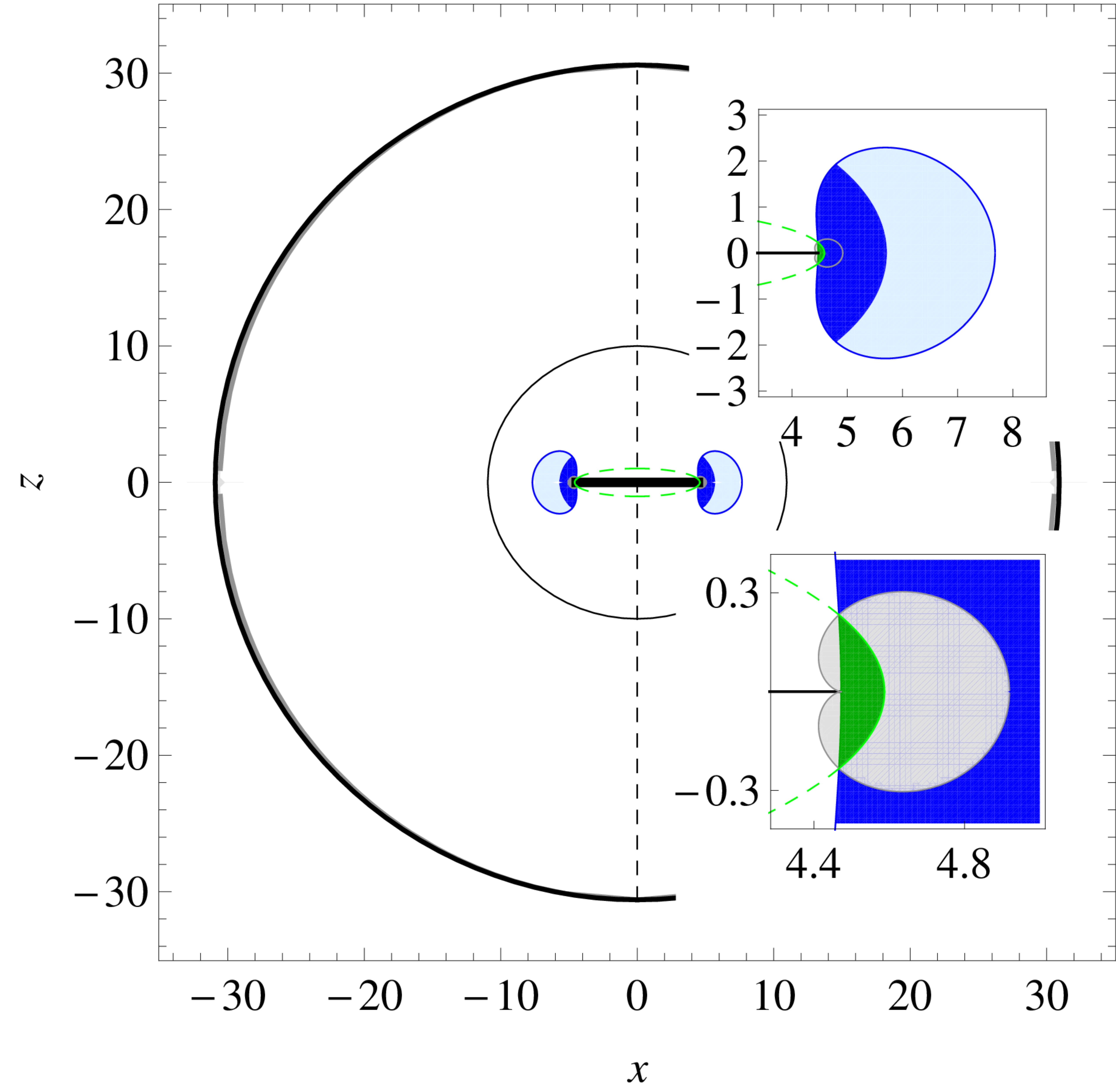}\\
		(e) Class IVb: $y=0.02,$ $a^2=1.5$ & (f) Class V: $y=0.001,$ $a^2=20$
	\end{tabular}
	\center (\textit{Figure continued})
\end{figure*}

\begin{figure*}[htbp]
	\centering
	\begin{tabular}{cc}
		
		\includegraphics[width=0.44\textwidth]{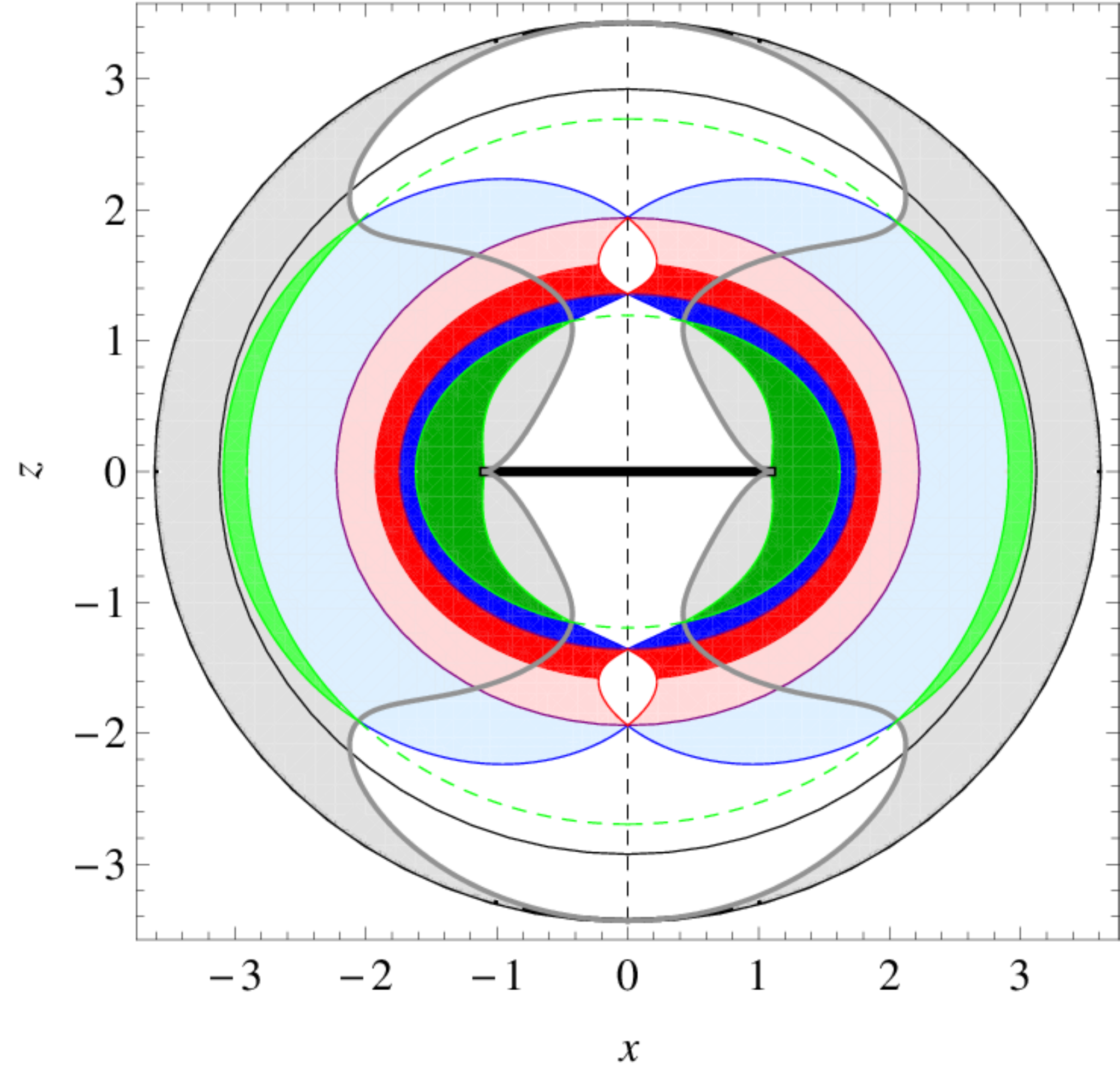} & \includegraphics[width=0.44\textwidth]{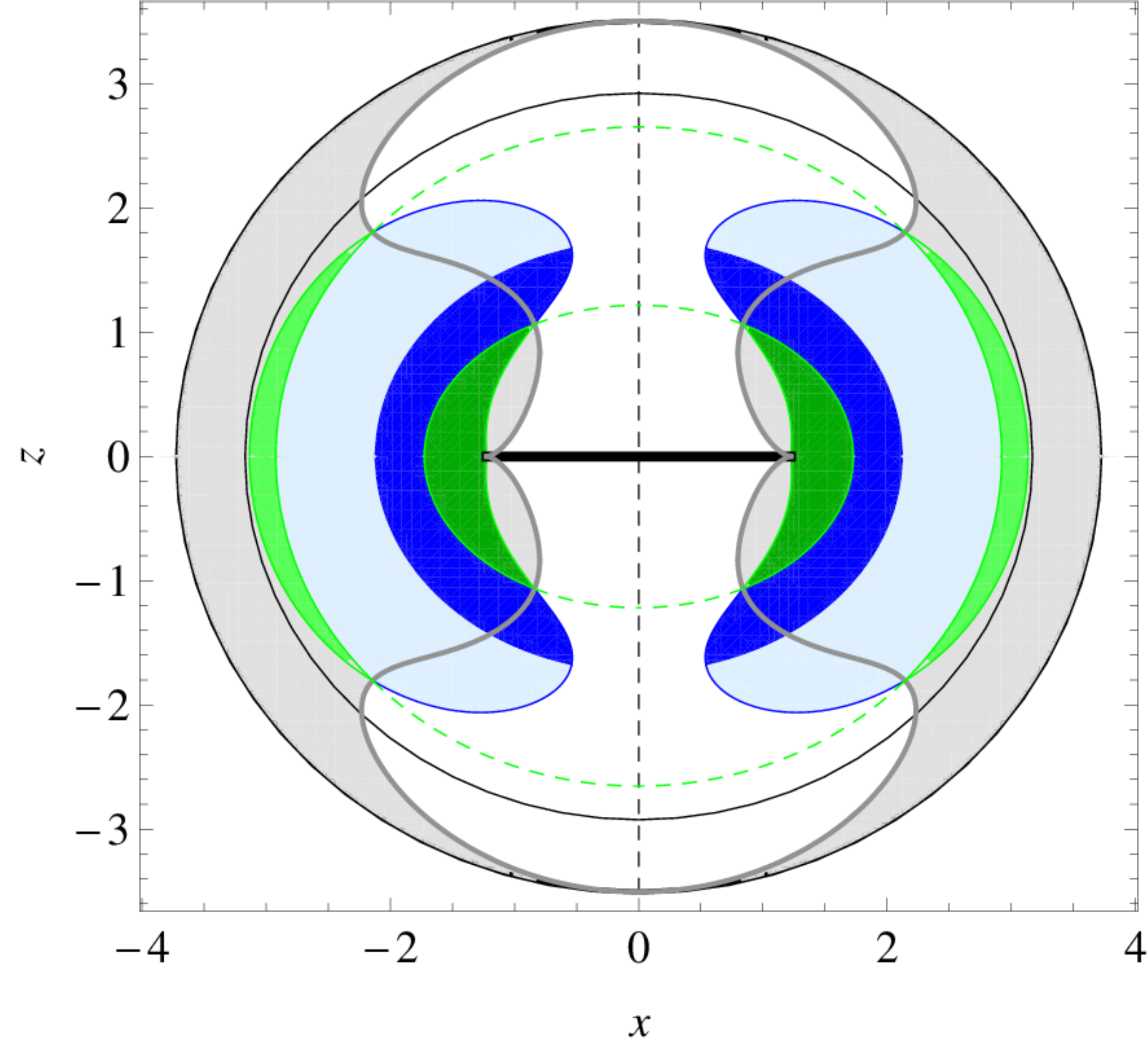}\\
		(g) Class VIa: $y=0.04,$ $a^2=1.2$ & (h)  Class VIb: $y=0.04,$ $a^2=1.5$\\
	\end{tabular}
	\begin{tabular}{c}
		\includegraphics[width=0.44\textwidth]{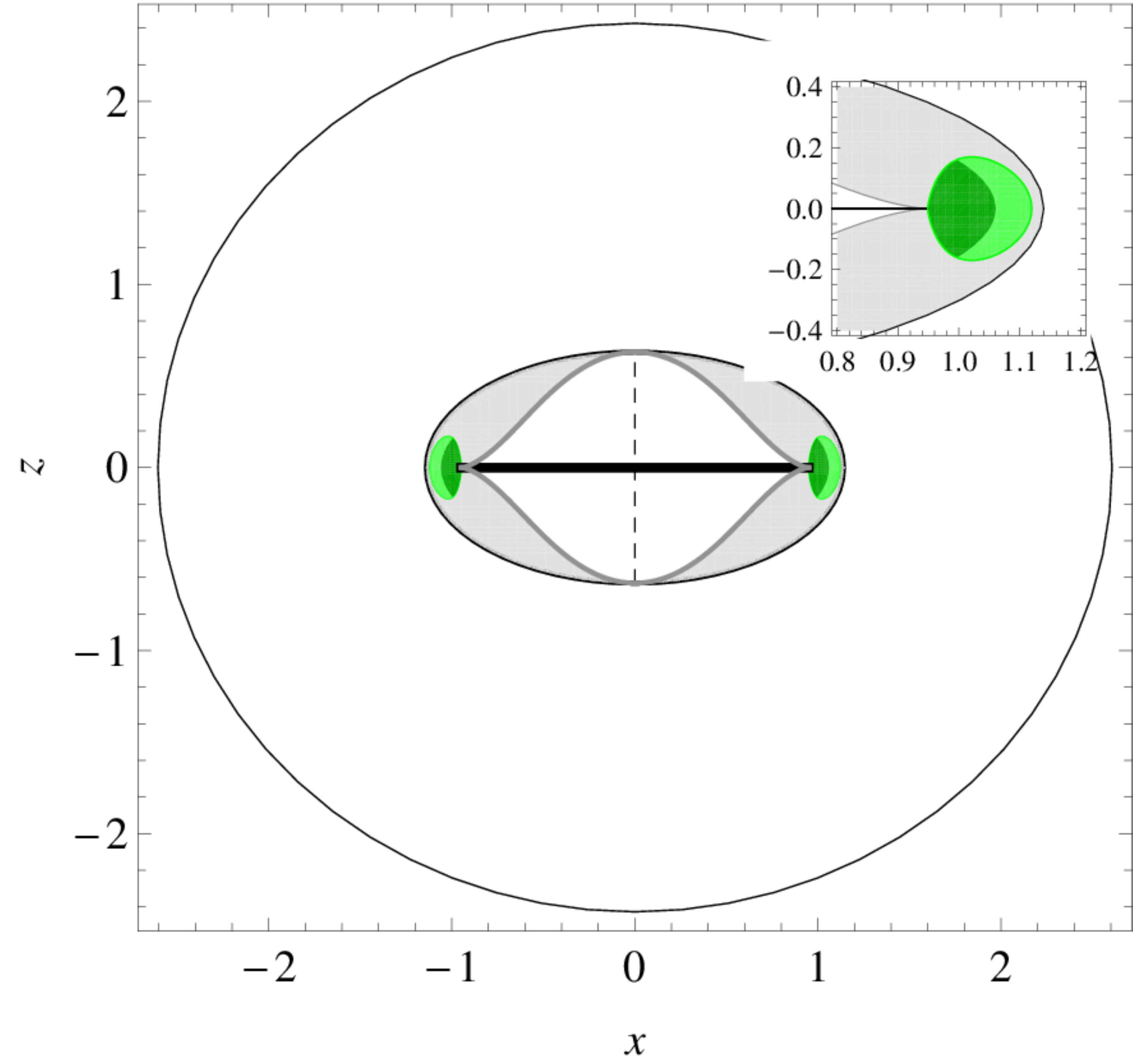} \\
		(i) Class VII: $y=0.07,$ $a^2=0.9$
	\end{tabular}	
	\caption{Spherical photon orbits in individual clases of \KdS\ spacetimes. The class VIII is not  presented, since the structure of the spherical orbits matches with the class VII, but it differs in character of the latitudinal motion.}\label{Classes}
\end{figure*}

\section{Conclusions}

We can summarize our results by the following concluding remarks.

	\begin{enumerate}
		\item In any kind of the black hole spacetimes, there are no radially bound null geodesics in the stationary region, i.e., the trajectory of a photon has at most one turning point in radial direction between the outer and cosmological horizon, or the photons can move freely between the outer black hole and the cosmological horizons. However, such bounded photon orbits exist in each naked singularity spacetime for photons with parameters $q>0$ and $X$ chosen appropriately.
		 
		\item No photons with $q>0$ can reach the ring singularity at $r=0$ in any of the \KdS\ spacetimes.
		
		\item In the \KdS\ spacetimes of classes I-VII, i.e., with the spacetime parameters satisfying the condition $y<1/a^2,$ there is a lower limit $q=-a^2$ of the parameter $q<0,$ for which the photon motion is allowed. The range of the allowed values of the impact parameter $X$ is then an interval given by the relations (\ref{qmin(X,y,a)}) - (\ref{q_l_plane1}). Photons with such tuned parameters have no turning point in radial direction, since the effective potential lies entirely in the forbidden region (Figs 10(e)-(g)). Further, by the results of Section 3, only such photons execute the vortical motion, or their trajectory lies completely on the cones of $\theta = constant.$ We can therefore reject possibility of existence of vortical photon motion of constant radius, or off-equatorial circular photon orbits.
		
		\item In the \KdS\ spacetimes of class VIII ($y>1/a^2$), the photon motion is allowed for any $q<0.$ The permissible values of the parameter $X$ are then two disjunct unlimited intervals determined by the relation (\ref{q_l_plane3}.) In the extreme case $y=1/a^2,$ it must be $q\ge -a^2$ again and for negative  $q,$ the parameter $X$ can take less than certain negative value given by (\ref{q_l_plane2}). The consequences for photon motion are then the same as in previous note (Figs 10($\gamma$)-($\delta$)).
		
		\item In the \KdS\ spacetimes with the divergent repulsive barrier of the radial photon motion, there exists a critical value $q_{max(ex)}(y,\:a^2),$ for which this barrier becomes impermeable between the outer black hole horizon and cosmological horizon, or, in naked singularity spacetimes, between the ring singularity and cosmological horizon, for photons with any impact parameter $X.$ In spacetimes with the restricted repulsive barrier of the radial photon motion, the height of this barrier slowly grows with increasing parameter $q,$ but stays finite for any $q>0$ (Figs 10(n), 10(y)).
        
        \item In the \KdS\ spacetimes of classes I-III, IVa, VIa, there exist spherical photon orbits, which can be both prograde or retrograde as seen by the family of locally non-rotating observers. Additionally, each of the two types can be stable or unstable with respect to radial perturbations. The regions of spherical orbits of different orientations are separated by the so called polar spherical orbit, at which photons cross the spacetime rotation axis alternately above both poles. In the naked singularity spacetimes of class IVa, VIa, there are two polar spherical orbits, the inner one being stable, the outer one being unstable.
        
        \item In the \KdS\  spacetimes of classes IVb, V ,VIb, VII, VIII there are no prograde or polar spherical orbits. 
        
        \item In each class of the \KdS\ spacetimes, there exist region, where the effective potential $X_{+}$ have positive values. Photons with impact parameter $X$ exceeding this values appear to move in retrograde direction as seen in LNRFs. This region must be located inside the ergosphere, and photons with such impact parameters must have negative energy, $E<0$. In the black hole spacetimes with the divergent barrier of the radial photon motion, the ergosphere has two parts above the black hole outer event horizon - the inner one, which is limited to the outer vicinity of the outer event horizon, and the outer one, limited to the inner vicinity of the cosmological horizon. In the black hole spacetimes with restricted repulsive barrier of the radial photon motion, the two regions of the ergosphere merge in the equatorial plane, and they spreads at any radii except for certain region in the vicinity of the rotation axis.

        \item In the LNRFs, trajectories of photons moving along any spherical orbit have no turning point of the azimuthal motion.  

	\end{enumerate}

We have thus demonstrated a variety of very extraordinary phenomena related to the photon motion in the KdS spacetimes, of both black hole and naked singularity types. Especially relevant effects are found in the case of spherical photon orbits that can be directly related to the observational phenomena. It is quite interesting that we could expect another interesting phenomena related to the charged Kerr-Newman or Kerr-Newman-de Sitter naked singularity spacetimes (with both the standard electric charge, or the tidal charge of the braneworld models), especially in the case of the so called mining Kerr-Newman spacetimes \cite{Bla-Stu:2016:PHYSR4:}, containing a special type of equatorial stable photon orbits.

\section*{Acknowledgments}

Z.S. acknowledges the Albert Einstein Centre for Gravitation and Astrophysics supported by the Czech
Science Foundation Grant No. 14-37086G. D.Ch. acknowledges the Silesian University in Opava Grant No. SGS/14/2016.

\bibliographystyle{abbrv}

\end{document}